\documentclass{aa}
\usepackage{graphicx}
\usepackage{txfonts}
\usepackage{lscape}
\begin{document}
   \title{HST/WFPC2 morphologies and bar structures of field galaxies at 0.4\,$<z<$\,1$^\star$}
   \author{X. Z. Zheng \inst{1,2}
          \and F. Hammer \inst{1}
          \and H. Flores \inst{1}
          \and F. Ass$\acute{\rm e}$mat \inst{1}
          \and A. Rawat \inst{3,1}
          }

   \offprints{zheng@mpia.de\\ 
$^\star$\,This work is based on observations  made with  the NASA/ESA  Hubble Space Telescope,  obtained from  the  data archive  at  the Space  Telescope
Institute. STScI  is operated by  the association of  Universities for
Research in Astronomy, Inc. under the NASA contract NAS 5-26555.}

   \institute{
              GEPI, Observatoire de Paris-Meudon, 92195 Meudon, France
          \and
              Max-Planck Institut f$\ddot{u}$r Astronomie, K$\ddot{o}$nigstuhl 17, D-69117 Heidelberg, Germany
          \and
              Inter-University Centre for Astronomy and Astrophysics, Post Bag 4, Ganeshkhind, Pune 411007, India
              }
   \date{Received 7 September 2004 / Accepted 27 January 2005}

   \abstract{
To address how the galaxy Hubble sequence is established and what physical processes are involved, we studied morphological properties and internal structures of field galaxies in the past (0.4\,$<z<$\,1). 
In addition to structural parameters derived from bulge+disk decomposition, Zheng et al. (\cite{Zheng}) introduced color maps in recognizing galaxies and properly classified morphologies of 36 luminous infrared galaxies (LIRGs, $L_{\rm IR}(8-1000\,\mu$m) $\geq$\,$10^{11}$\,$L_\odot$). Here we presented morphological classification of a parallel 75 non-LIRG sample. Our examination revealed that a significant fraction of the galaxies shows remarkable morphological evolution, most likely related to the present-day spiral galaxies. 
Comparison of the morphological properties between LIRGs and non-LIRGs shows that the LIRGs contain a higher fraction of ongoing major mergers and systems with signs of merging/interaction. This suggests that the merging process is one of the major mechanisms to trigger star formation. We found that spiral LIRGs probably host much fewer bars than spiral non-LIRGs, suggesting that a bar is not efficient in triggering violent star formation. Differing from Abraham et al. (\cite{Abraham99}), no dramatic change of the bar frequency is detected up to redshift $z\sim$\,0.8. 
The bar frequency of the distant spirals is similar to (and may be higher than) the present-day spirals in the rest-frame $B$ band. We conclude that bar-driven secular evolution is not a major mechanism to drive morphological evolution of field galaxies, especially their bulge formation, which is more likely related to multiple intense star formation episodes during which the galaxies appear as LIRGs (Hammer et al. \cite{Hammer04a}).
  }
   \titlerunning{Morphologies and bar structures of field galaxies at 0.4 $<z<$ 1}
   \maketitle

\section{Introduction}

The origin of the Hubble sequence is still a challenge for cosmological studies. With HST/WFPC2 deep observations, morphological properties can be derived for field galaxies up to redshift $z\sim$\,1. Over the last decade, many works contributed to addressing the morphological evolution of the field galaxies. Lilly et al.(\cite{Lilly98}) found that massive disk galaxies show a constant number density from $z\sim$\,1. Elliptical galaxies also exhibit little evolution up to $z\sim$\,1 in their number density (Schade et al. \cite{Schade}) although non-negligible star formation at $z<$\,1 was detected in a substantial sub-population (Menanteau et al. \cite{Menanteau01}; Treu et al. \cite{Treu02}). Morphological investigations based on large galaxy samples revealed that the number density of irregular galaxies increases with redshift (Brinchmann et al. \cite{Brinchmann}, hereafter B98; van den Bergh et al. \cite{Bergh00}). A significant evolution of irregular galaxies was suggested to associate with transformation into regular galaxies (Brinchmann \& Ellis \cite{BrinchmannEllis}).

However, the formation of massive spiral galaxies is still under debate (Hammer et al. \cite{Hammer04b}), while cumulate evidence suggests that most massive elliptical galaxies in the field are in place prior to $z$\,=\,1 (e.g. Daddi et al. \cite{Daddi}; Cimatti et al. \cite{Cimatti}; but see also van Dokkum \& Ellis \cite{vanDokkum}; Bell et al. \cite{Bell}; Drory et al. \cite{Drory}). 
Indeed, field galaxies at intermediate redshifts 0.4\,$<z<$\,1 are dominated by emission line galaxies (Hammer et al. \cite{Hammer97}). Spectroscopic studies showed that the emission line galaxies are metal poorer than their local counterparts by $\sim$0.3 dex, suggesting that they had formed a significant fraction of their stars since $z$\,=\,1 (Kobulnicky et al. \cite{Kobulnicky}; Liang et al. \cite{Liang}). 
By analyzing stellar populations of $\sim 10^5$ nearby galaxies in the {\it Sloan Digital Sky Survey} (SDSS), Heavens et al. (\cite{Heavens}) found that the cosmic star formation rate peaks at $z\simeq$\,0.6, mostly due to the galaxies with stellar mass in the range 3\,-\,30\,$\times$\,10$^{10} M_\odot$.  From a direct investigation of such galaxies at 0.4\,$<z<$\,1, Hammer et al. (\cite{Hammer04a}) suggested that they are related to spiral galaxies in formation through multiple violent starburst episodes, appearing as LIRGs. This is also consistent with the results derived from deep ISOCAM observations at 15\,$\mu$m showing that the infrared emission from LIRGs at $z\sim$\,0.8 dominates the cosmic infrared background (Elbaz \& Cesarsky \cite{Elbaz03}). 

 The epoch from $z$\,=\,1 to $z$\,=\,0.4 is crucial for the assembly of spiral galaxies (Hammer et al. \cite{Hammer04a} and references therein). It still remains to be answered how their morphological structures were formed and what physical processes fundamentally drove the assembly. While disks can be built up by accreting cooling gas in a dark halo (e.g. Fall \& Efstathiou \cite{Fall}), the formation of bulges may be regulated by either external or internal processes (see Combes \cite{Combes00} for a review). The bulges can be formed via either primordial collapses at an early time (e.g. Eggen et al. \cite{Eggen}) or hierarchical mergers (e.g. Kauffman \cite{Kauffmann}). In the later case, violent star formation is expected to be ignited and the merging systems are mostly associated with LIRGs (Sanders \& Mirabel \cite{Sanders}). Comparison of morphological properties between the LIRGs and non-LIRGs will essentially contribute to unveiling whether the merging process dominantly regulates bulge formation. 
Alternatively, bulges may be formed within preexisting disks through internal secular evolution (e.g. Combes \cite{Combes00}). 
In the secular processes, bars are mostly suggested to be efficient at driving gas into the galactic center and changing star orbits to form a central bulge-like structure (Combes et al. \cite{Combes}; Debattista et al. \cite{Debattista}). However, observations have not been made to see if the bar-driven secular process is important to trigger star formation in field galaxies at 0.4\,$<z<$\,1 which on average contain richer gas than the local ones. 

Compared with the galaxy distribution along the Hubble sequence in the local universe, a morphological census of field galaxies at 0.4\,$<z<$\,1 will provide clues to understand their global morphological transformation and evolution. 
We have performed a morphological investigation of a sample of 36 LIRGs (Zheng et al. \cite{Zheng}, hereafter Z04). Here, a similar morphological investigation is applied to a parallel non-LIRG sample. With HST $F814W$ imaging, strong bars can be still detected up to $z\sim$\,0.7 (van den Bergh et al. \cite{Bergh02}). We examine the bar phenomena in both samples to see the roles of the bars in governing bulge formation.

This paper is organized as follows: In Section 2, we briefly describe our sample. Section 3 presents the methodology and results of our morphological classification. In Section 4, we describe the examination of bars in spiral galaxies. Our results are discussed in Section 5 and our conclusions summarized in Section 6. Throughout this paper we adopt a cosmology with $H_0$\,=\,70\,km\,s$^{-1}$\,Mpc$^{-1}$,  $\Omega_{\rm  M}$\,=\,0.3 and $\Omega_\Lambda$\,=\,0.7 unless specially stated.

\section{HST sample}

We collected 17 deep HST/WFPC2 observations through the $F814W$ and $F606W$ (or $F450W$) filters in the Canada-France Redshift Survey (CFRS) fields 0300+00 and 1415+52, which have also been deeply surveyed at 15\,$\mu$m by the ISOCAM on board the {\it Infrared Space Observatory} (ISO). Parameters of the HST/WFPC2 observations and data properties are described in detail in Z04. Here we present a brief description. 
The primary sample was drawn from 87 square arcminute sky areas, including 169 galaxies with magnitude $I_{\rm AB}<$\,22.5 and known spectroscopic redshift from the CFRS survey or literature (mostly from the DEEP I survey; see Weiner et al.~\cite{Weiner}). The DEEP I survey is complete by CFRS selection criteria. Thus, the primary sample is still consistent with the original CFRS sample. Of the 169 objects, 111 are in the redshift range 0.4 to $\sim$\,1.1. Of the 111 objects, 36 are LIRGs. The other 75 objects make up our non-LIRG sample.  Note that additional efforts were made to identify redshift of the ISO-detected sources following the CFRS strategy (Lilly et al.~\cite{Lilly95}), aimed at having a large LIRG sample. Cross-correlating the CFRS redshift catalog with ISO-detected source catalog of two $10\arcmin\times 10\arcmin$ CFRS fields 0300+00 and 1415+52, we estimated the frequency of LIRGs as 15\% in field galaxies at 0.4\,$<z\leq$\,1 (see Hammer et al.~\cite{Hammer04a} for more details). Except for two LIRGs at redshift $z\sim$\,1.1, all targets are indeed at 0.4\,$<z\leq$\,1. We refer our sample to the redshift range 0.4 to 1.

Table~\ref{table} lists the parameters of the non-LIRG sample, including CFRS ID (Col. 1), redshift (Col. 2), CFRS redshift confidence (Col. 3), HST photometry (Col. 4-6). Almost all  objects (70 of 75) show a high redshift confidence (class 3, 4, 8 and 9). The aperture used in our photometry is  3$\arcsec$. We refer $B_{450}$, $V_{606}$ and  $I_{814}$ to HST filters $F450W$, $F606W$ and $F814W$, respectively.  The Vega system is adopted for our photometry, which can be converted into the AB system using equations: $B_{450}$\,=\,$B_{AB}(450)-$0.11, $V_{606}$\,=\,$V_{AB}(606)$+0.10 and $I_{814}$\,=\,$I_{AB}(814)$+0.42. To be consistent with published CFRS papers, the absolute $B$ band magnitude (Col. 7) and the $K$ band magnitude  (Col. 8) are given in the AB system,  using isophotal magnitudes from ground-based imaging (Lilly et al.~\cite{Lilly95}). The K-correction is calculated based on the ground-based $B, V, I$ and $K$ band CFRS photometry (see Hammer et al. \cite{Hammer01}  for details).

\begin{figure} \centering
   \includegraphics[width=0.4\textwidth]{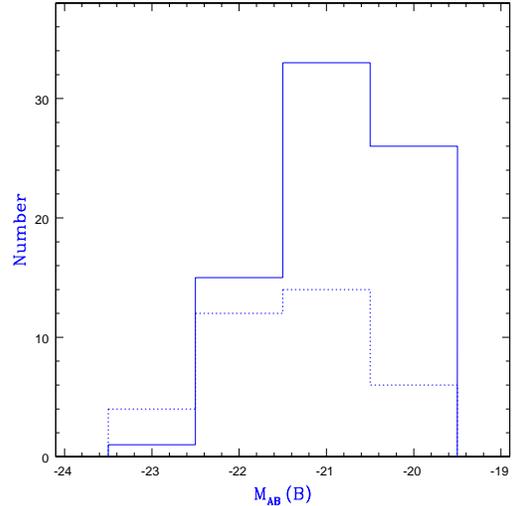}
   \caption{The histograms of the absolute $B$ band magnitude for 75 non-LIRGs ({\it solid line}) and 36 LIRGs ({\it dotted line}).}
   \label{hisMB}
\end{figure}

\begin{landscape}
\begin{table}[h]
  \caption[]{Catalog of field galaxies in CFRS 0300+00 and 1415+52 fields}
  \label{table}
  \begin{tabular}{cccccccccccccclll}
  \hline
  \noalign{\smallskip}
         & & & & & &    &   &  \multicolumn{2}{c}{\hrulefill $B_{450}$\hrulefill} & \multicolumn{2}{c}{\hrulefill $V_{606}$ \hrulefill} & \multicolumn{2}{c}{\hrulefill $I_{814}$ \hrulefill}  &  & & \\
  CFRS ID & $z$ & $Q_z^{\mathrm{a}}$ & $m_{450}$ & $m_{606}$ & $m_{814}$ & $M_{\rm AB}(B)$ & $M_{\rm AB}(K)$ & $B/T$ & $\chi^2$ & $B/T$ &  $\chi^2$ &  $B/T$ &  $\chi^2$ & Type$^{\mathrm{b}}$ & $Q^{\mathrm{c}}$ & Int/M$^{\mathrm{d}}$ \\
(1) & (2) & (3) & (4) & (5) & (6) & (7) & (8) & (9) & (10) & (11) & (12) & (13) & (14) & (15) & (16) & (17) \\
  \noalign{\smallskip}
  \hline
  \noalign{\smallskip}

03.0032 & 0.5220 & 3 &  --   & 22.08 & 20.62 & $-$20.19 & $-$22.53 &           --            &  --   &  0.11$^{-0.01}_{+0.01}$ & 1.117 & 0.17$^{-0.02}_{+0.16}$ & 1.192& Sab & 1 &     \\
03.0046 & 0.5120 & 3 &  --   & 21.73 & 20.58 & $-$20.44 & $-$22.10 &           --            &  --   &  0.03$^{-0.01}_{+0.01}$ & 2.681 & 0.13$^{-0.05}_{+0.04}$ & 1.566& Sbc & 1 &     \\
03.0327 & 0.6064 & 3 & 23.63 &  --   & 21.45 & $-$20.05 & $-$21.33 &  0.01$^{-0.01}_{+0.11}$ & 1.030 &           --            &  --   & 0.47$^{-0.04}_{+0.07}$ & 1.339& C   & 1 & R   \\
03.0350 & 0.6950 & 3 & 24.00 &  --   & 20.62 & $-$21.18 & $-$22.84 &  0.44$^{-0.26}_{+0.30}$ & 0.906 &           --            &  --   & 0.54$^{-0.02}_{+0.02}$ & 1.029& S0  & 1 &     \\
03.0508 & 0.4642 & 4 &  --   & 21.92 & 21.16 & $-$19.66 & $-$20.31 &           --            &  --   &  0.29$^{-0.01}_{+0.01}$ & 1.525 & 0.30$^{-0.04}_{+0.04}$ & 1.250& C   & 2 & R  \\
03.0528 & 0.7140 & 3 &  --   & 22.53 & 20.96 & $-$21.01 & $-$22.36 &           --            &  --   &  0.38$^{-0.05}_{+0.04}$ & 1.084 & 0.51$^{-0.04}_{+0.04}$ & 1.089& Sab & 2 &     \\
03.0560 & 0.6968 & 3 &  0.00 &  --   & 20.92 & $-$20.91 &    --    &           --            &  --   &           --            &  --   & 0.80$^{-0.06}_{+0.04}$ & 0.928& E/C & 1 &     \\
03.0589 & 0.7160 & 3 &  --   & 22.85 & 21.65 & $-$20.25 & $-$21.02 &           --            &  --   &  0.13$^{-0.04}_{+0.05}$ & 1.290 & 0.36$^{-0.11}_{+0.04}$ & 1.216& C   & 2 &     \\
03.0619 & 0.4854 & 3 &  --   & 21.42 & 20.36 & $-$20.93 & $-$22.06 &           --            &  --   &  0.11$^{-0.02}_{+0.03}$ & 1.369 & 0.16$^{-0.02}_{+0.03}$ & 1.165& Sab & 2 &     \\
03.0645 & 0.5275 & 4 &  --   & 21.60 & 20.78 & $-$20.45 & $-$21.33 &           --            &  --   &  0.68$^{-0.02}_{+0.03}$ & 1.907 & 0.61$^{-0.10}_{+0.05}$ & 1.463& C & 3 & M2  \\
03.0717 & 0.6070 & 3 & 23.09 &  --   & 20.47 & $-$20.92 &    --    &           --            &  --   &           --            &   --  & 0.04$^{-0.00}_{+0.03}$ & 1.011& Sbc & 2 & R?  \\
03.0728 & 0.5200 & 3 &  --   & 23.02 & 21.45 & $-$19.16 &    --    &           --            &  --   &  0.82$^{-0.09}_{+0.10}$ & 1.167 & 0.49$^{-0.04}_{+0.04}$ & 1.075& S0/C & 1 &     \\
03.0767 & 0.6690 & 3 &  --   & 23.07 & 21.37 & $-$20.40 & $-$22.62 &           --            &  --   &  0.32$^{-0.08}_{+0.10}$ & 1.044 & 0.67$^{-0.06}_{+0.09}$ & 1.022& S0  & 1 &     \\
03.0837 & 0.8215 & 3 &  --   & 23.58 & 21.57 & $-$20.98 & $-$23.39 &           --            &  --   &  0.49$^{-0.49}_{+0.31}$ & 0.926 & 0.70$^{-0.15}_{+0.11}$ & 1.159& Irr & 2 & R   \\
03.0844 & 0.8220 & 3 &  --   & 23.32 & 21.35 & $-$20.91 & $-$22.62 &           --            &  --   &  0.48$^{-0.10}_{+0.14}$ & 1.212 & 0.63$^{-0.08}_{+0.10}$ & 0.951& S0  & 1 &     \\
03.1318 & 0.8575 & 9 &  --   & 23.33 & 21.93 & $-$20.69 & $-$20.94 &           --            &  --   &           --            &  --   &          --            &  --  &     & 4 & M1  \\
03.1345 & 0.6167 & 3 &  --   & 22.38 & 20.87 & $-$20.48 & $-$22.49 &           --            &  --   &  0.14$^{-0.02}_{+0.03}$ & 1.037 & 0.40$^{-0.11}_{+0.16}$ & 0.819& Sab & 1 & I1  \\
03.1353 & 0.6340 & 2 &  --   & 22.30 & 20.98 & $-$20.68 & $-$22.40 &           --            &  --   &  0.11$^{-0.01}_{+0.01}$ & 1.197 & 0.18$^{-0.02}_{+0.03}$ & 1.097& Sab & 1 &     \\
03.1531 & 0.7148 & 3 &  --   & 23.32 & 21.70 & $-$20.52 & $-$22.33 &           --            &  --   &  0.00$^{-0.00}_{+0.04}$ & 0.933 & 0.00$^{-0.00}_{+0.02}$ & 0.920& Sd  & 2 &     \\
03.1541 & 0.6895 & 4 &  0.00 &  --   & 21.41 & $-$20.52 &    --    &           --            &  --   &           --            &  --   & 0.06$^{-0.04}_{+0.10}$ & 0.927& Sbc & 1 &     \\
14.0207 & 0.5460 & 4 & 22.78 &  --   & 19.21 & $-$22.01 & $-$23.88 &  0.92$^{-0.27}_{+0.04}$ & 0.823 &           --            &  --   & 0.93$^{-0.02}_{+0.02}$ & 0.943& E   & 1 &     \\
14.0400 & 0.6760 & 4 &  --   & 21.93 & 20.84 & $-$20.90 & $-$21.76 &           --            &  --   &  0.03$^{-0.03}_{+0.05}$ & 1.051 & 0.25$^{-0.11}_{+0.08}$ & 1.023& Sab & 1 &     \\
14.0411 & 0.8538 & 4 &  --   & 21.93 & 20.98 & $-$21.38 &    --    &           --            &  --   &  0.11$^{-0.02}_{+0.03}$ & 1.226 & 0.26$^{-0.04}_{+0.05}$ & 1.329& C   & 2 & R?  \\
14.0422 & 0.4210 & 2 &  --   & 21.18 & 19.97 & $-$20.39 & $-$22.30 &           --            &  --   &  0.38$^{-0.01}_{+0.01}$ & 1.238 & 0.53$^{-0.02}_{+0.01}$ & 1.272& S0  & 2 &     \\
14.0485 & 0.6545 & 3 & 23.63 &  --   & 21.69 & $-$20.18 & $-$21.02 &  0.01$^{-0.01}_{+0.14}$ & 0.938 &           --            &  --   & 0.01$^{-0.01}_{+0.04}$ & 1.006& Sbc & 2 & I2  \\
14.0557 & 0.8800 & 4 &  --   & 22.12 & 21.00 & $-$21.43 &    --    &           --            &  --   &  0.58$^{-0.09}_{+0.11}$ & 1.812 & 0.68$^{-0.02}_{+0.02}$ & 1.507& C   & 1 &     \\
14.0580 & 0.7440 & 3 &  --   & 22.25 & 20.82 & $-$21.83 &    --    &           --            &  --   &  0.05$^{-0.00}_{+0.00}$ & 1.113 & 0.37$^{-0.02}_{+0.09}$ & 1.983& Sab & 1 &     \\
14.0651 & 0.5480 & 1 &  --   & 22.91 & 21.44 & $-$19.57 & $-$21.47 &           --            &  --   &  0.81$^{-0.02}_{+0.01}$ & 1.014 & 0.82$^{-0.02}_{+0.02}$ & 1.039& S0/C& 1 &     \\
14.0697 & 0.8271 & 9 &  --   & 22.52 & 21.59 & $-$20.77 &    --    &           --            &  --   &  0.41$^{-0.03}_{+0.04}$ & 1.282 & 0.72$^{-0.09}_{+0.13}$ & 1.158& C   & 1 &     \\
14.0700 & 0.6526 & 4 &  --   & 21.56 & 19.96 & $-$21.44 &    --    &           --            &  --   &  0.92$^{-0.02}_{+0.02}$ & 1.003 & 0.92$^{-0.02}_{+0.02}$ & 1.005& S0/C& 1 &     \\
14.0727 & 0.4638 & 3 &  --   & 21.36 & 20.45 & $-$20.76 &    --    &           --            &  --   &  0.00$^{-0.00}_{+0.00}$ & 1.517 & 0.14$^{-0.03}_{+0.03}$ & 1.275& Sbc & 1 &     \\
14.0746 & 0.6750 & 3 &  --   & 22.60 & 20.92 & $-$20.50 & $-$22.41 &           --            &  --   &  0.73$^{-0.03}_{+0.06}$ & 0.995 & 0.72$^{-0.03}_{+0.02}$ & 1.018& S0  & 1 &     \\
14.0820 & 0.9800 & 4 &  --   & 23.39 & 21.25 & $-$21.83 & $-$23.87 &           --            &  --   &  0.58$^{-0.12}_{+0.10}$ & 0.989 & 0.75$^{-0.07}_{+0.06}$ & 1.037& S0  & 1 &     \\
14.0846 & 0.9900 & 9 & 23.90 & 22.88 & 21.47 & $-$21.69 & $-$22.71 &           --            &  --   &  0.01$^{-0.01}_{+0.03}$ & 1.063 & 0.08$^{-0.03}_{+0.04}$ & 1.132& Sbc & 3 & I2  \\
14.0854 & 0.9920 & 4 &  --   & 23.13 & 21.16 & $-$22.08 & $-$23.79 &           --            &  --   &  0.93$^{-0.13}_{+0.07}$ & 1.046 & 0.64$^{-0.04}_{+0.05}$ & 1.005& S0  & 2 & I2  \\
14.0899 & 0.8750 & 9 &  --   & 21.75 & 21.21 & $-$21.40 &    --    &           --            &  --   &  0.00$^{-0.00}_{+0.00}$ & 1.091 & 0.00$^{-0.00}_{+0.00}$ & 1.040& Sd  & 1 &     \\
14.0913 & 1.0058 & 4 &  --   & 23.47 & 21.55 & $-$21.57 &    --    &           --            &  --   &  0.61$^{-0.06}_{+0.06}$ & 0.978 & 0.71$^{-0.12}_{+0.03}$ & 0.943& S0/C& 2 &     \\
14.0937 & 1.0097 & 4 &  --   & 23.51 & 21.41 & $-$21.45 &    --    &           --            &  --   &  0.64$^{-0.03}_{+0.03}$ & 1.014 & 0.70$^{-0.01}_{+0.02}$ & 1.024& S0  & 1 &     \\
14.0962 & 0.7617 & 4 &  --   & 22.29 & 20.48 & $-$21.61 & $-$23.03 &           --            &  --   &  0.45$^{-0.04}_{+0.04}$ & 1.008 & 0.44$^{-0.01}_{+0.02}$ & 1.034& Sab & 2 &     \\
14.0964 & 0.4351 & 4 &  --   & 21.95 & 21.19 & $-$19.57 &    --    &           --            &  --   &  0.01$^{-0.00}_{+0.00}$ & 2.069 & 0.03$^{-0.01}_{+0.01}$ & 1.405& Sbc & 1 &     \\
  \noalign{\smallskip}
  \hline
  \end{tabular}
\end{table}
\end{landscape}

\addtocounter{table}{-1}
\begin{landscape}
\begin{table}
  \caption[]{Catalog of field galaxies in CFRS 0300+00 and 1415+52 fields -- continued}
  \begin{tabular}{cccccccccccccclll}
  \hline
  \noalign{\smallskip}
         & & & & & &    &   &  \multicolumn{2}{c}{\hrulefill $B_{450}$\hrulefill} & \multicolumn{2}{c}{\hrulefill $V_{606}$ \hrulefill} & \multicolumn{2}{c}{\hrulefill $I_{814}$ \hrulefill}  &  & & \\
  CFRS ID & $z$ & $Q_z^{\mathrm{a}}$ & $m_{450}$ & $m_{606}$ & $m_{814}$ & $M_{\rm AB}(B)$ & $M_{\rm AB}(K)$ & $B/T$ & $\chi^2$ & $B/T$ &  $\chi^2$ &  $B/T$ &  $\chi^2$ & Type$^{\mathrm{b}}$ & $Q^{\mathrm{c}}$ & Int/M$^{\mathrm{d}}$ \\
(1) & (2) & (3) & (4) & (5) & (6) & (7) & (8) & (9) & (10) & (11) & (12) & (13) & (14) & (15) & (16)  & (17)\\
  \noalign{\smallskip}
  \hline
  \noalign{\smallskip}

14.0972 & 0.6809 & 4 & 22.34 & 21.43 & 20.58 & $-$21.25 & $-$21.93 &  0.24$^{-0.22}_{+0.15}$ & 1.003 &  0.60$^{-0.02}_{+0.02}$ & 1.388 & 0.49$^{-0.02}_{+0.03}$ & 1.350& C   & 1 & R   \\
14.1008 & 0.4362 & 4 &  --   & 21.62 & 20.18 & $-$20.29 &    --    &           --            &  --   &  0.25$^{-0.01}_{+0.00}$ & 1.849 & 0.26$^{-0.00}_{+0.01}$ & 2.198& Sab & 1 &     \\
14.1012 & 0.4815 & 3 & 22.84 & 21.87 & 20.96 & $-$20.07 & $-$21.03 &  0.86$^{-0.03}_{+0.0.02}$ & 0.882 &  0.56$^{-0.04}_{+0.03}$ & 1.082 & 0.59$^{-0.14}_{+0.09}$ & 1.070& C   & 1 &     \\
14.1028 & 0.9876 & 3 & 24.93 & 23.10 & 21.07 & $-$21.96 & $-$23.56 &           --            &  --   &  0.46$^{-0.10}_{+0.07}$ & 0.999 & 0.44$^{-0.21}_{+0.06}$ & 1.087& Sab & 2 &     \\
14.1037 & 0.5489 & 3 & 22.03 & 21.87 & 20.31 & $-$20.32 & $-$21.32 &  0.00$^{-0.00}_{+0.05}$ & 0.874 &  0.00$^{-0.00}_{+0.00}$ & 1.136 & 0.00$^{-0.00}_{+0.00}$ & 1.133& Irr & 2 &     \\
14.1041 & 0.4351 & 4 &  --   & 21.80 & 21.07 & $-$20.04 &    --    &           --            &  --   &  0.00$^{-0.00}_{+0.00}$ & 2.237 & 0.00$^{-0.00}_{+0.00}$ & 1.644& Irr & 3 & M2  \\
14.1043 & 0.6516 & 4 & 22.73 & 21.25 & 19.75 & $-$22.02 & $-$23.69 &  0.04$^{-0.04}_{+0.06}$ & 0.862 &  0.10$^{-0.00}_{+0.00}$ & 1.382 & 0.19$^{-0.00}_{+0.00}$ & 1.270& Sab & 1 &     \\
14.1079 & 0.9130 & 9 &  --   & 22.78 & 21.37 & $-$21.22 & $-$22.21 &           --            &  --   &  0.00$^{-0.00}_{+0.01}$ & 1.290 & 0.03$^{-0.01}_{+0.00}$ & 1.192& Irr & 3 & M2  \\
14.1087 & 0.6595 & 3 &  --   & 22.41 & 21.35 & $-$20.31 & $-$20.95 &           --            &  --   &  0.55$^{-0.35}_{+0.20}$ & 1.144 & 0.90$^{-0.13}_{+0.10}$ & 1.020& C   & 1 & R   \\
14.1136 & 0.6404 & 3 & 23.09 & 22.30 & 21.49 & $-$20.71 & $-$20.16 &  0.74$^{-0.04}_{+0.04}$ & 0.845 &  0.74$^{-0.05}_{+0.04}$ & 1.122 & 0.69$^{-0.16}_{+0.06}$ & 0.926& C   & 1 & R   \\
14.1146 & 0.7437 & 3 &  --   & 22.39 & 21.16 & $-$20.84 &    --    &           --            &  --   &  0.25$^{-0.05}_{+0.05}$ & 1.217 & 0.13$^{-0.03}_{+0.04}$ & 1.222& C   & 1 & R   \\
14.1164 & 0.6773 & 4 &  --   & 22.26 & 21.29 & $-$20.67 &    --    &           --            &  --   &  0.00$^{-0.00}_{+0.00}$ & 1.210 & 0.00$^{-0.00}_{+0.01}$ & 1.140& Sd  & 2 &     \\
14.1179 & 0.4345 & 4 &  --   & 22.23 & 20.96 & $-$19.45 &    --    &           --            &  --   &  0.66$^{-0.11}_{+0.09}$ & 0.996 & 0.53$^{-0.06}_{+0.07}$ & 1.035& S0  & 1 &     \\
14.1189 & 0.7526 & 3 & 23.47 &  --   & 21.57 & $-$20.51 & $-$21.28 &  0.07$^{-0.07}_{+0.24}$ & 0.923 &           --            &  --   & 0.01$^{-0.01}_{+0.03}$ & 1.035& Sbc & 2 &     \\
14.1213 & 0.7631 & 4 &  --   & 22.48 & 20.95 & $-$21.63 &    --    &           --            &  --   &  0.00$^{-0.00}_{+0.00}$ & 1.091 & 0.00$^{-0.00}_{+0.00}$ & 1.155&     & 4 & M1  \\
14.1223 & 0.6936 & 4 &  --   & 22.48 & 21.38 & $-$20.39 &    --    &           --            &  --   &  0.16$^{-0.04}_{+0.05}$ & 1.042 & 0.15$^{-0.02}_{+0.03}$ & 1.036& Sab & 2 &     \\
14.1232 & 0.7660 & 4 &  --   & 23.29 & 22.67 & $-$20.61 & $-$22.23 &           --            &  --   &  0.42$^{-0.03}_{+0.03}$ & 1.389 & 0.70$^{-0.04}_{+0.05}$ & 1.150&     & 4 & M1  \\
14.1246 & 0.6486 & 2 &  --   & 23.37 & 21.92 & $-$19.89 &    --    &           --            &  --   &  0.18$^{-0.11}_{+0.17}$ & 1.137 & 0.13$^{-0.05}_{+0.07}$ & 1.122& Irr & 3 & M2  \\
14.1277 & 0.8100 & 4 &  --   & 22.74 & 20.88 & $-$21.70 &    --    &           --            &  --   &  0.22$^{-0.02}_{+0.01}$ & 1.050 & 0.45$^{-0.02}_{+0.02}$ & 1.020& Sab & 2 & R?  \\
14.1311 & 0.8065 & 3 &  --   & 21.99 & 20.11 & $-$22.72 &    --    &           --            &  --   &  0.51$^{-0.09}_{+0.07}$ & 1.061 & 0.59$^{-0.04}_{+0.03}$ & 1.039& S0  & 2 &     \\
14.1326 & 0.8111 & 4 &  --   & 22.63 & 21.46 & $-$20.55 &    --    &           --            &  --   &  0.02$^{-0.01}_{+0.01}$ & 1.006 & 0.01$^{-0.01}_{+0.02}$ & 0.955& Sbc & 2 & M2  \\
14.1355 & 0.4801 & 4 &  --   & 21.87 & 21.01 & $-$20.01 & $-$20.84 &           --            &  --   &  0.00$^{-0.00}_{+0.00}$ & 1.697 & 0.00$^{-0.00}_{+0.00}$ & 1.256& Sd  & 1 &     \\
14.1356 & 0.8307 & 3 &  --   & 22.91 & 21.70 & $-$20.81 &    --    &           --            &  --   &  0.41$^{-0.06}_{+0.06}$ & 0.912 & 0.72$^{-0.11}_{+0.13}$ & 0.963& S0/C& 2 &     \\
14.1395 & 0.5301 & 4 &  --   & 22.20 & 21.39 & $-$19.90 &    --    &           --            &  --   &  0.02$^{-0.01}_{+0.00}$ & 1.071 & 0.01$^{-0.01}_{+0.02}$ & 1.059& Sbc & 2 &     \\
14.1412 & 0.9122 & 4 &  --   & 23.05 & 21.57 & $-$21.17 &    --    &           --            &  --   &  0.00$^{-0.00}_{+0.01}$ & 1.075 & 0.13$^{-0.02}_{+0.03}$ & 1.083& Sbc & 2 &     \\
14.1427 & 0.5379 & 4 &  --   & 22.23 & 21.11 & $-$19.83 &    --    &           --            &  --   &  0.03$^{-0.03}_{+0.04}$ & 1.085 & 0.02$^{-0.02}_{+0.02}$ & 1.100& Sbc & 1 &     \\
14.1464 & 0.4620 & 2 &  --   & 21.93 & 20.55 & $-$19.77 &    --    &           --            &  --   &  0.41$^{-0.04}_{+0.05}$ & 0.985 & 0.41$^{-0.02}_{+0.02}$ & 1.042& Sab & 1 &     \\
14.1477 & 0.8191 & 4 &  --   & 22.86 & 21.01 & $-$21.25 &    --    &           --            &  --   &  0.29$^{-0.04}_{+0.07}$ & 0.988 & 0.76$^{-0.21}_{+0.23}$ & 1.082& S0  & 1 &     \\
14.1496 & 0.9163 & 3 &  --   & 22.94 & 21.59 & $-$21.32 &    --    &           --            &  --   &  0.60$^{-0.04}_{+0.03}$ & 1.032 & 0.87$^{-0.05}_{+0.06}$ & 1.027& C   & 2 &     \\
14.1501 & 1.0018 & 4 &  --   & 22.70 & 21.55 & $-$21.78 &    --    &           --            &  --   &  0.00$^{-0.00}_{+0.01}$ & 1.260 & 0.00$^{-0.00}_{+0.00}$ & 1.349& Sd  & 2 &     \\
14.1506 & 1.0126 & 4 &  --   & 23.46 & 21.65 & $-$21.64 &    --    &           --            &  --   &  0.32$^{-0.05}_{+0.08}$ & 1.015 & 0.84$^{-0.06}_{+0.08}$ & 1.036& E   & 2 &     \\
14.1524 & 0.4297 & 4 &  --   & 20.33 & 19.37 & $-$21.50 &    --    &           --            &  --   &  0.00$^{-0.00}_{+0.00}$ & 1.787 & 0.03$^{-0.01}_{+0.00}$ & 1.581& Sbc & 2 &     \\
14.1554 & 0.9115 & 4 &  --   & 22.78 & 21.40 & $-$21.26 &    --    &           --            &  --   &  0.00$^{-0.00}_{+0.03}$ & 1.001 & 0.00$^{-0.00}_{+0.02}$ & 1.014& Sd  & 2 &     \\
14.1601 & 0.5370 & 4 &  --   & 21.66 & 20.84 & $-$20.31 &    --    &           --            &  --   &  0.00$^{-0.00}_{+0.01}$ & 1.200 & 0.01$^{-0.01}_{+0.01}$ & 1.192& Sd  & 2 &     \\
14.1620 & 0.9182 & 4 &  --   & 23.03 & 20.98 & $-$21.77 &    --    &           --            &  --   &  0.66$^{-0.11}_{+0.08}$ & 0.989 & 0.54$^{-0.05}_{+0.06}$ & 0.998& S0  & 1 &     \\
                      
  \noalign{\smallskip}
  \hline
  \end{tabular}
  \begin{list}{}{}
  \item[$^{\mathrm{a}}$]
$Q_z$ CFRS redshift confidence class --- 4,8,9: 100\%; 3: 97\%; 2: 80\%; 1: 50\%.
  \item[$^{\mathrm{b}}$]
Galaxy type --- E/S0: 0.8\,$<B/T\leq$\,1; S0: 0.5\,$<B/T\leq$\,0.8; Sab: 0.15\,$<B/T\leq$\,0.5; Sbc: 0\,$<B/T\leq$\,0.15; Sd: $B/T$\,=\,0; C: compact; Irr: irregular. 
  \item[$^{\mathrm{c}}$]
$Q$ quality factor --- 1: secure; 2: possibly secure; 3: insecure; 4: undetermined.
  \item[$^{\mathrm{d}}$]
Interaction/Merging --- M1: obvious merging; M2: possible merging; I1: obvious interaction; I2: possible interaction; R: relics of fusion/interactions.
  \end{list}
\end{table}
\end{landscape}

\begin{figure*} \centering

\includegraphics[height=0.22\textwidth,clip]{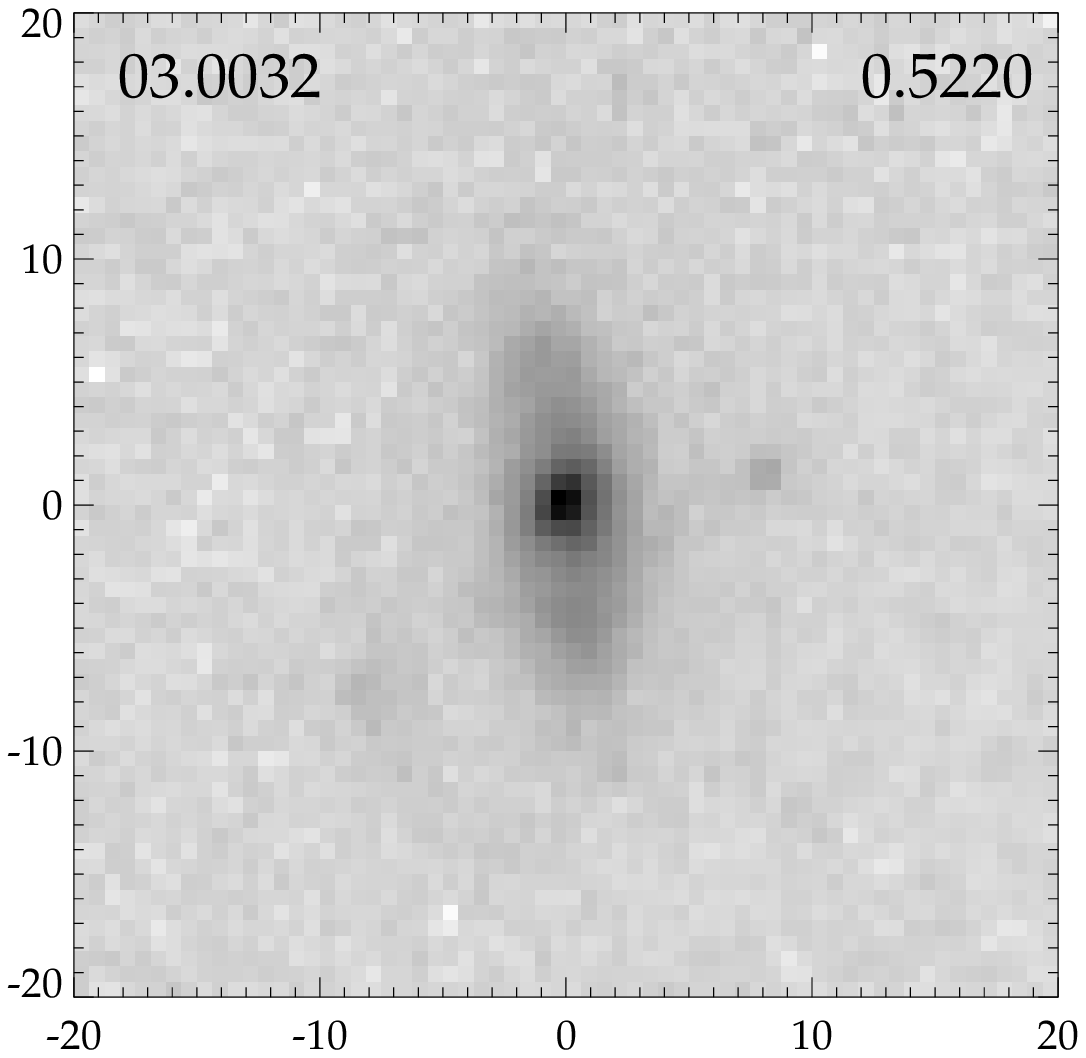} \includegraphics[height=0.22\textwidth,clip]{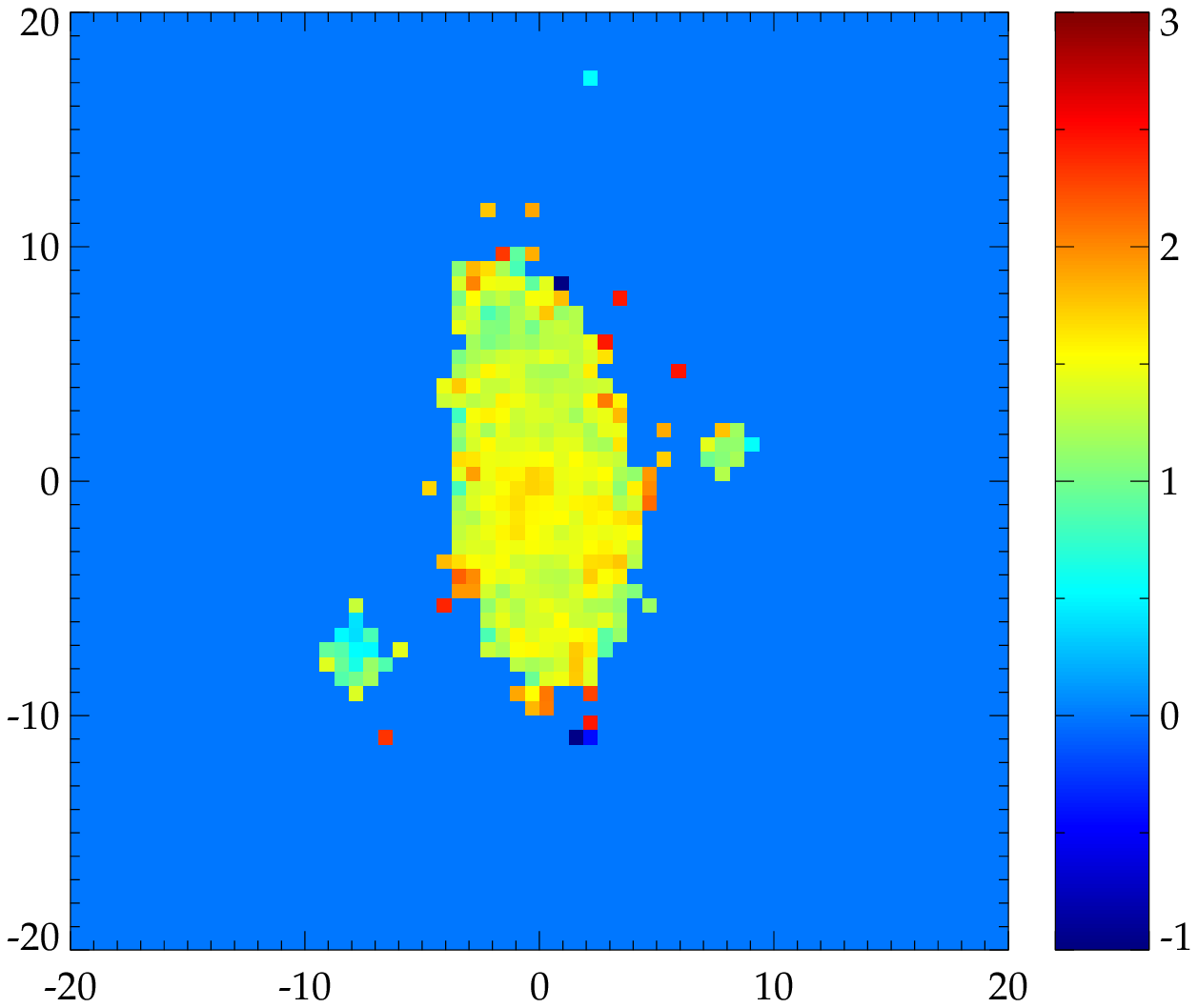}
\includegraphics[height=0.22\textwidth,clip]{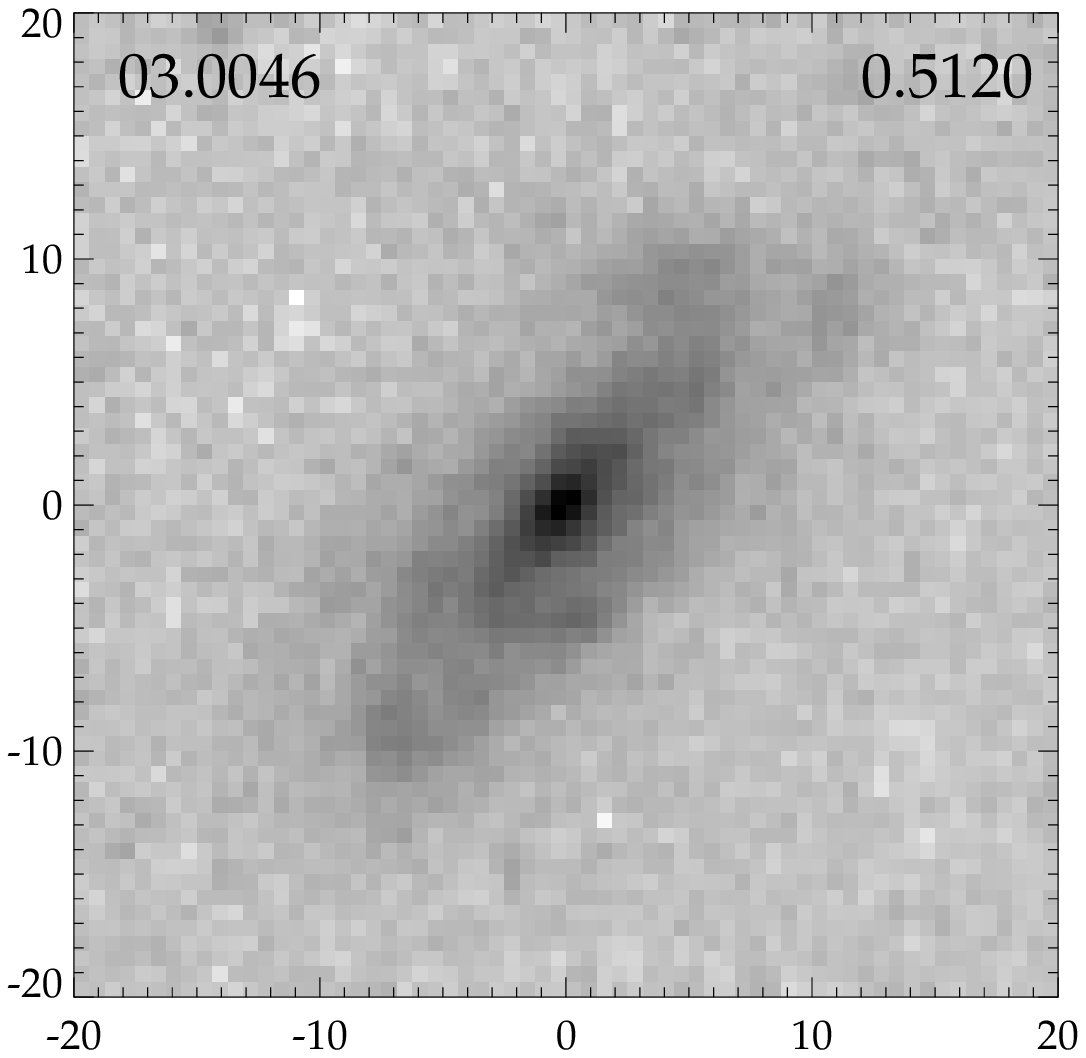} \includegraphics[height=0.22\textwidth,clip]{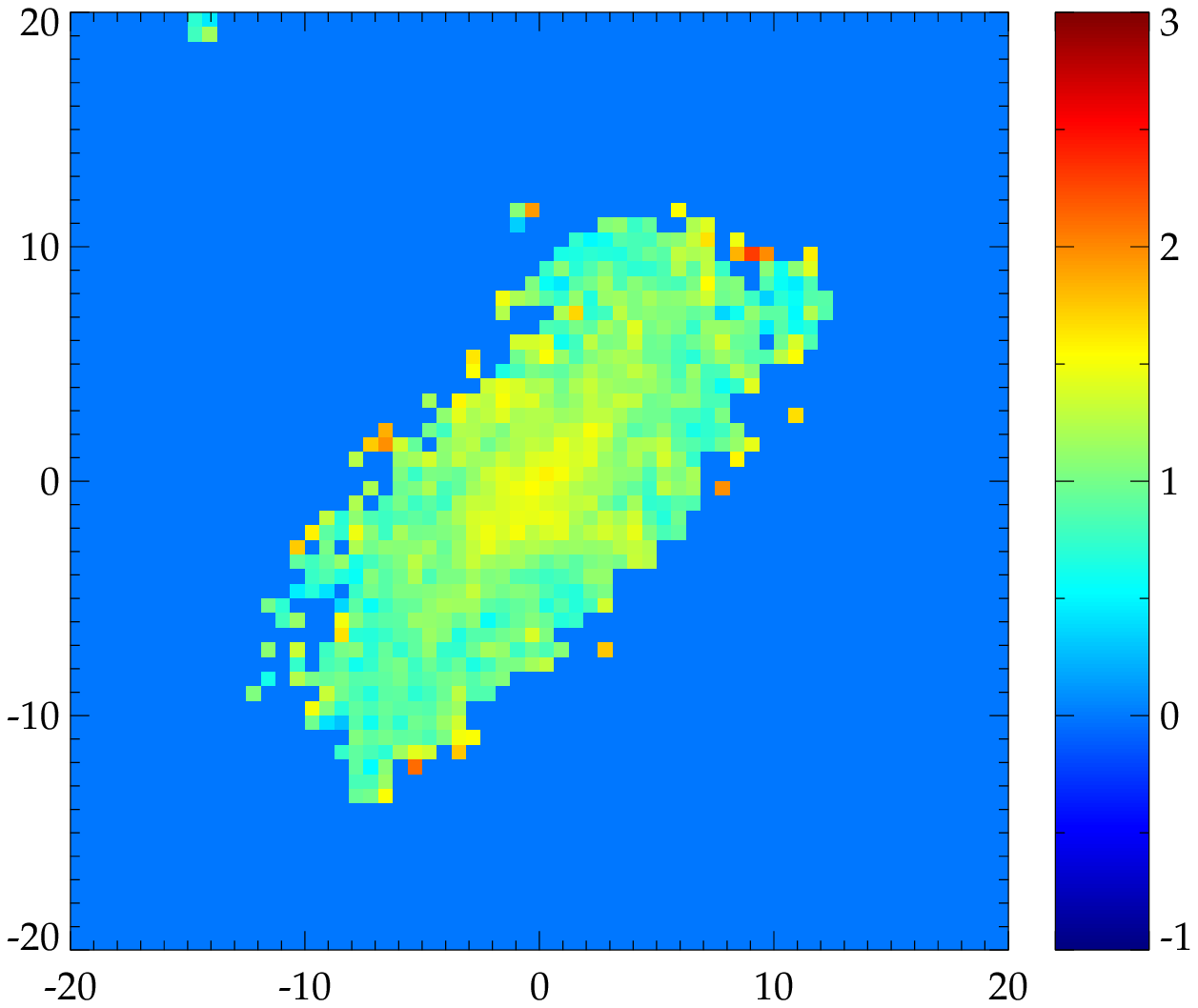}
\includegraphics[height=0.22\textwidth,clip]{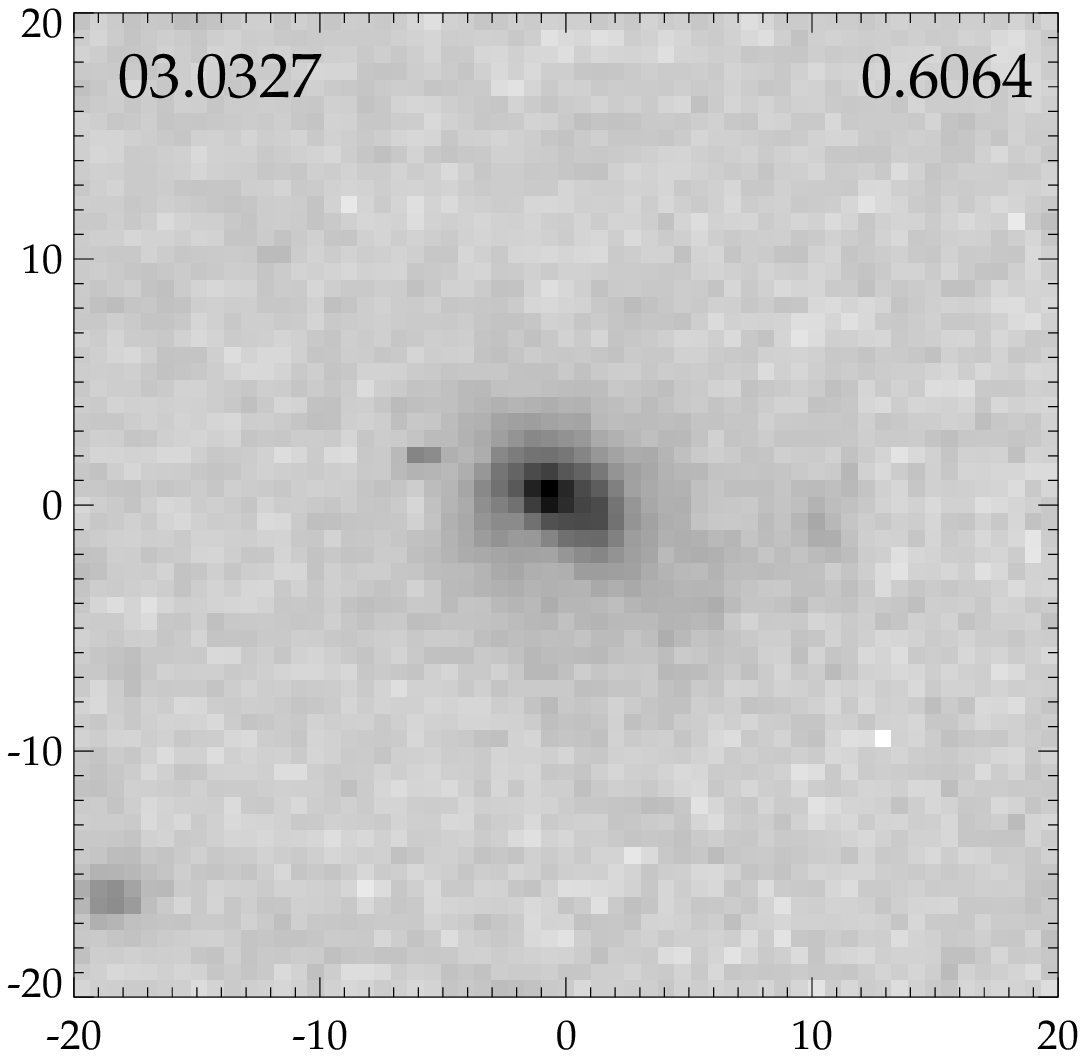} \includegraphics[height=0.22\textwidth,clip]{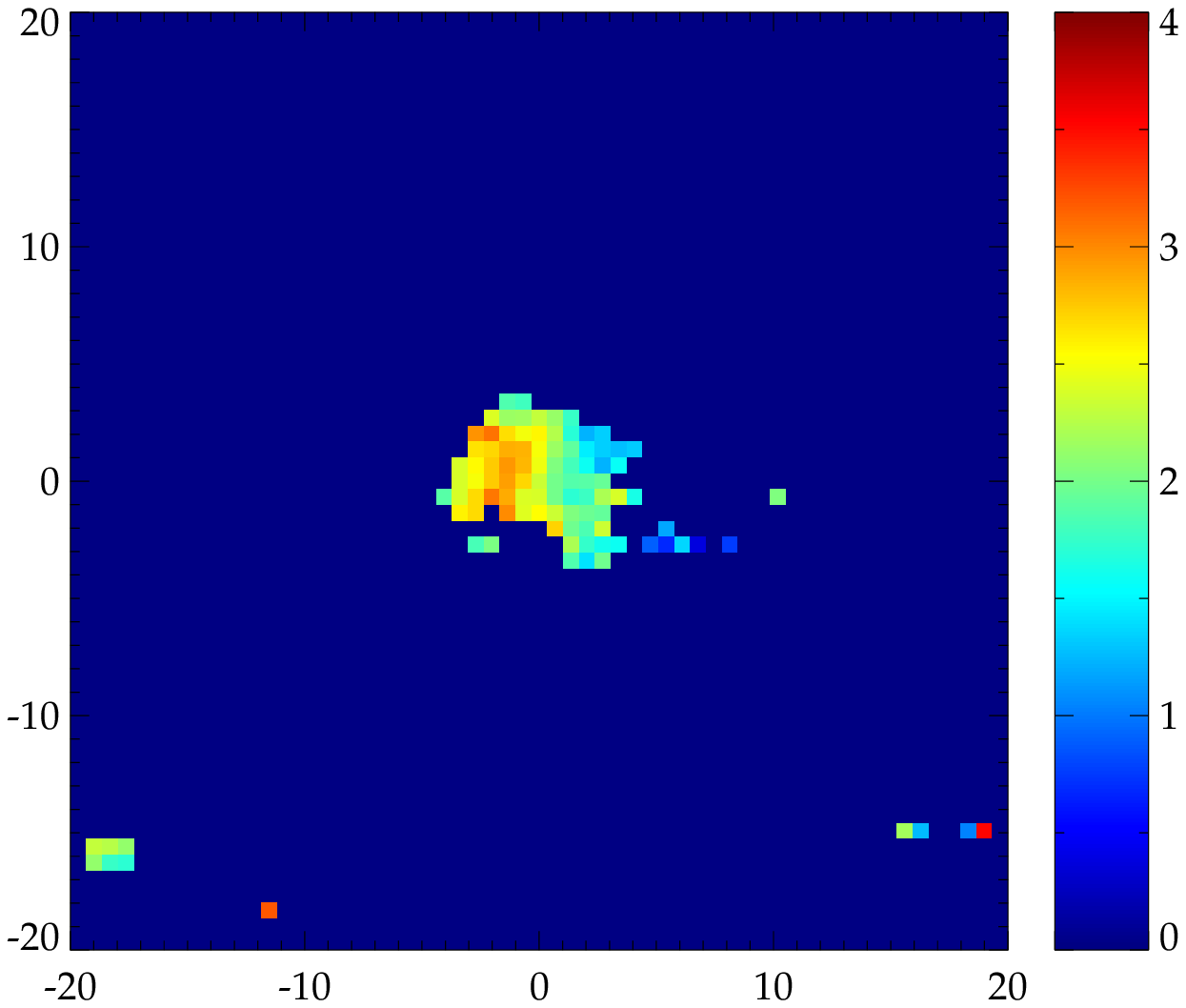}
\includegraphics[height=0.22\textwidth,clip]{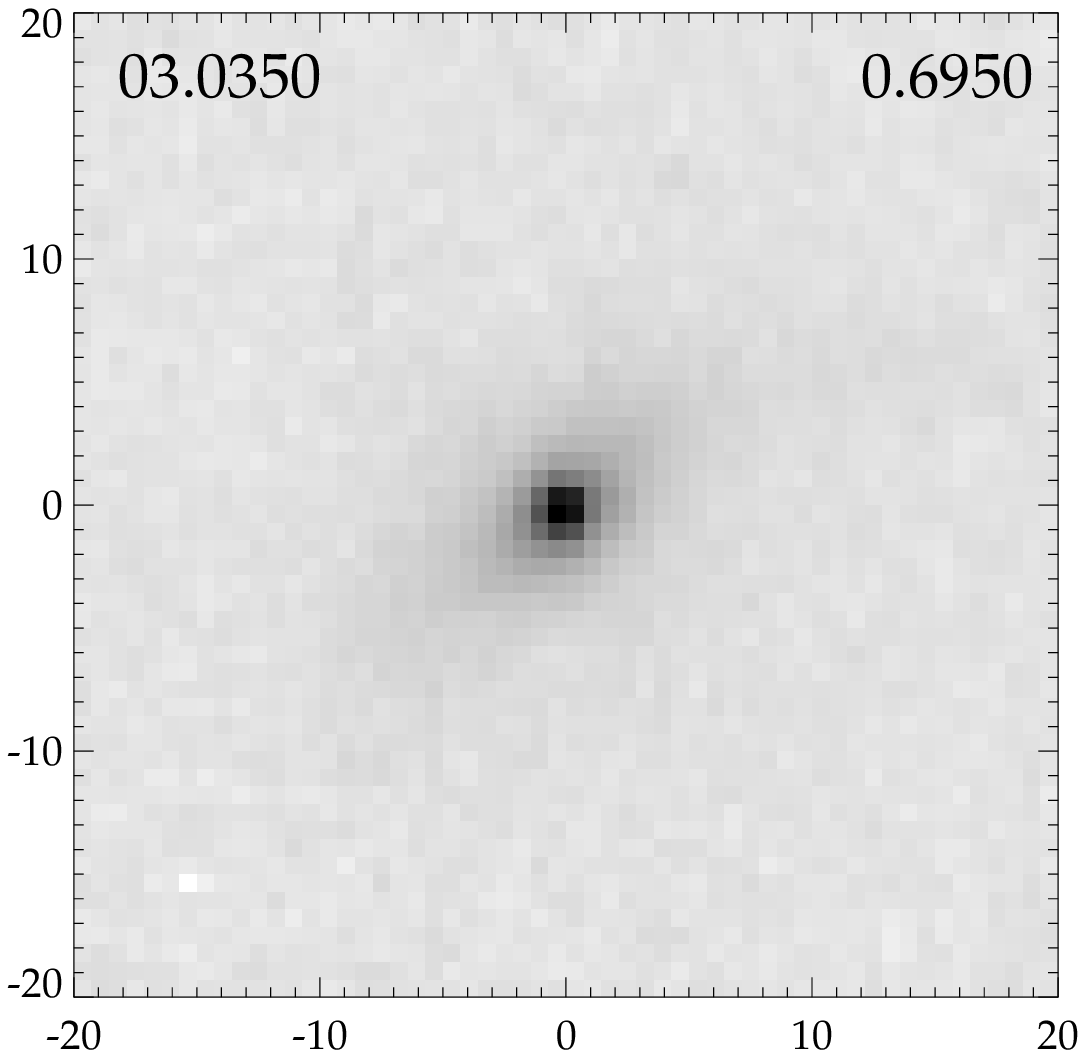} \includegraphics[height=0.22\textwidth,clip]{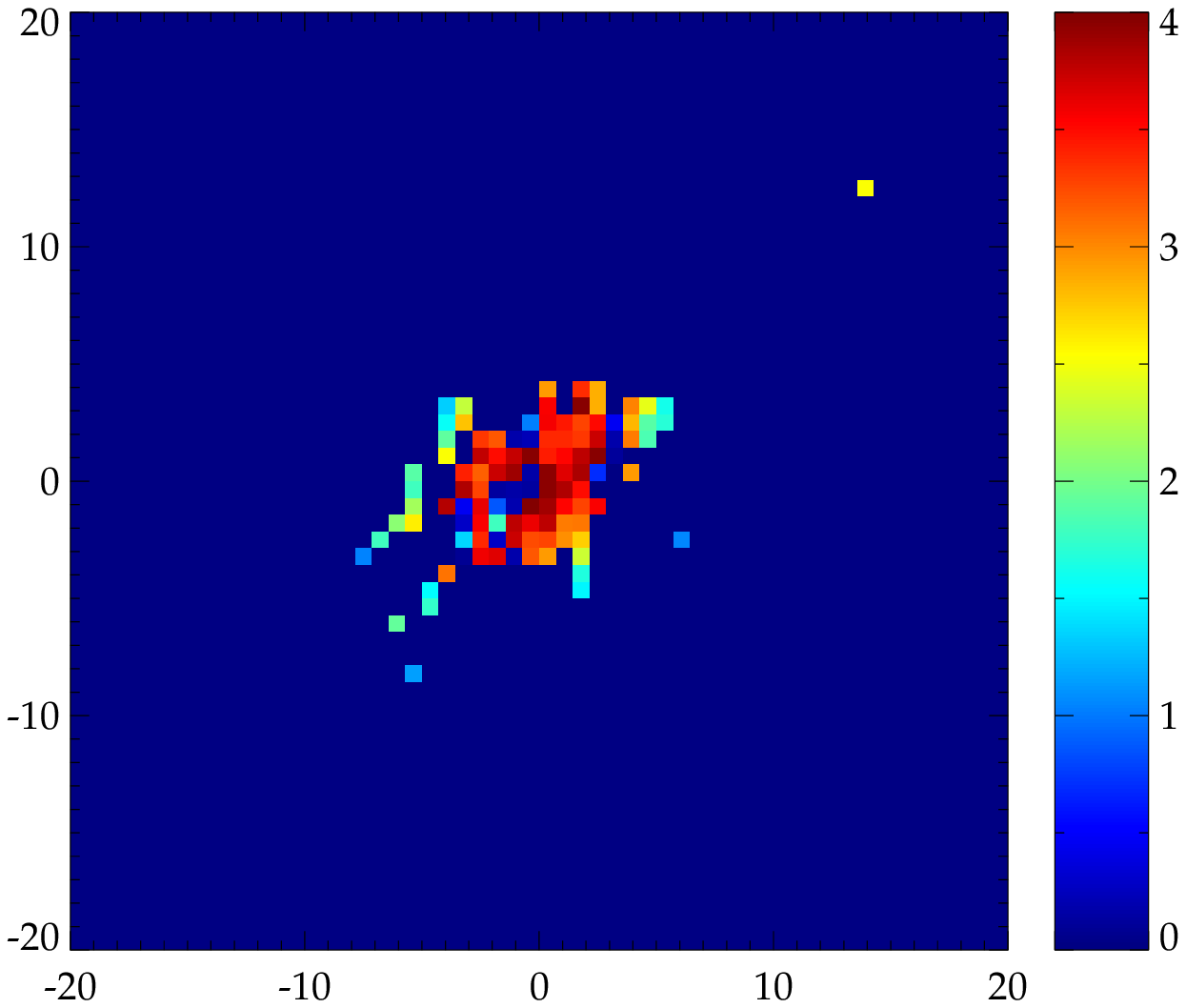}
\includegraphics[height=0.22\textwidth,clip]{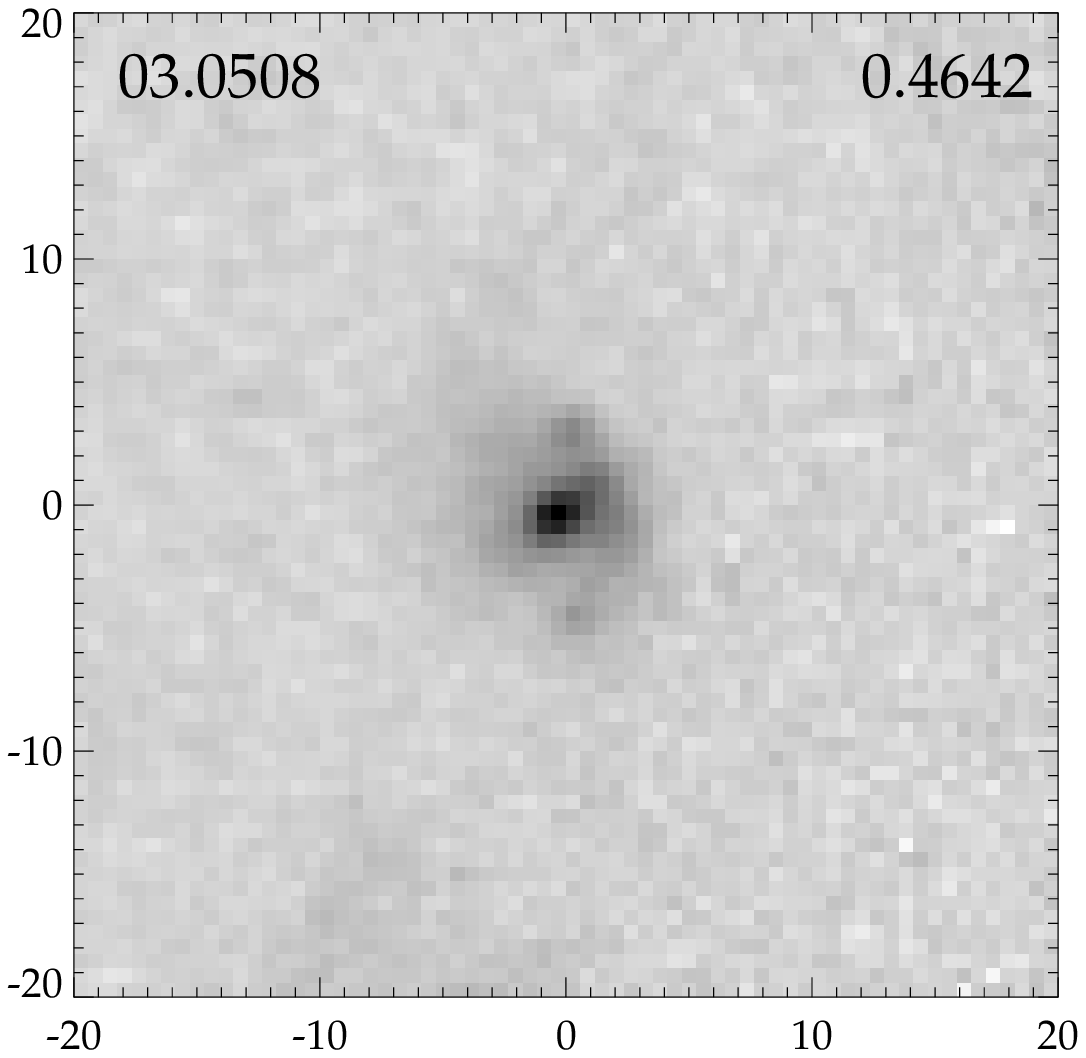} \includegraphics[height=0.22\textwidth,clip]{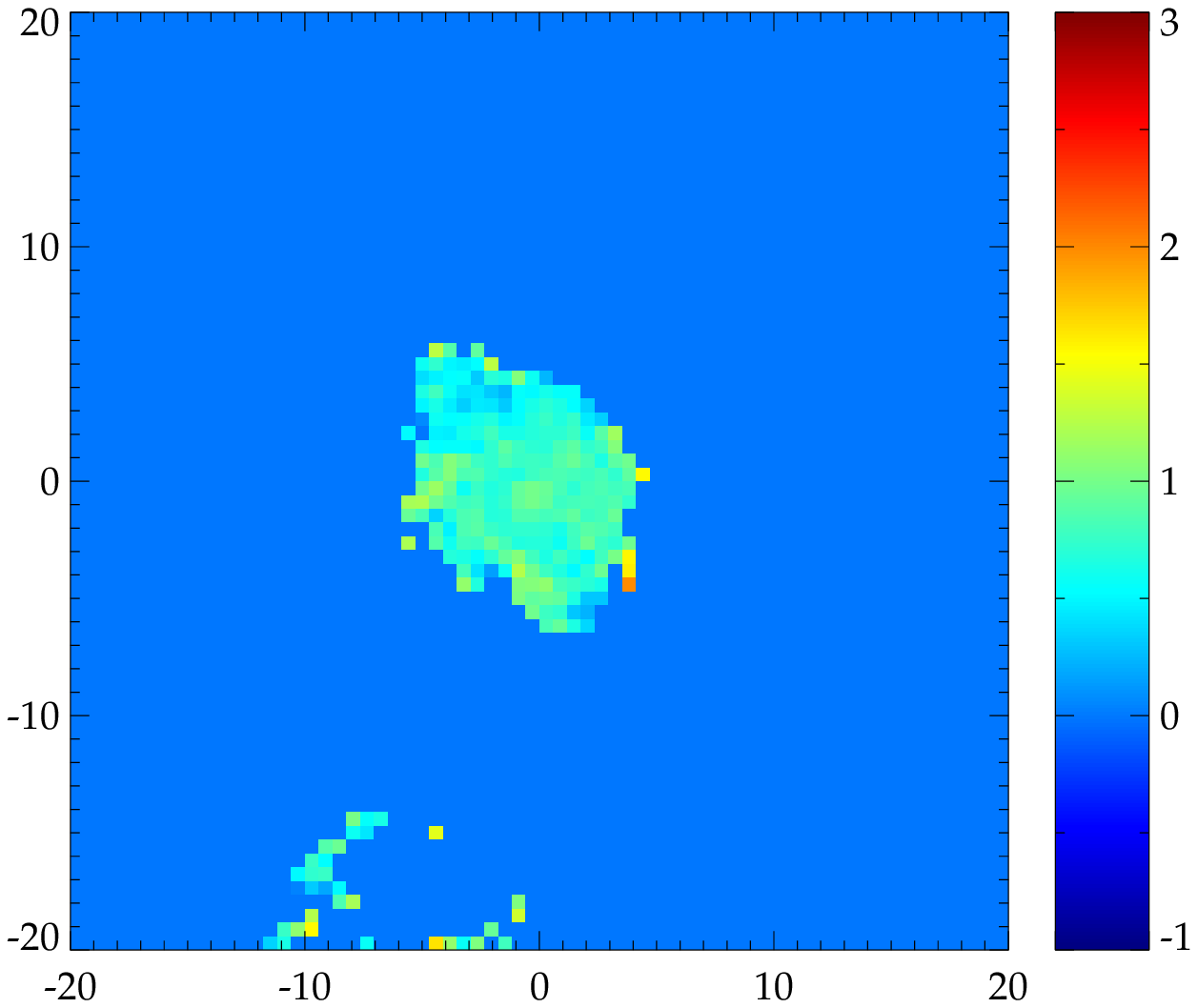}
\includegraphics[height=0.22\textwidth,clip]{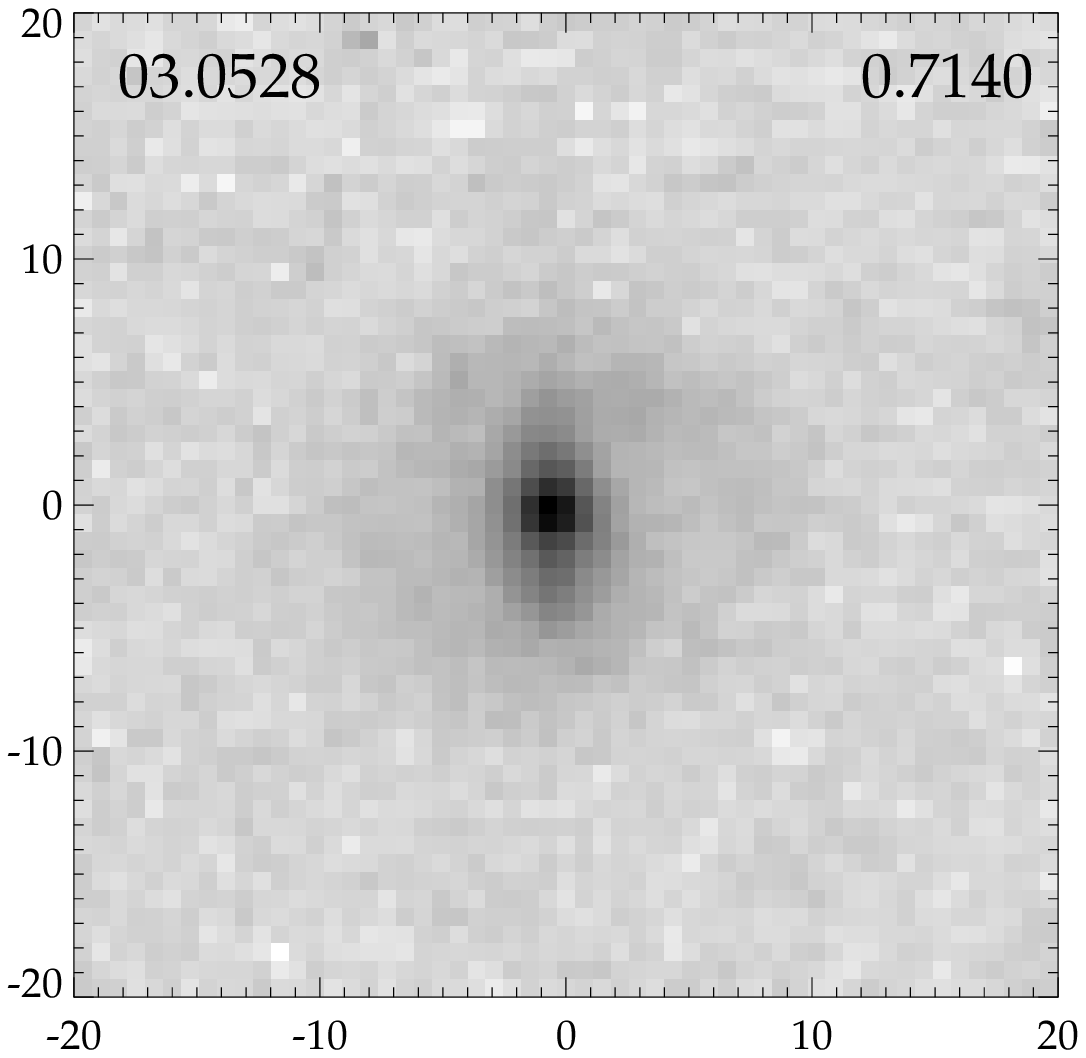} \includegraphics[height=0.22\textwidth,clip]{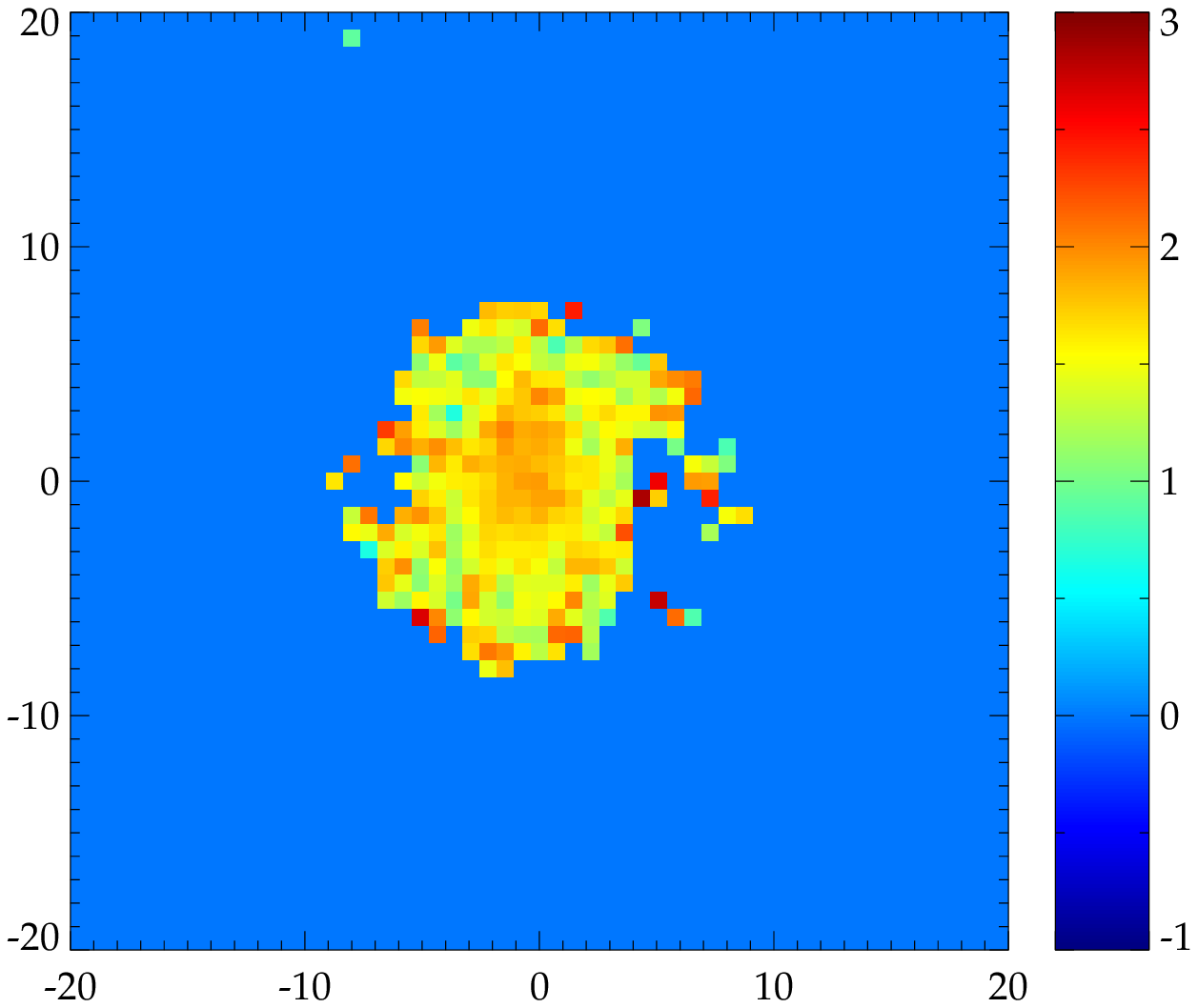}
\includegraphics[height=0.22\textwidth,clip]{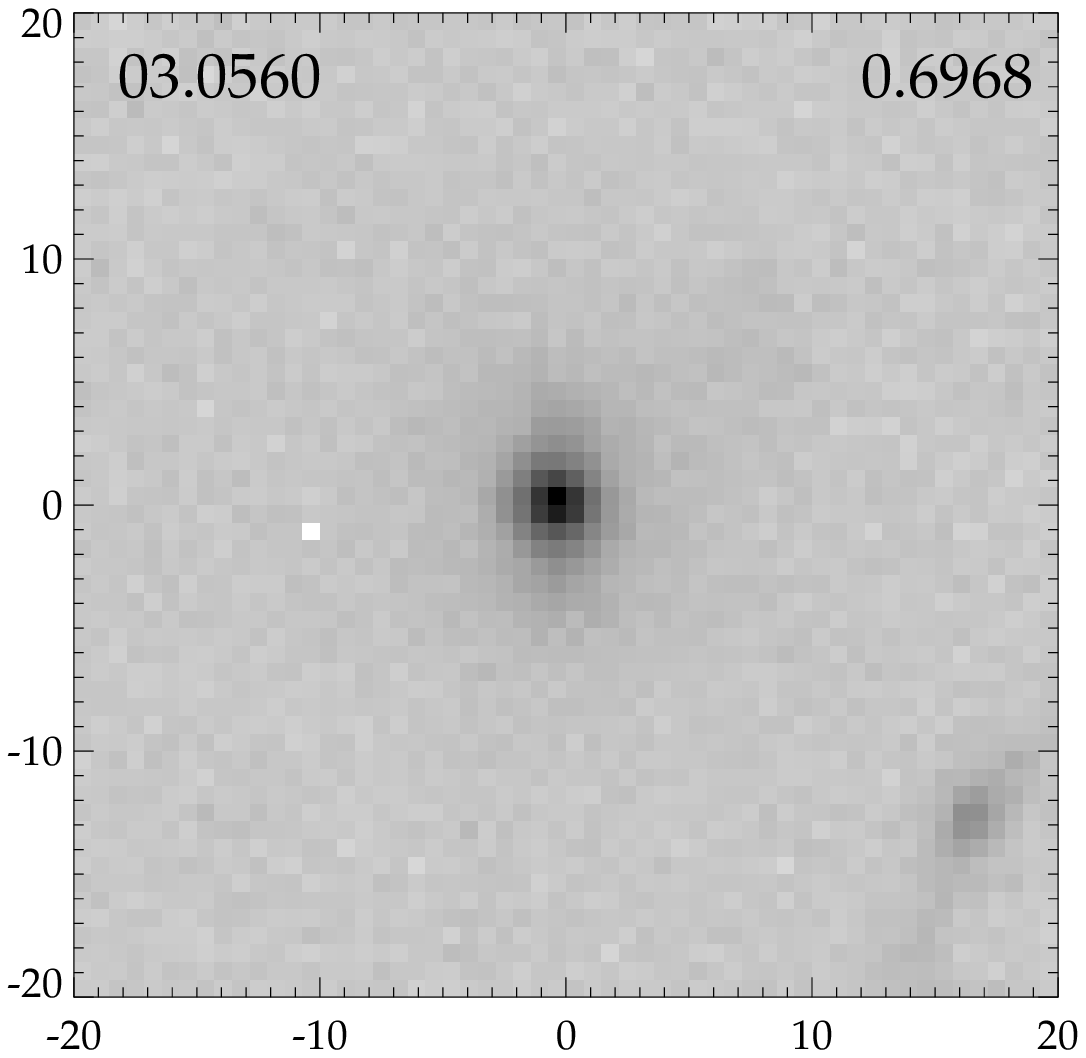} \includegraphics[height=0.22\textwidth,clip]{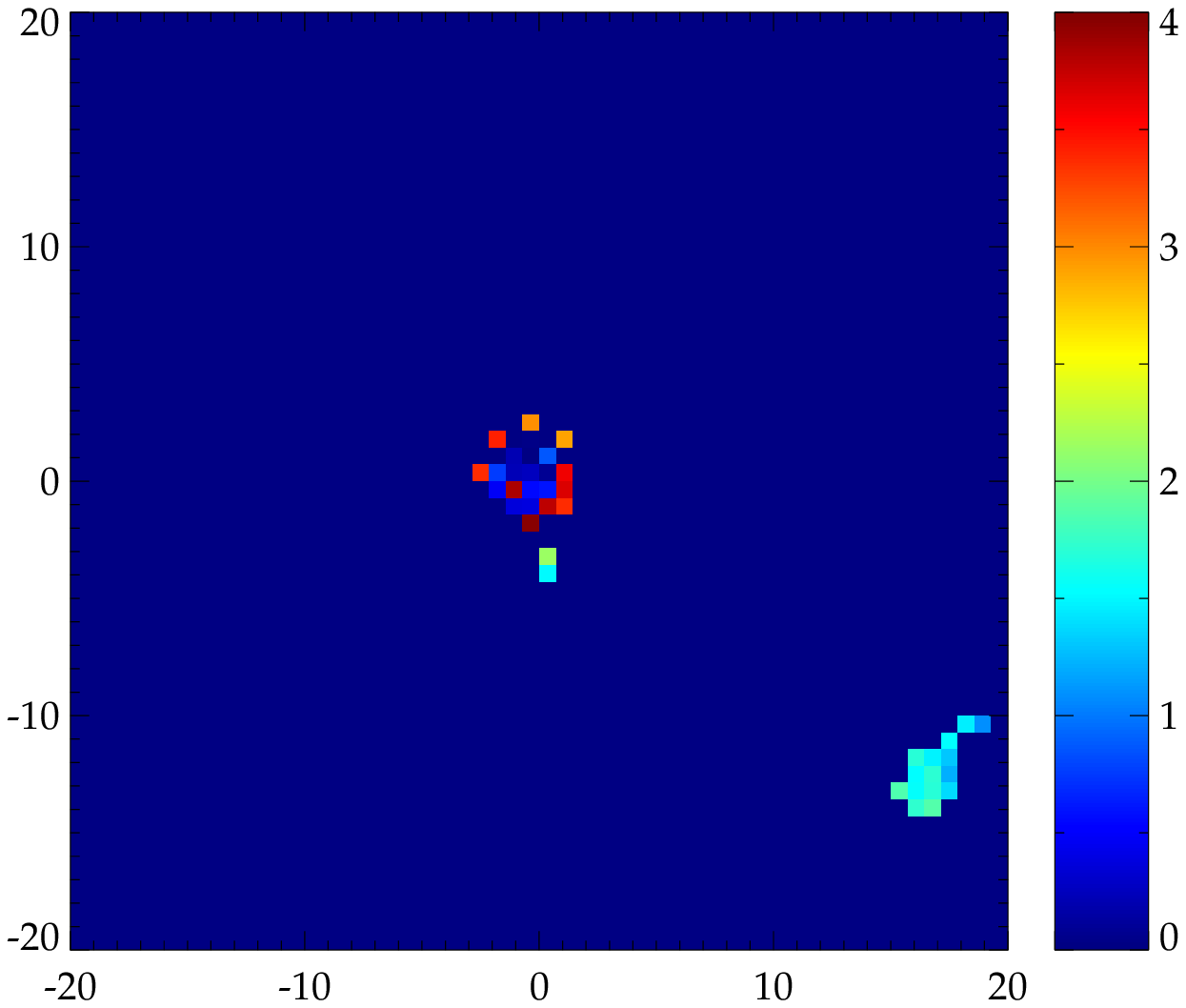}
\includegraphics[height=0.22\textwidth,clip]{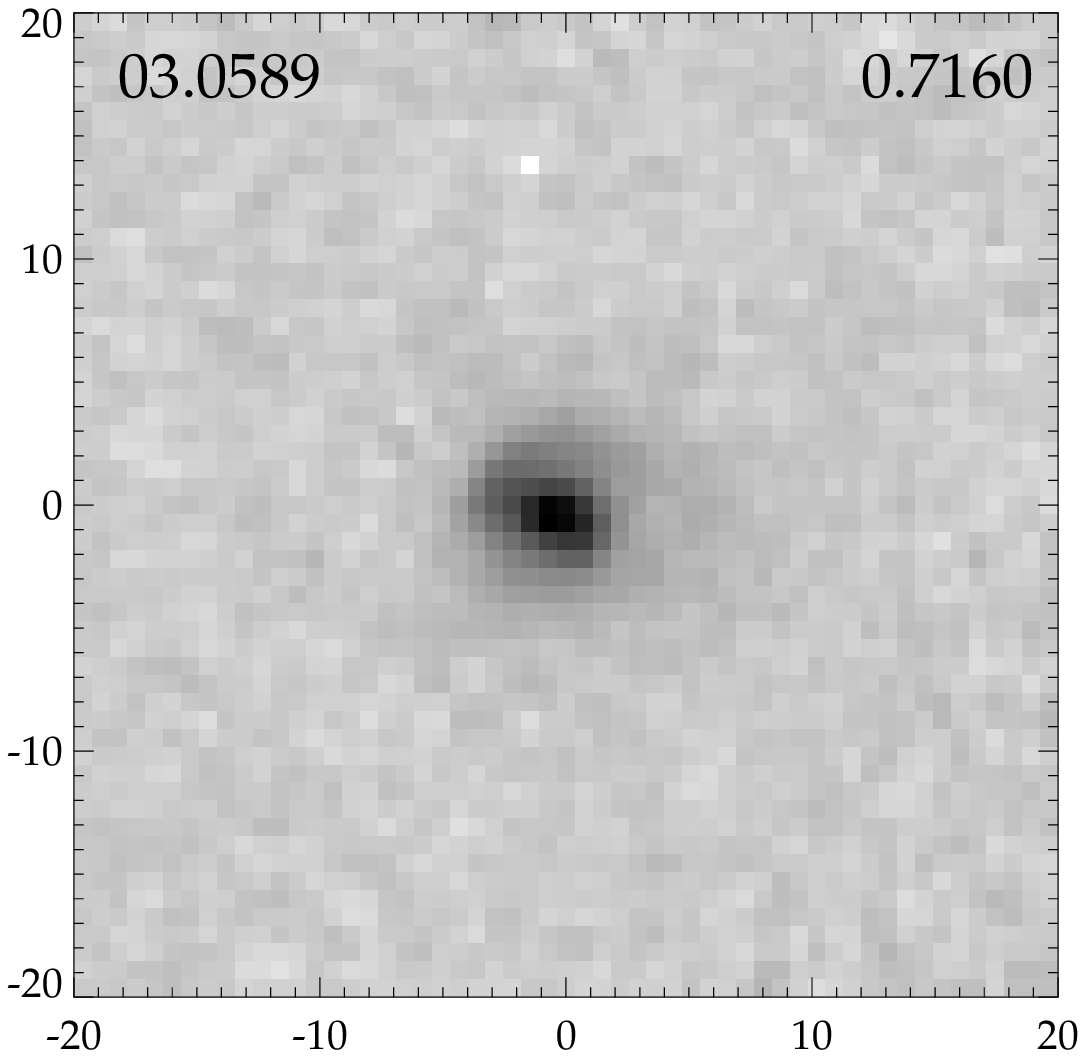} \includegraphics[height=0.22\textwidth,clip]{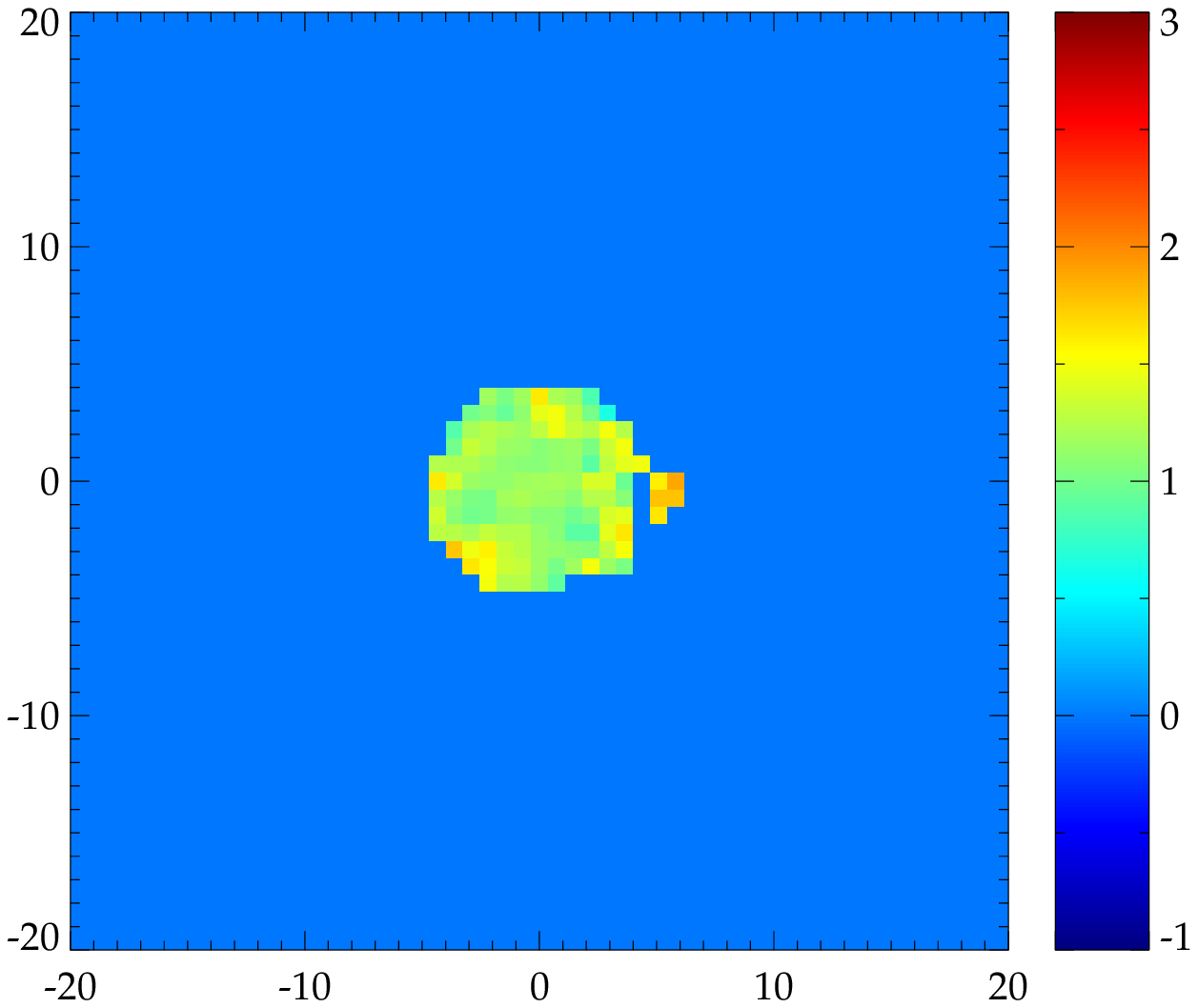}
\includegraphics[height=0.22\textwidth,clip]{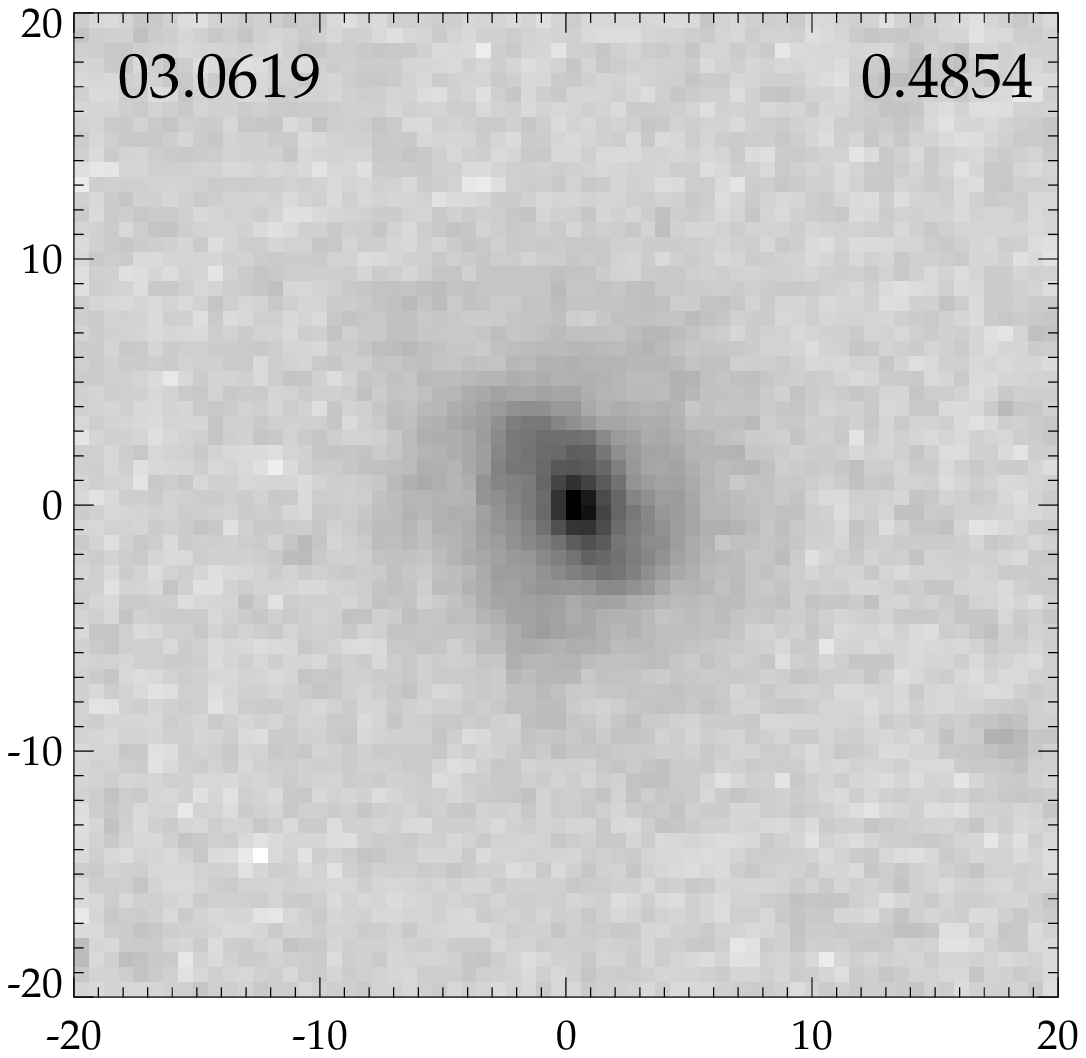} \includegraphics[height=0.22\textwidth,clip]{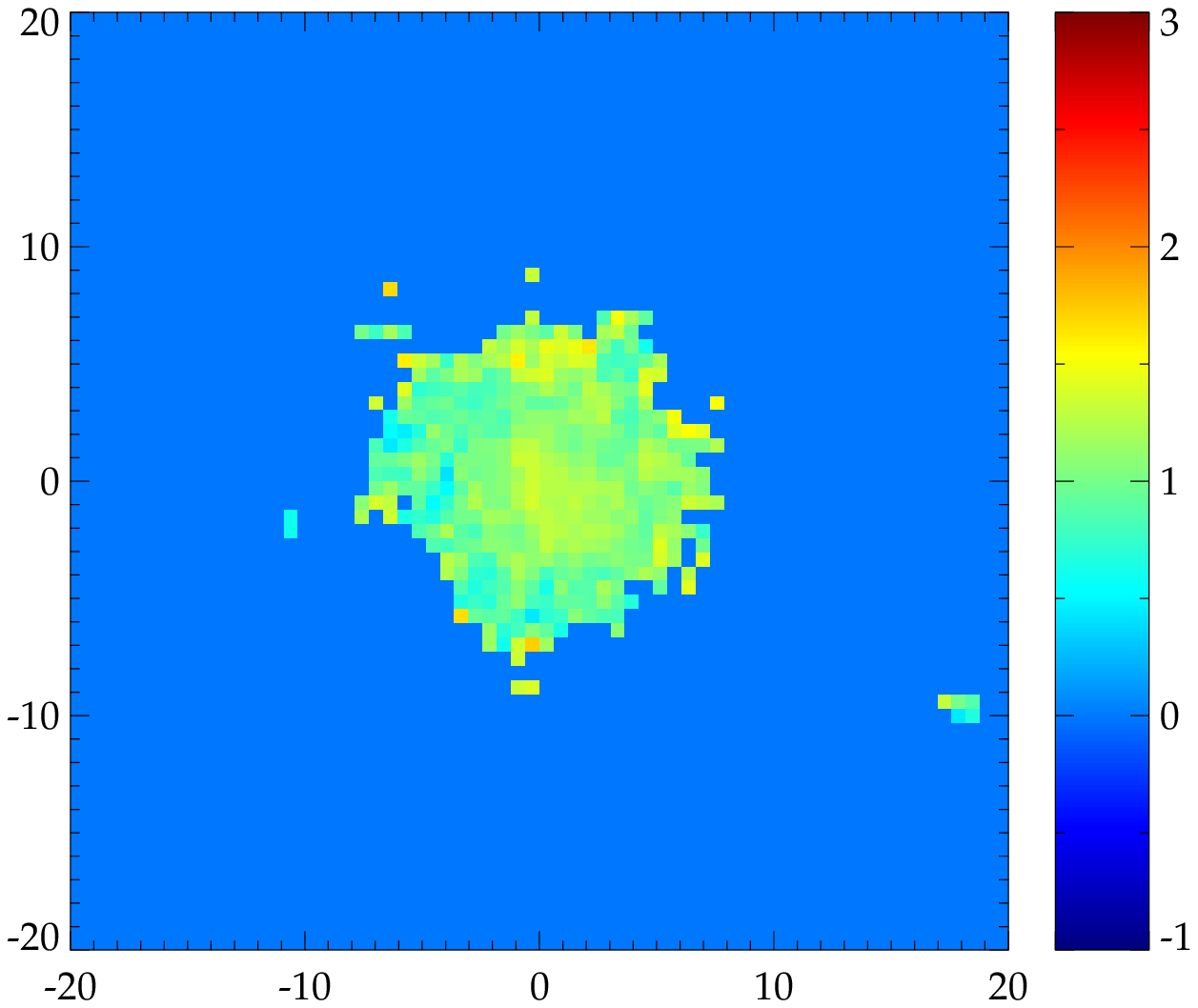}
\includegraphics[height=0.22\textwidth,clip]{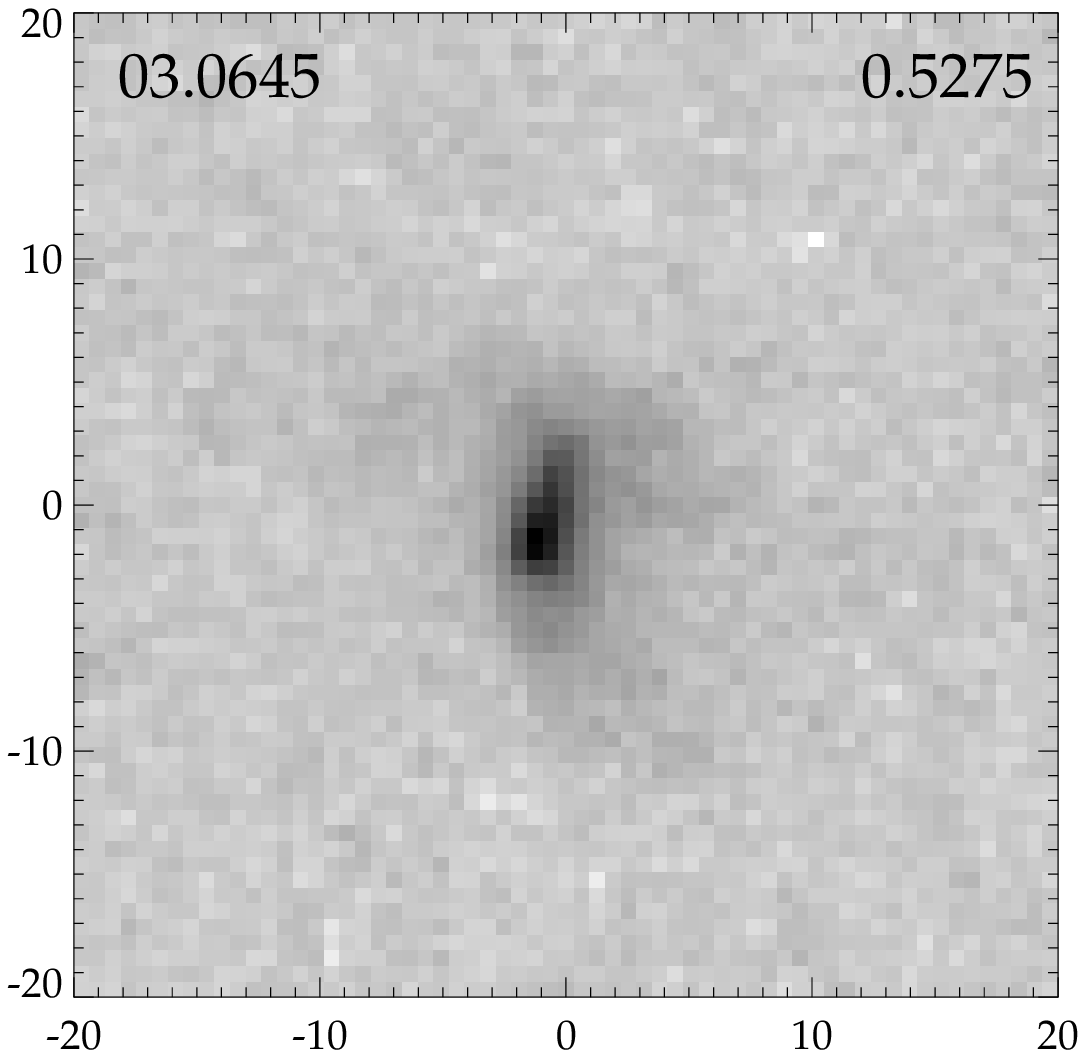} \includegraphics[height=0.22\textwidth,clip]{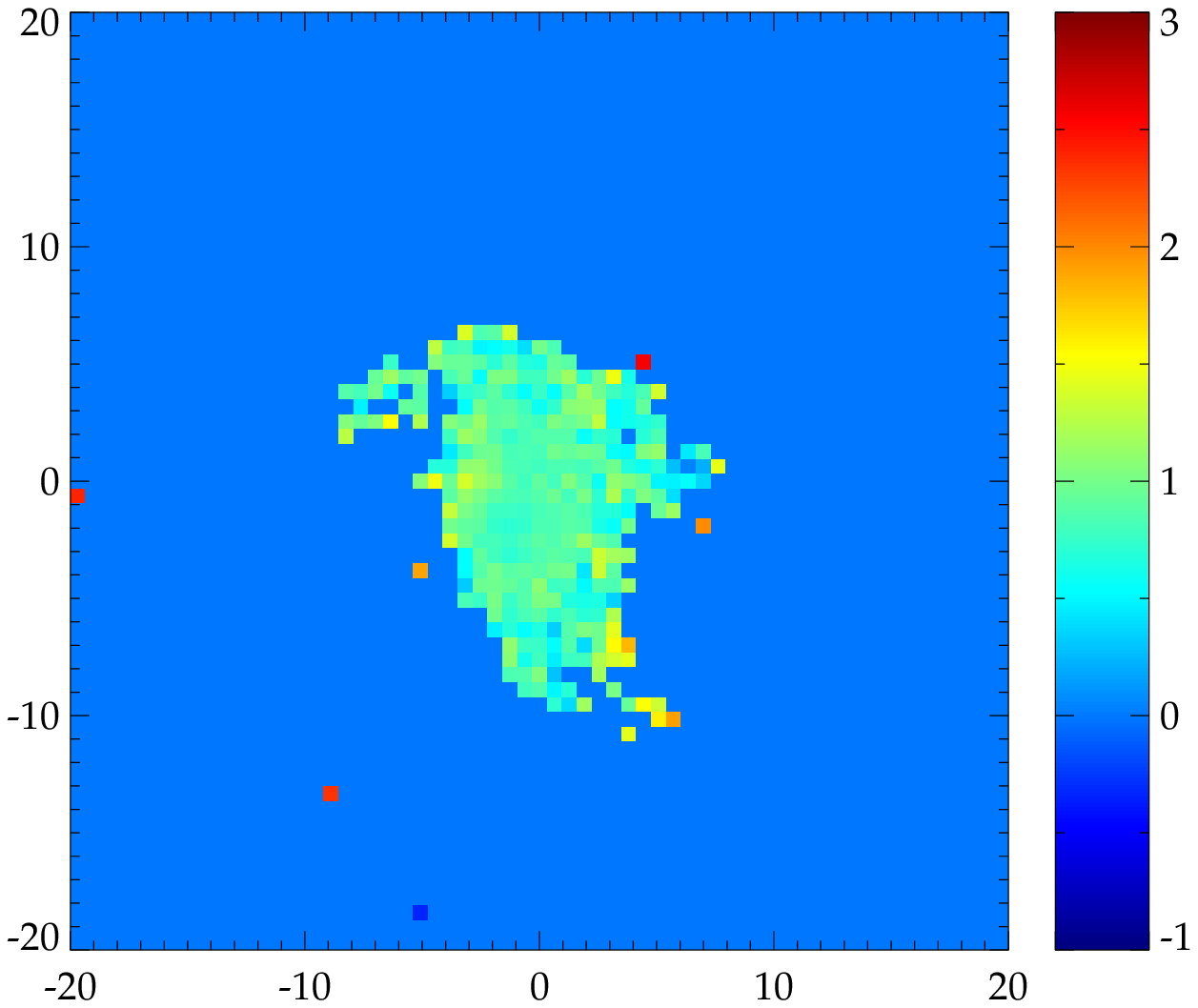}
\includegraphics[height=0.22\textwidth,clip]{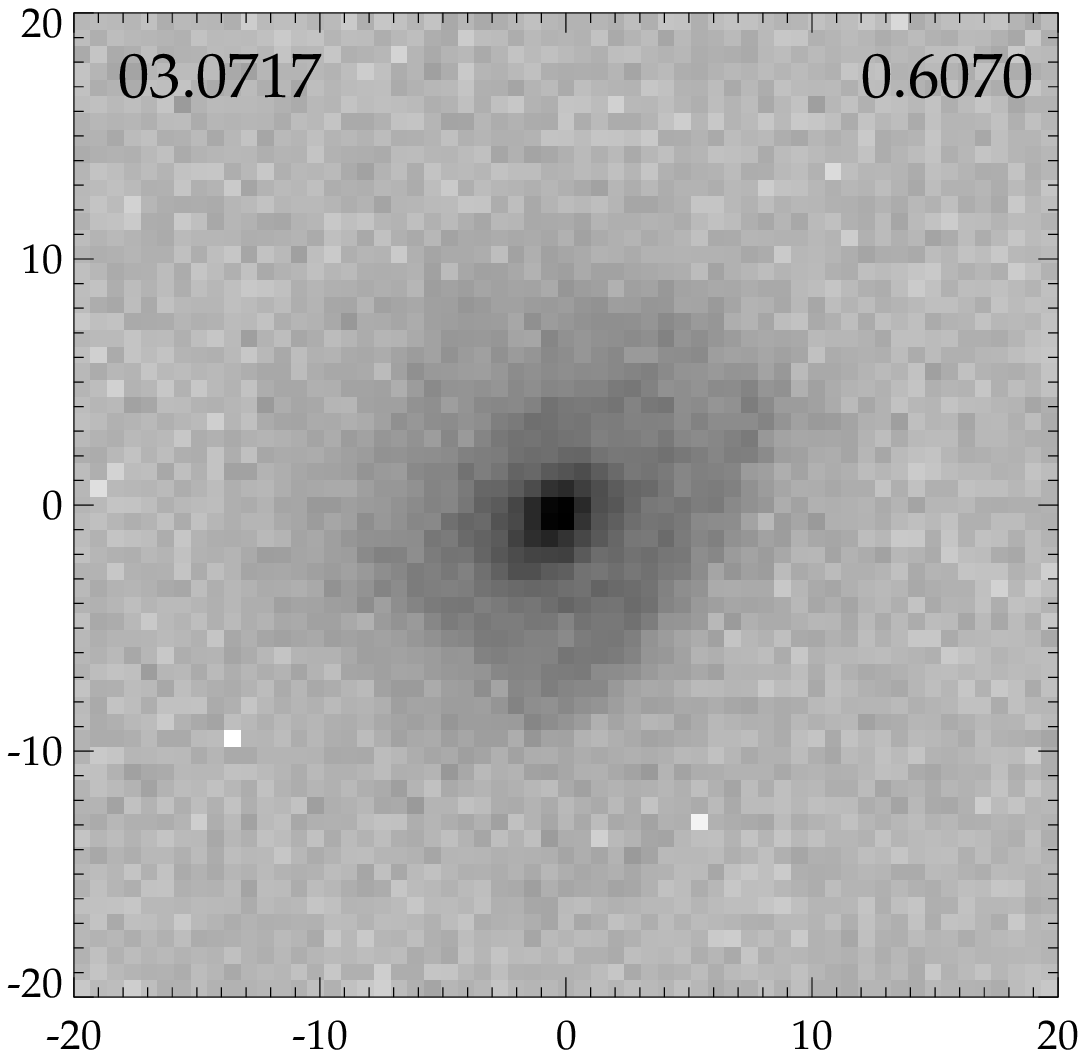} \includegraphics[height=0.22\textwidth,clip]{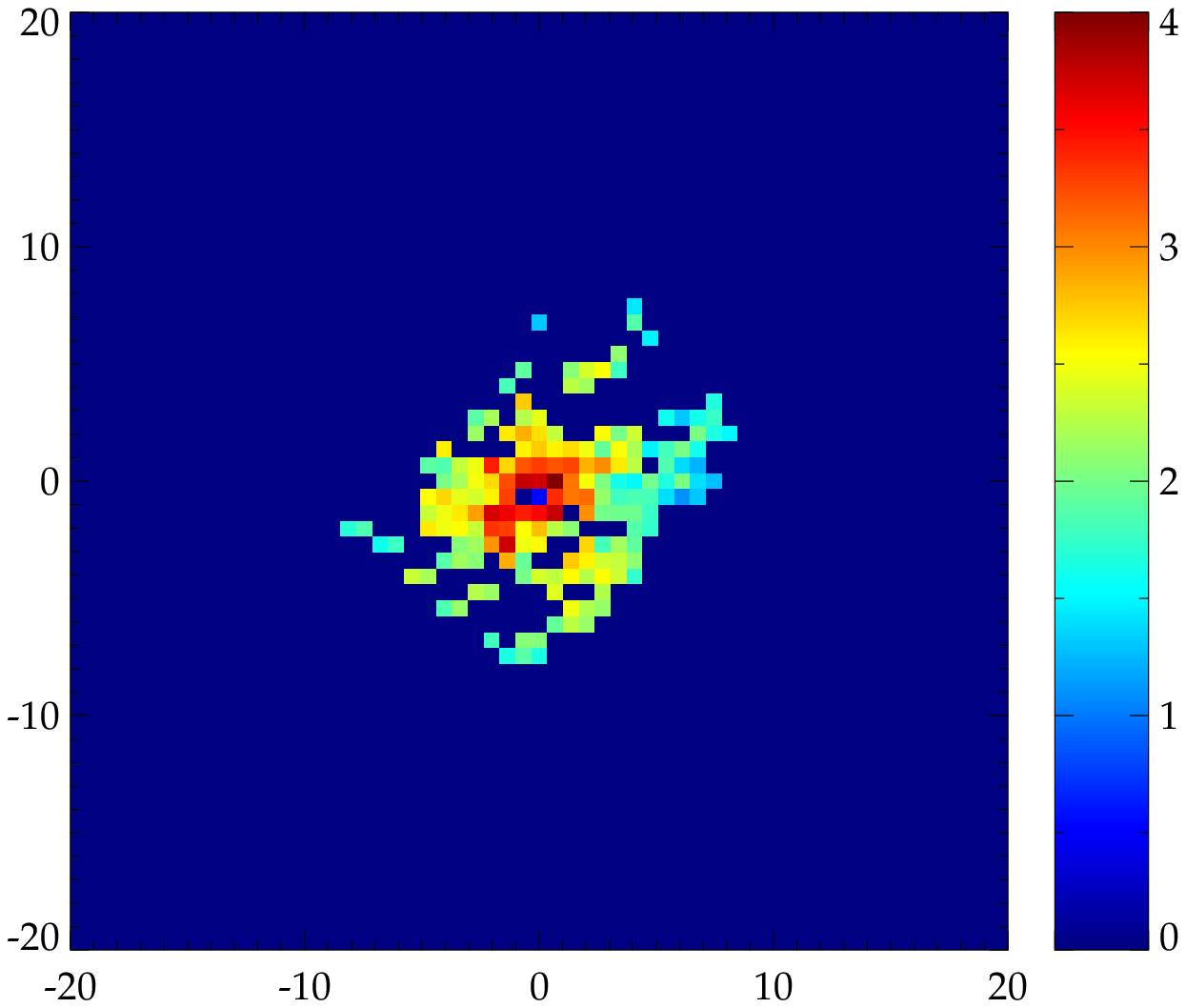}
\includegraphics[height=0.22\textwidth,clip]{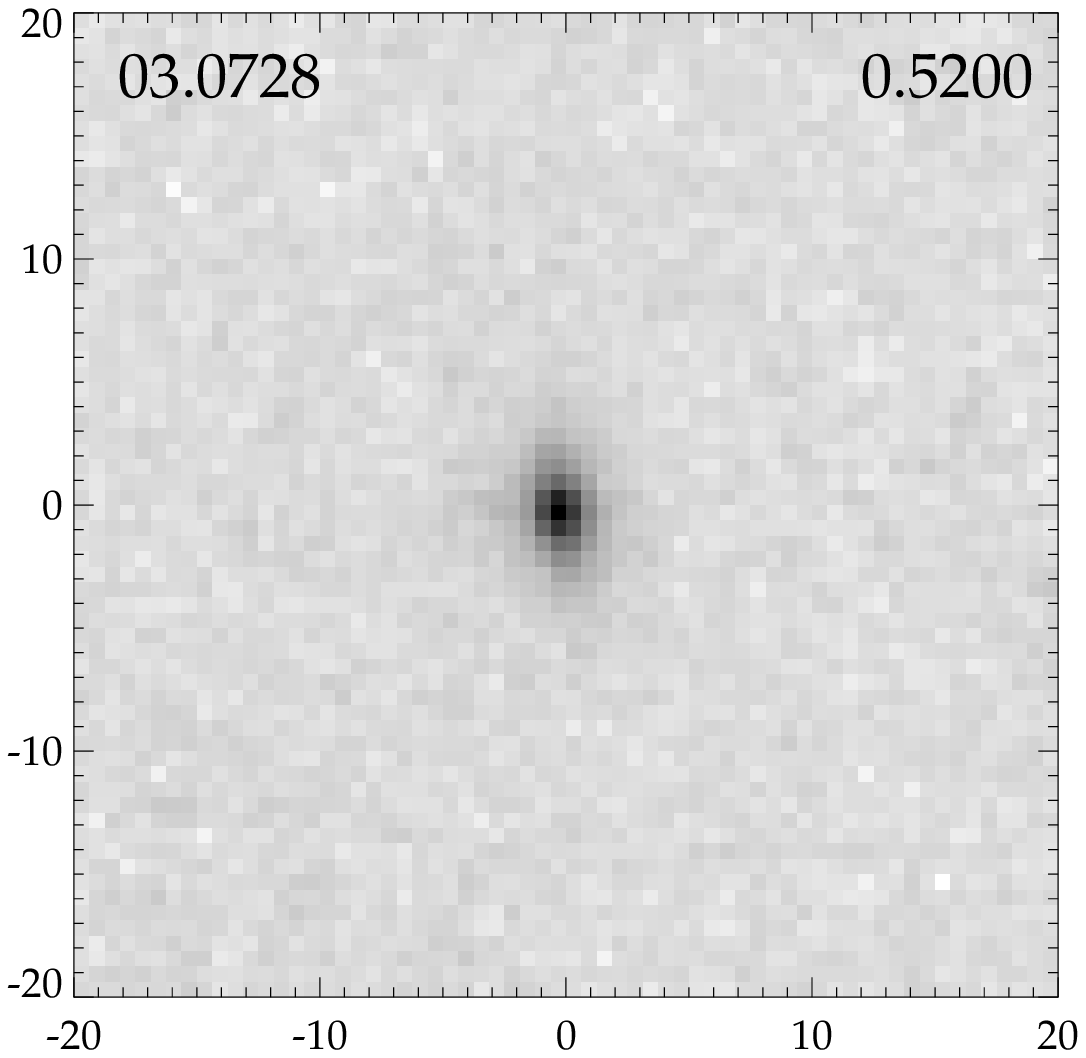} \includegraphics[height=0.22\textwidth,clip]{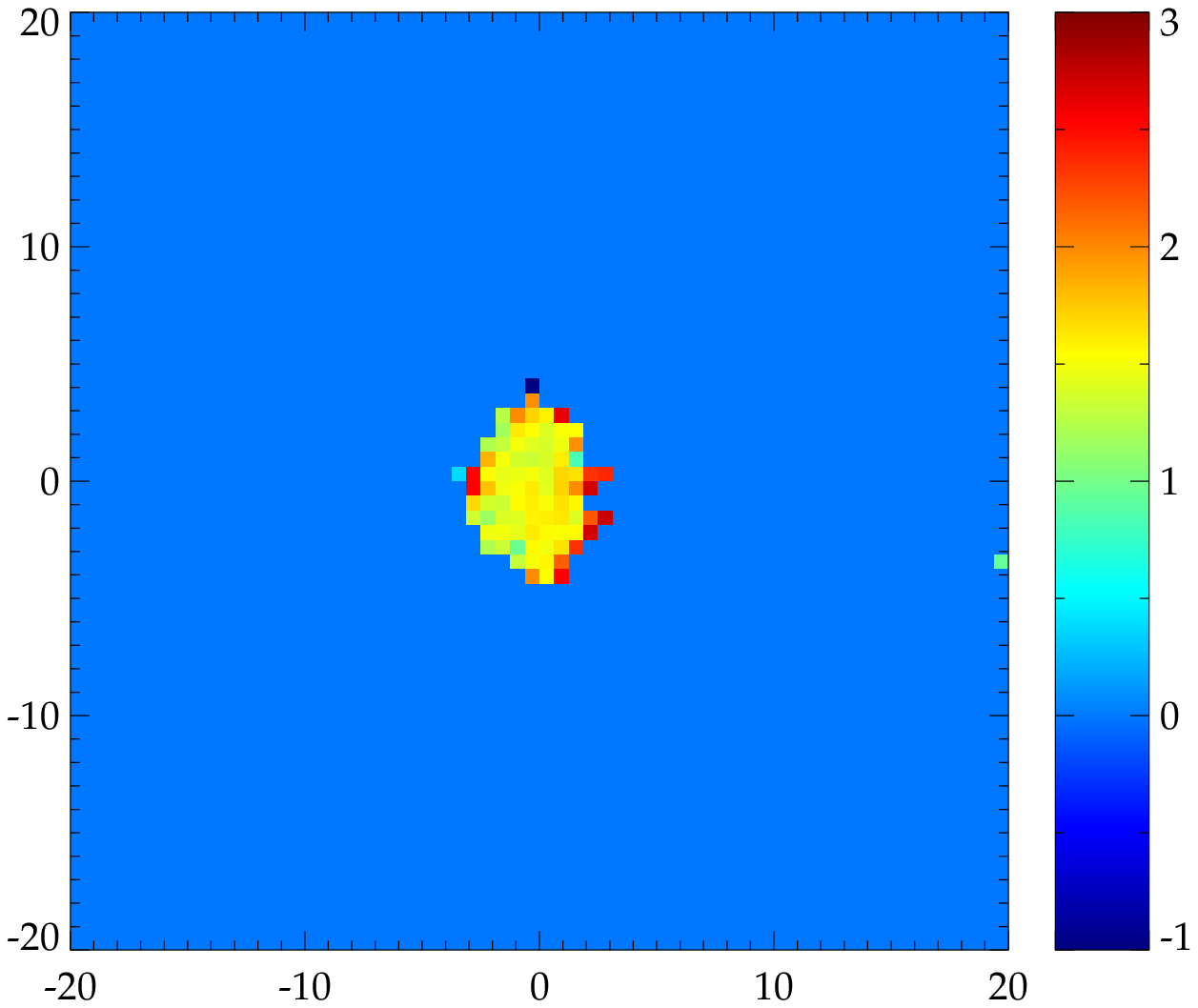}

\caption{$I_{814}$ and color map images. Explanation is given at the end of this figure.} \label{colormap} \end{figure*}

\addtocounter{figure}{-1}
\begin{figure*} \centering

\includegraphics[height=0.22\textwidth,clip]{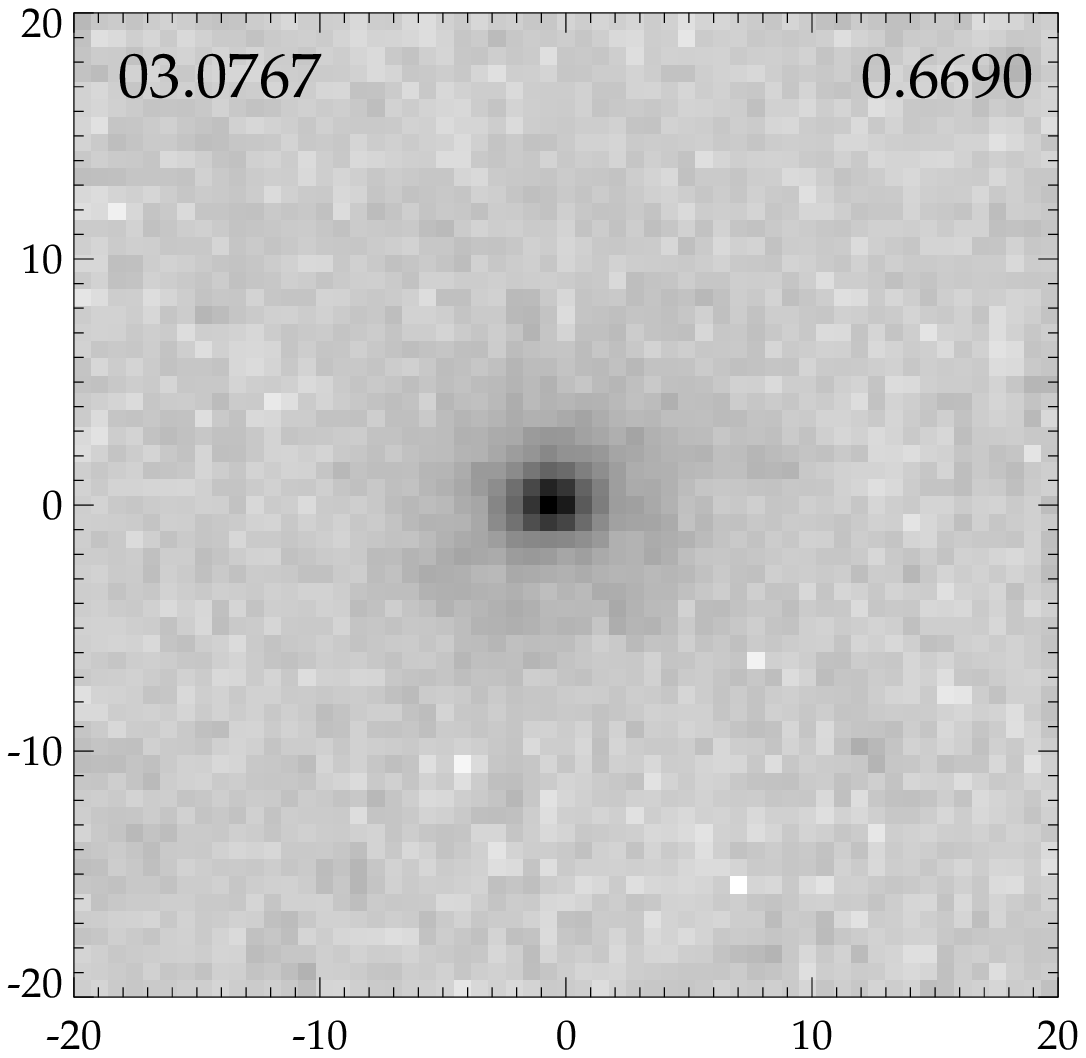} \includegraphics[height=0.22\textwidth,clip]{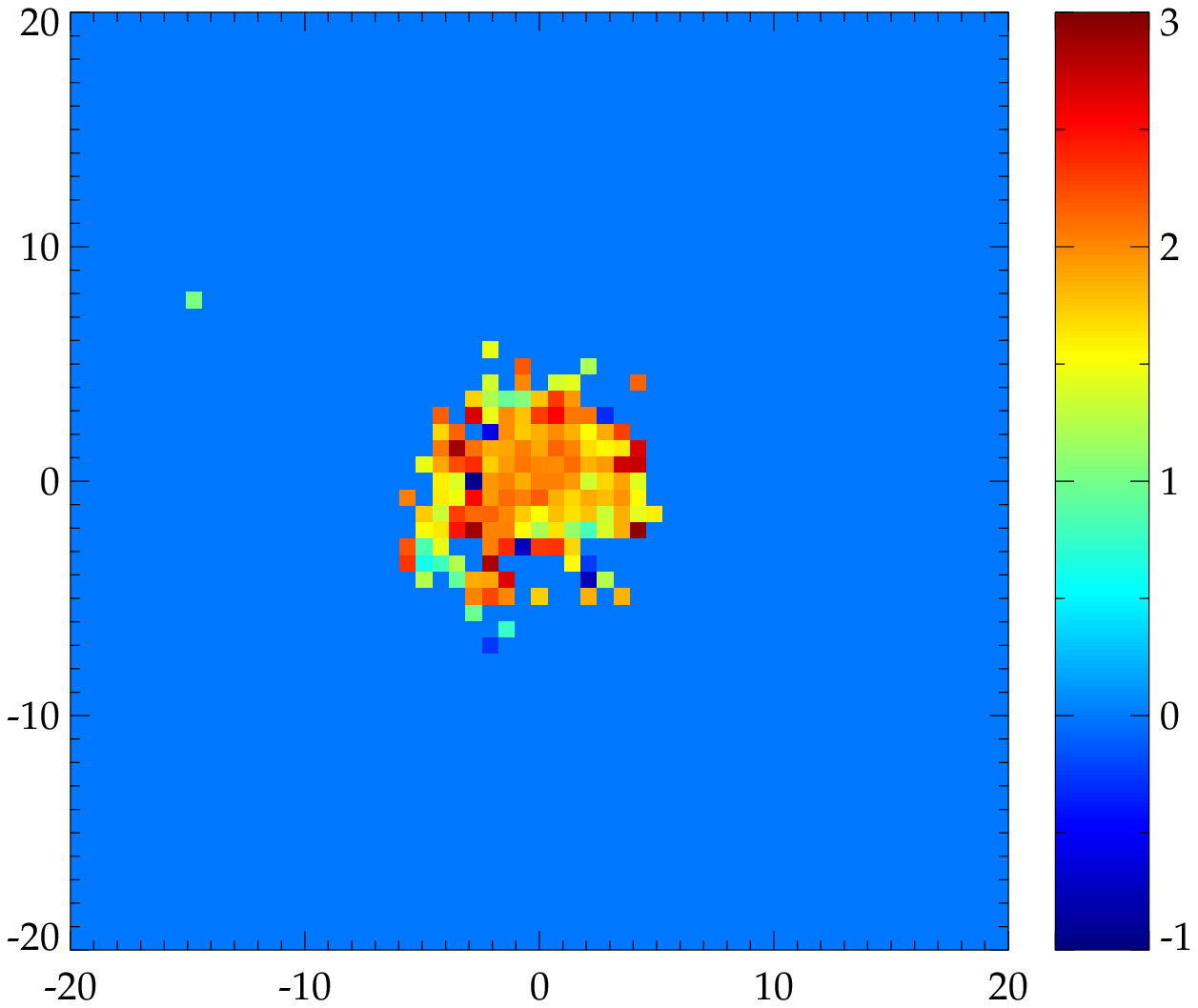}
\includegraphics[height=0.22\textwidth,clip]{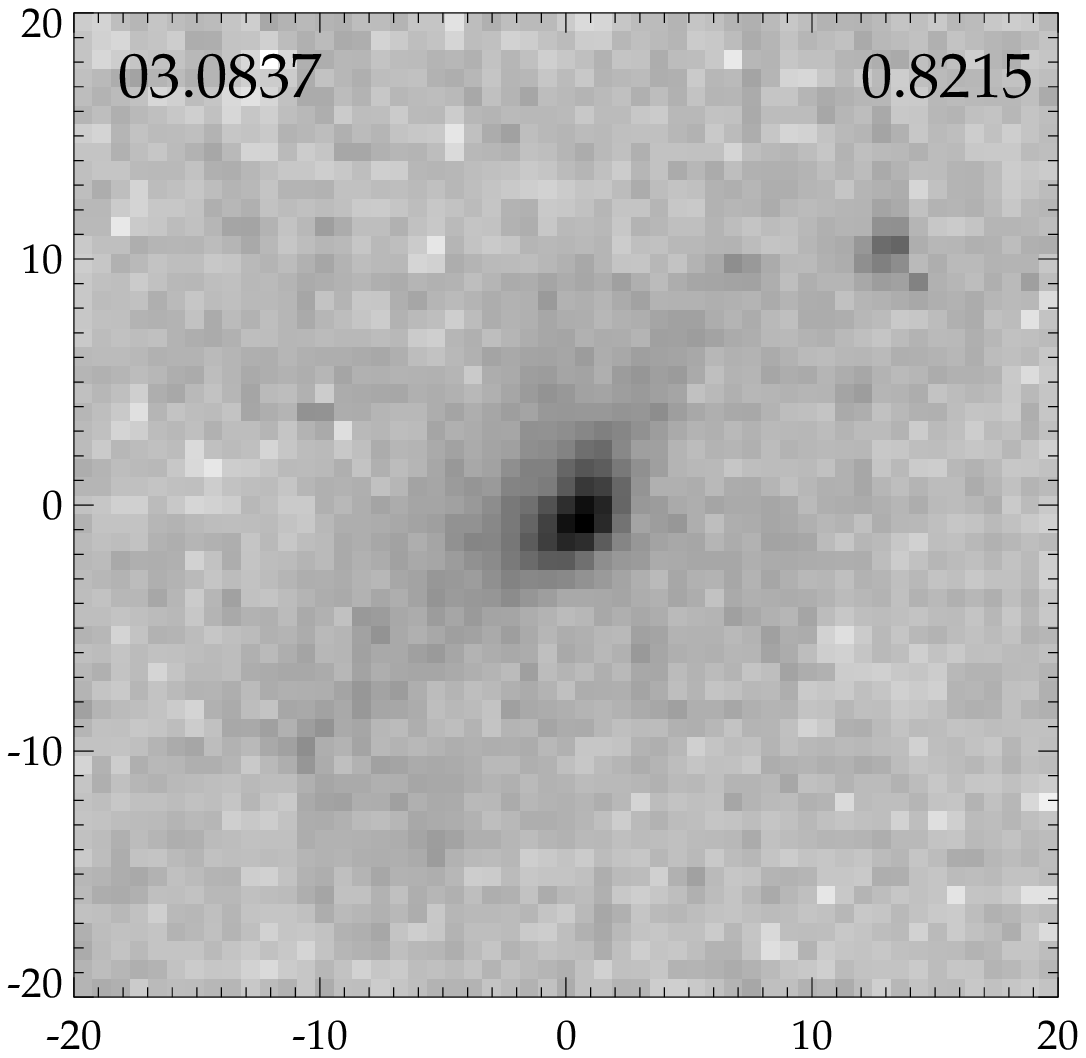} \includegraphics[height=0.22\textwidth,clip]{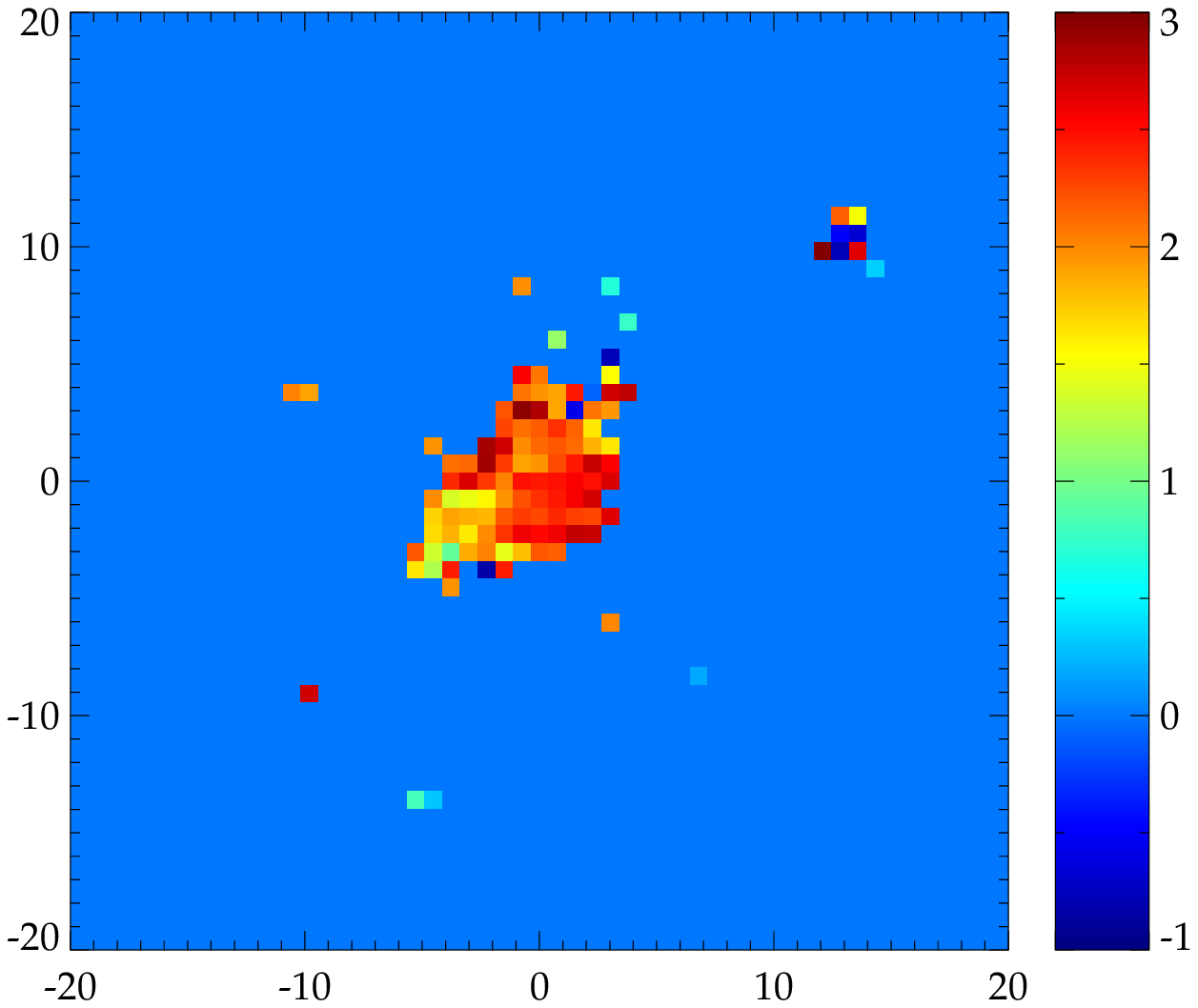}
\includegraphics[height=0.22\textwidth,clip]{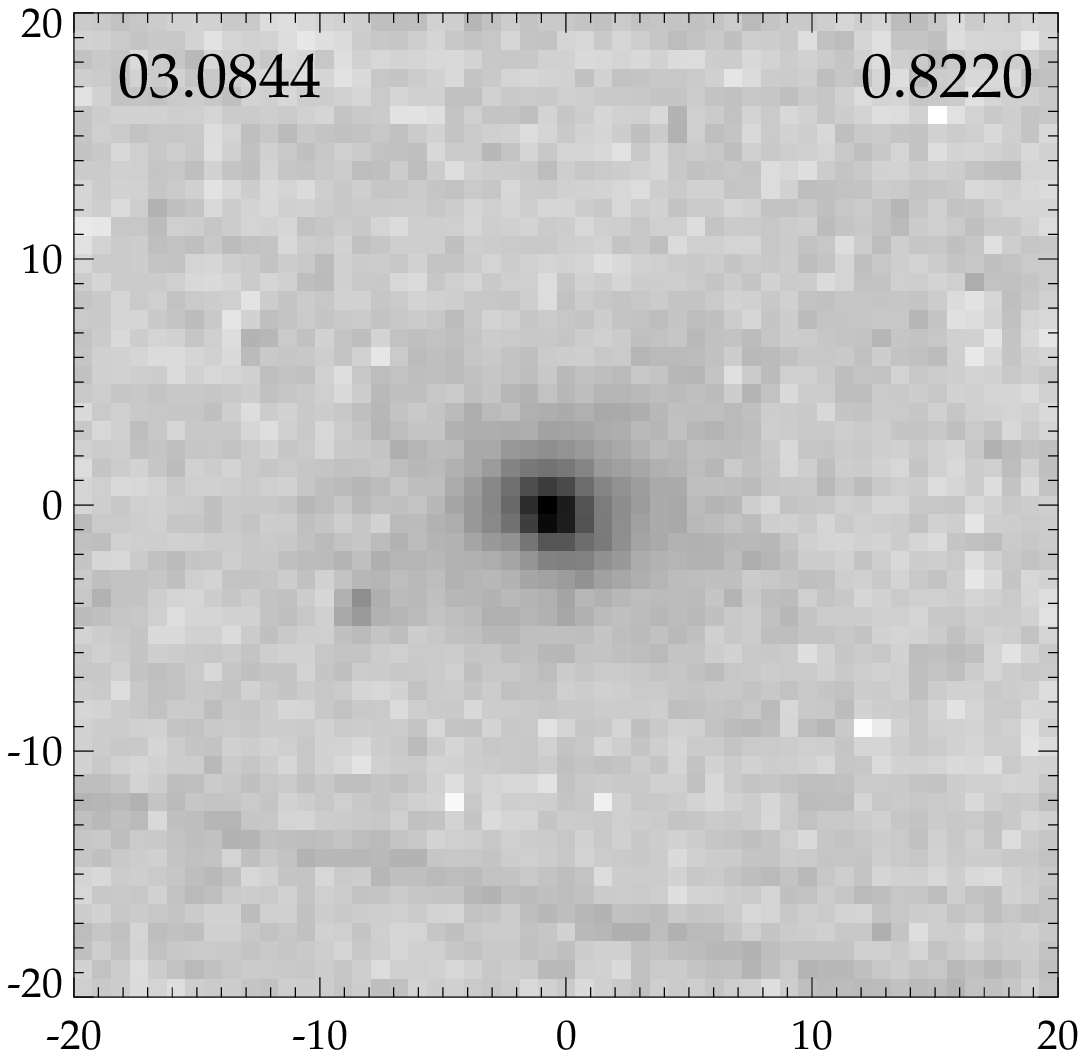} \includegraphics[height=0.22\textwidth,clip]{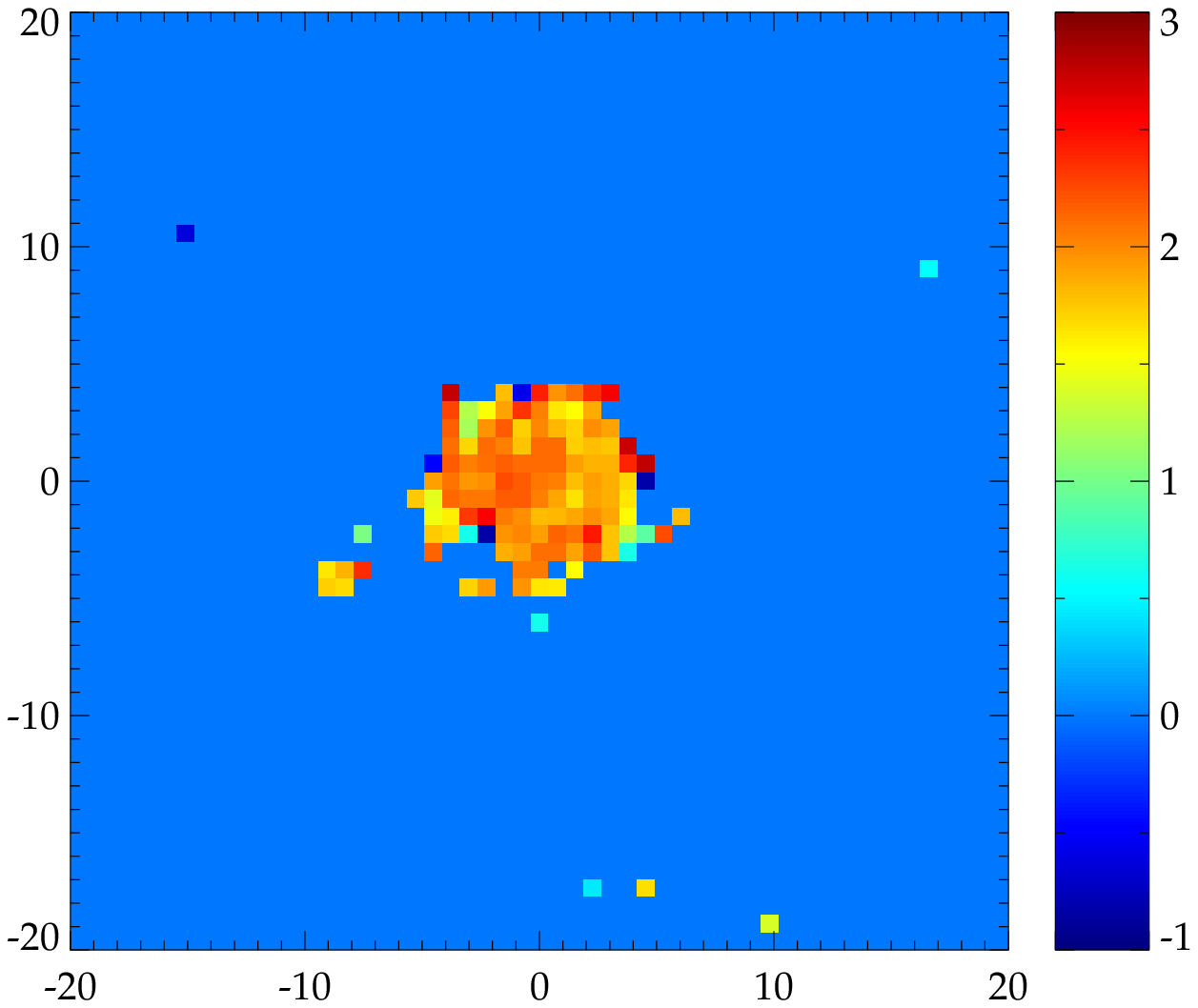}
\includegraphics[height=0.22\textwidth,clip]{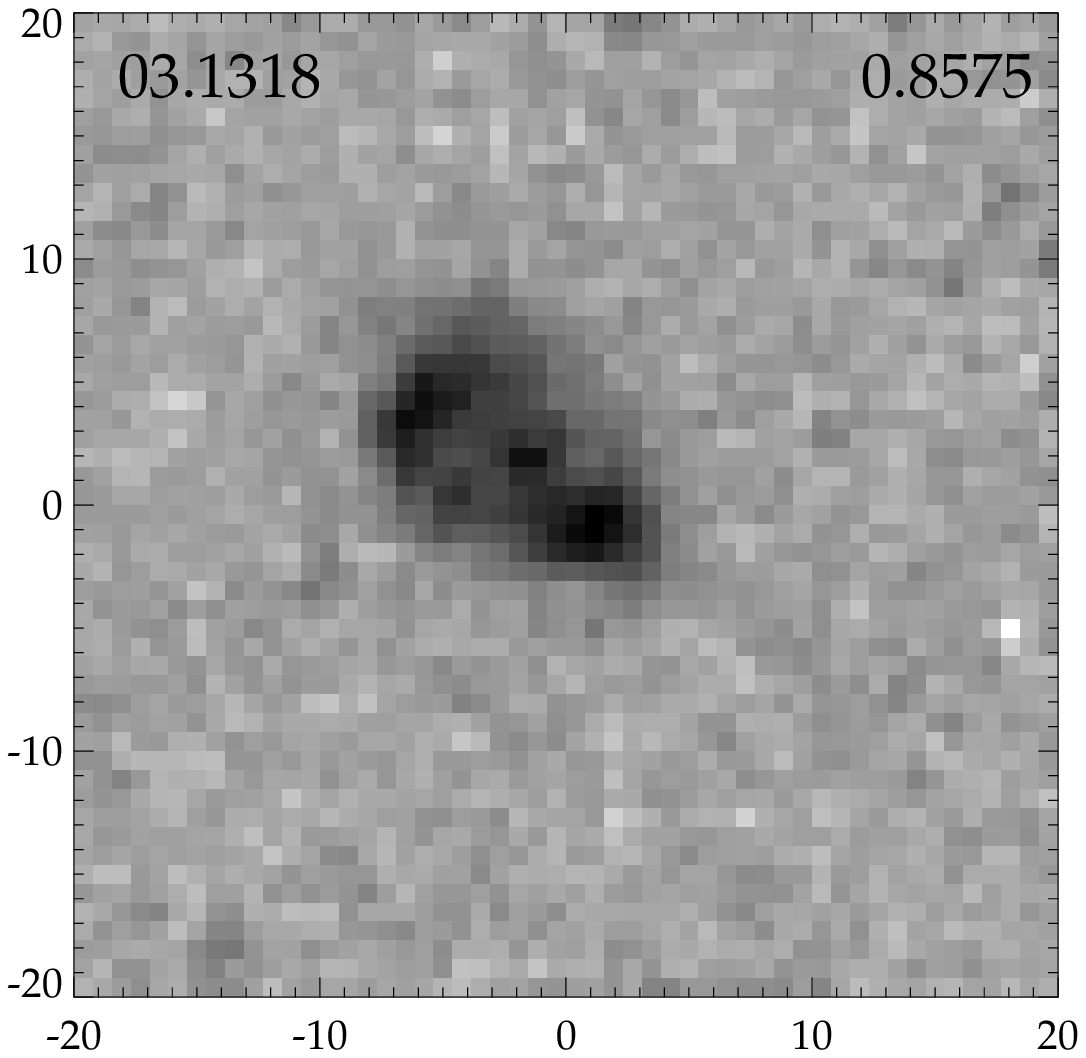} \includegraphics[height=0.22\textwidth,clip]{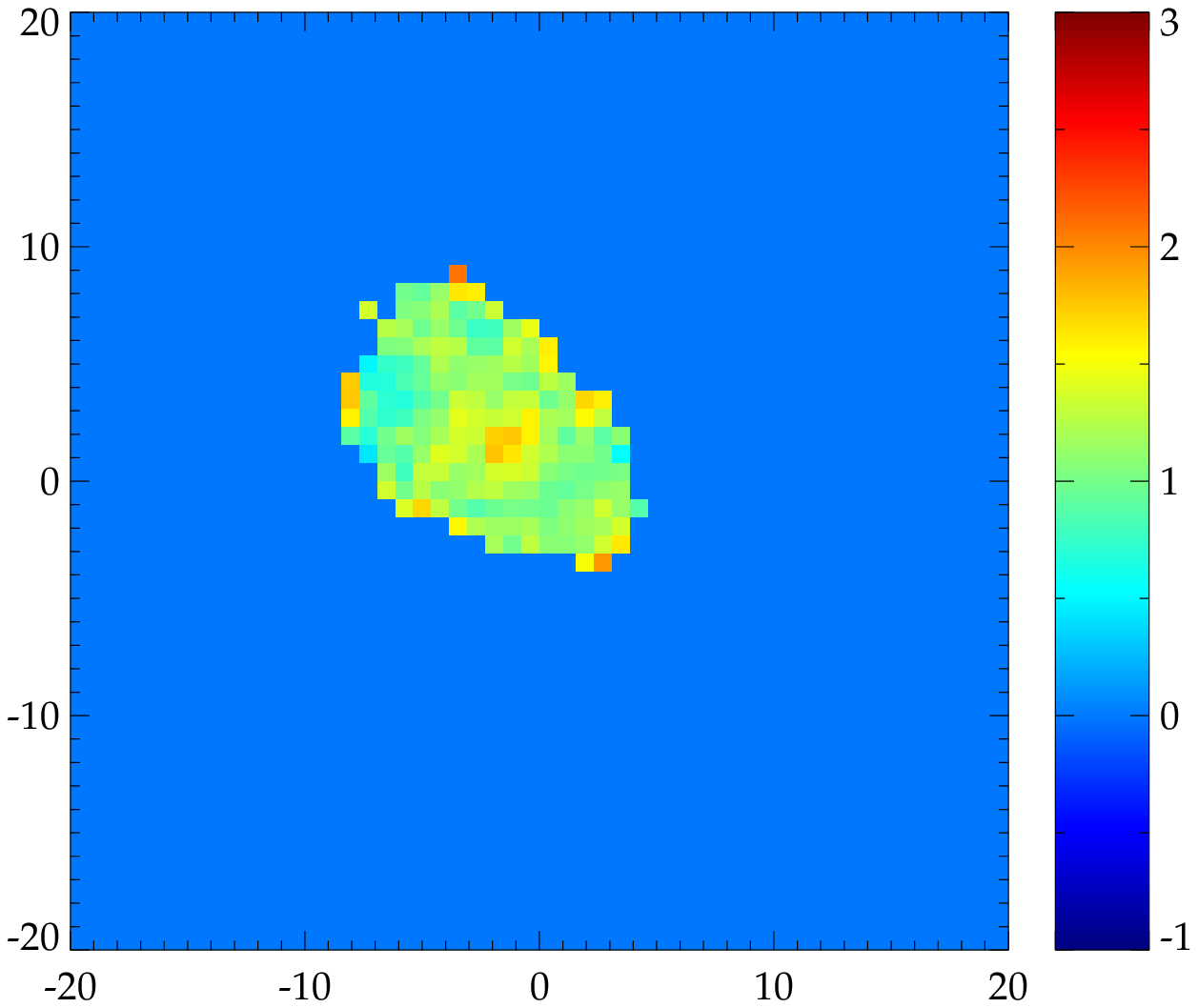}
\includegraphics[height=0.22\textwidth,clip]{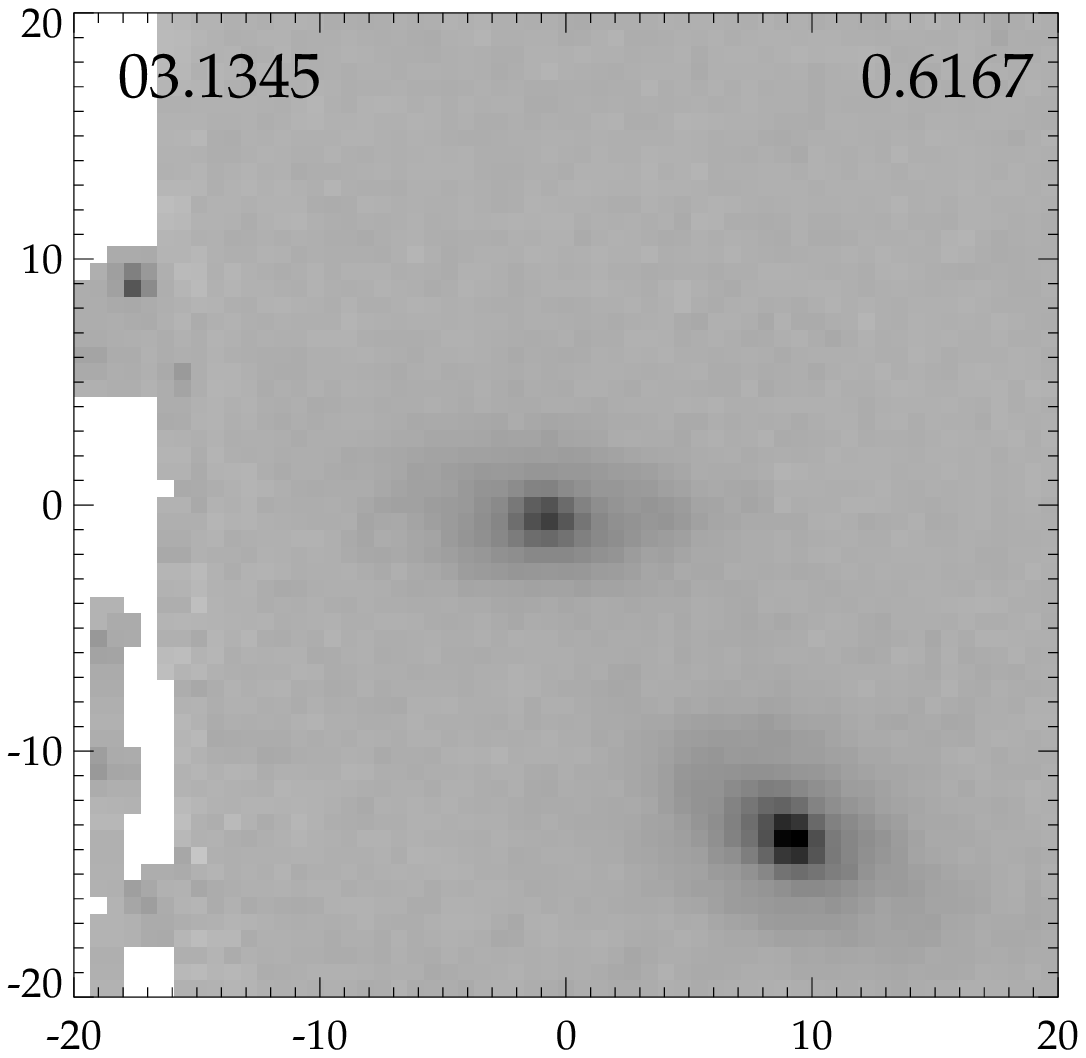} \includegraphics[height=0.22\textwidth,clip]{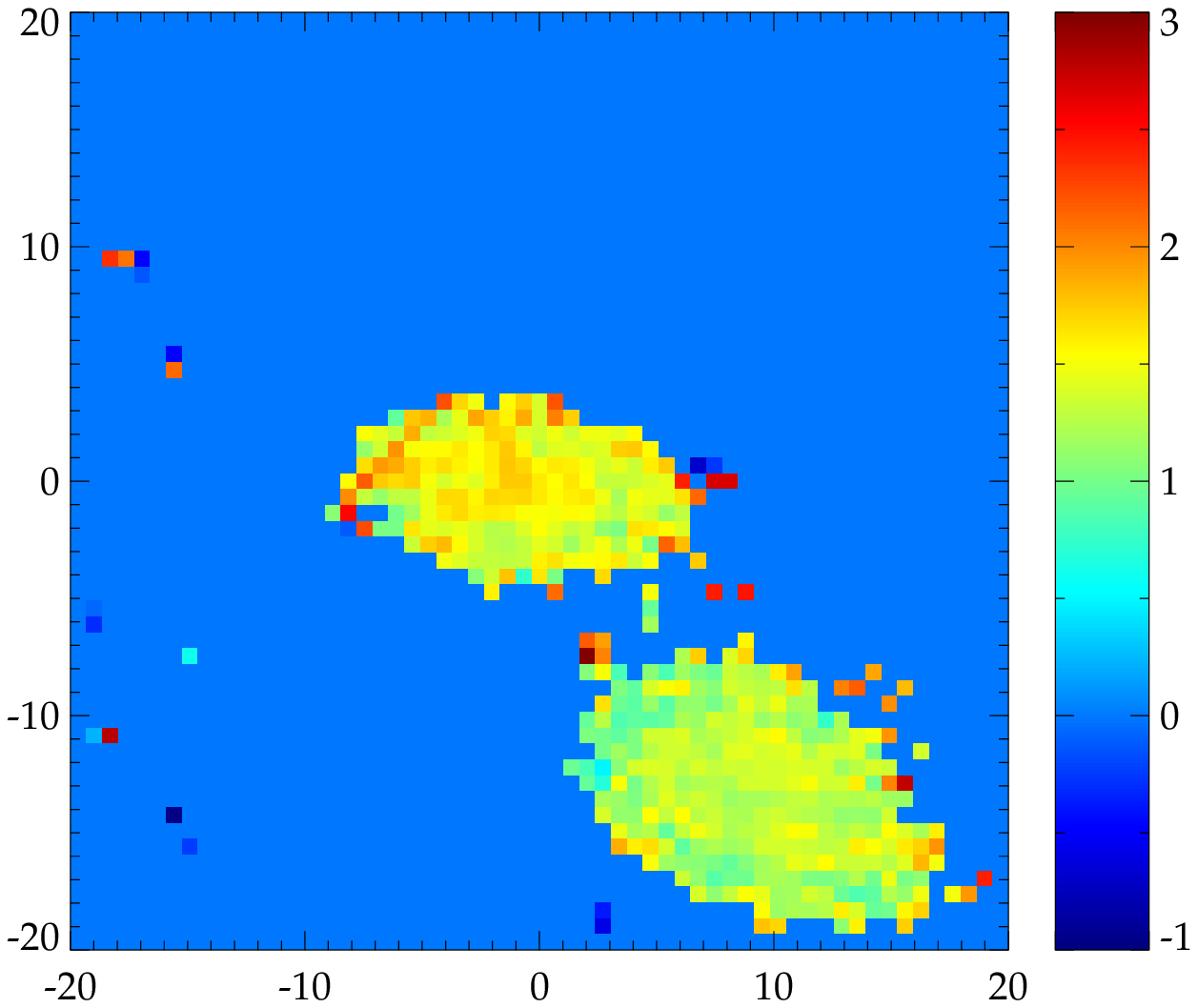}
\includegraphics[height=0.22\textwidth,clip]{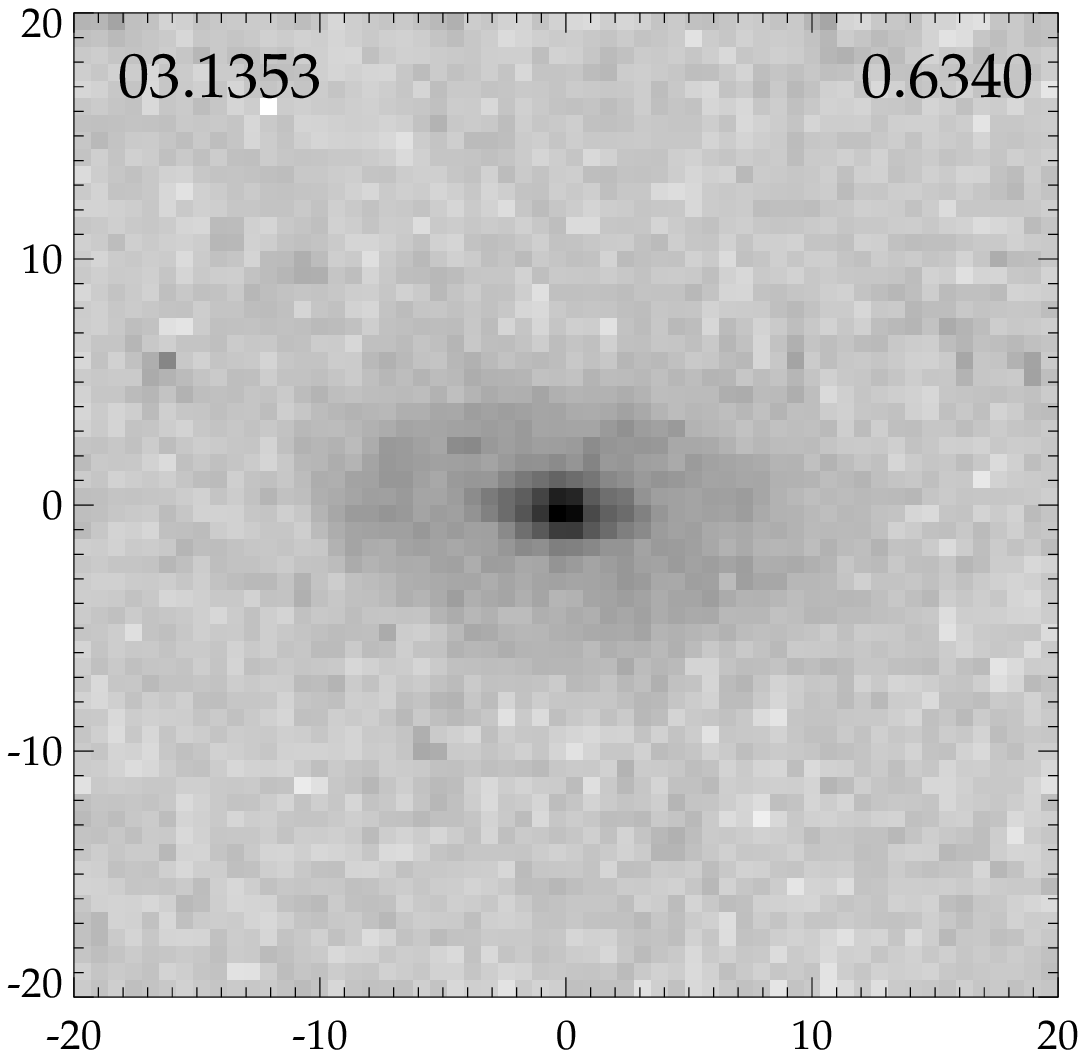} \includegraphics[height=0.22\textwidth,clip]{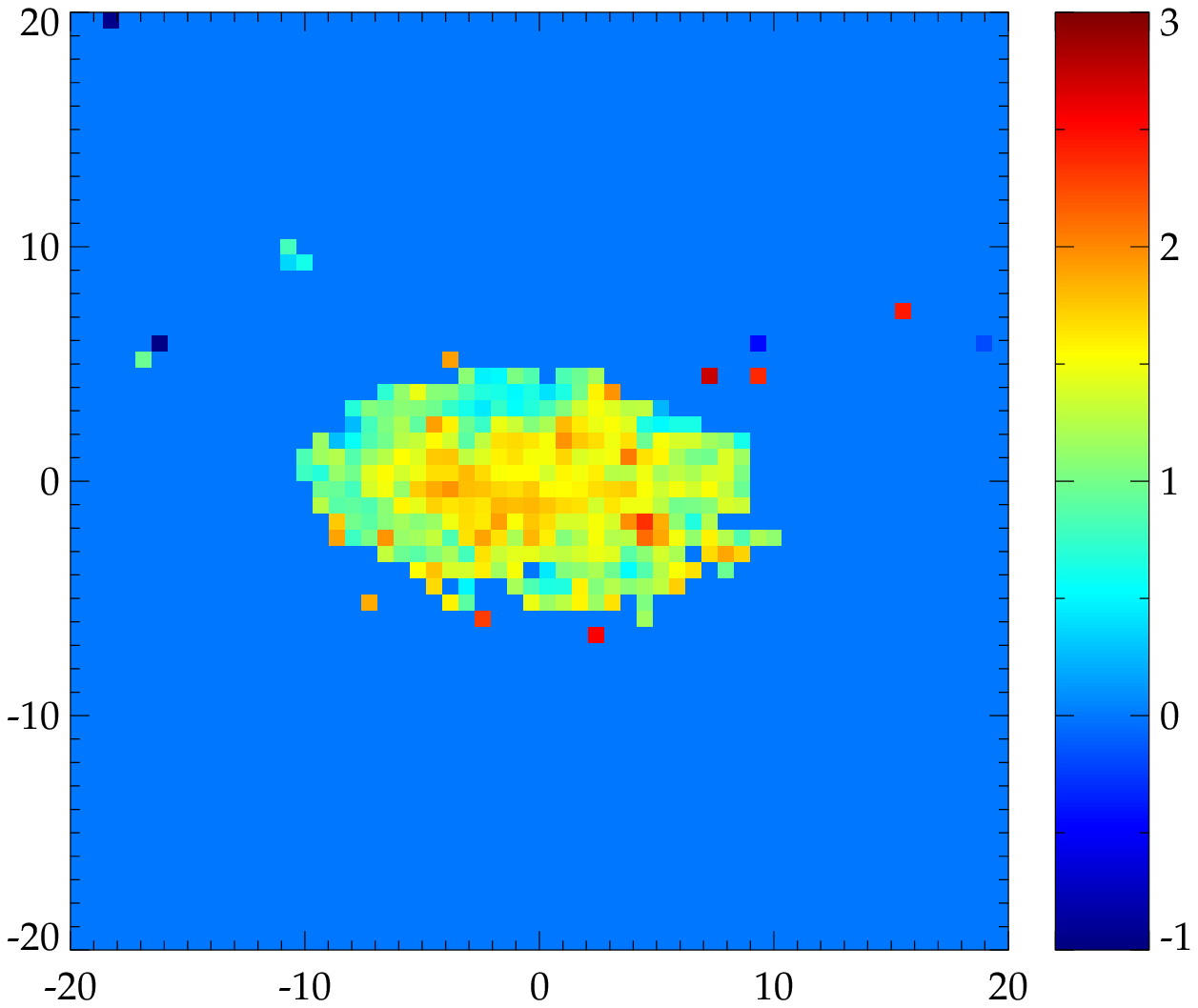}
\includegraphics[height=0.22\textwidth,clip]{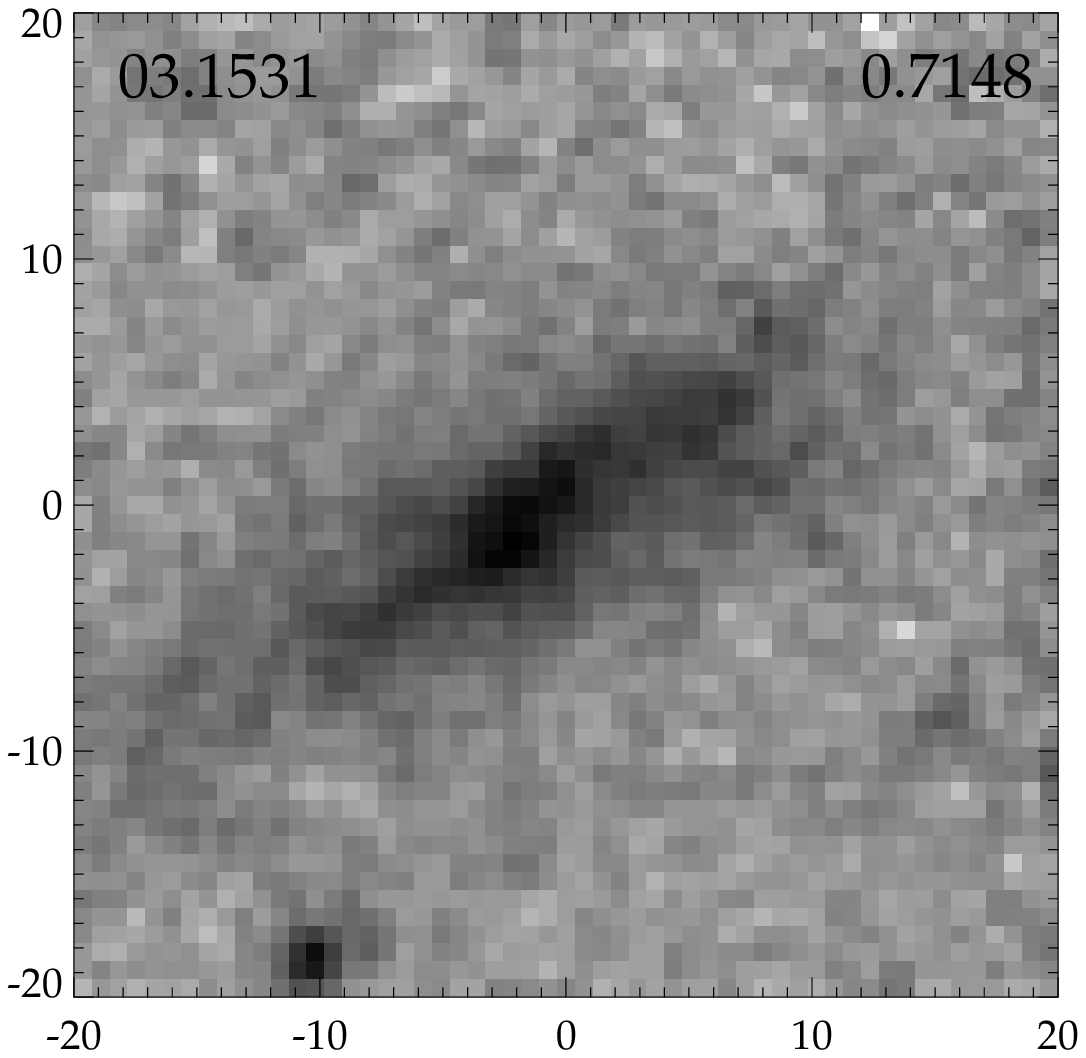} \includegraphics[height=0.22\textwidth,clip]{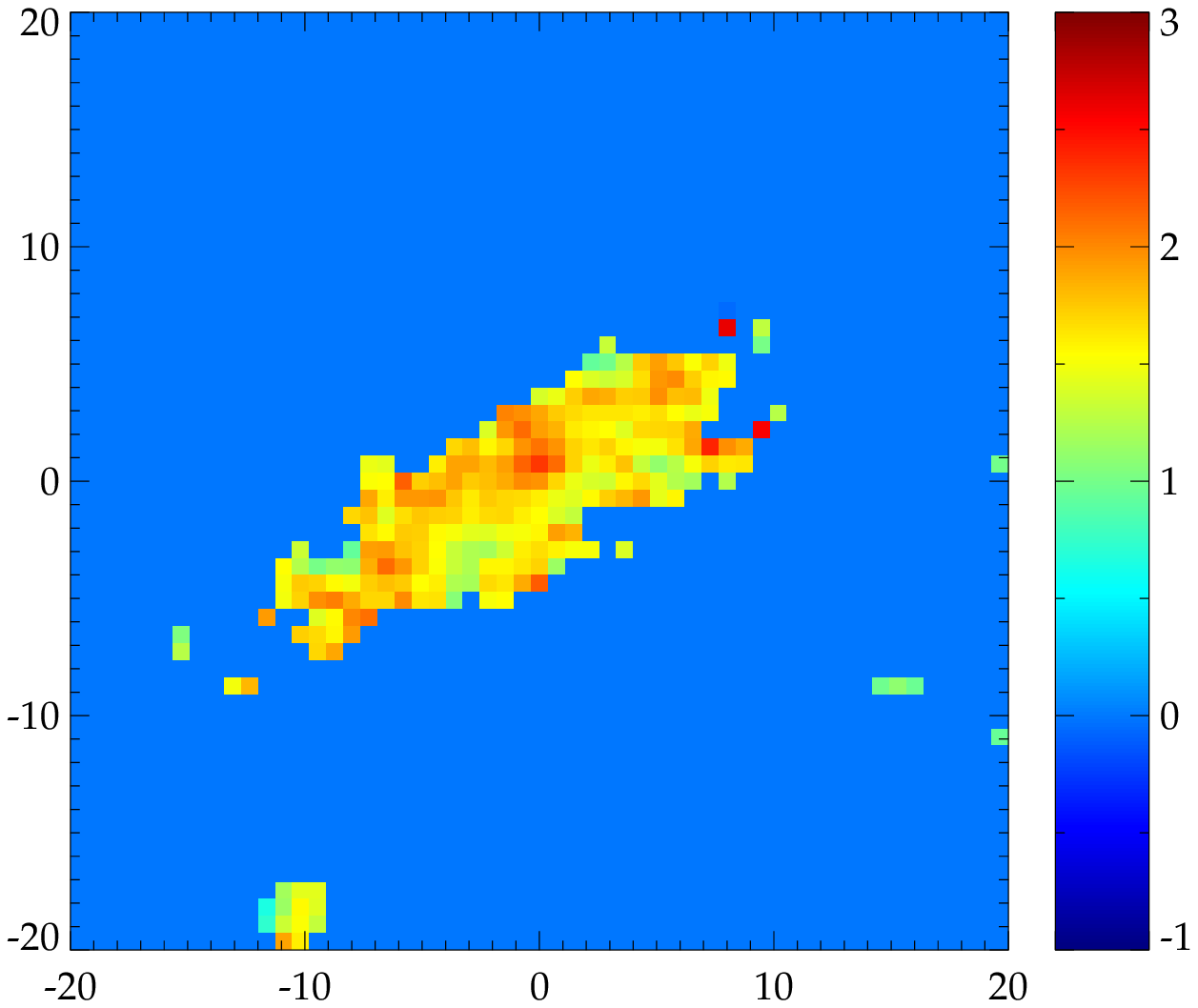}
\includegraphics[height=0.22\textwidth,clip]{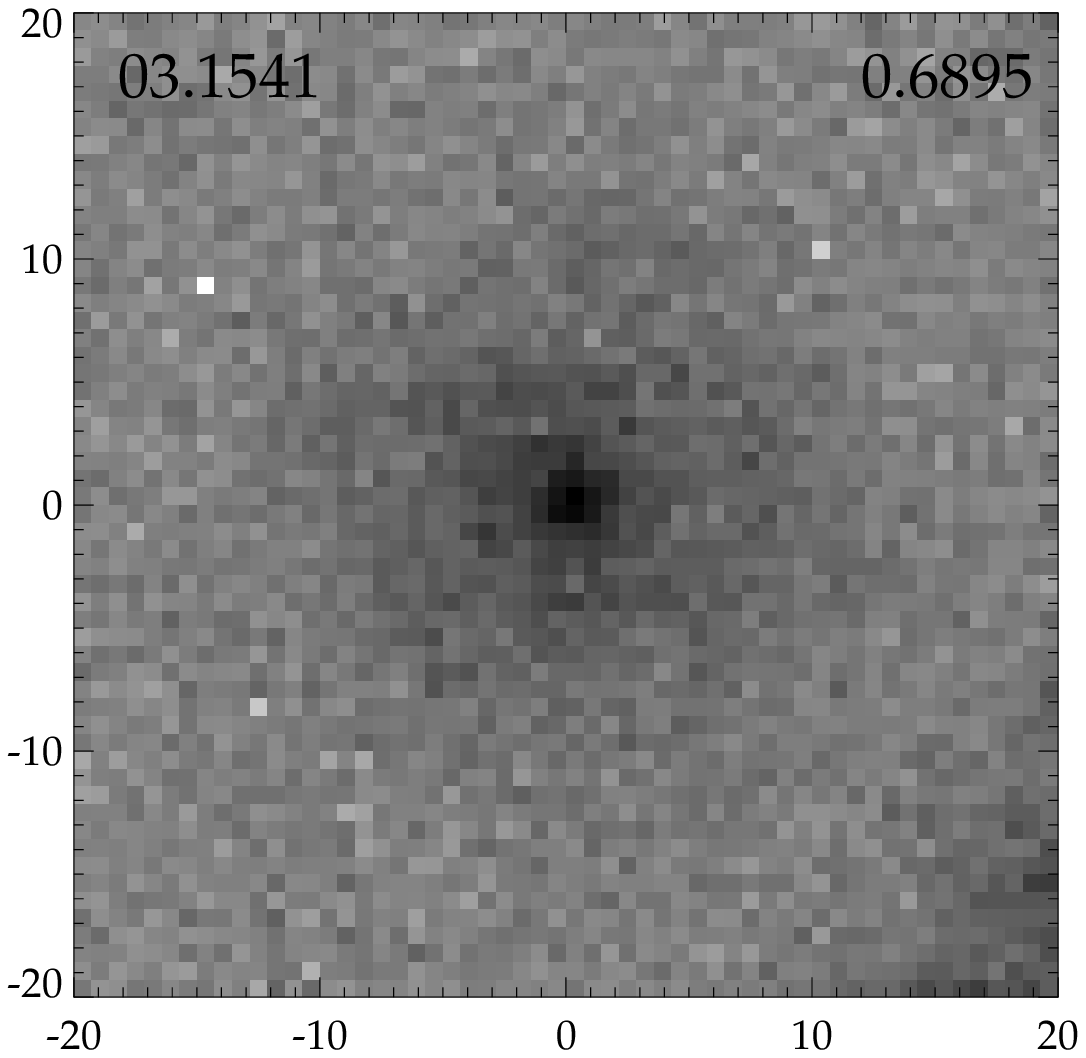} \includegraphics[height=0.22\textwidth,clip]{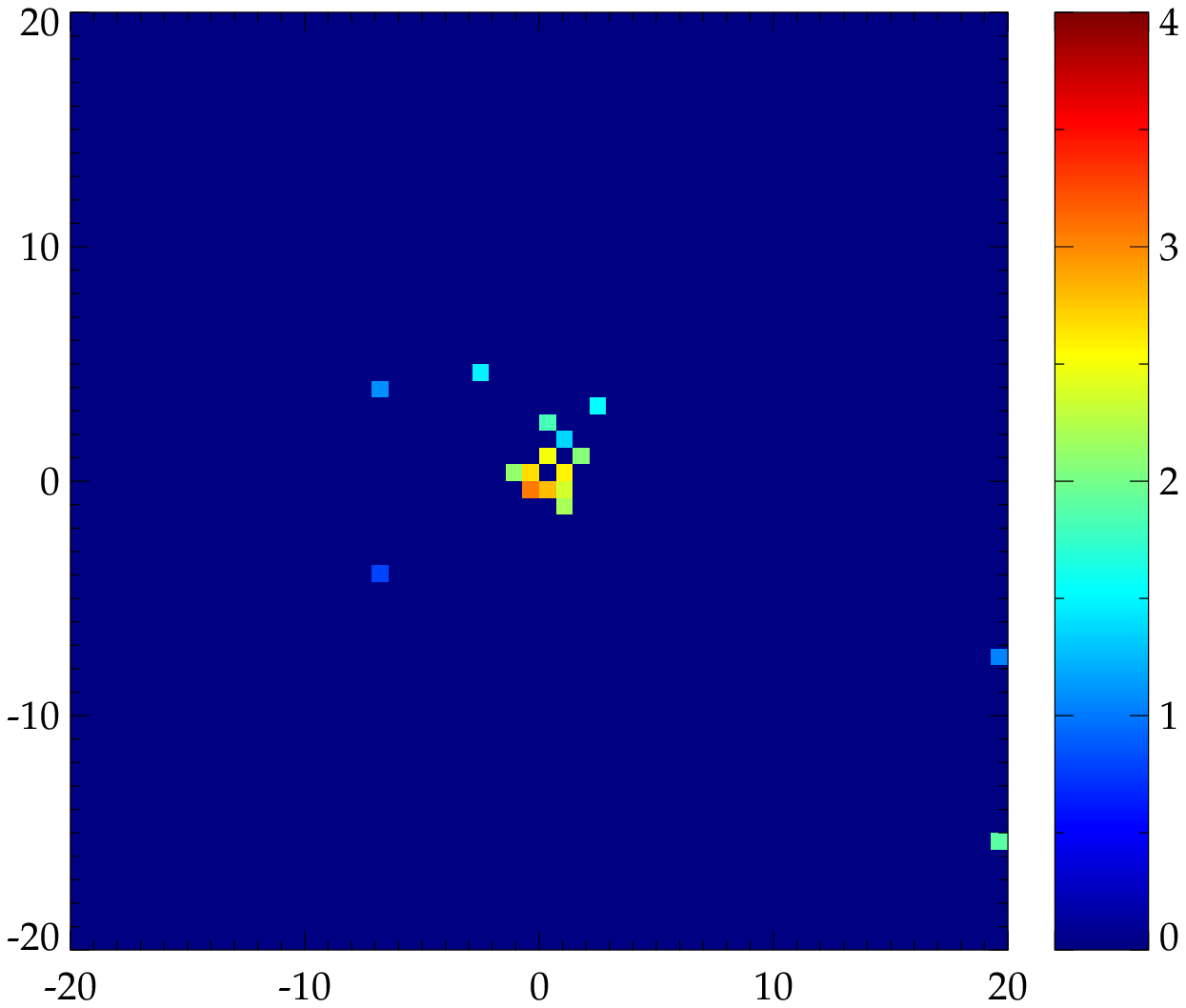}
\includegraphics[height=0.22\textwidth,clip]{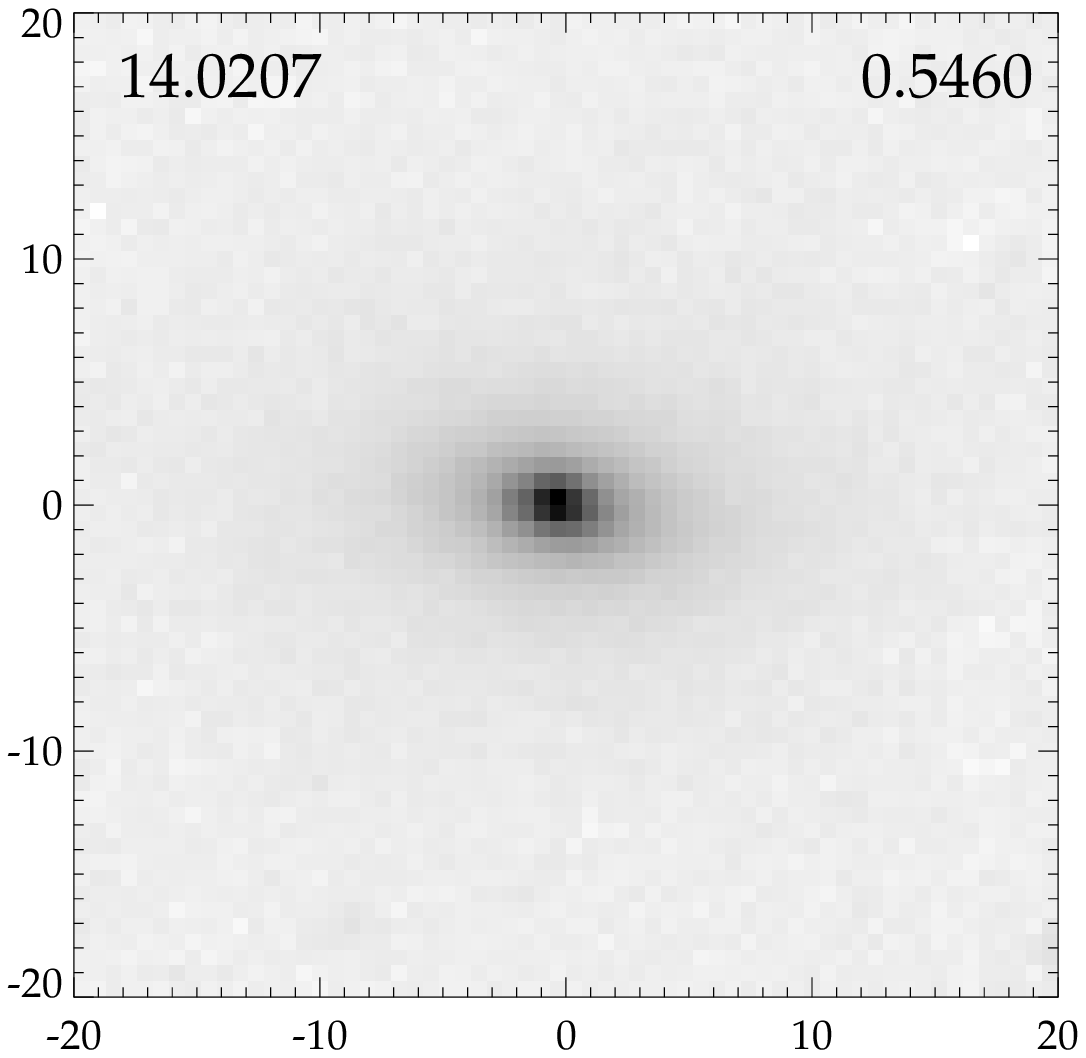} \includegraphics[height=0.22\textwidth,clip]{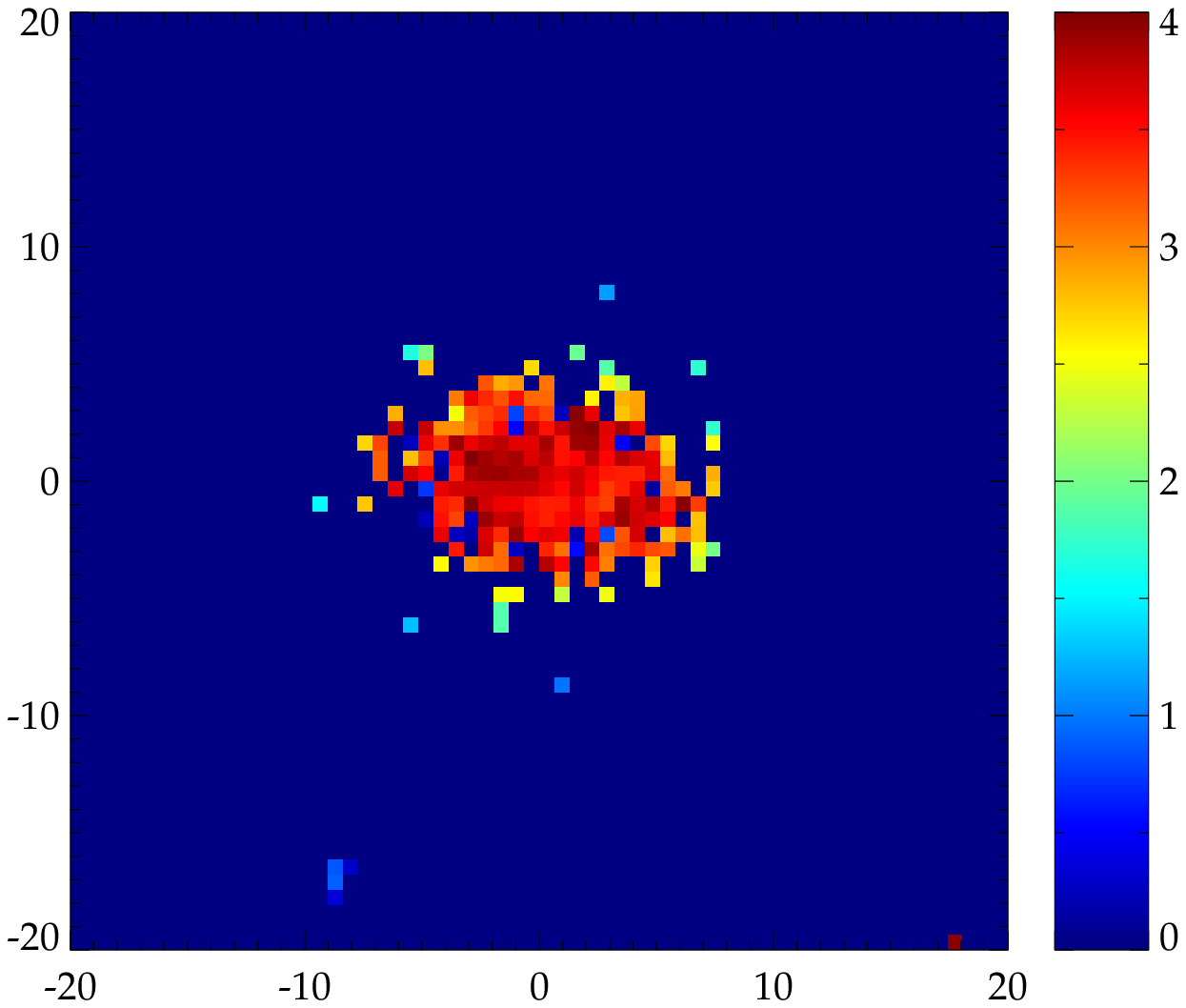}
\includegraphics[height=0.22\textwidth,clip]{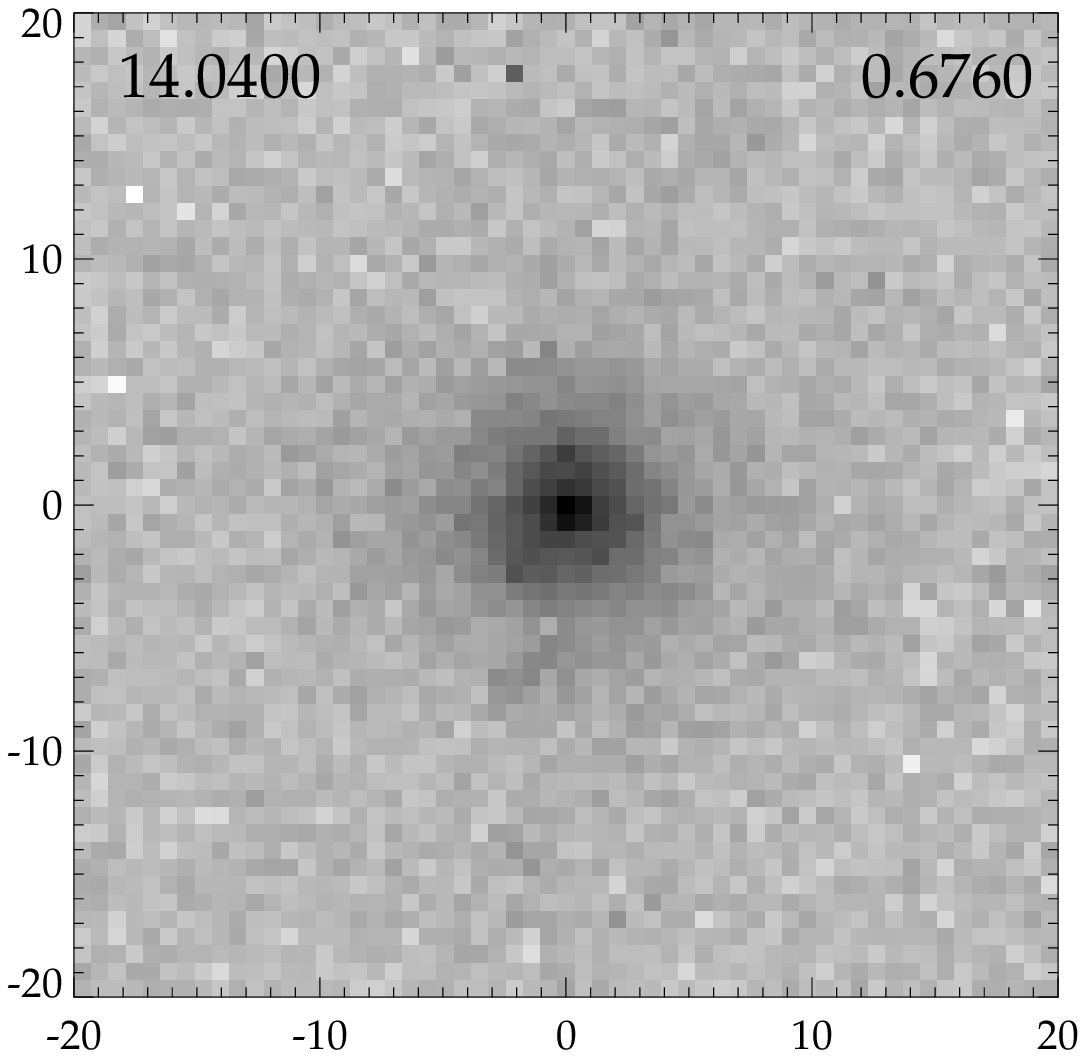} \includegraphics[height=0.22\textwidth,clip]{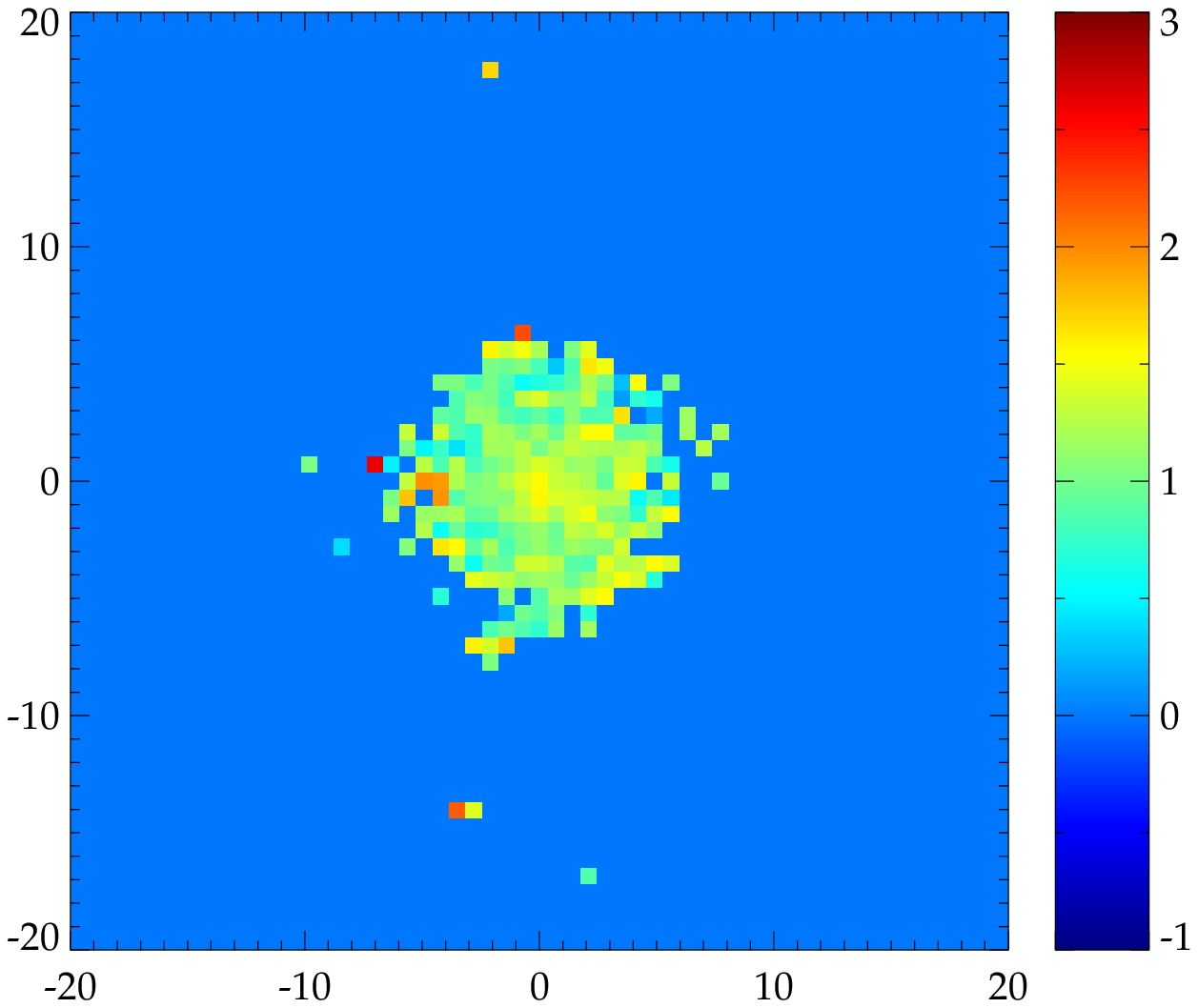}
\includegraphics[height=0.22\textwidth,clip]{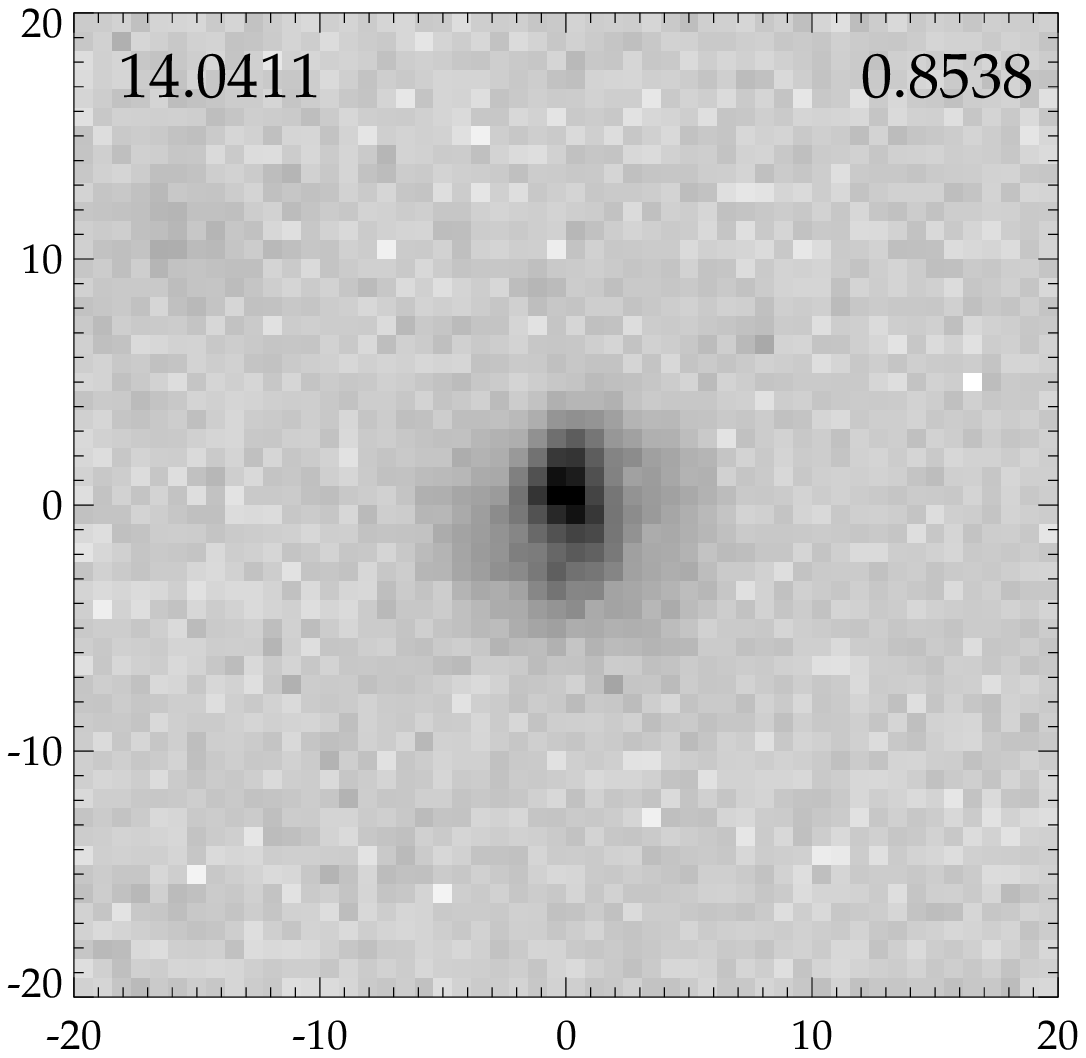} \includegraphics[height=0.22\textwidth,clip]{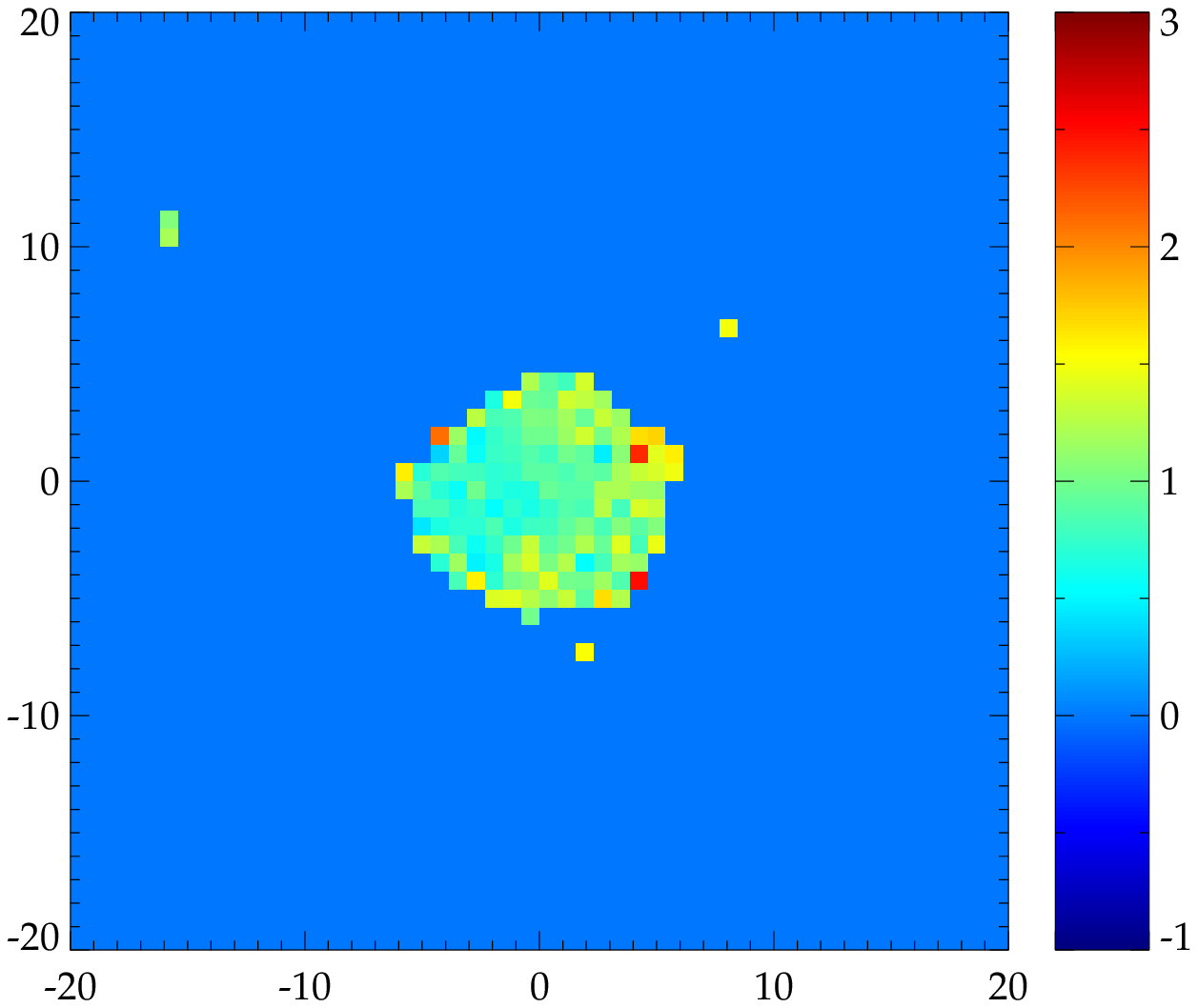}
\includegraphics[height=0.22\textwidth,clip]{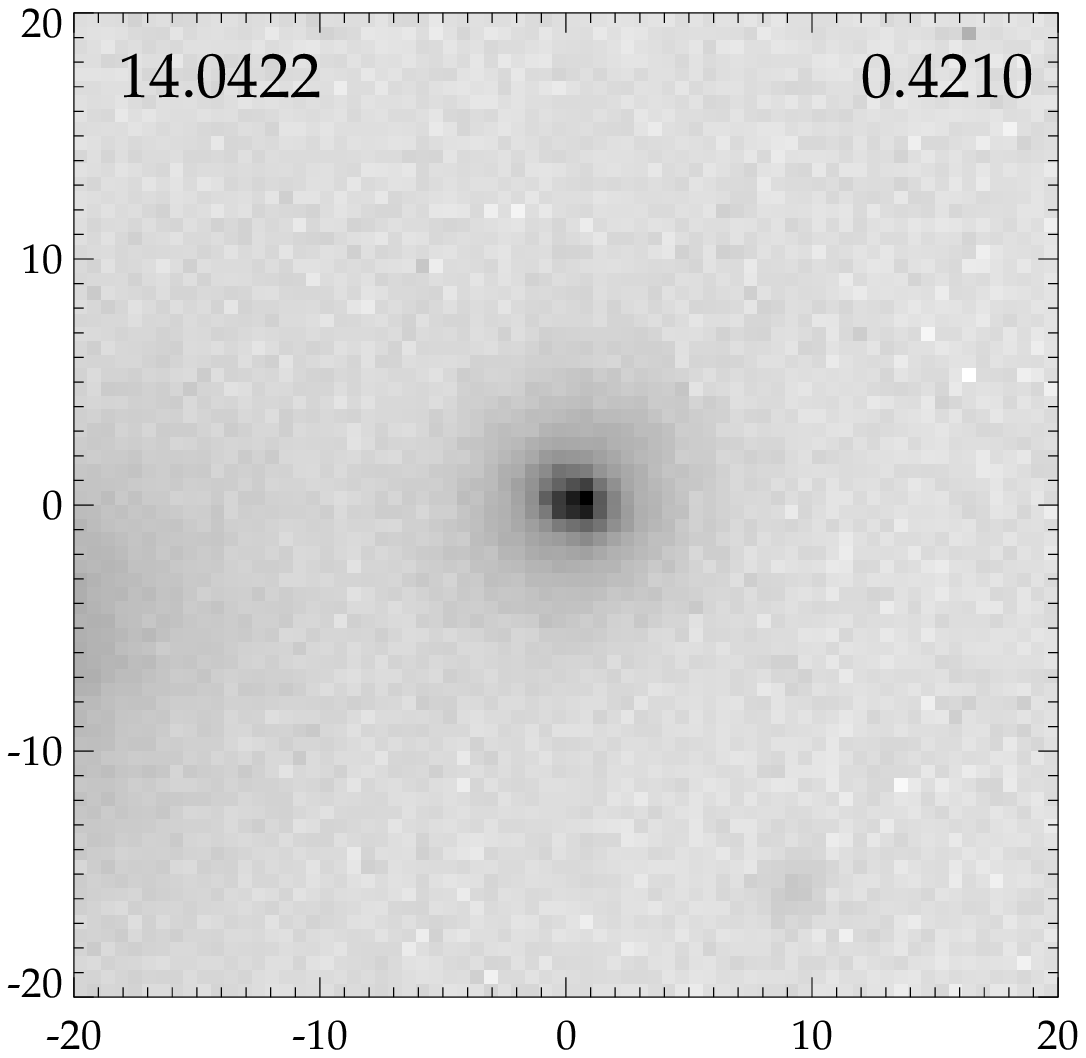} \includegraphics[height=0.22\textwidth,clip]{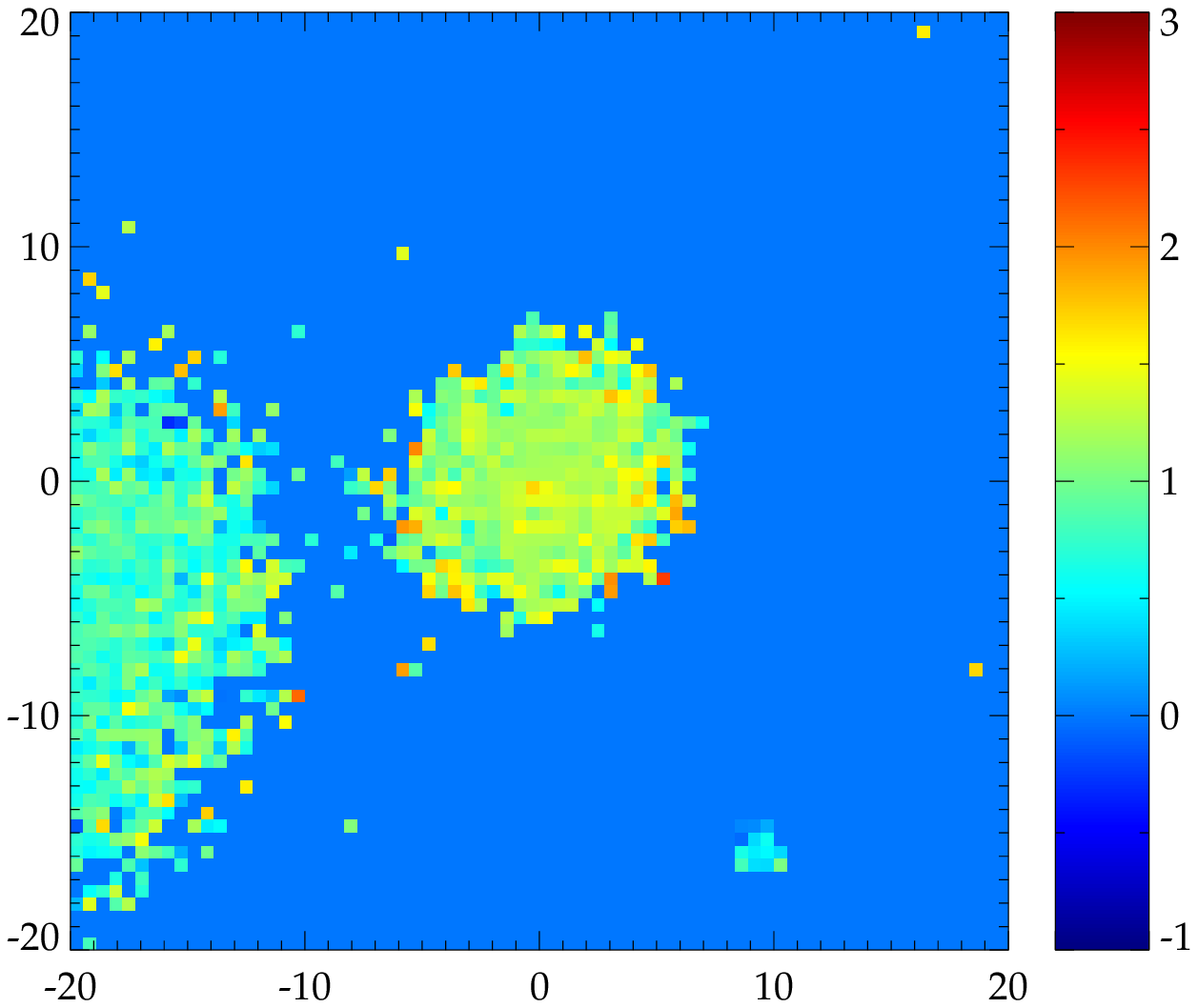}
\caption{Continued.} \end{figure*}

\addtocounter{figure}{-1}
\begin{figure*} \centering

\includegraphics[height=0.22\textwidth,clip]{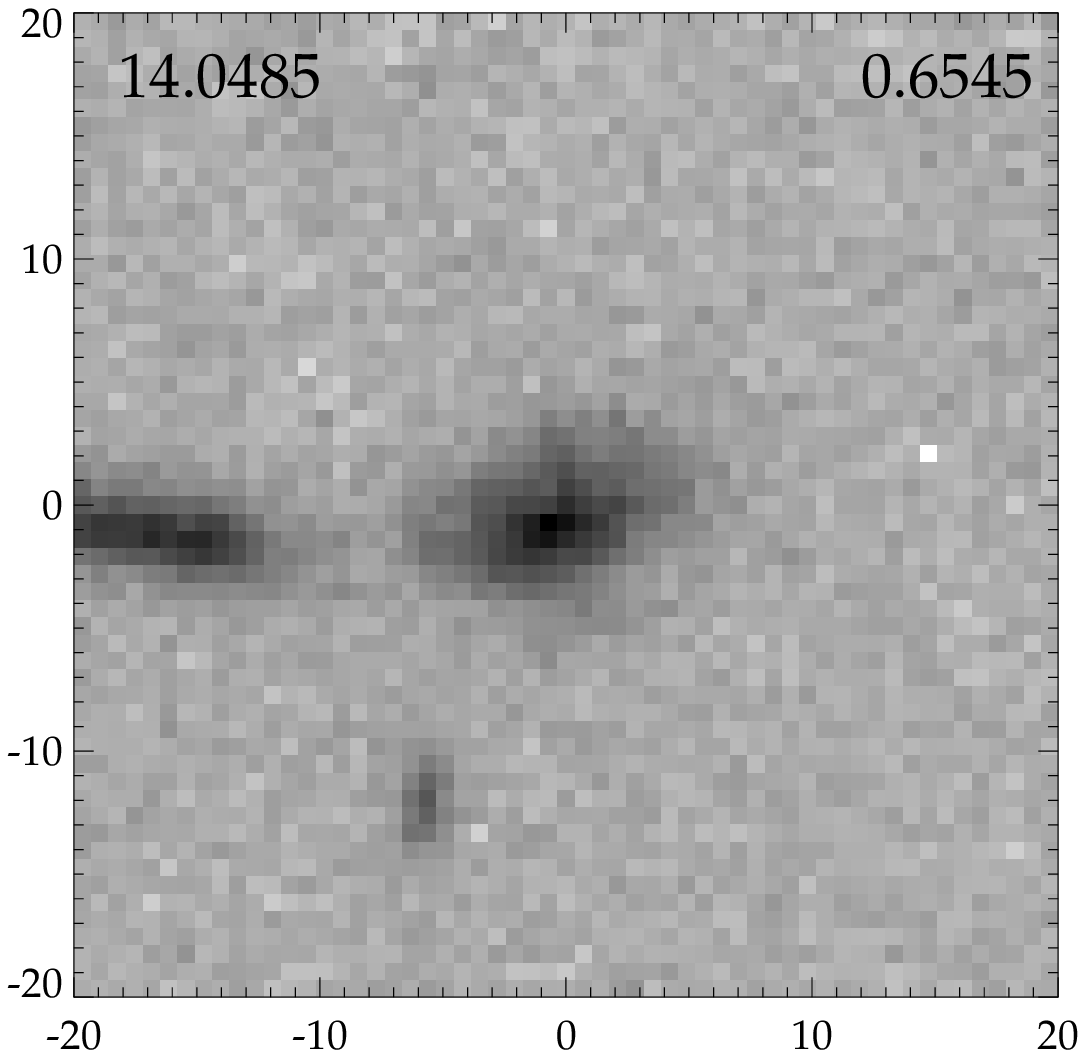} \includegraphics[height=0.22\textwidth,clip]{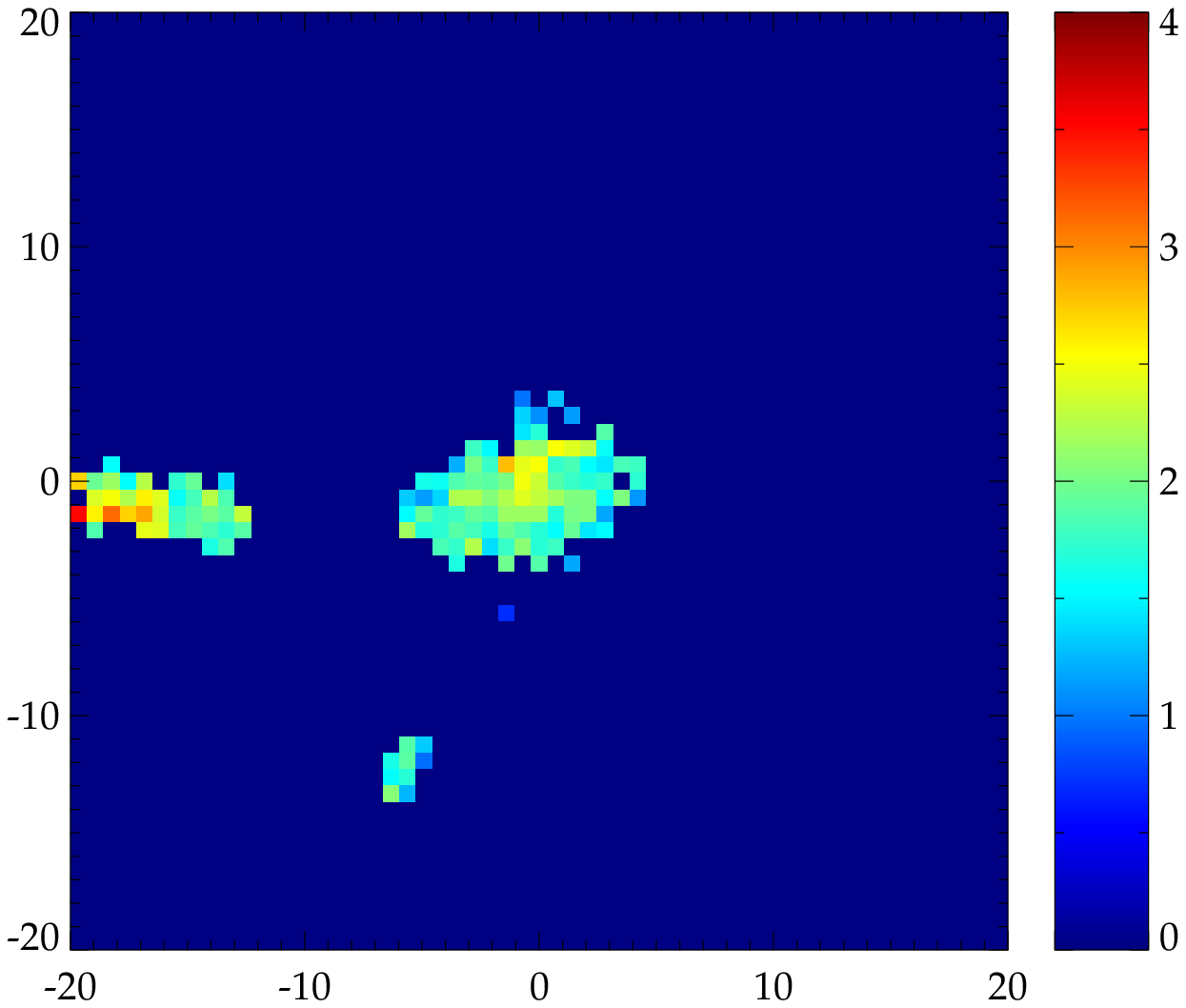}
\includegraphics[height=0.22\textwidth,clip]{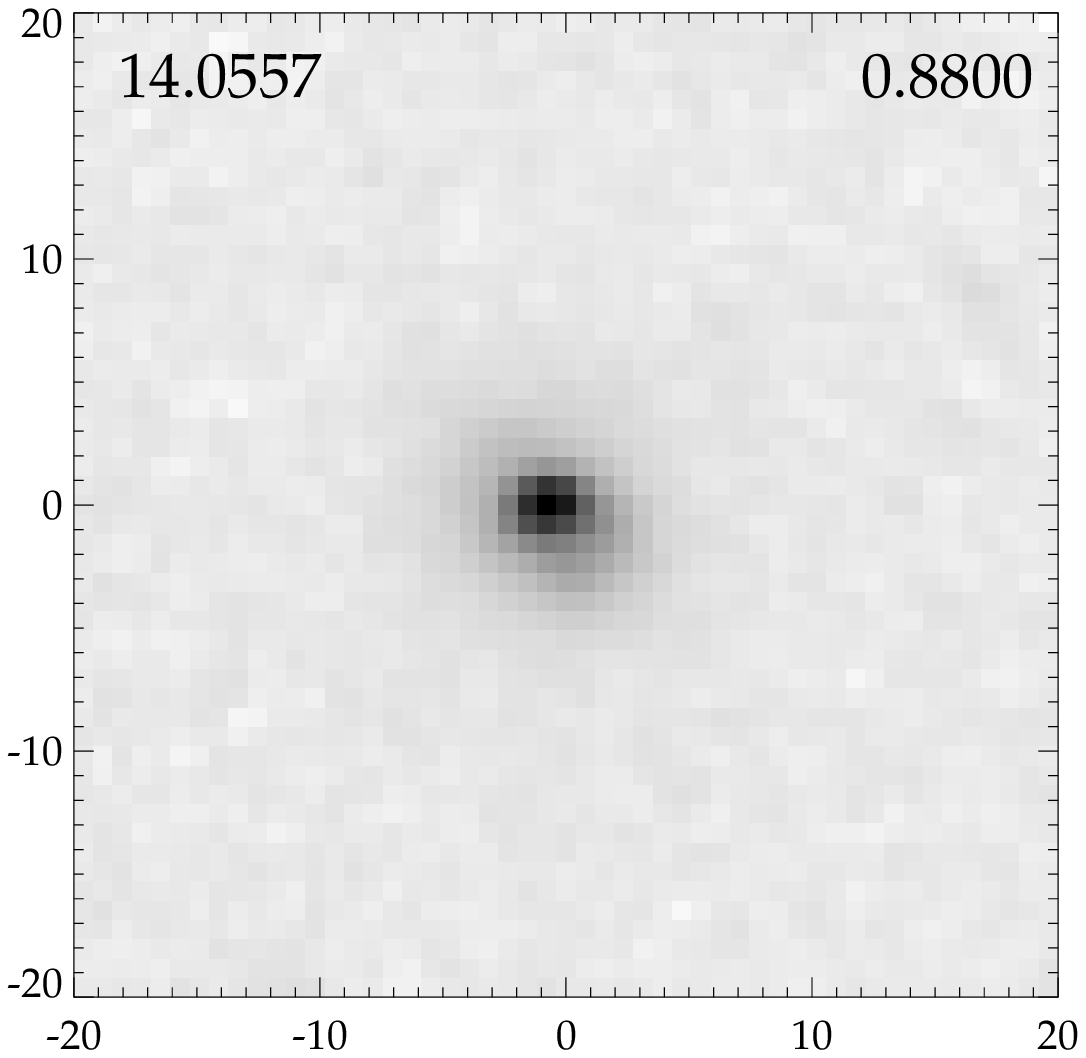} \includegraphics[height=0.22\textwidth,clip]{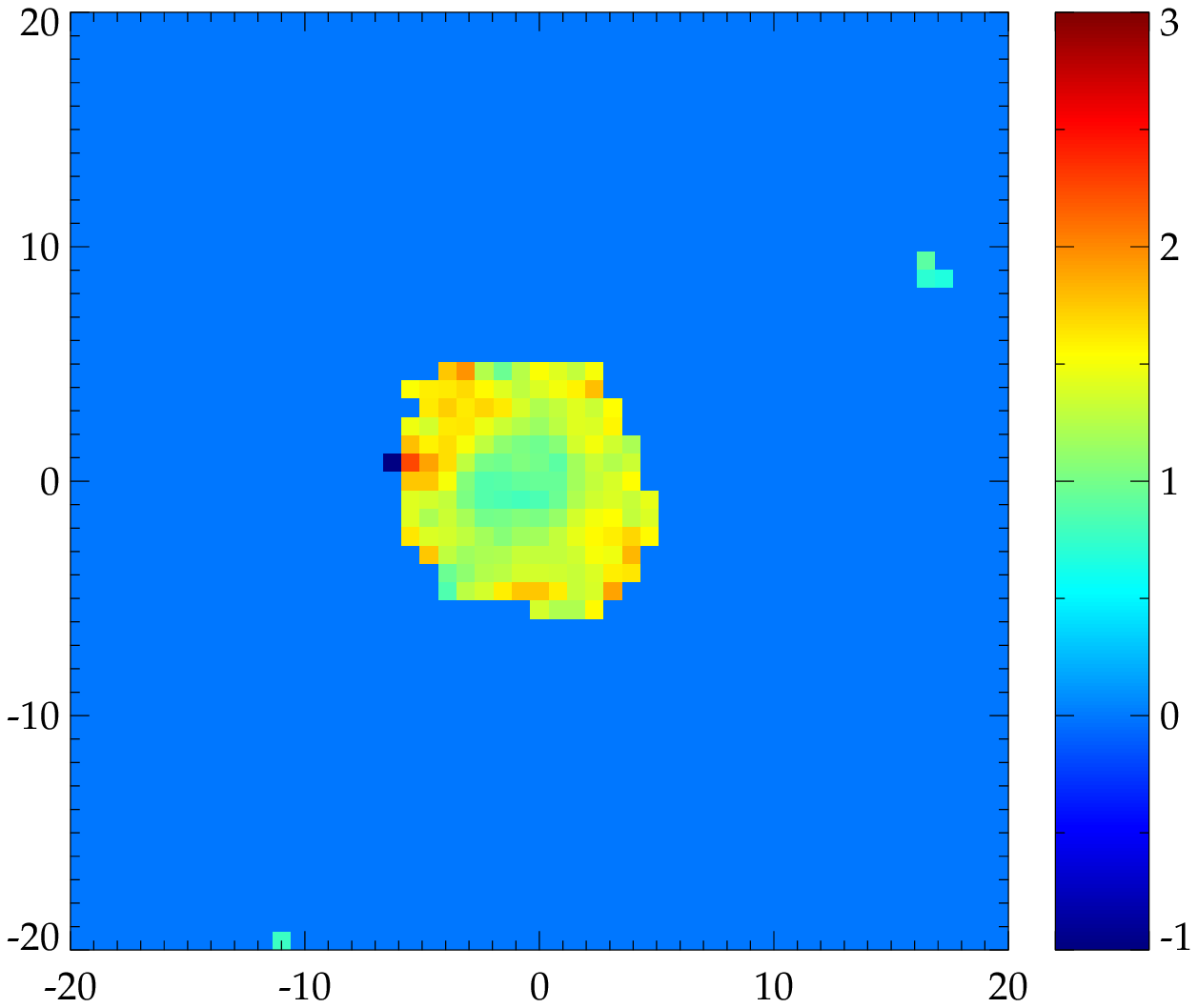}
\includegraphics[height=0.22\textwidth,clip]{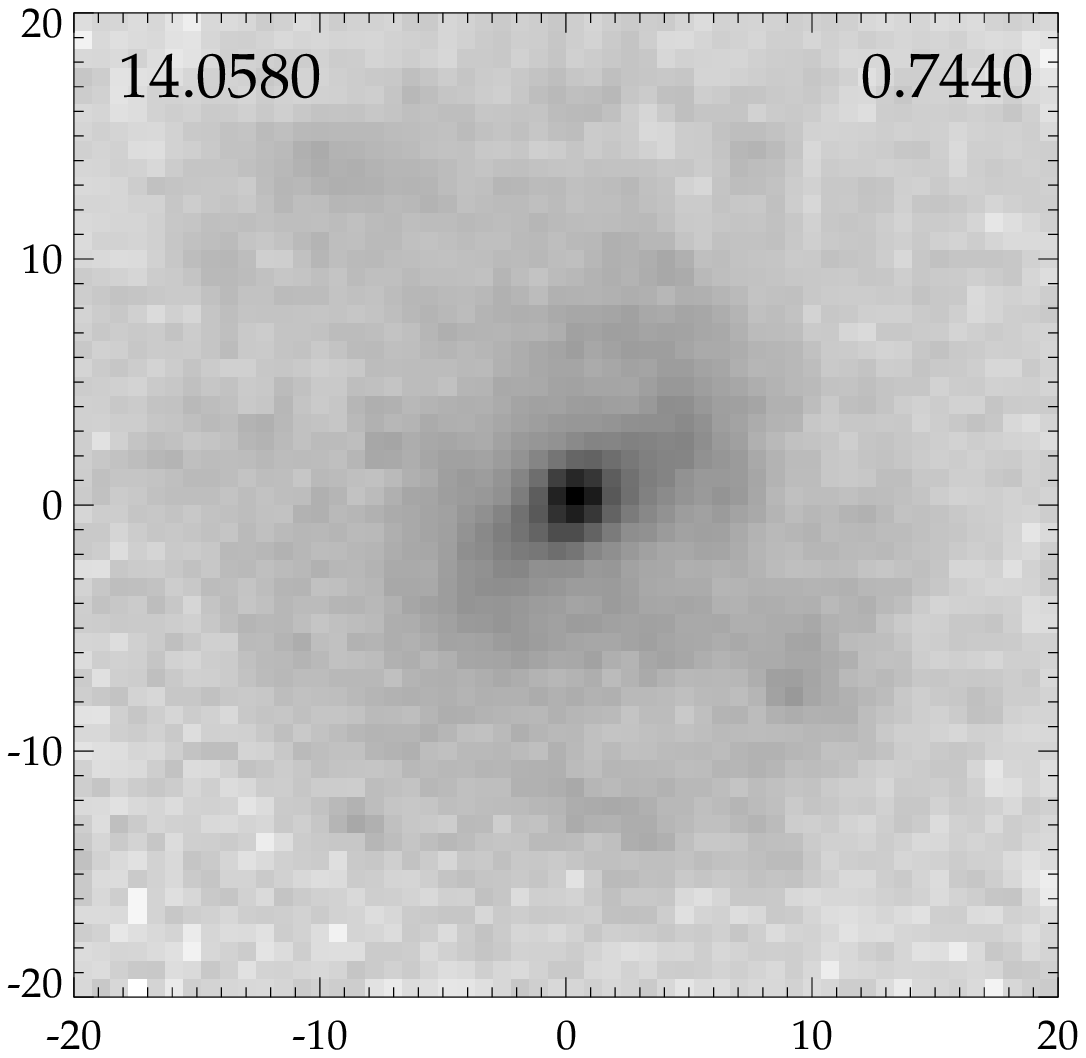} \includegraphics[height=0.22\textwidth,clip]{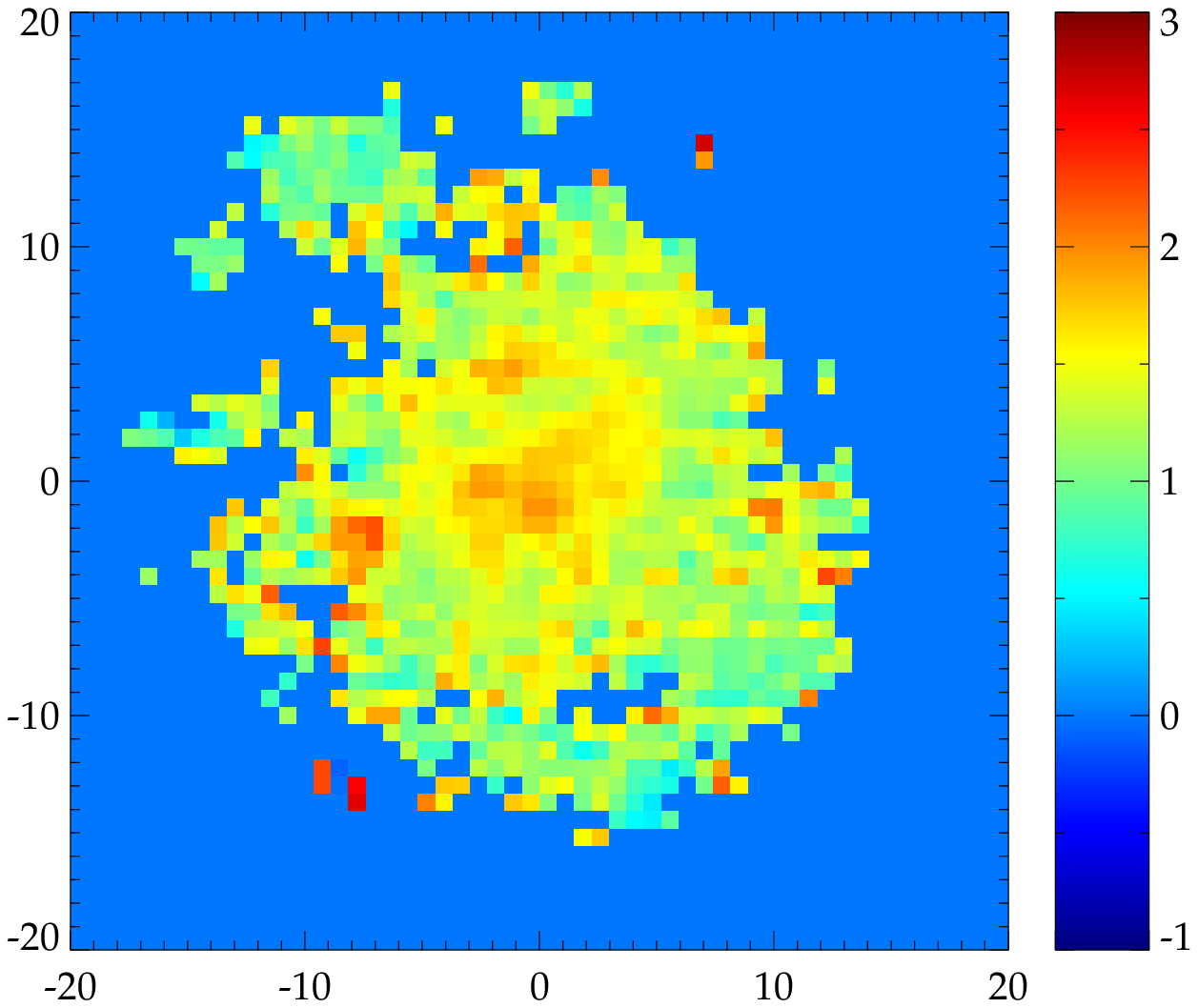}
\includegraphics[height=0.22\textwidth,clip]{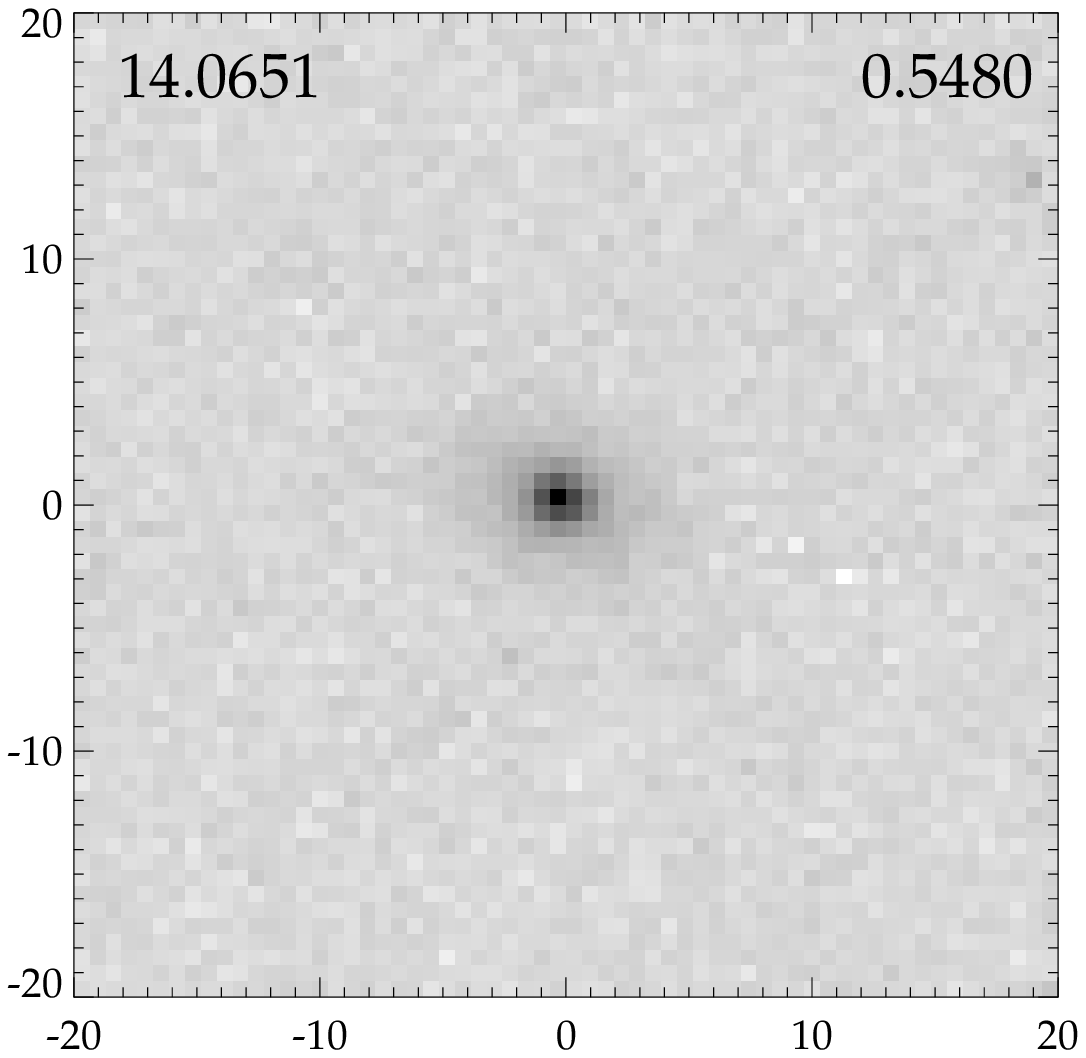} \includegraphics[height=0.22\textwidth,clip]{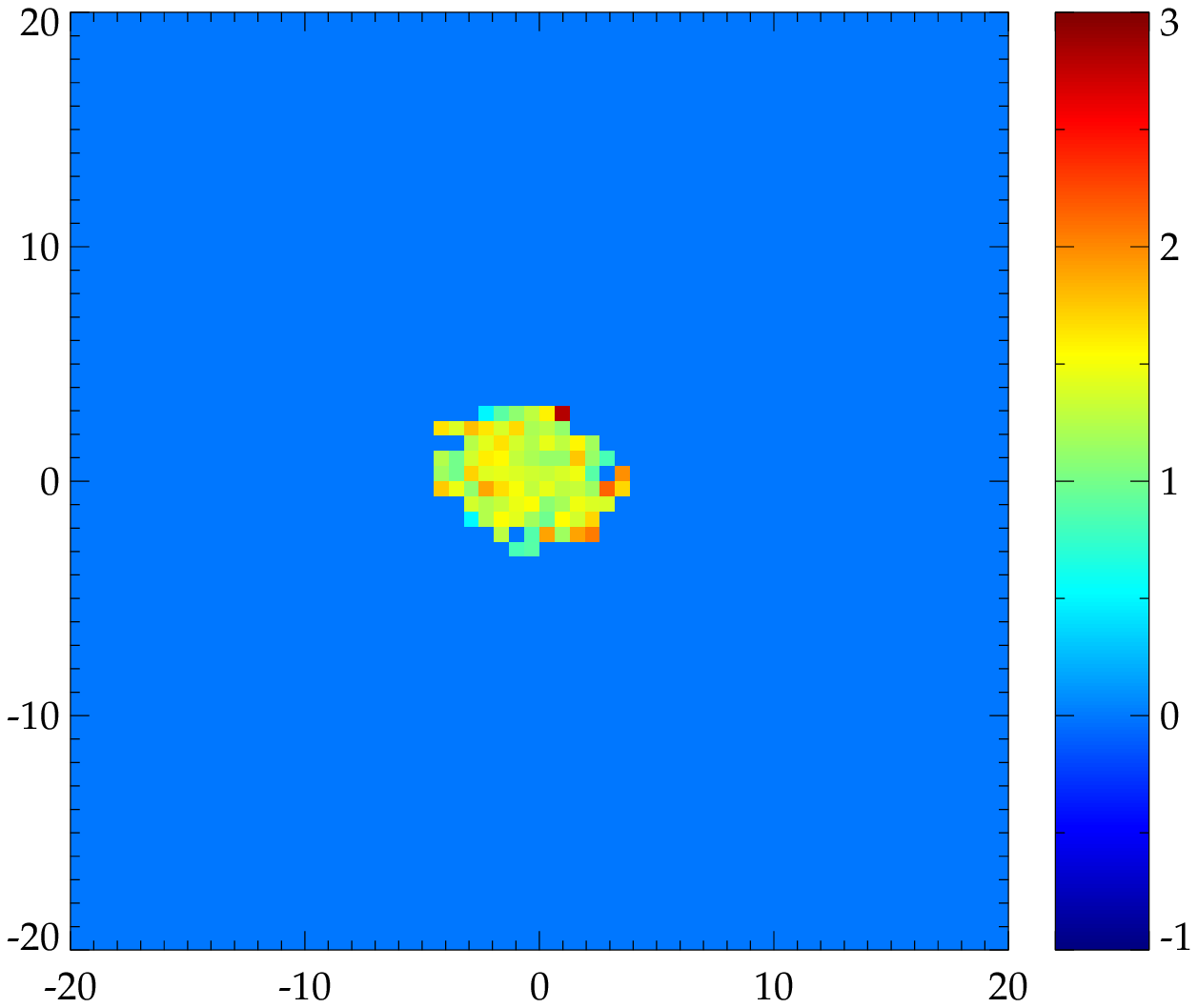}
\includegraphics[height=0.22\textwidth,clip]{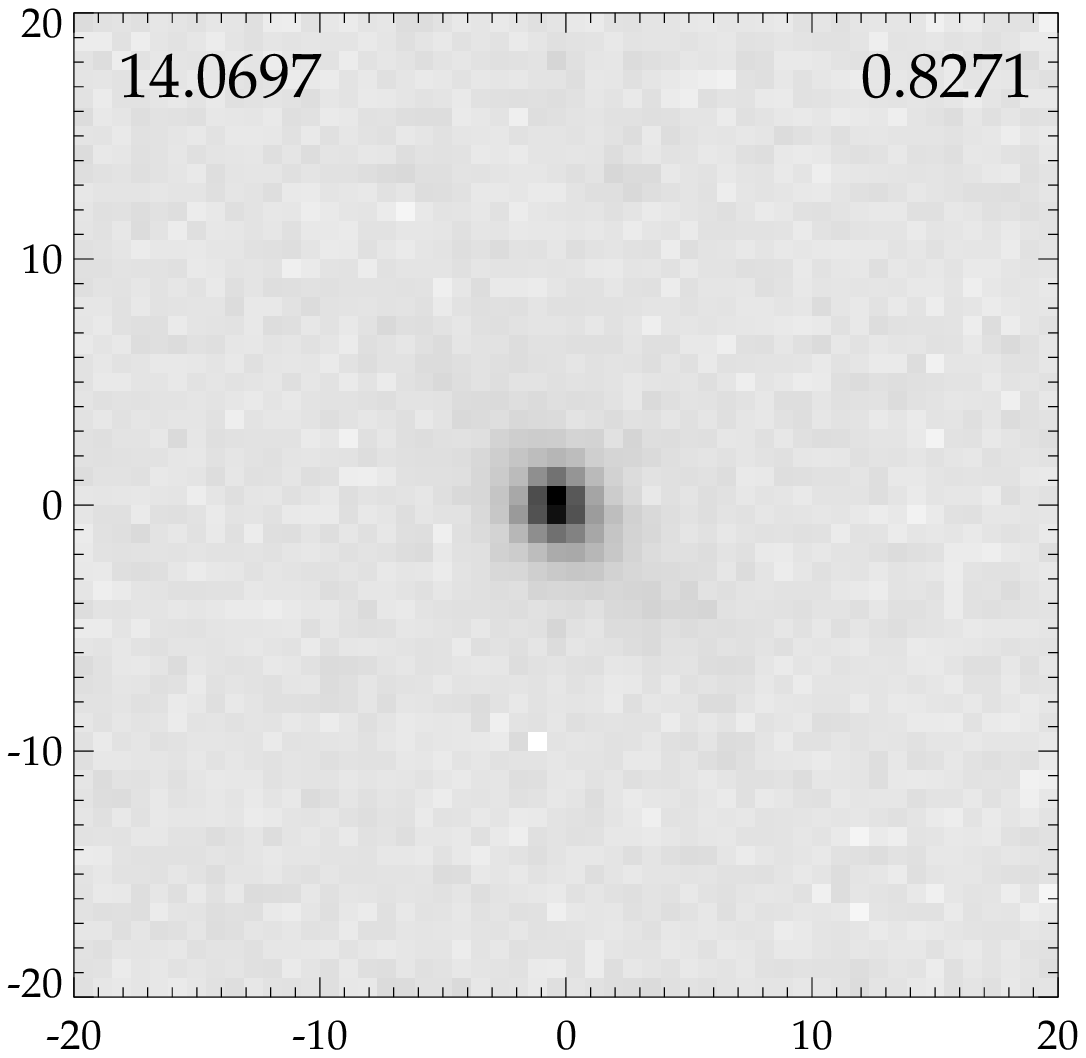} \includegraphics[height=0.22\textwidth,clip]{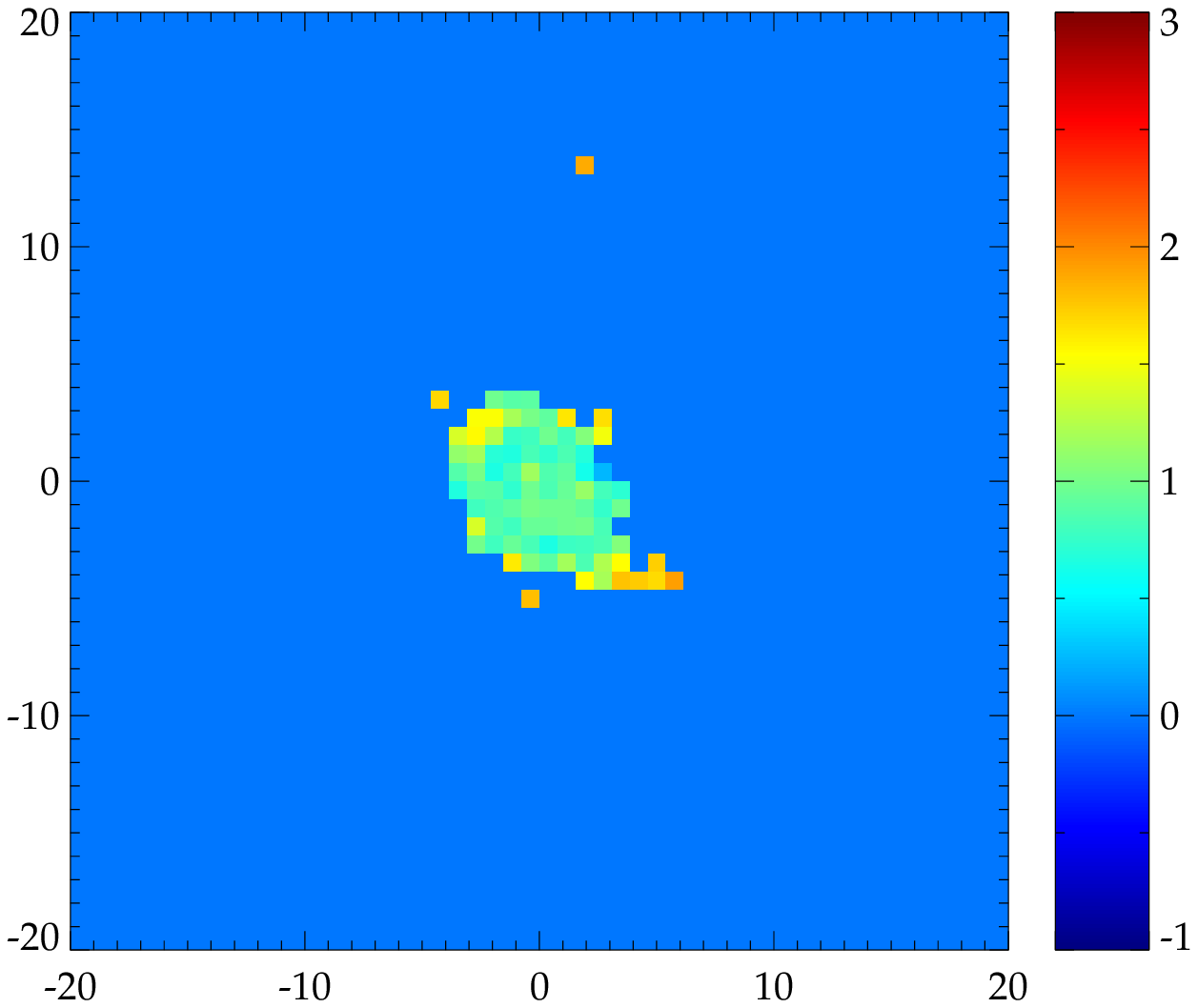}
\includegraphics[height=0.22\textwidth,clip]{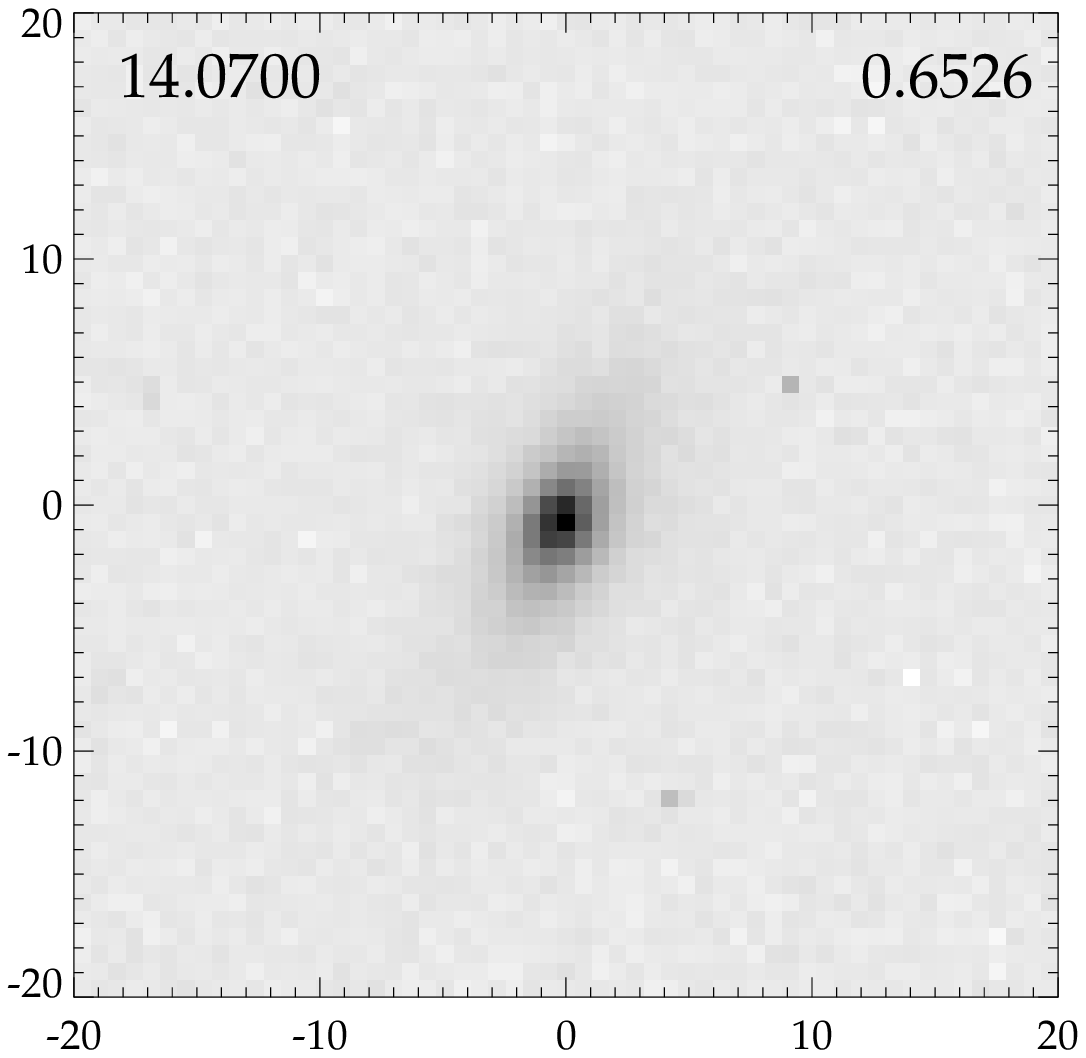} \includegraphics[height=0.22\textwidth,clip]{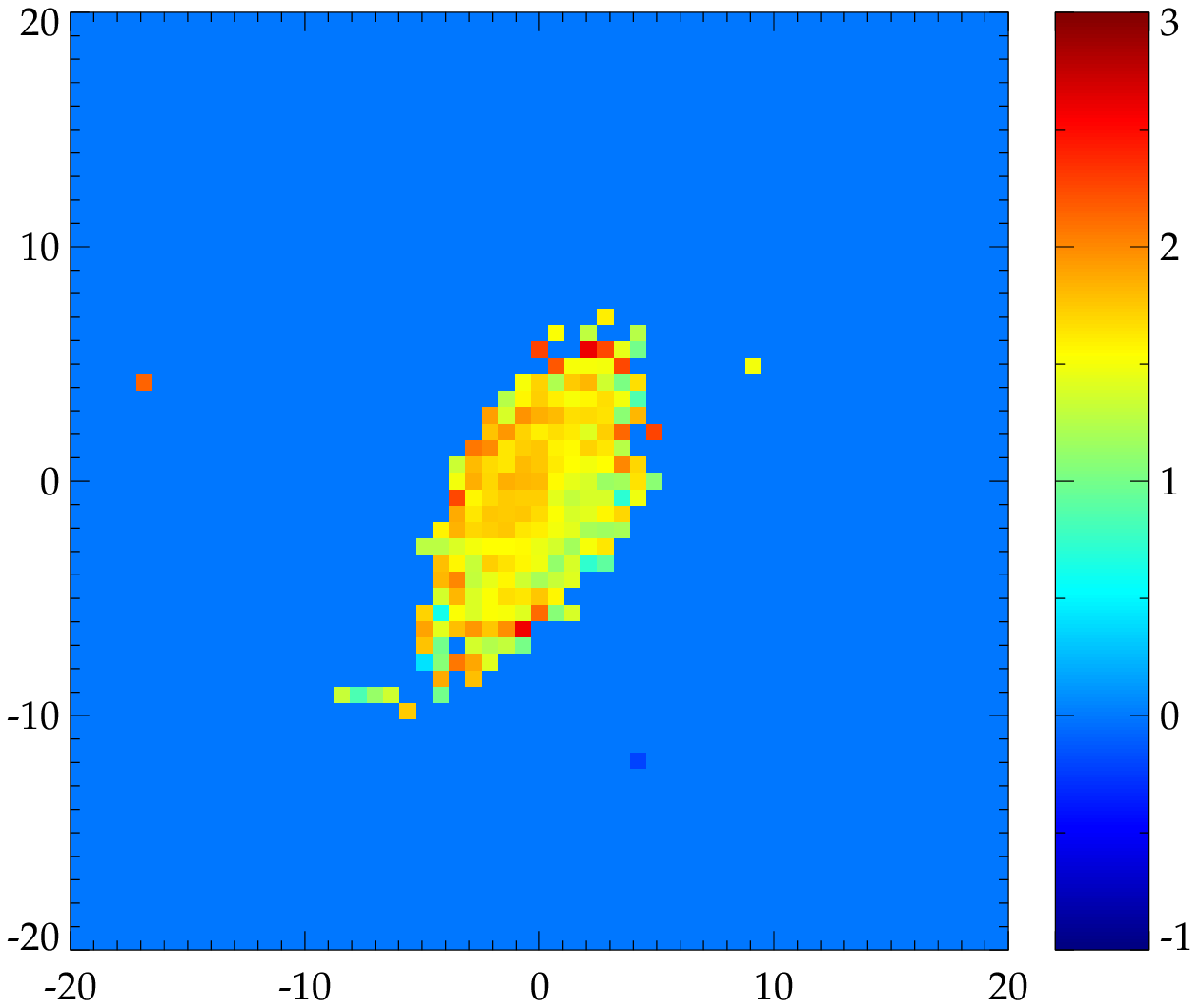}
\includegraphics[height=0.22\textwidth,clip]{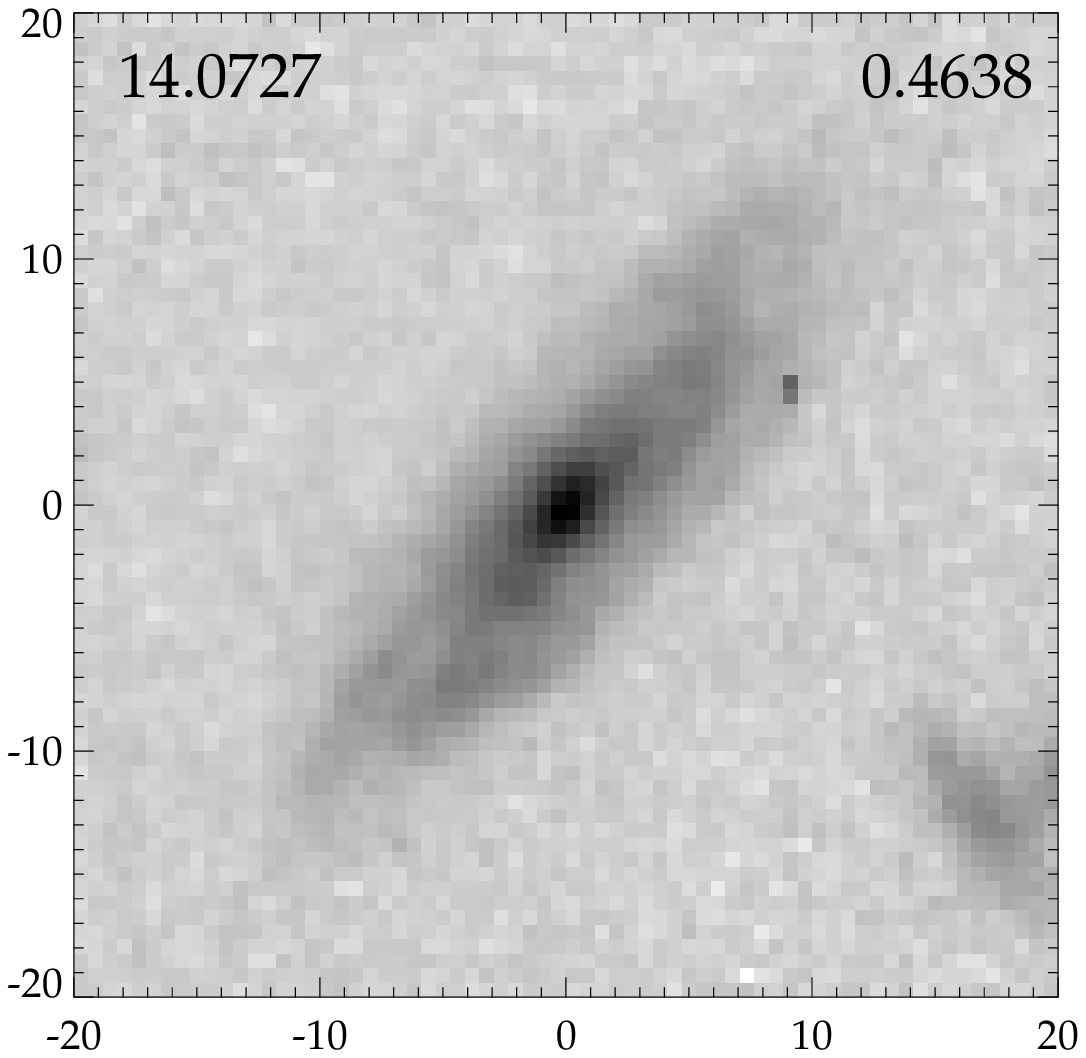} \includegraphics[height=0.22\textwidth,clip]{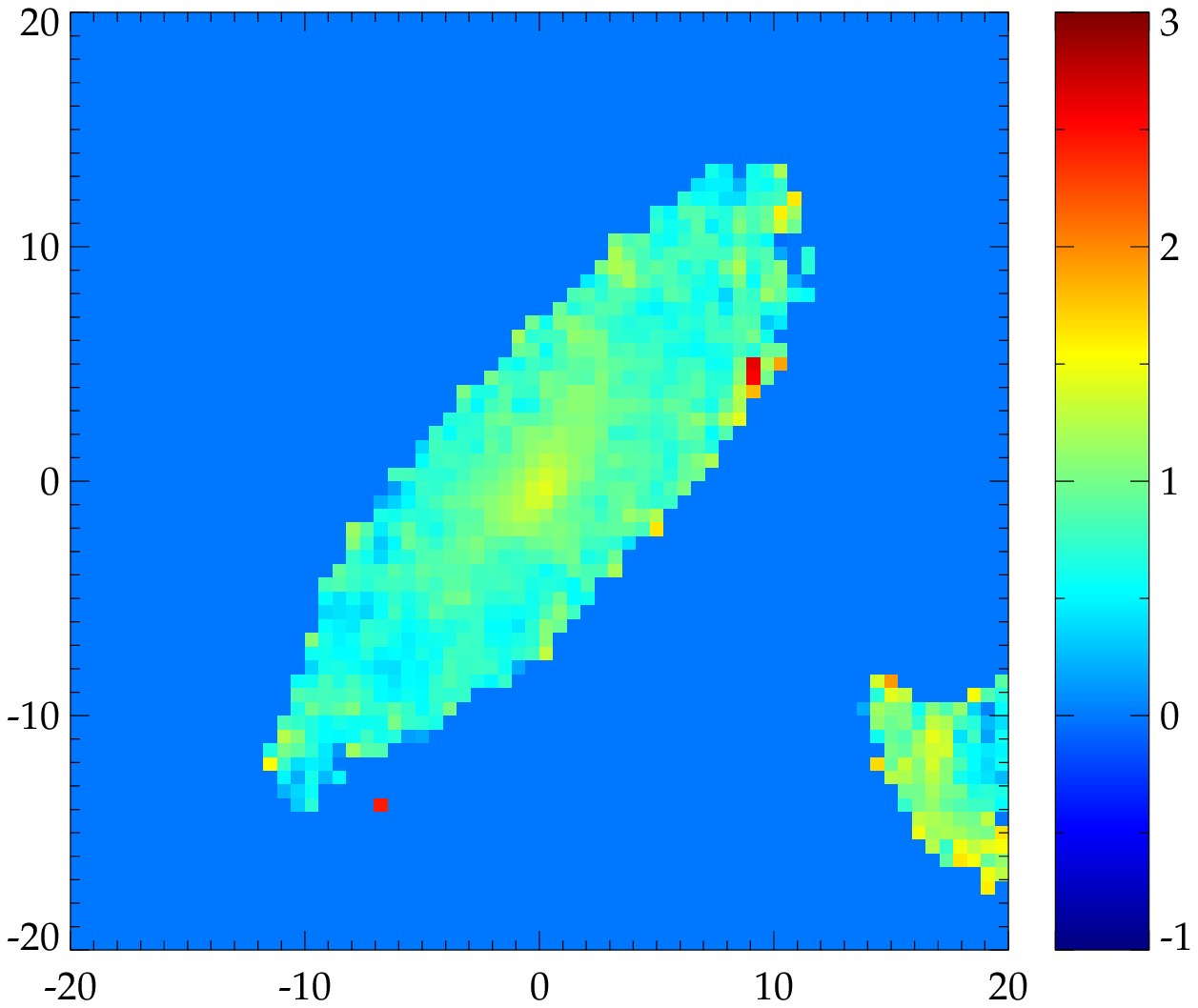}
\includegraphics[height=0.22\textwidth,clip]{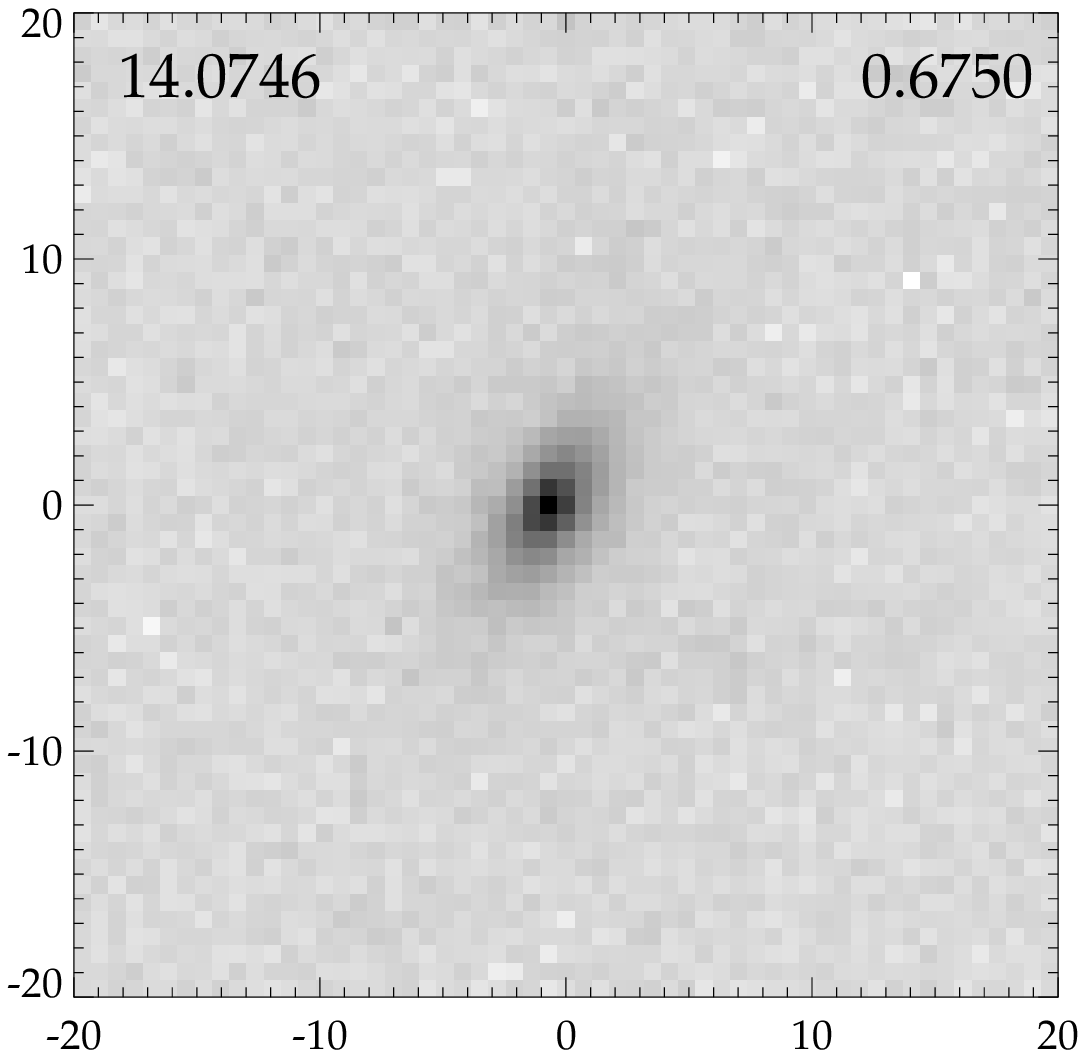} \includegraphics[height=0.22\textwidth,clip]{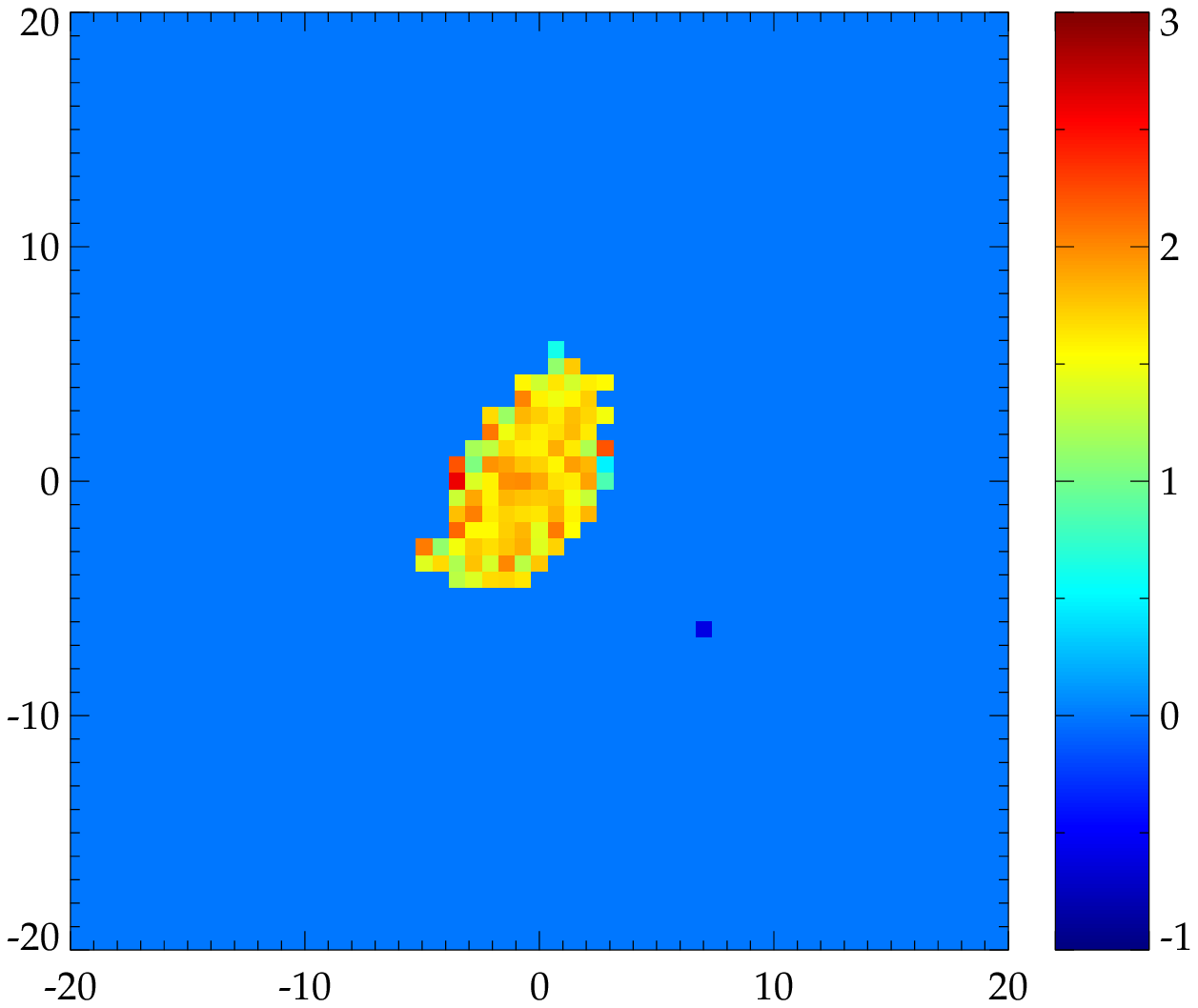}
\includegraphics[height=0.22\textwidth,clip]{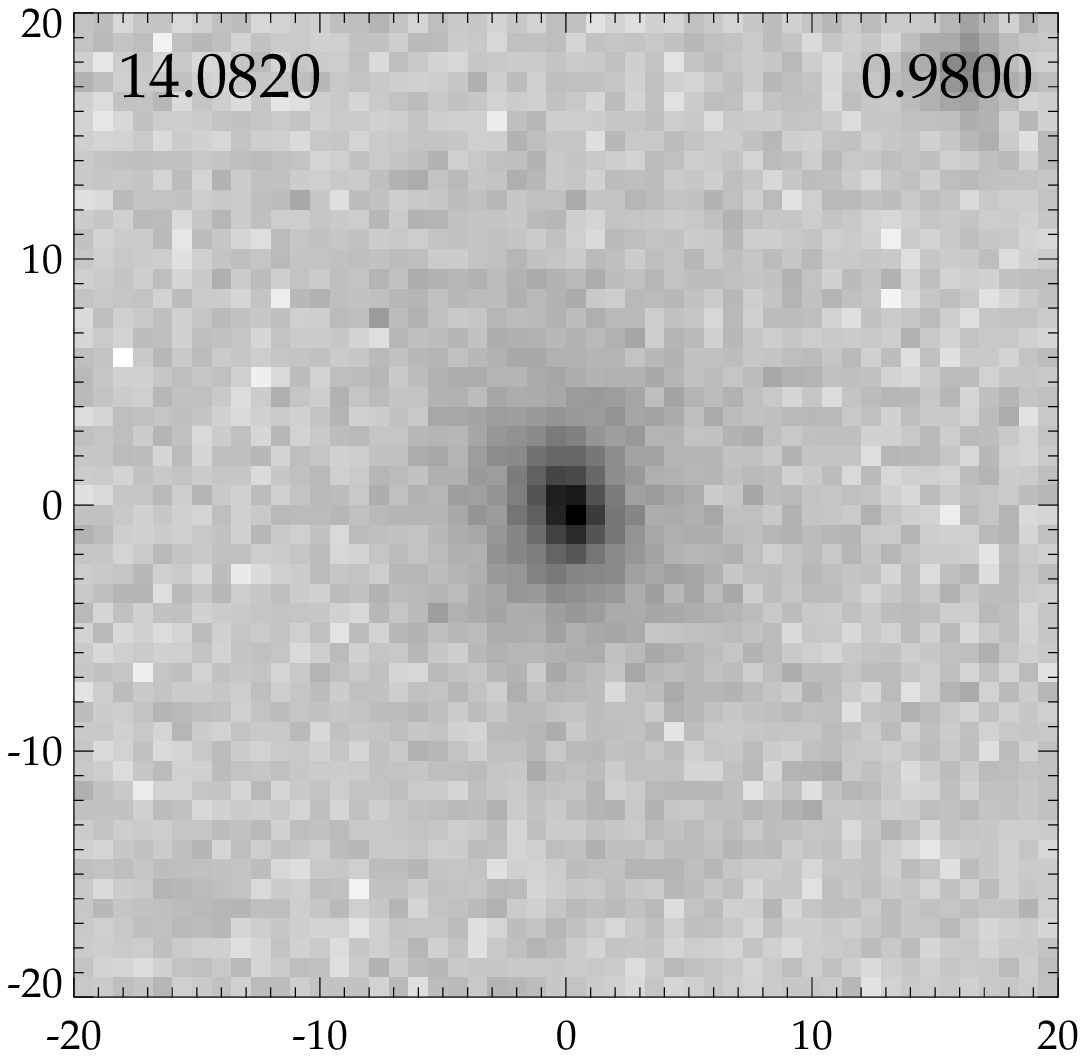} \includegraphics[height=0.22\textwidth,clip]{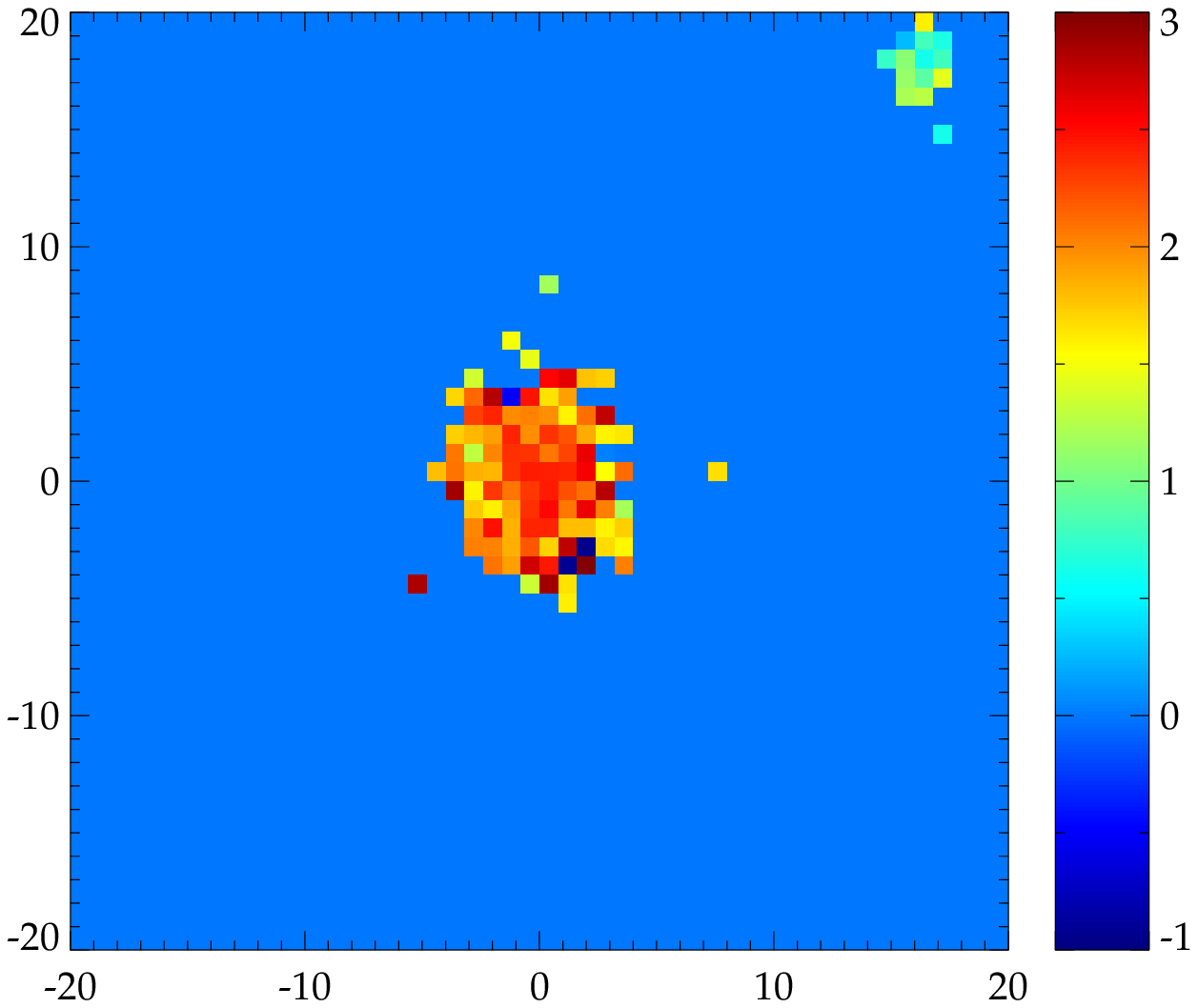}
\includegraphics[height=0.22\textwidth,clip]{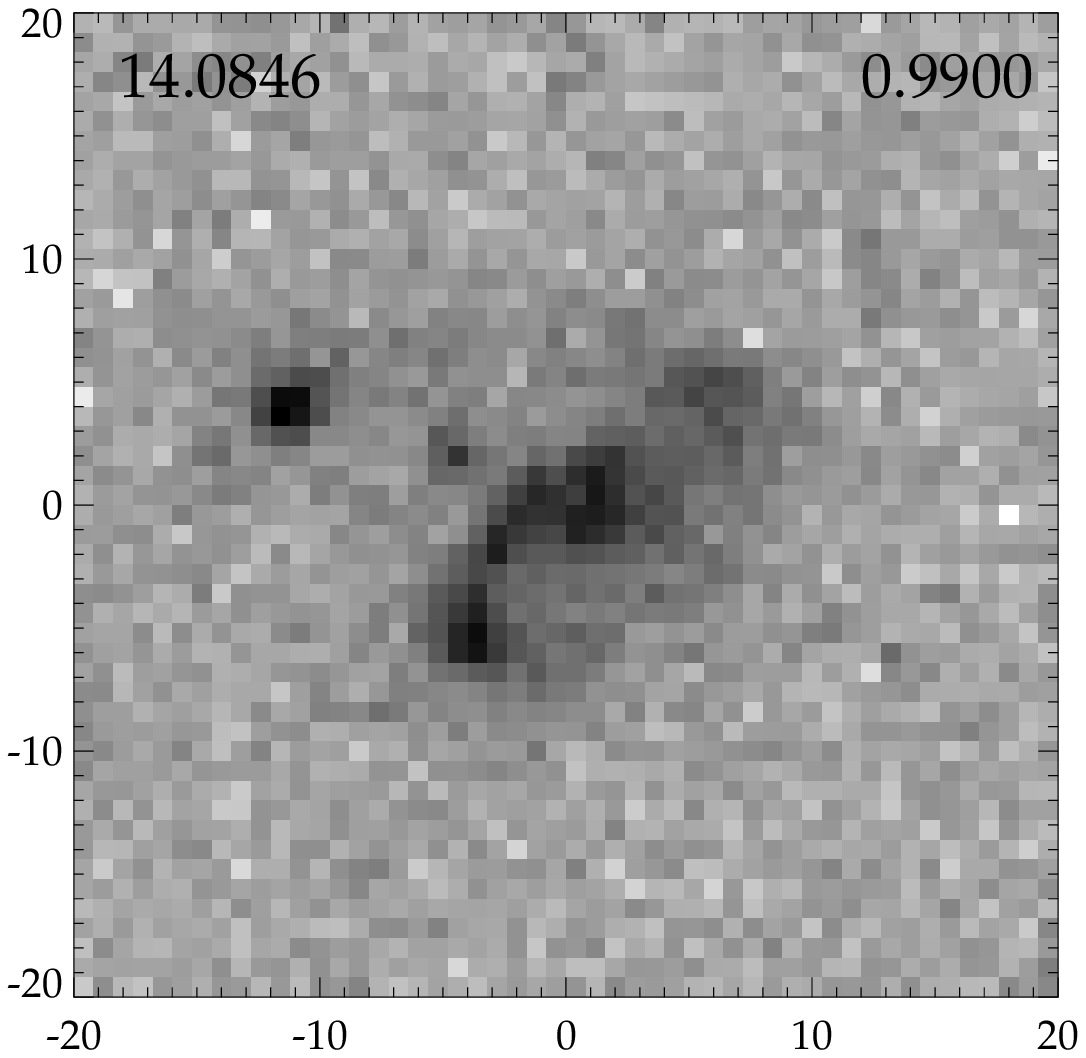} \includegraphics[height=0.22\textwidth,clip]{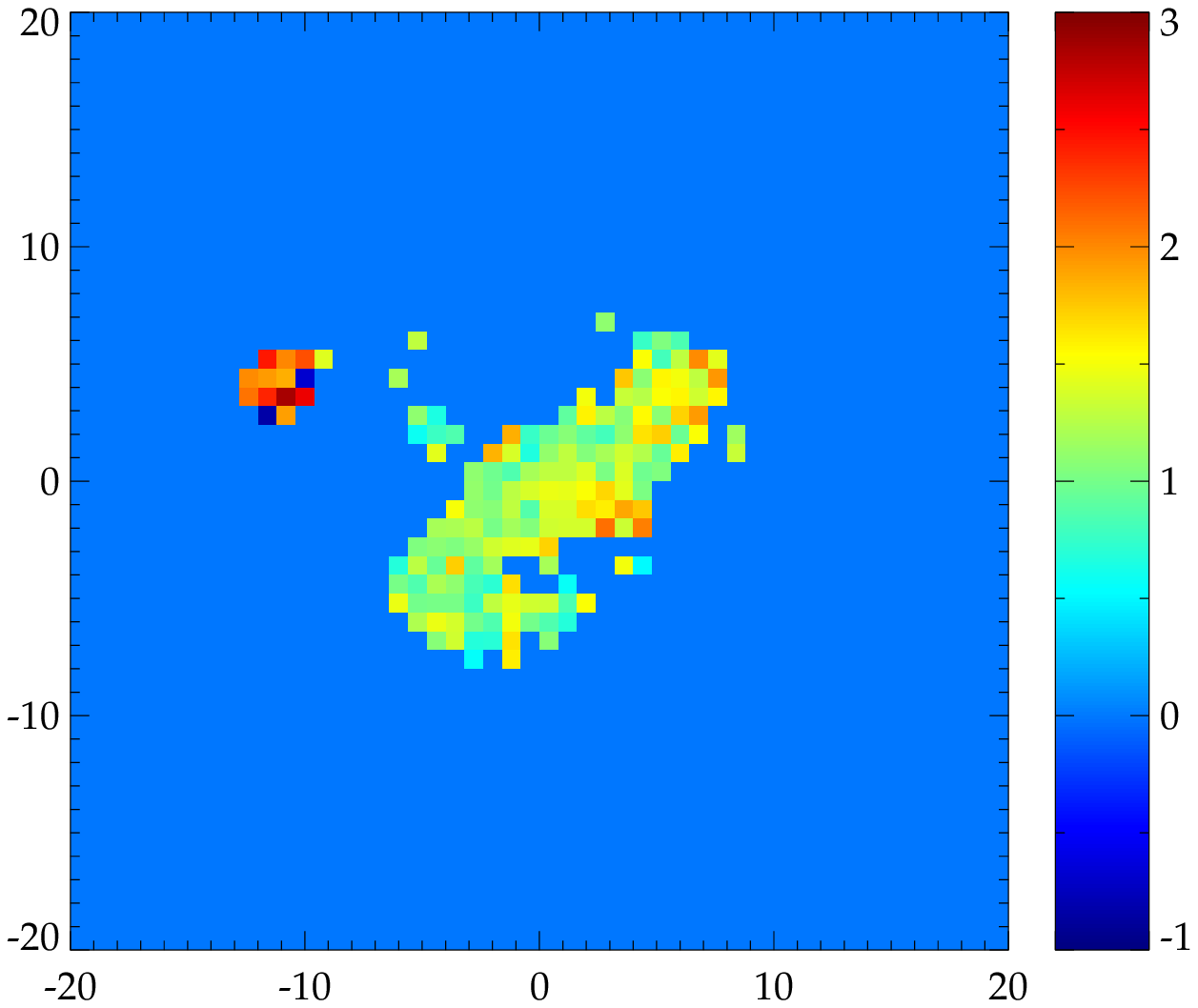}
\includegraphics[height=0.22\textwidth,clip]{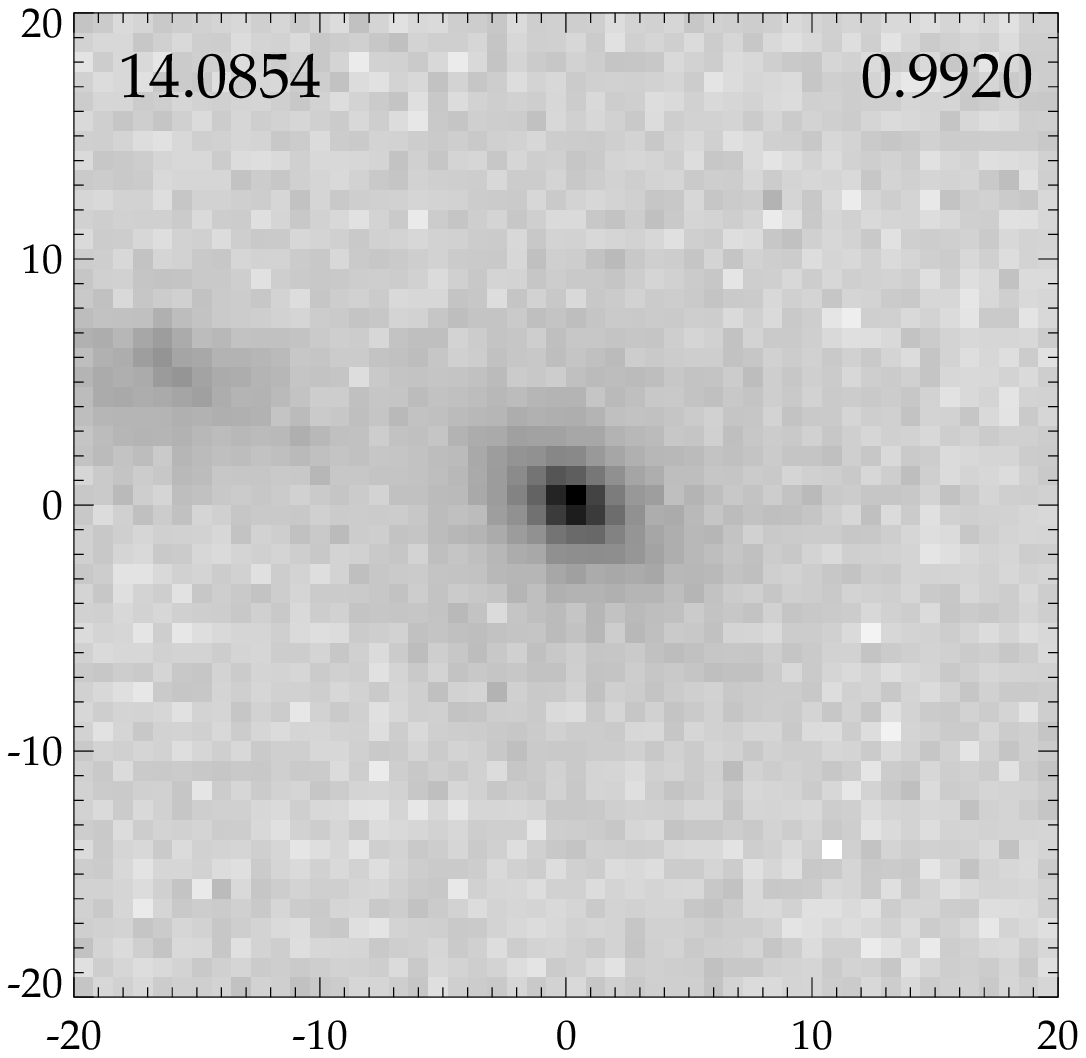} \includegraphics[height=0.22\textwidth,clip]{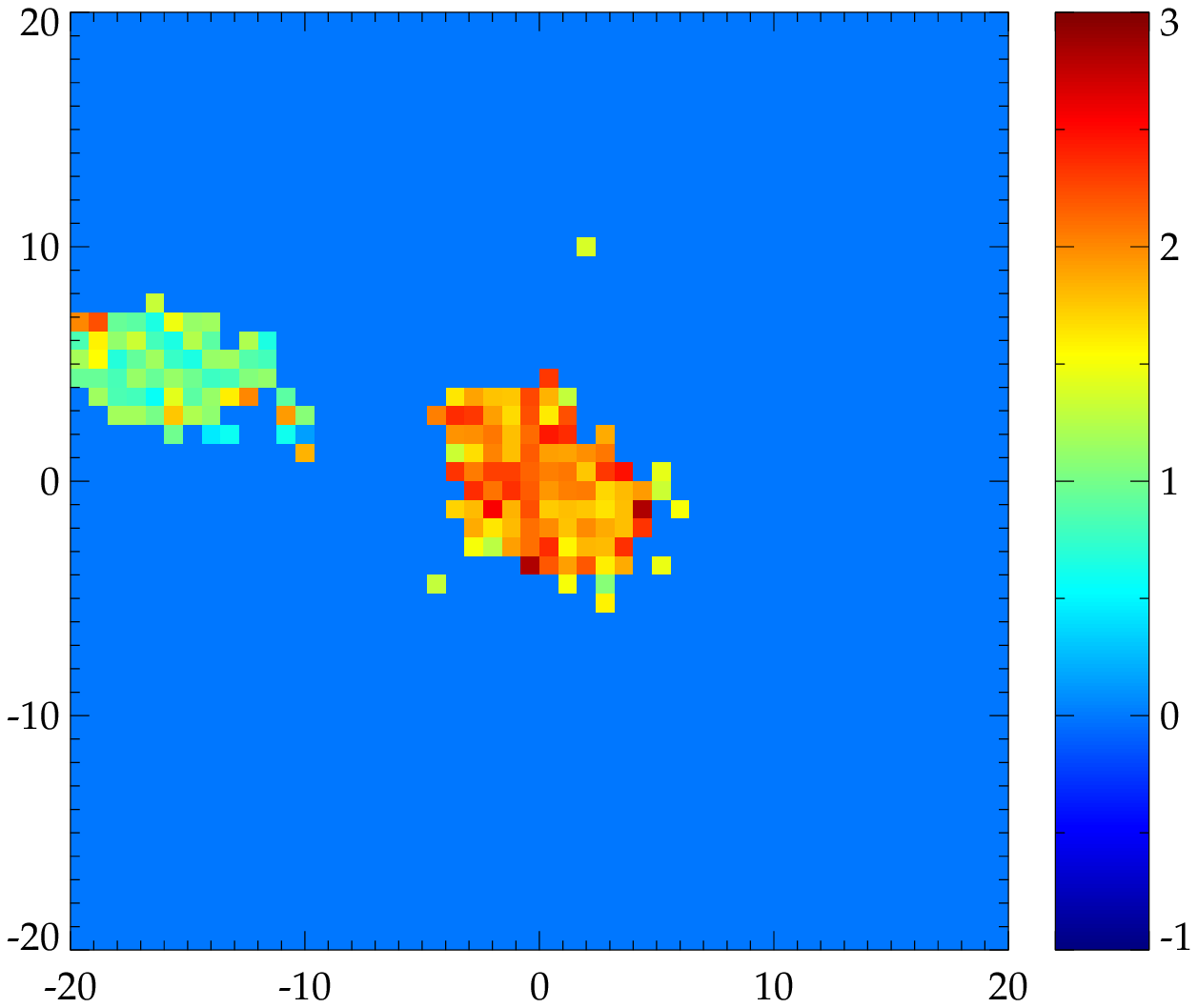}
\includegraphics[height=0.22\textwidth,clip]{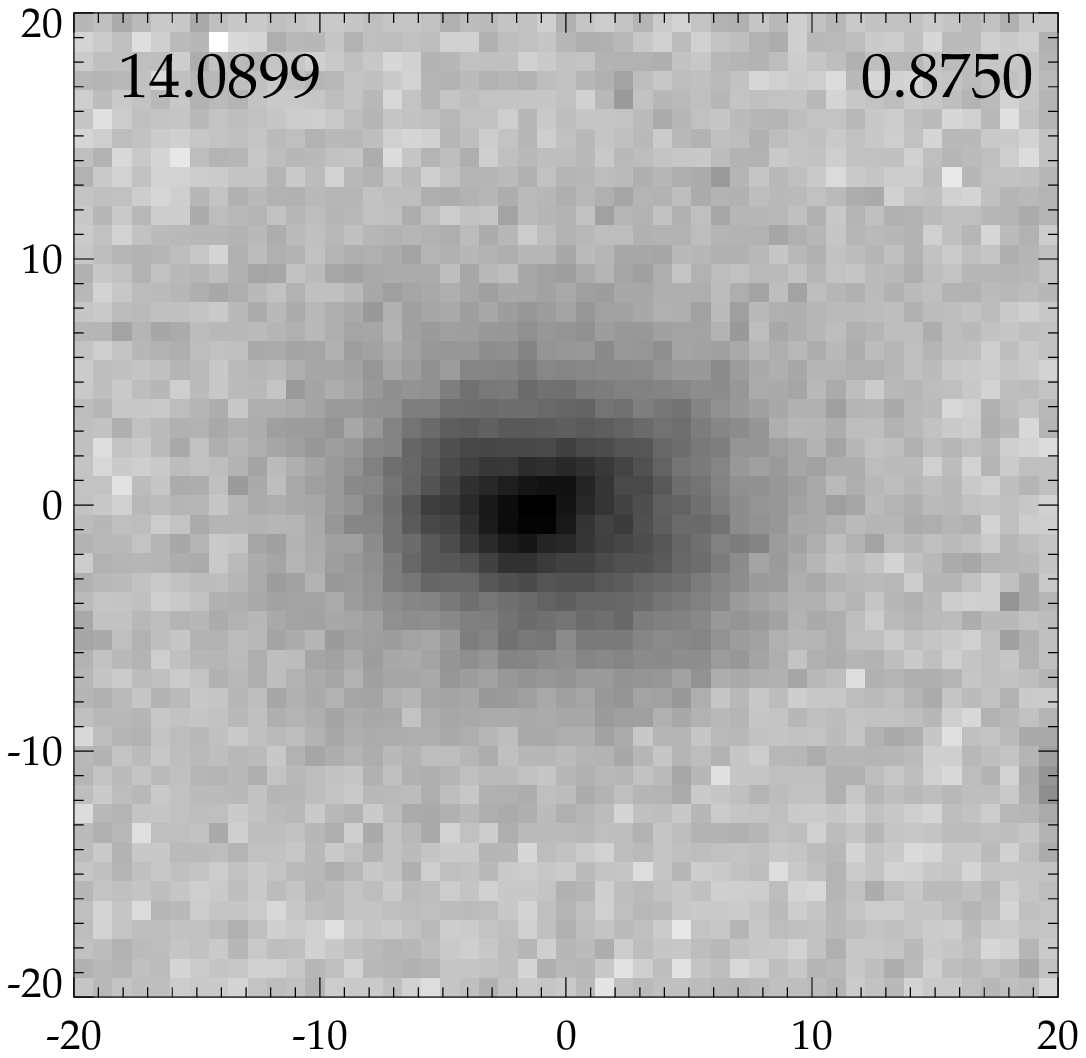} \includegraphics[height=0.22\textwidth,clip]{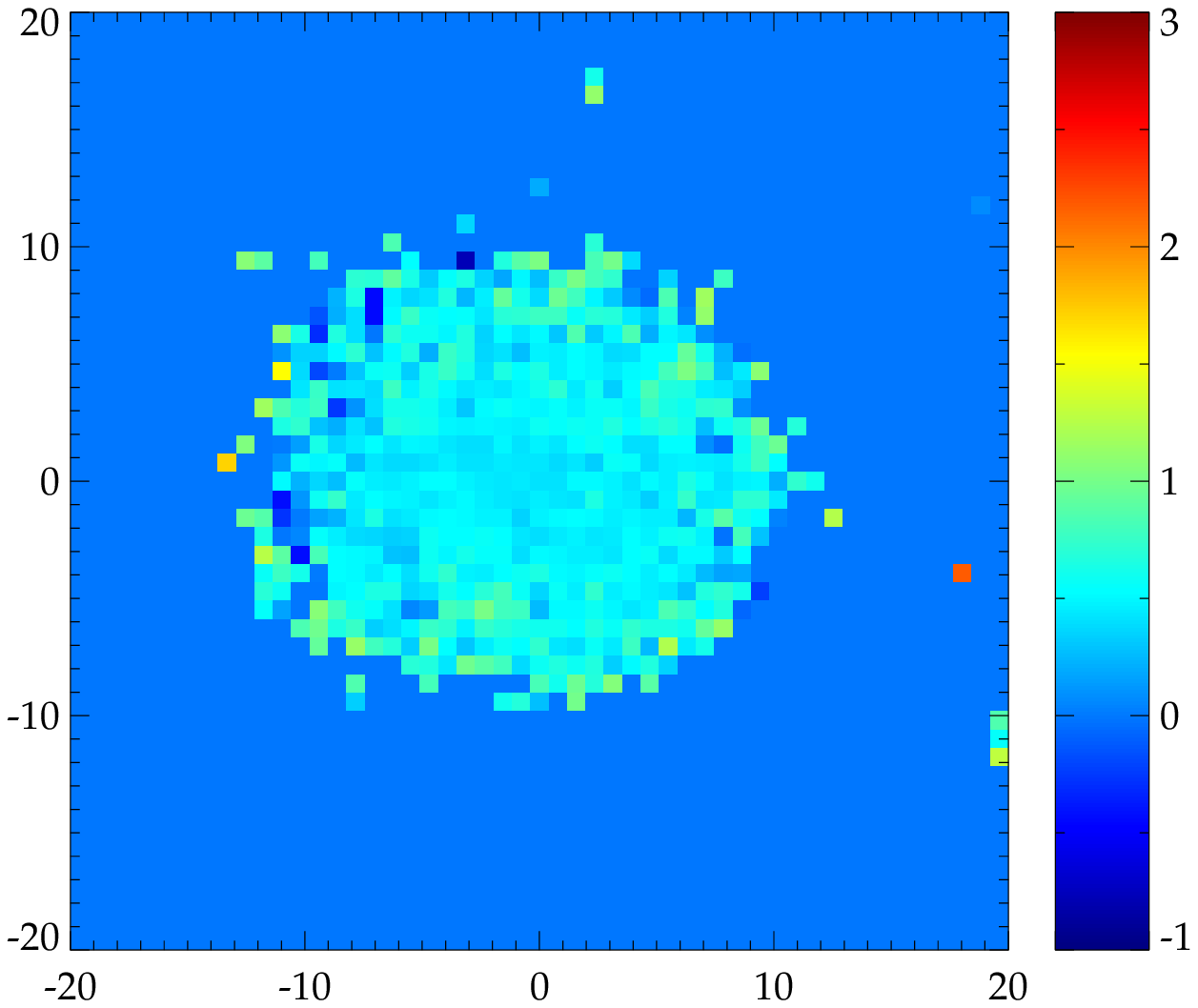}
\caption{Continued.} \end{figure*}

\addtocounter{figure}{-1}
\begin{figure*} \centering

\includegraphics[height=0.22\textwidth,clip]{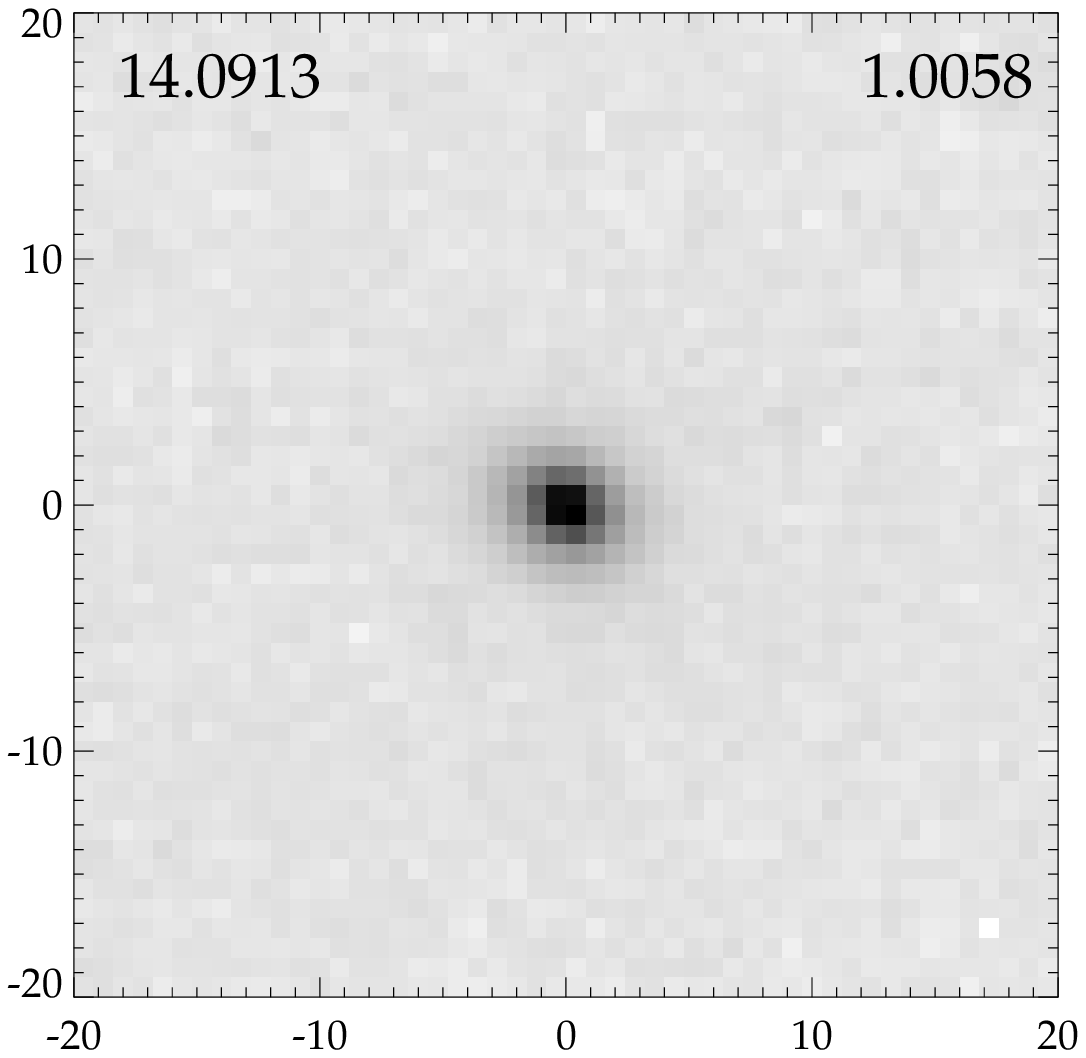} \includegraphics[height=0.22\textwidth,clip]{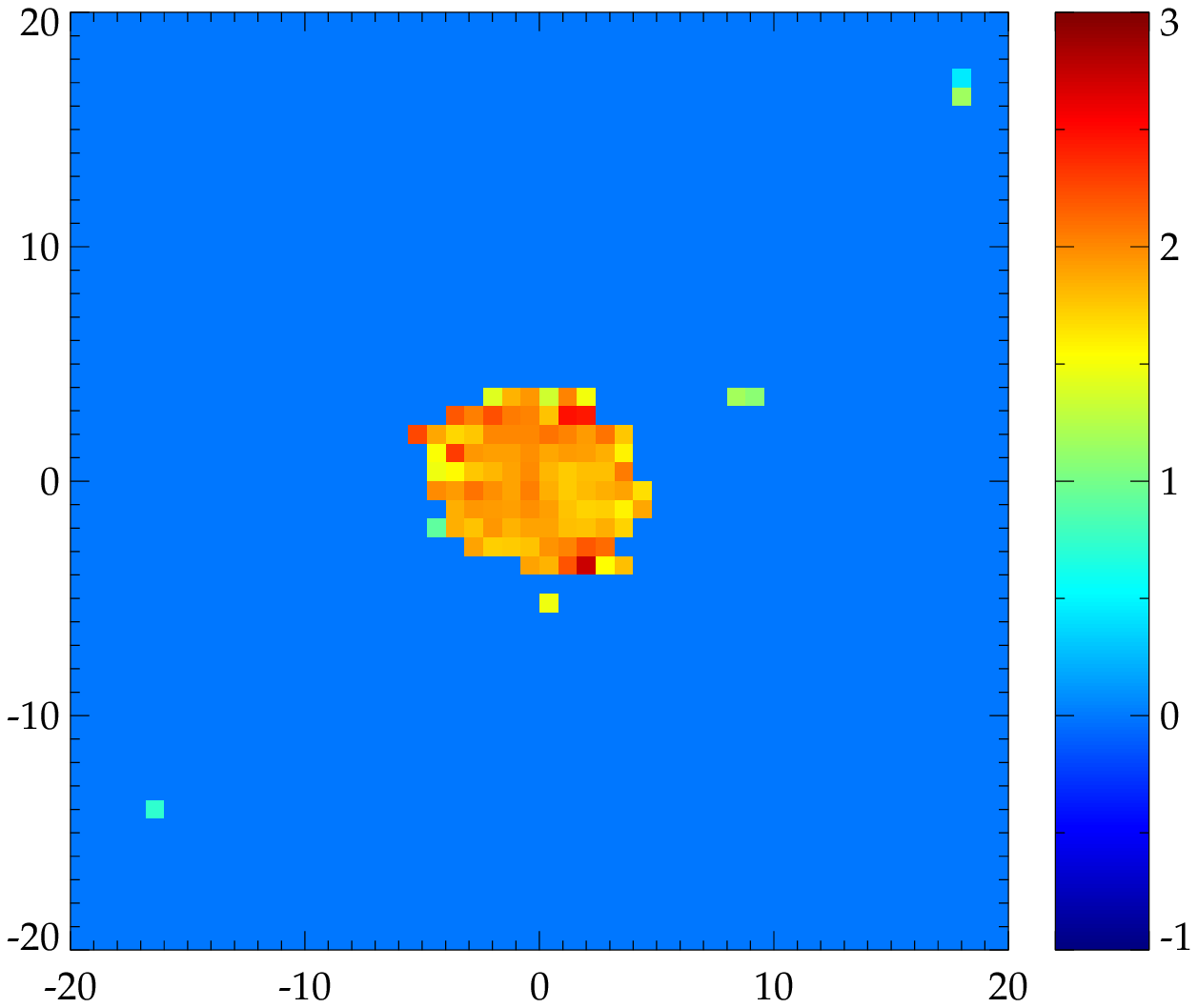}
\includegraphics[height=0.22\textwidth,clip]{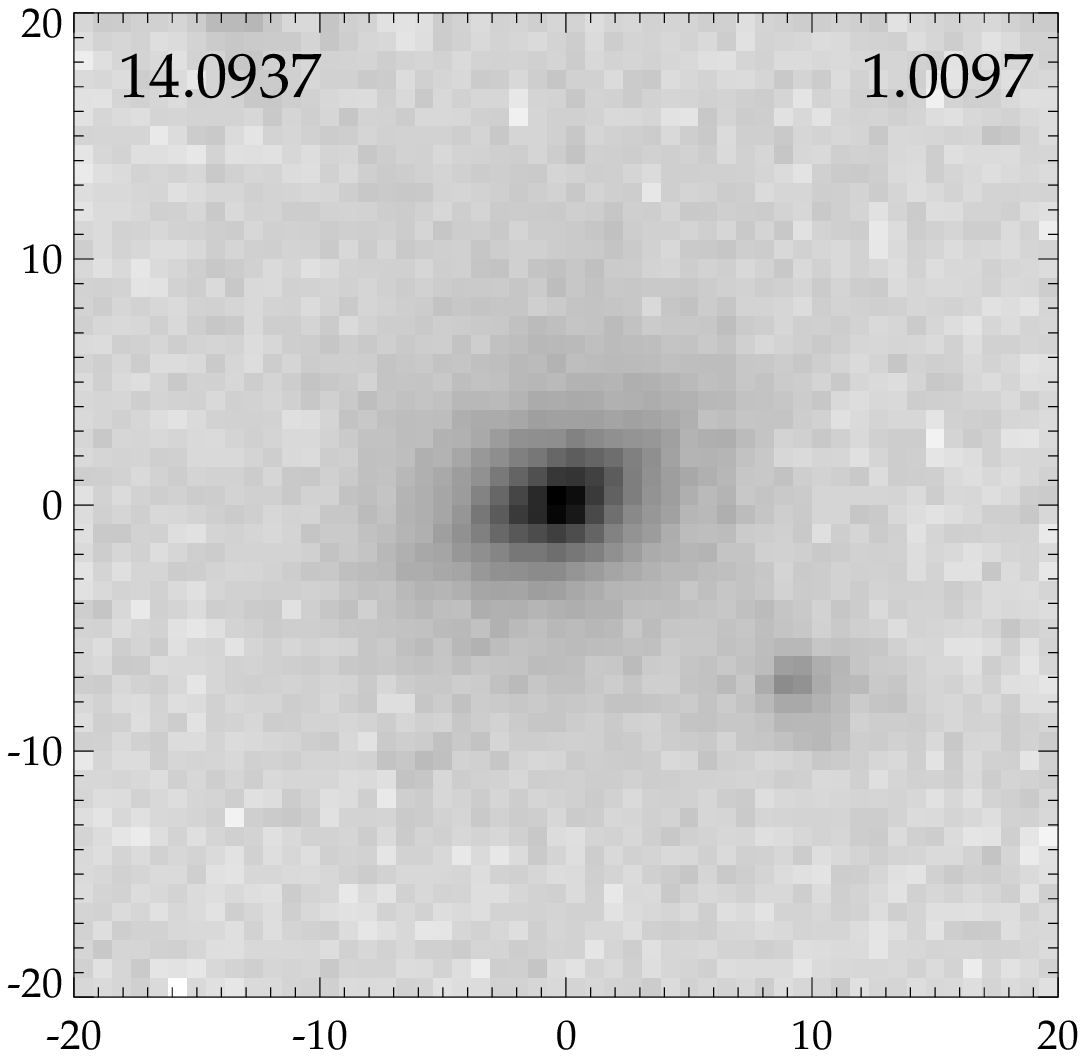} \includegraphics[height=0.22\textwidth,clip]{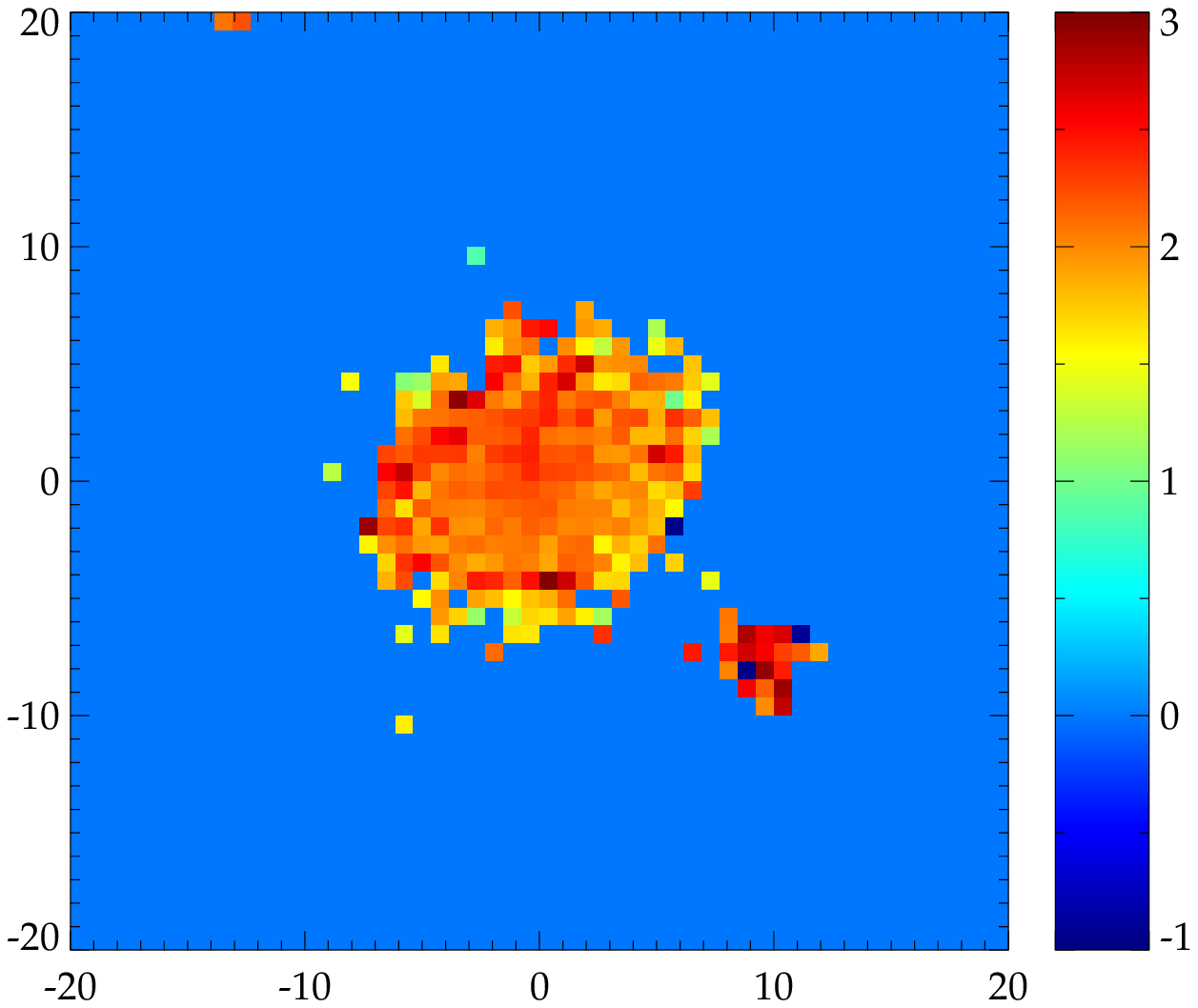}
\includegraphics[height=0.22\textwidth,clip]{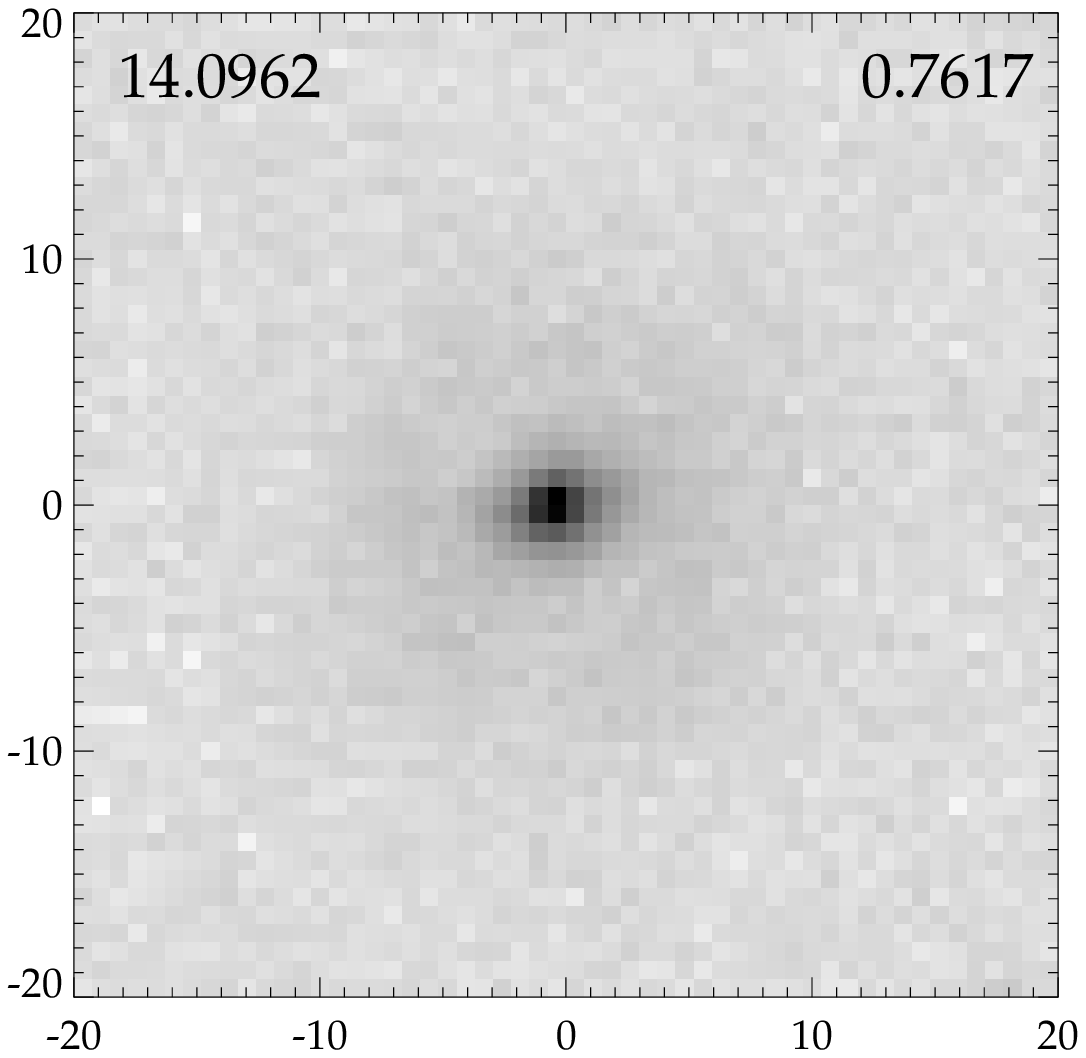} \includegraphics[height=0.22\textwidth,clip]{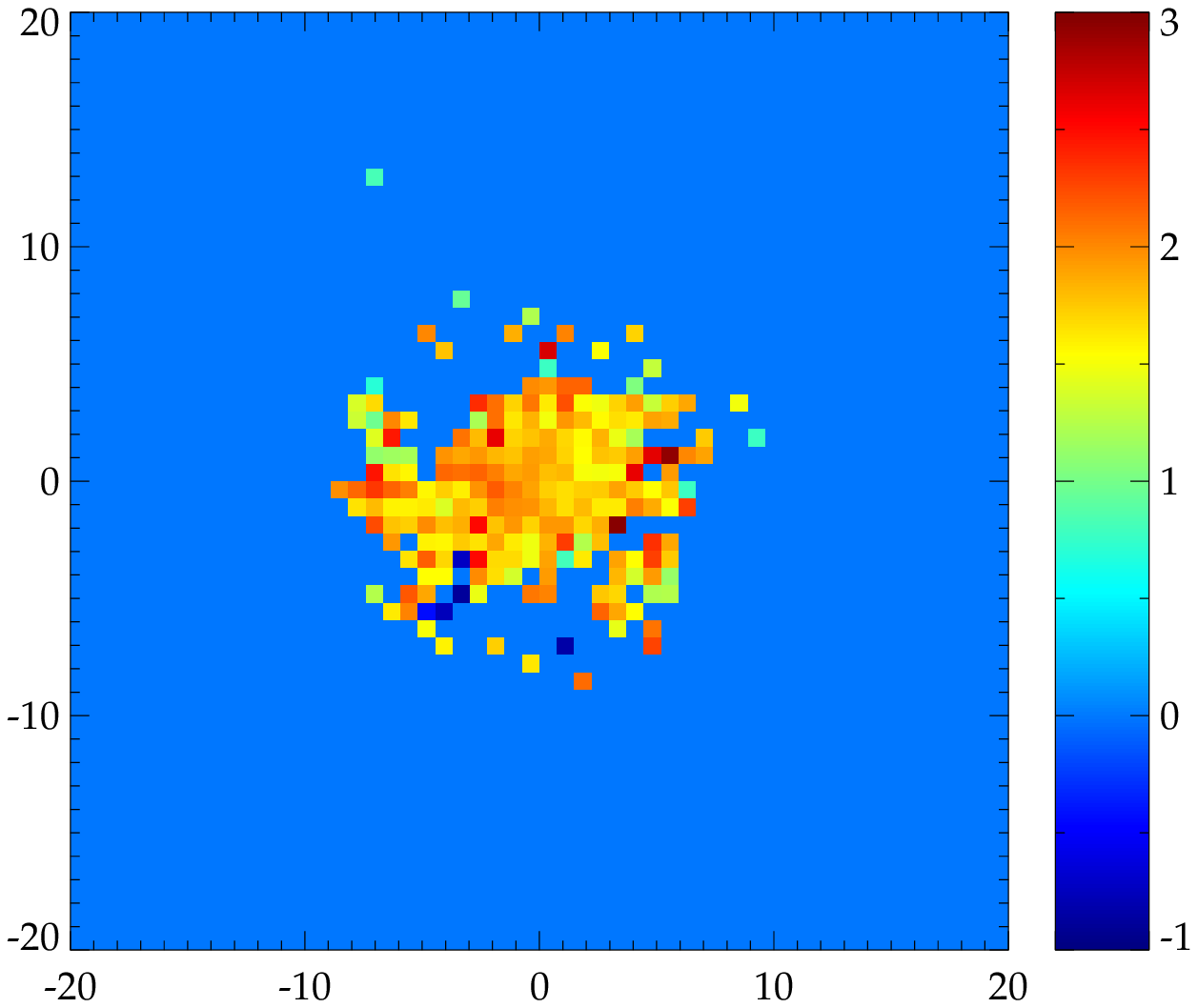}
\includegraphics[height=0.22\textwidth,clip]{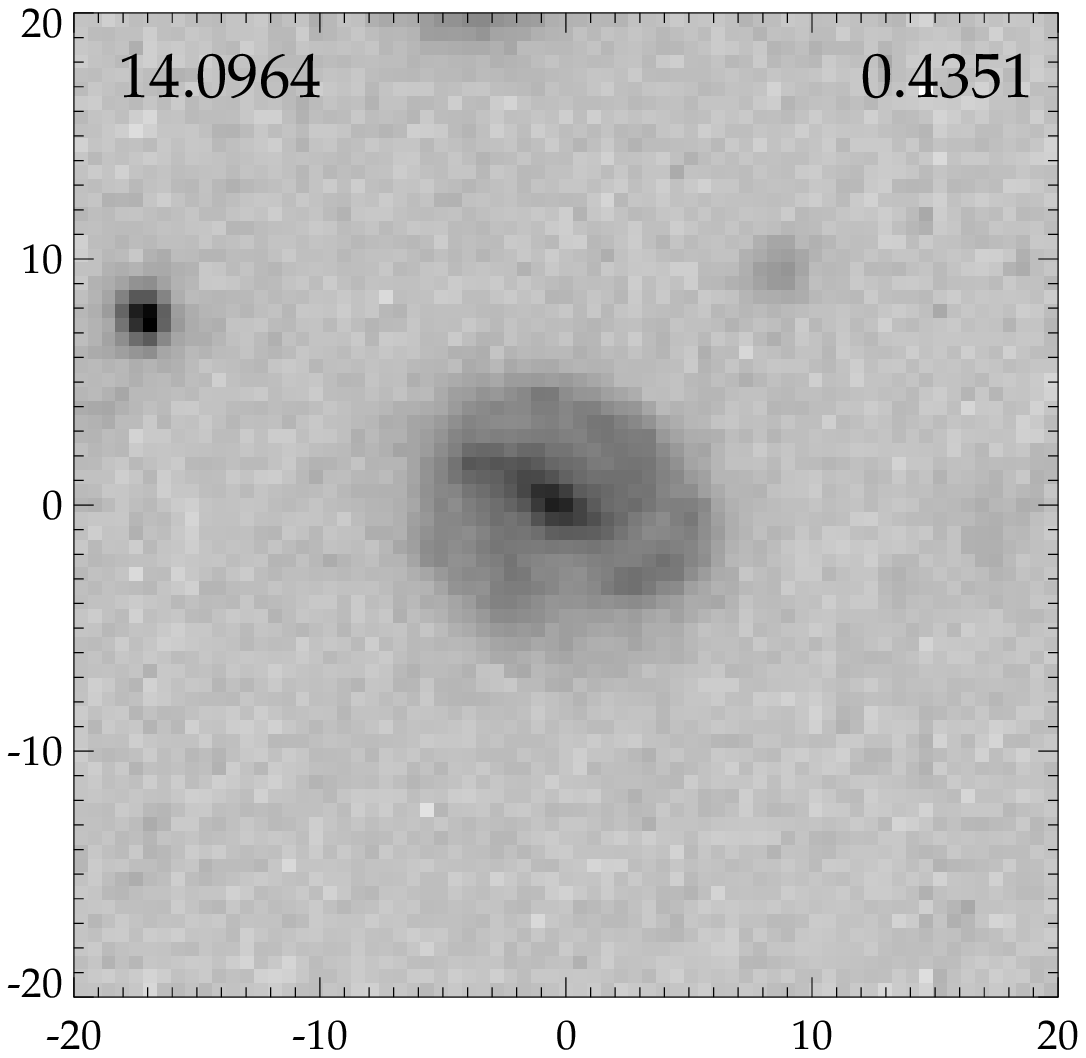} \includegraphics[height=0.22\textwidth,clip]{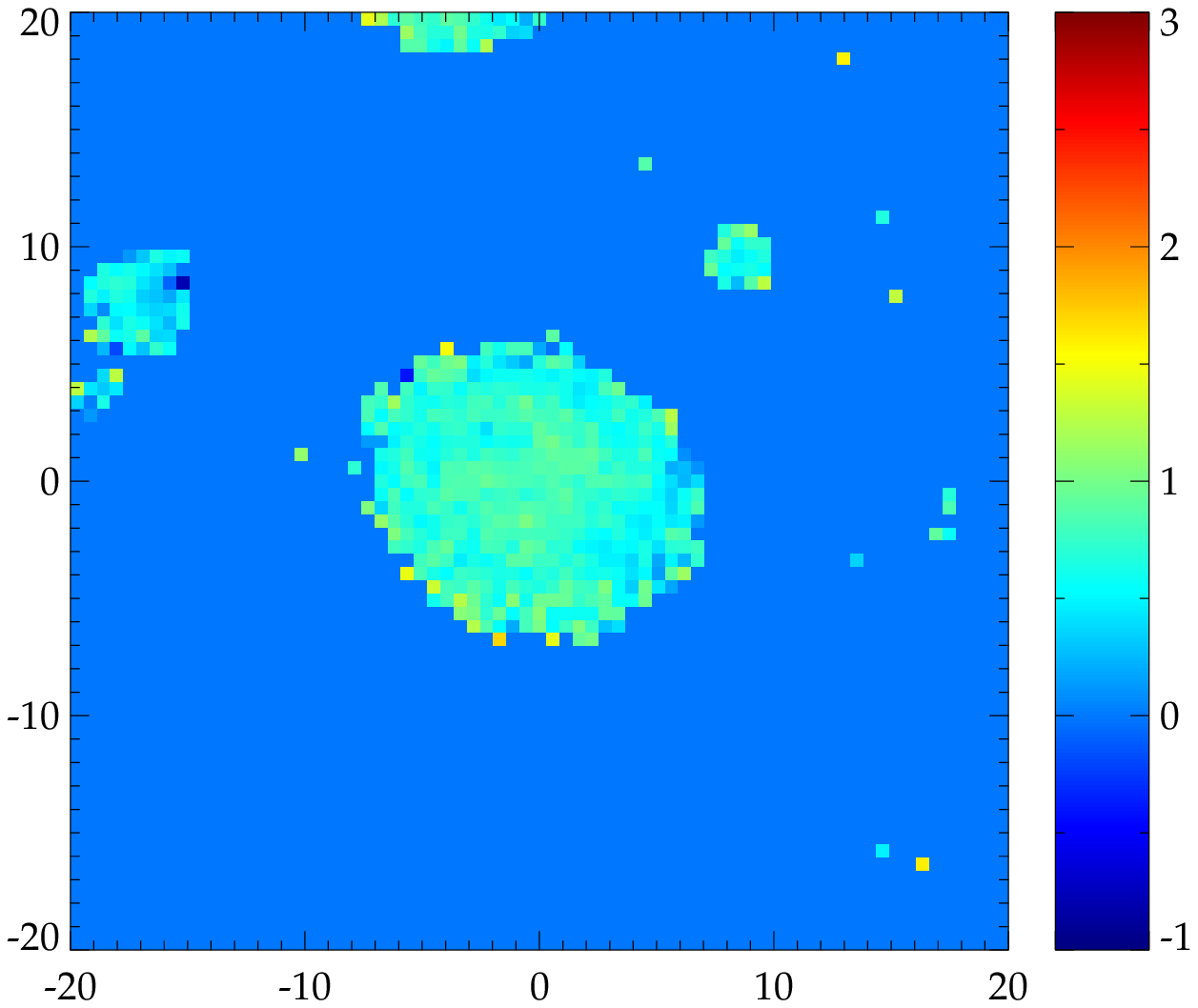}
\includegraphics[height=0.22\textwidth,clip]{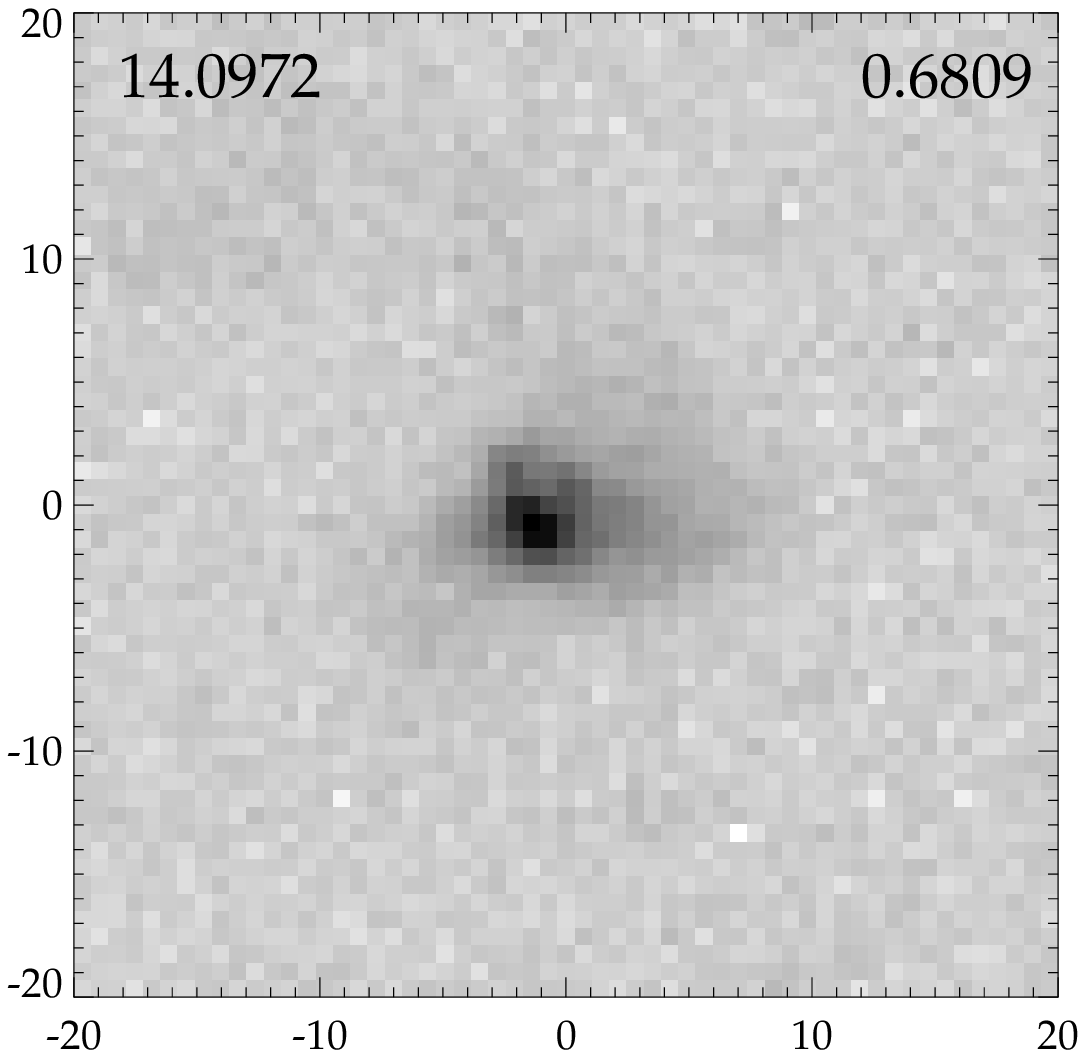} \includegraphics[height=0.22\textwidth,clip]{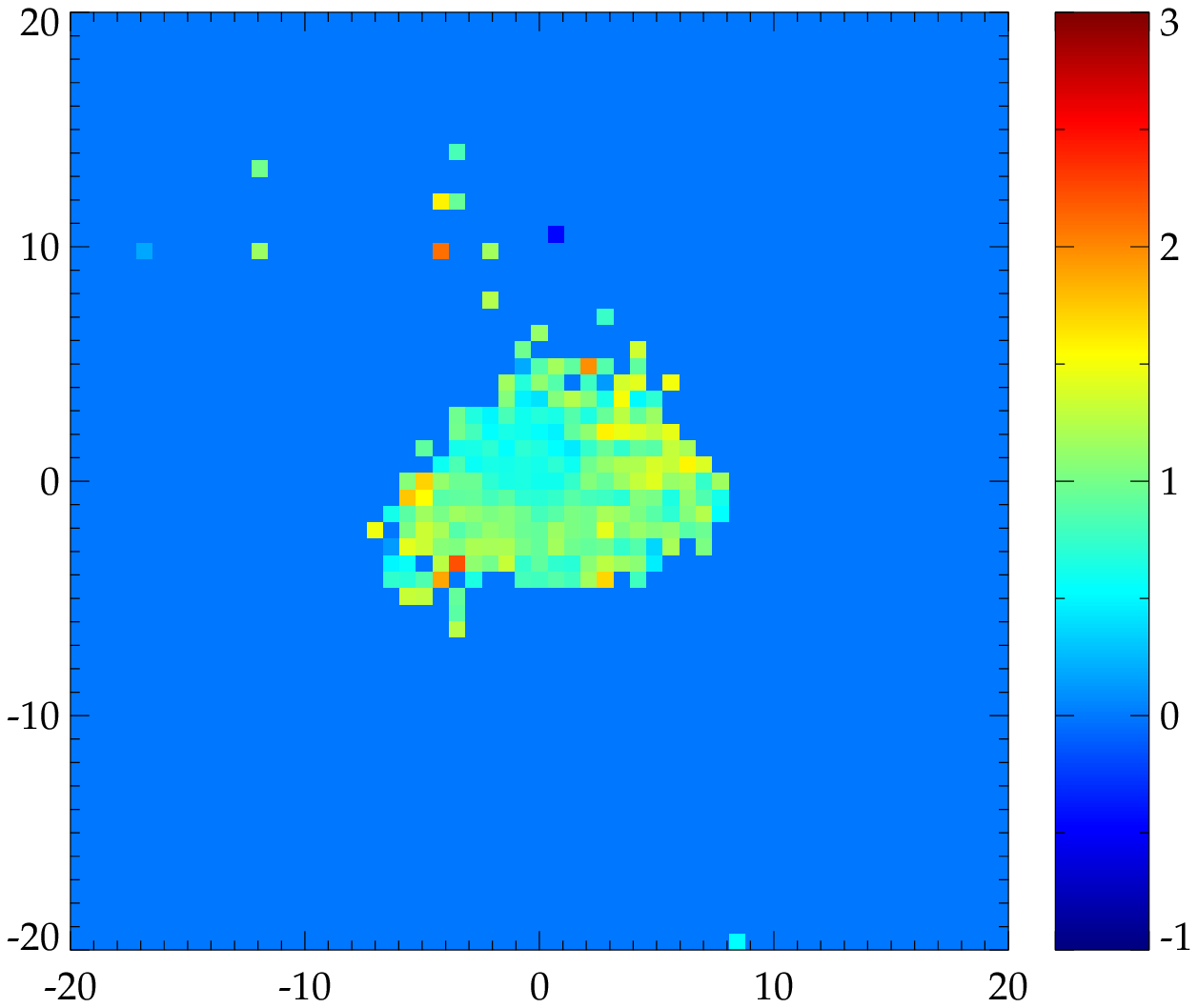}
\includegraphics[height=0.22\textwidth,clip]{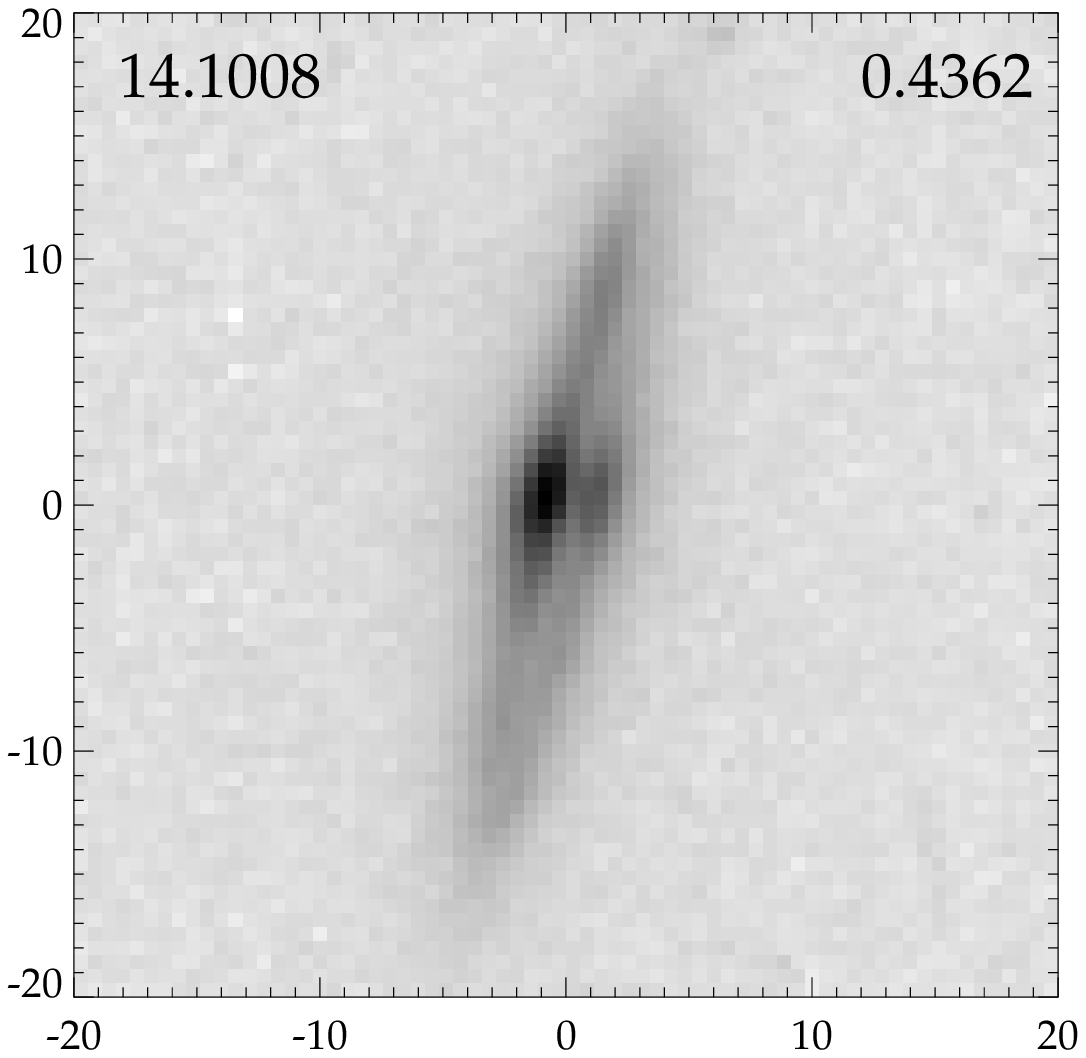} \includegraphics[height=0.22\textwidth,clip]{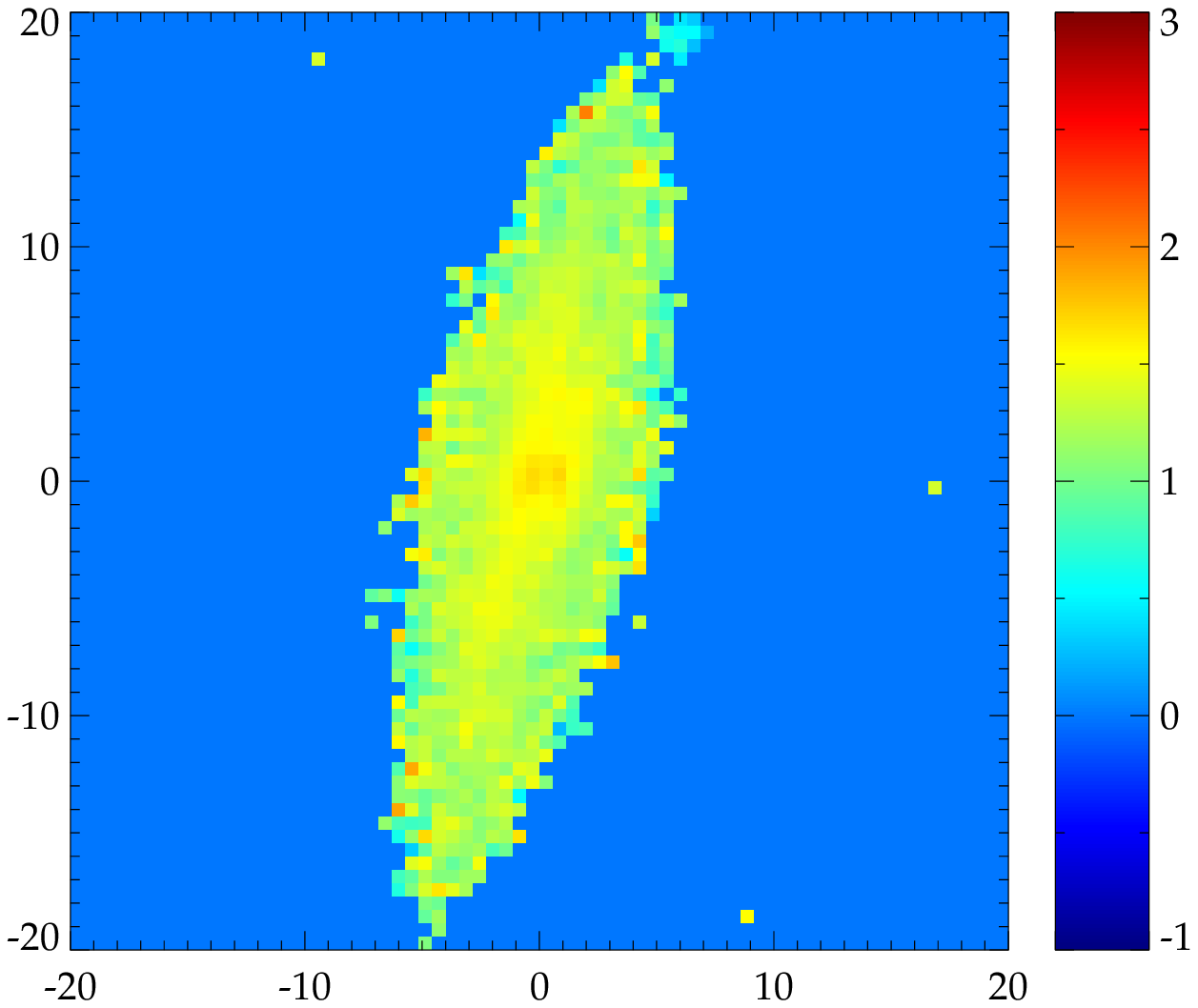}
\includegraphics[height=0.22\textwidth,clip]{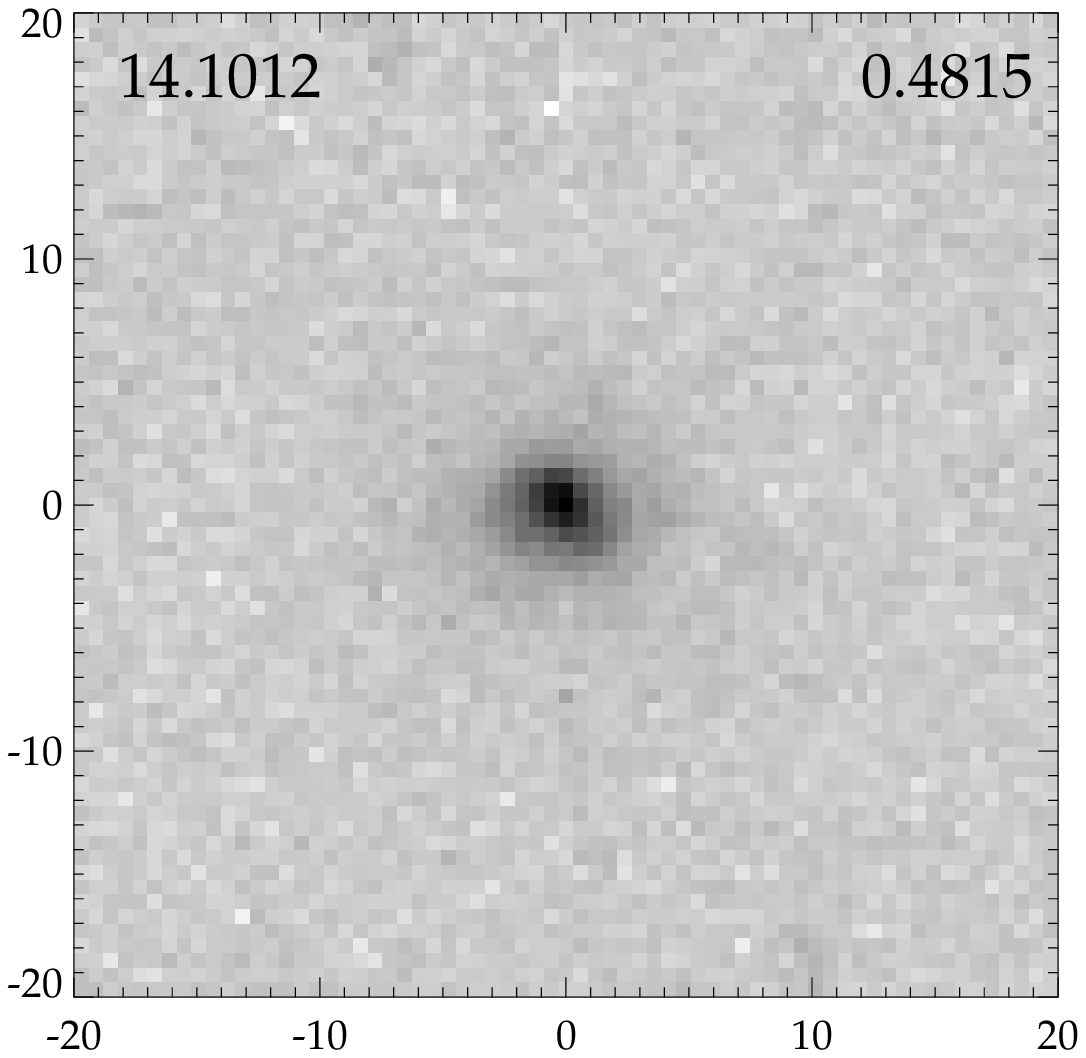} \includegraphics[height=0.22\textwidth,clip]{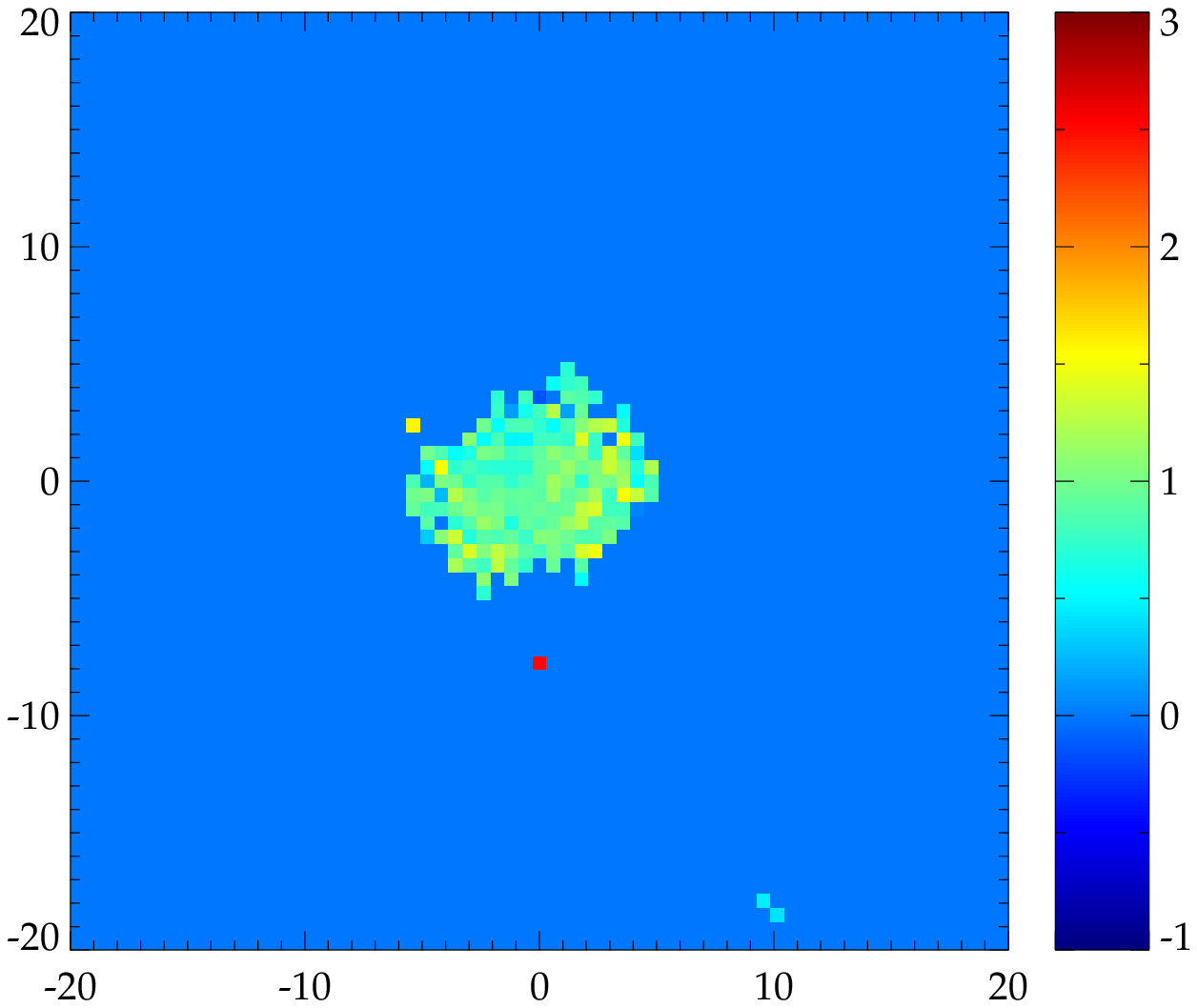}
\includegraphics[height=0.22\textwidth,clip]{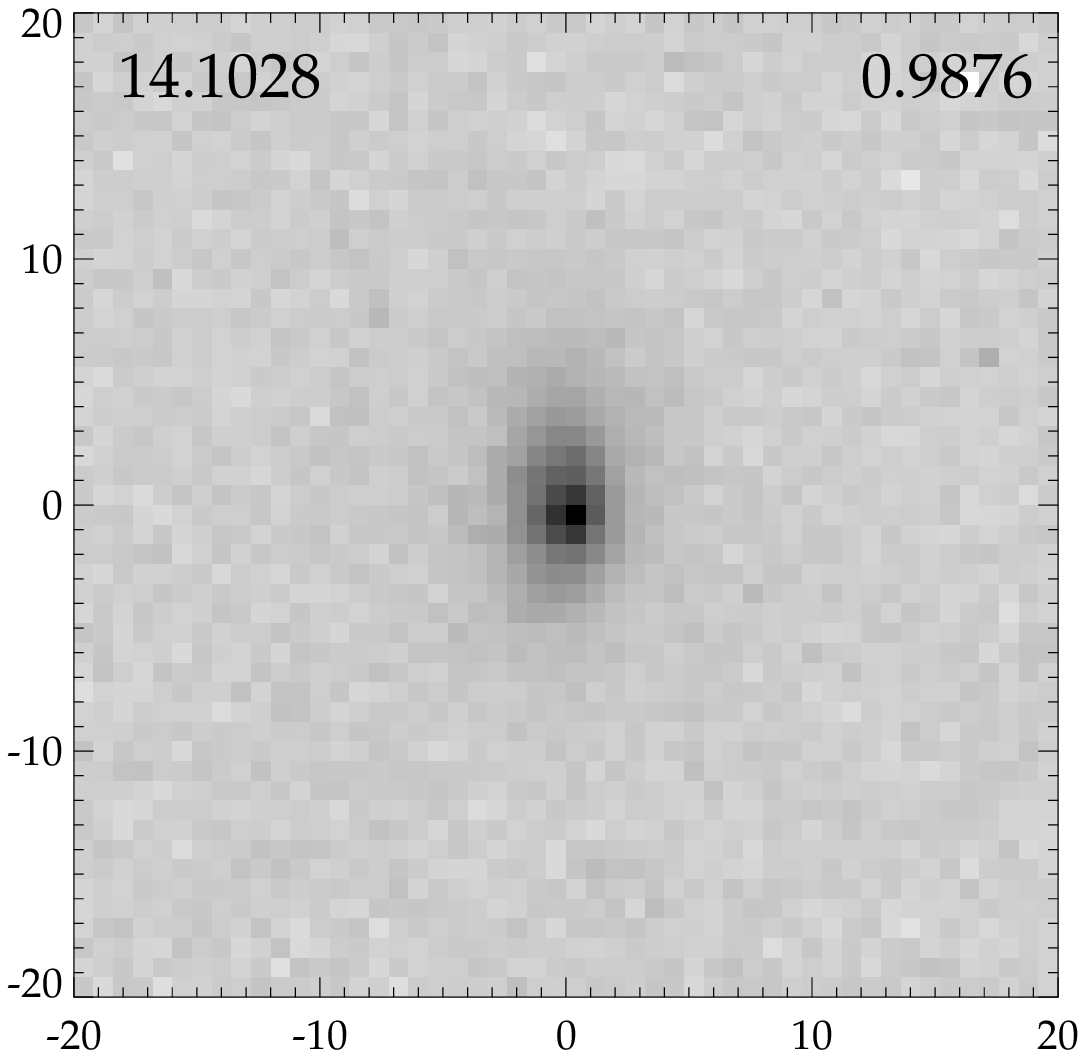} \includegraphics[height=0.22\textwidth,clip]{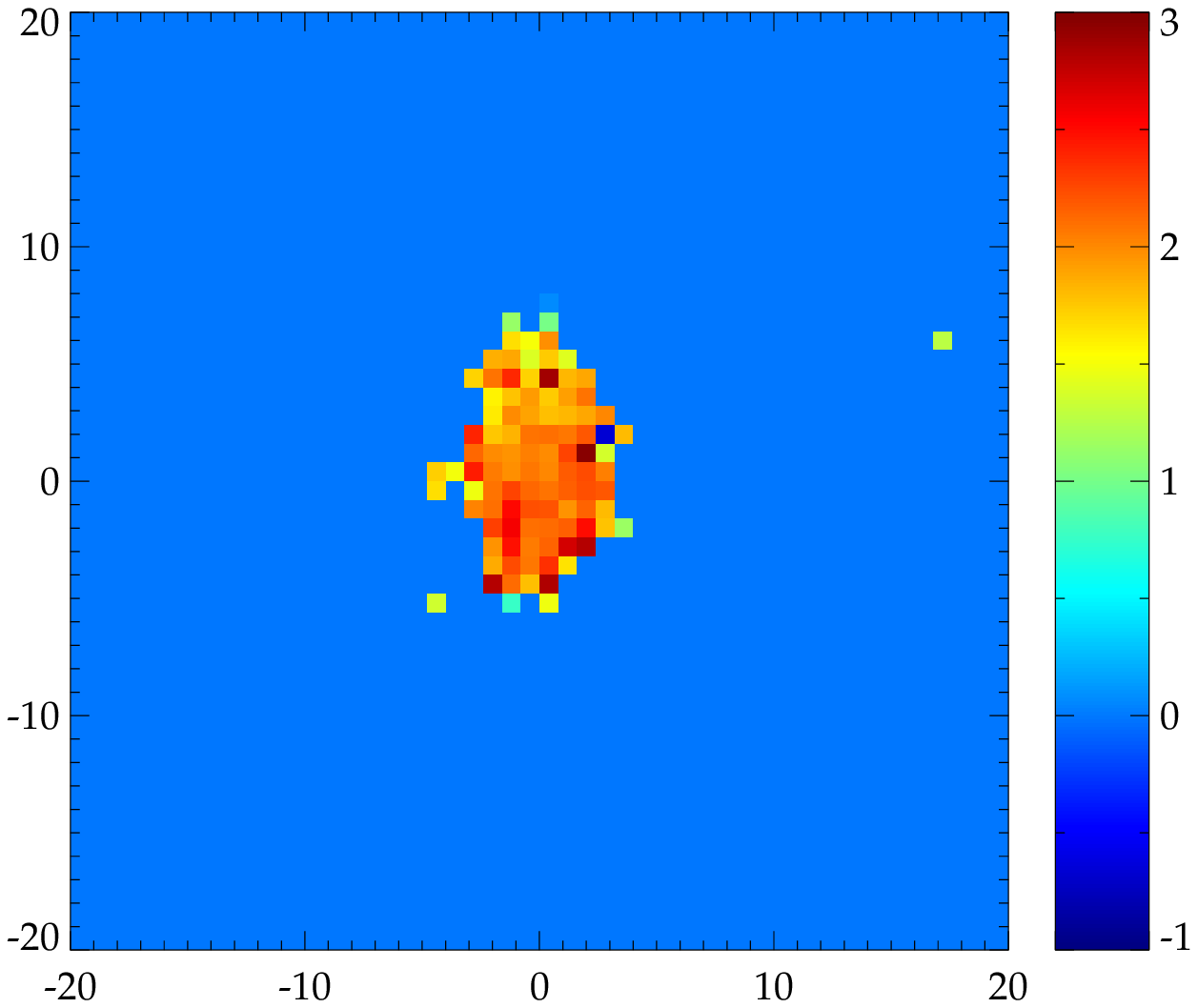}
\includegraphics[height=0.22\textwidth,clip]{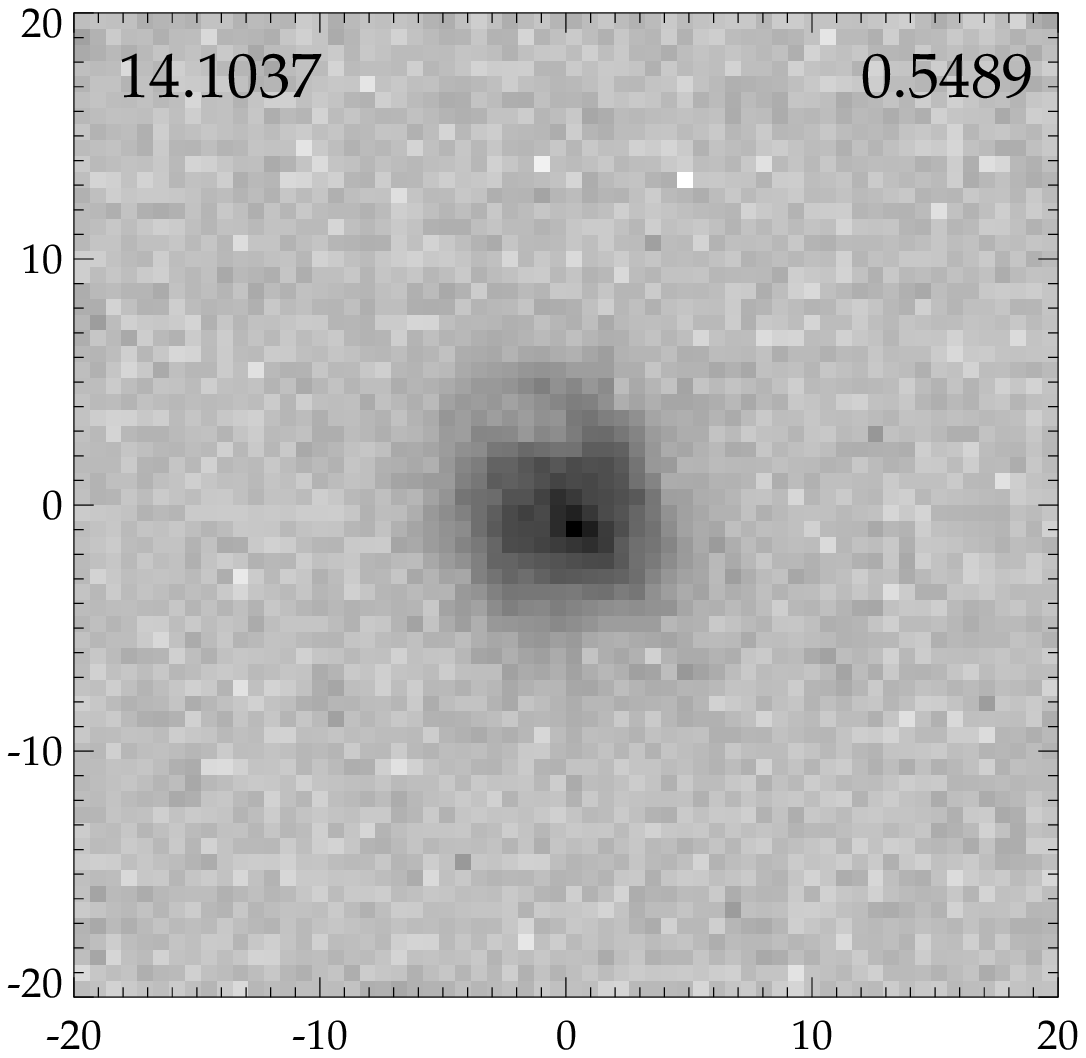} \includegraphics[height=0.22\textwidth,clip]{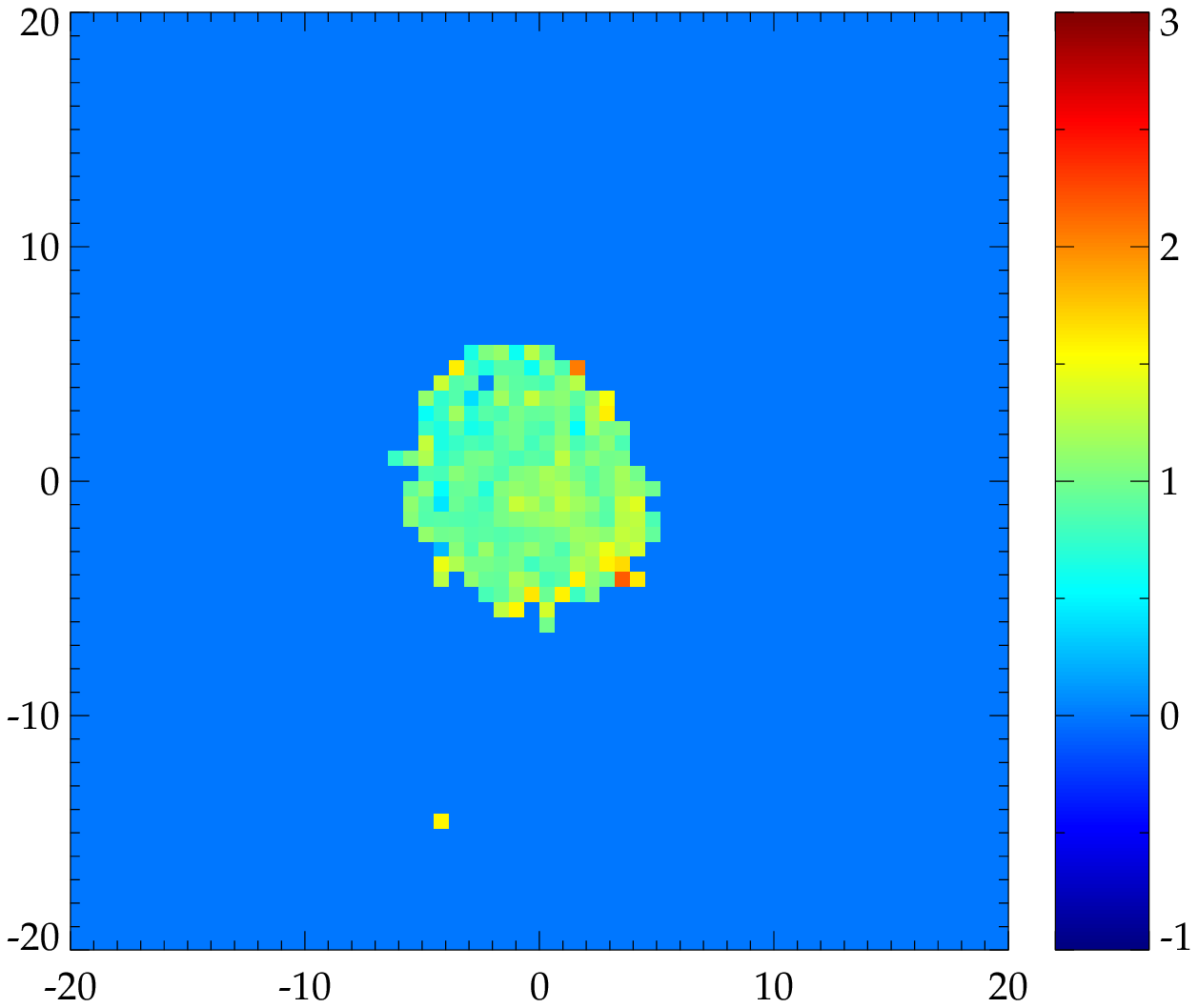}
\includegraphics[height=0.22\textwidth,clip]{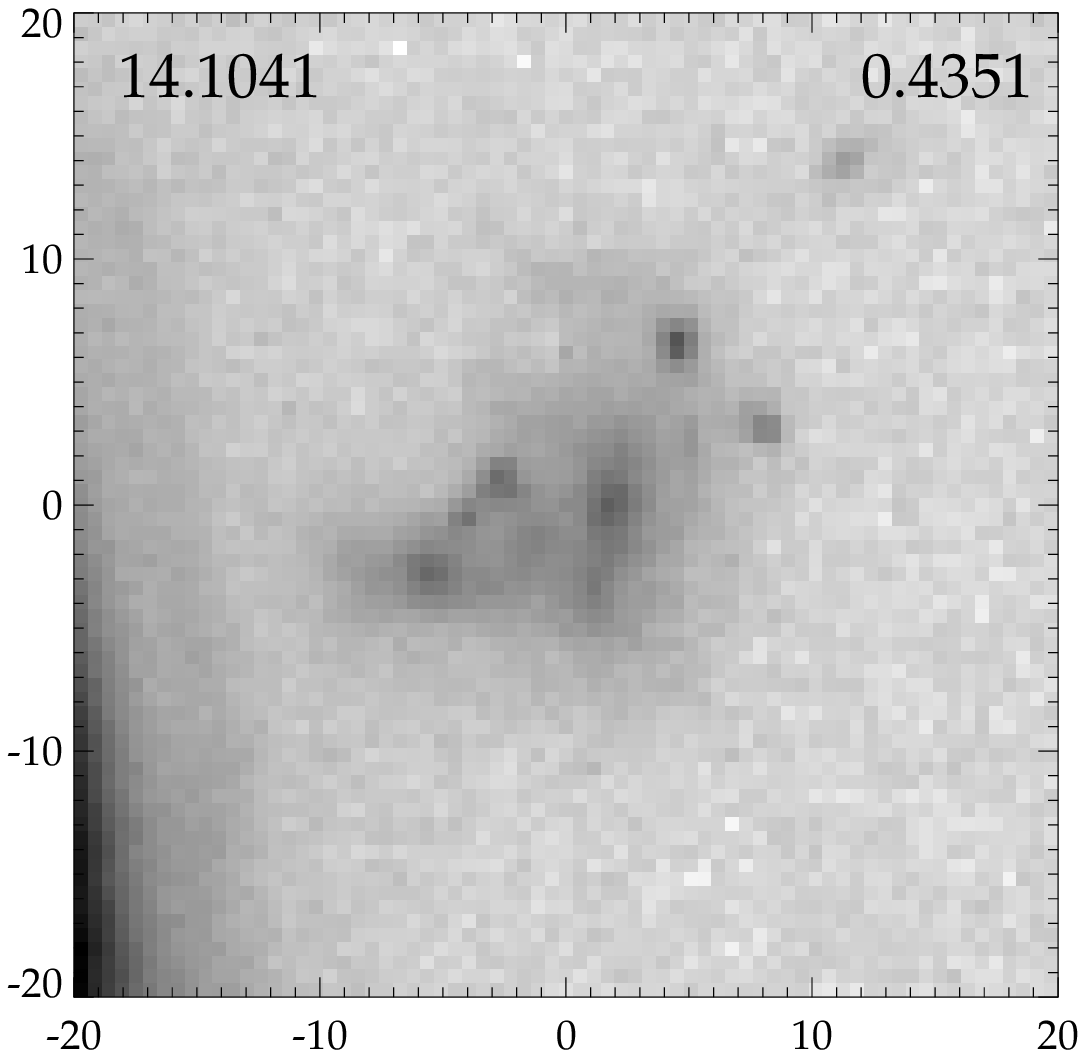} \includegraphics[height=0.22\textwidth,clip]{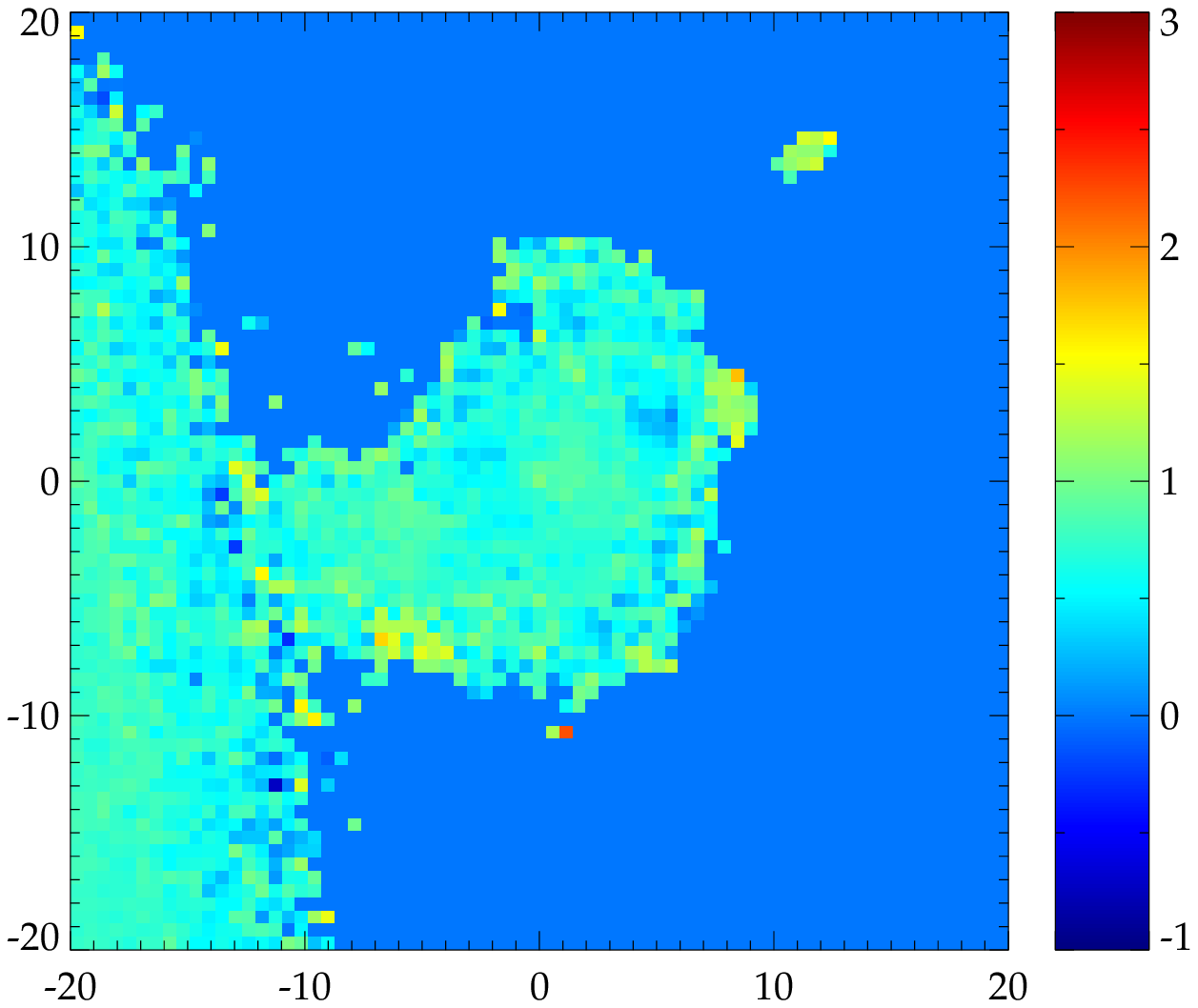}
\includegraphics[height=0.22\textwidth,clip]{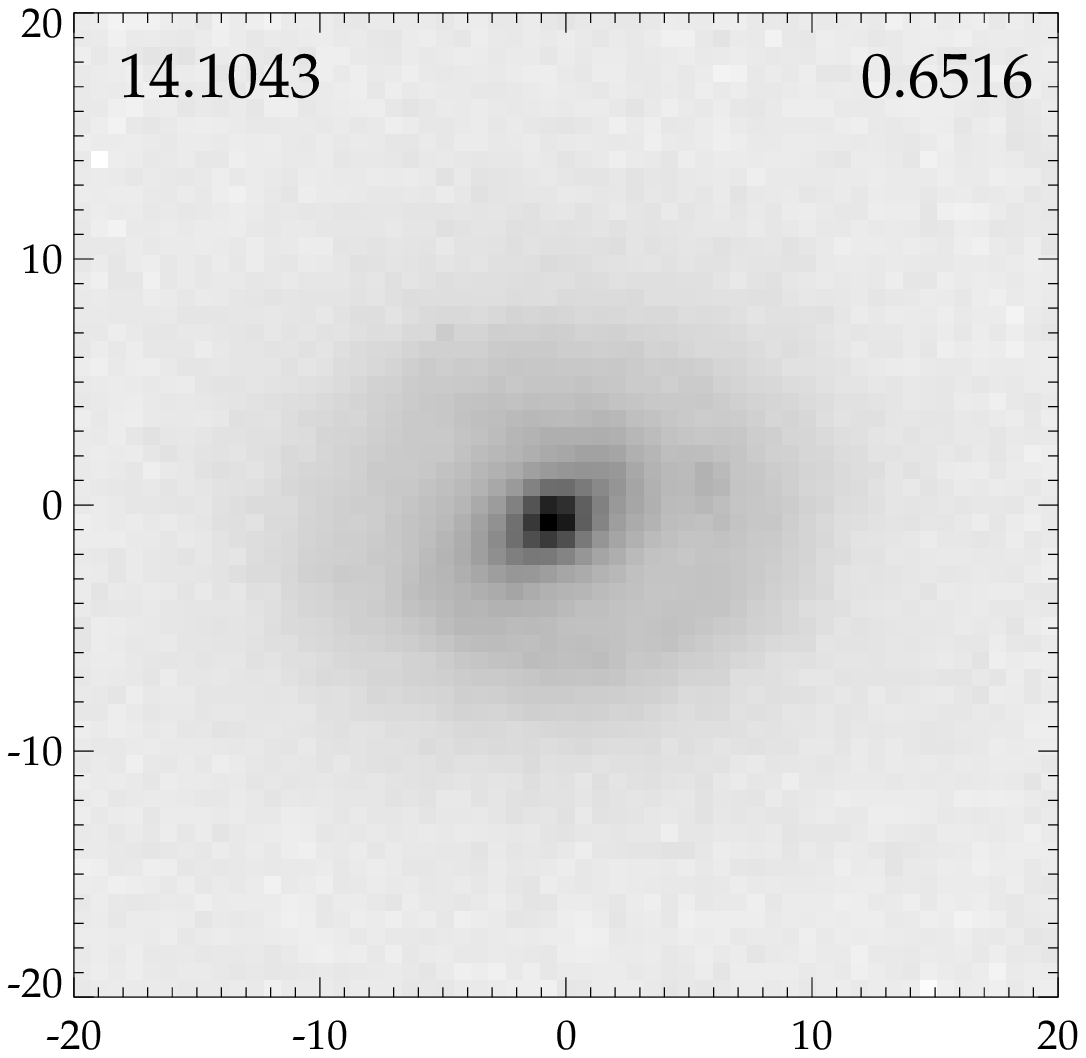} \includegraphics[height=0.22\textwidth,clip]{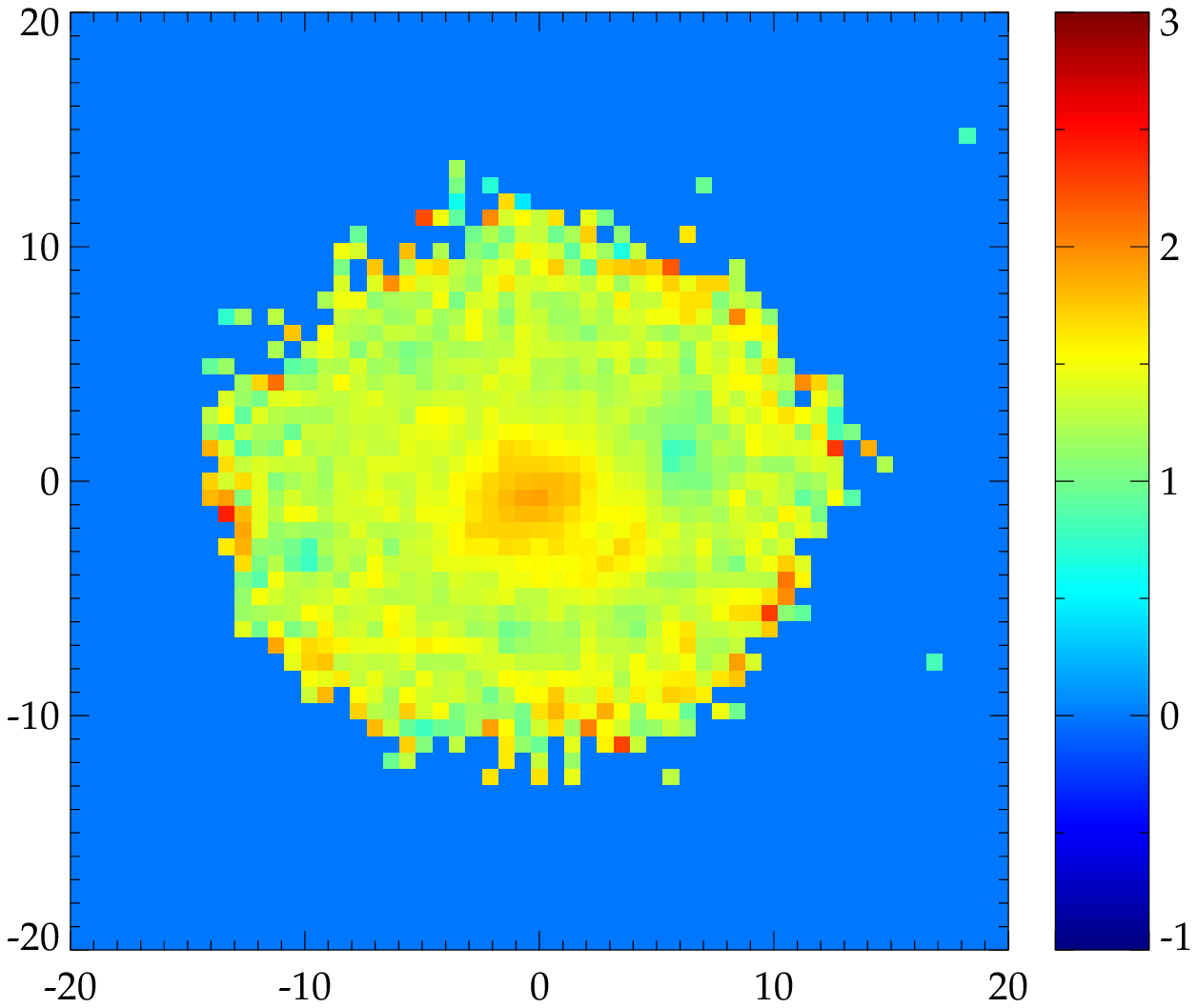}
\includegraphics[height=0.22\textwidth,clip]{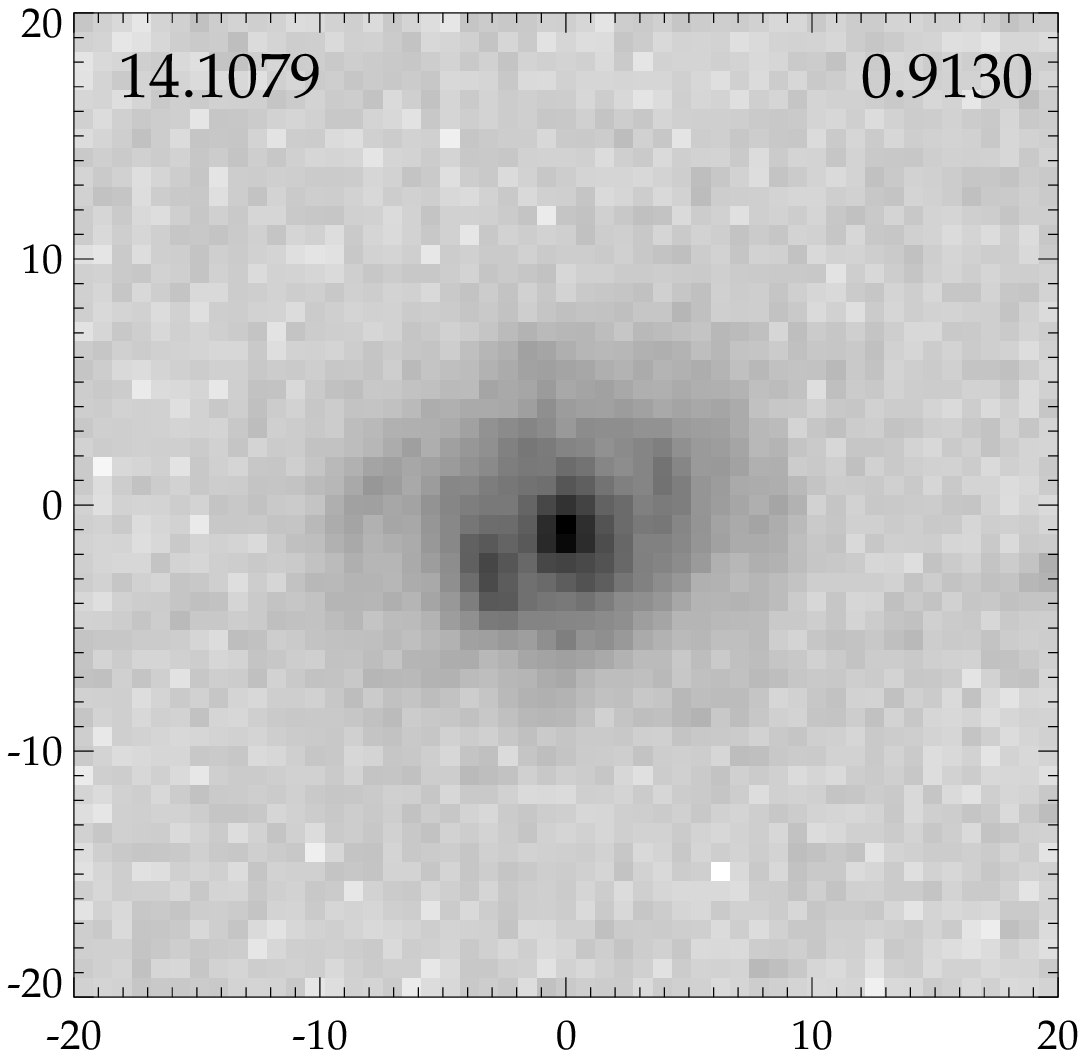} \includegraphics[height=0.22\textwidth,clip]{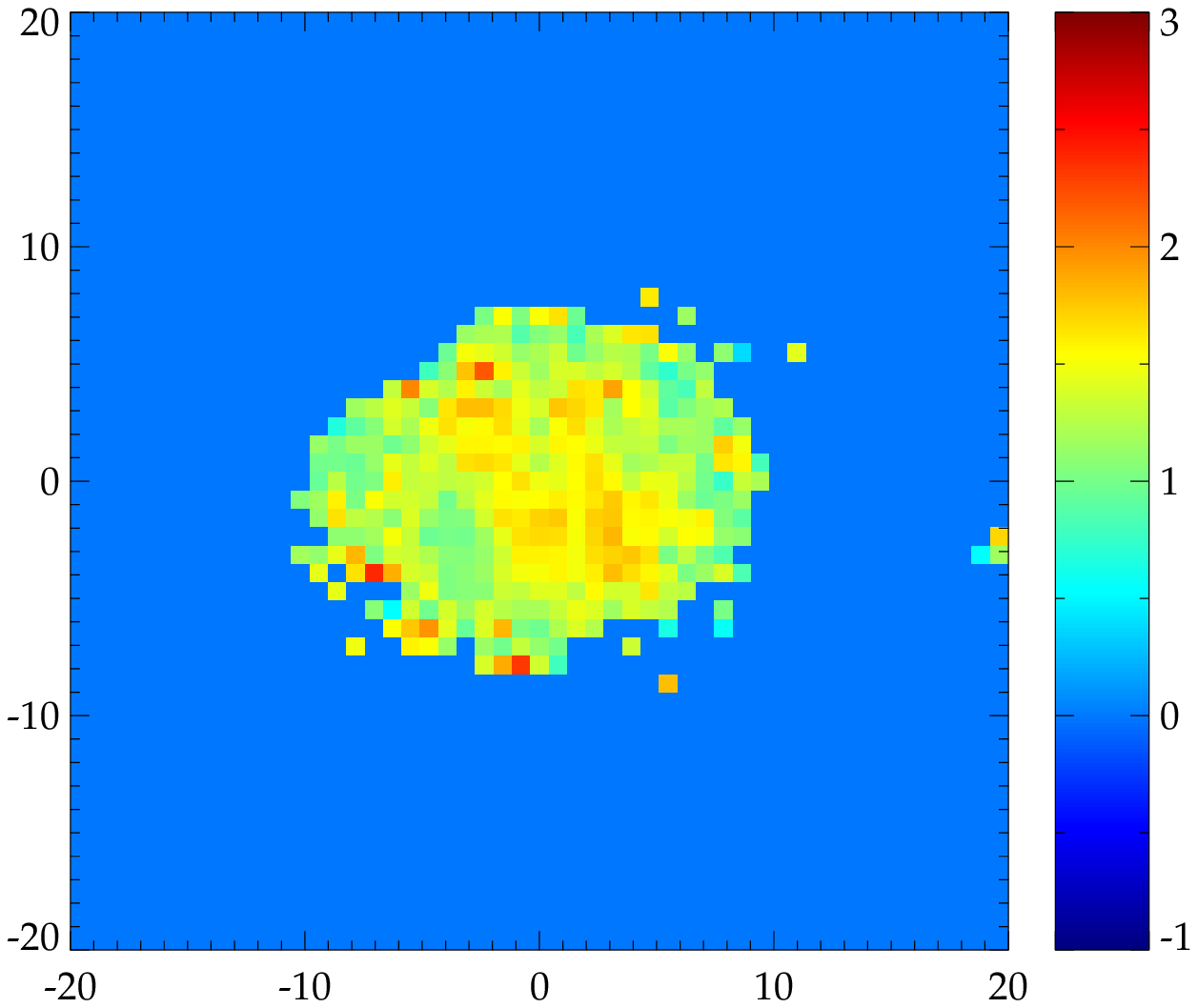}
\caption{Continued.} \end{figure*}

\addtocounter{figure}{-1}
\begin{figure*} \centering

\includegraphics[height=0.22\textwidth,clip]{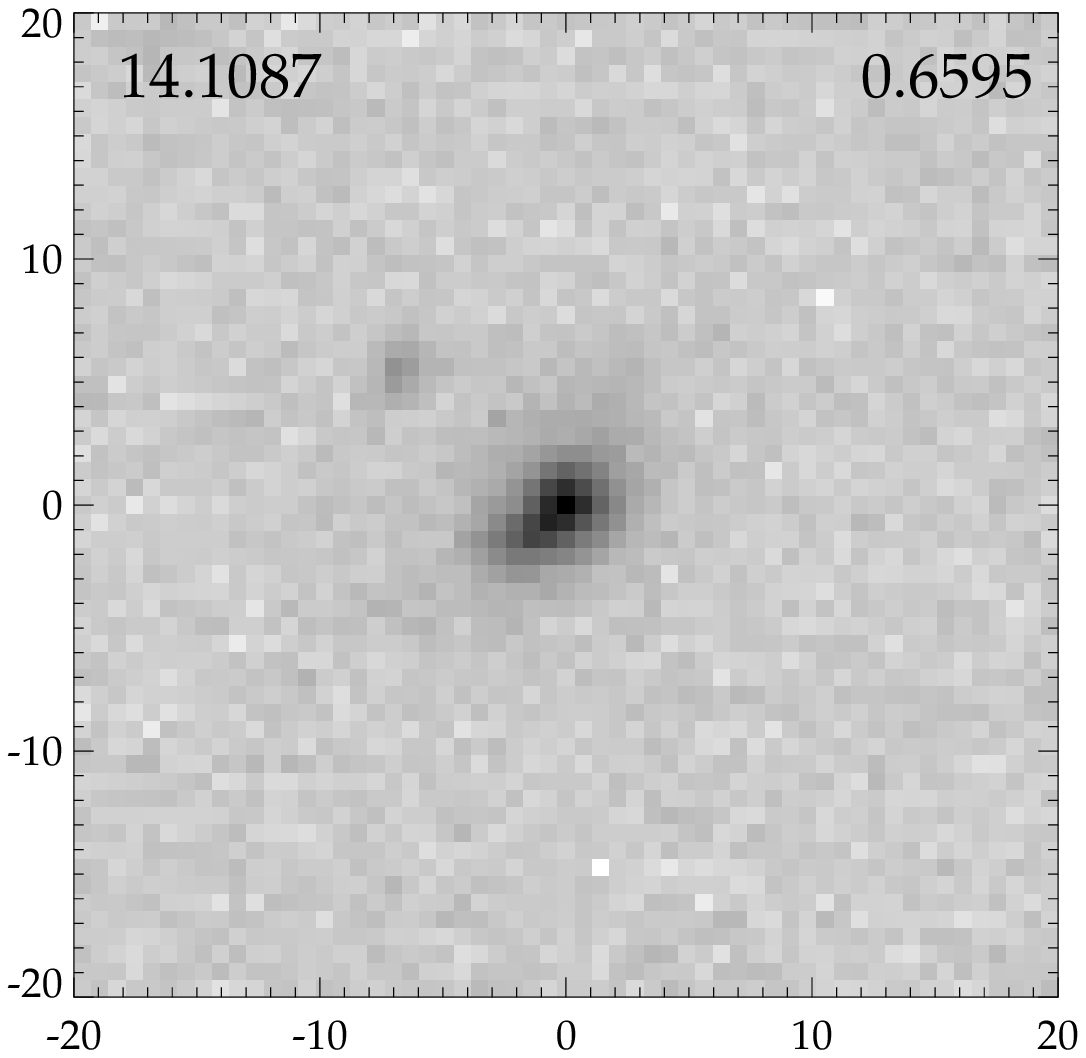} \includegraphics[height=0.22\textwidth,clip]{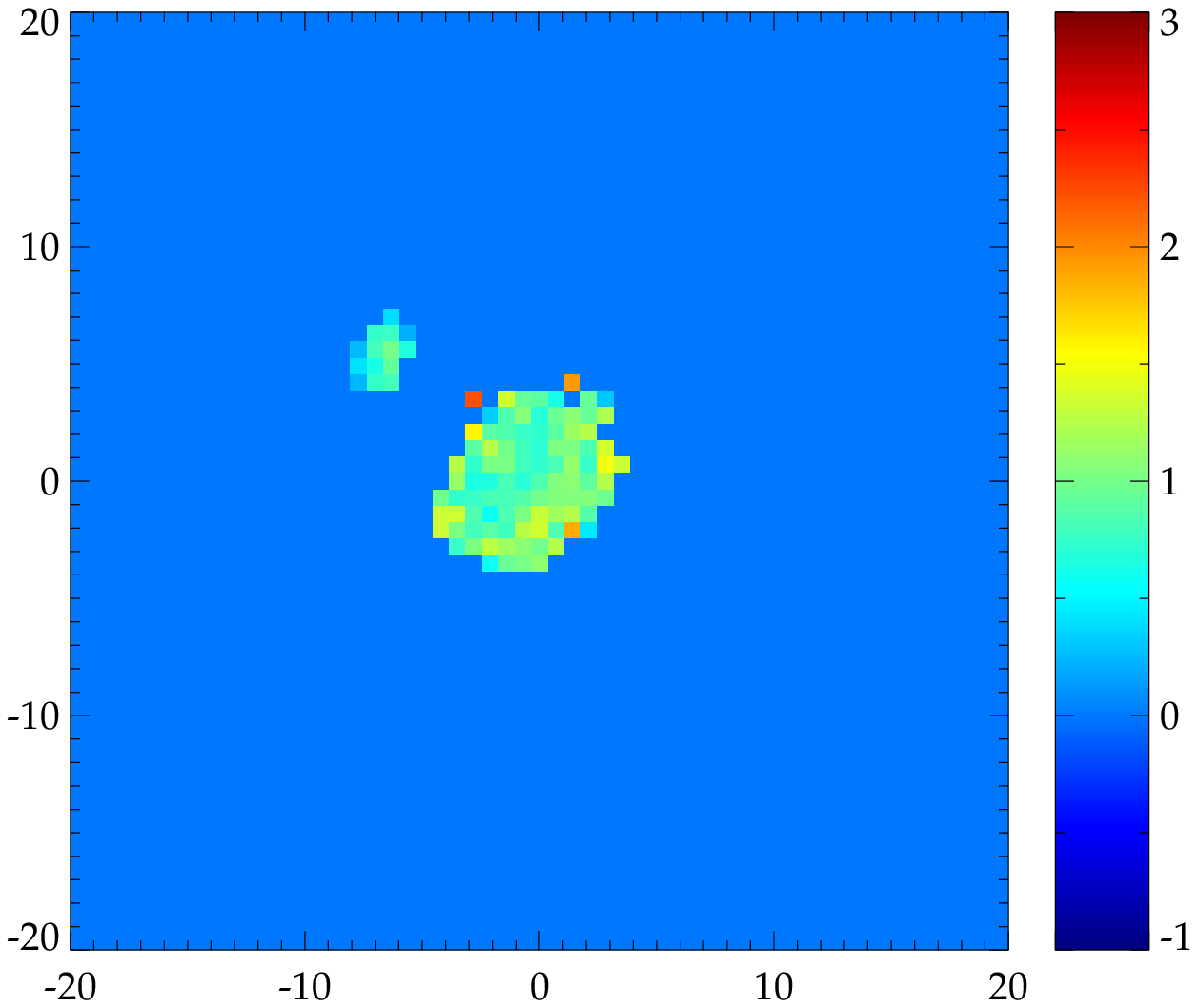}
\includegraphics[height=0.22\textwidth,clip]{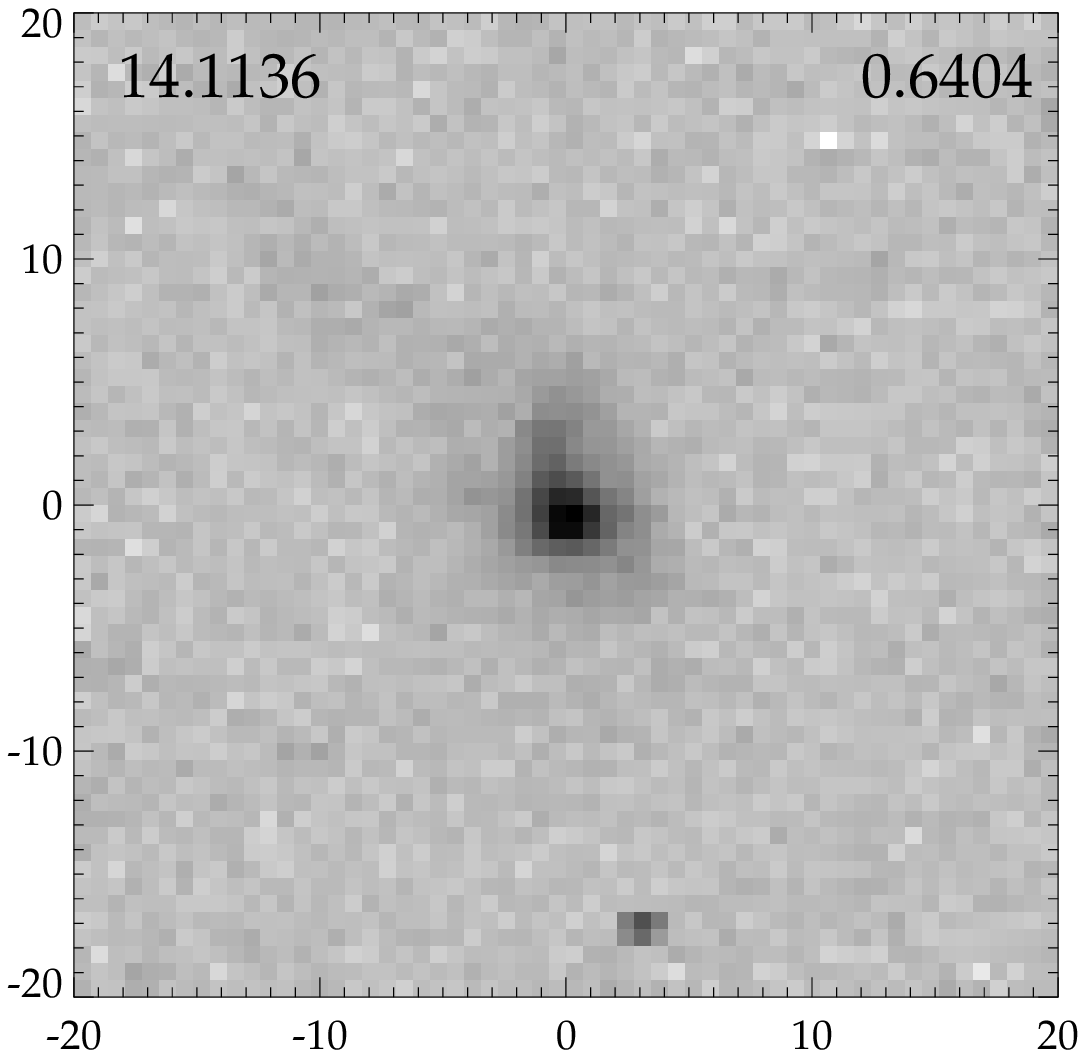} \includegraphics[height=0.22\textwidth,clip]{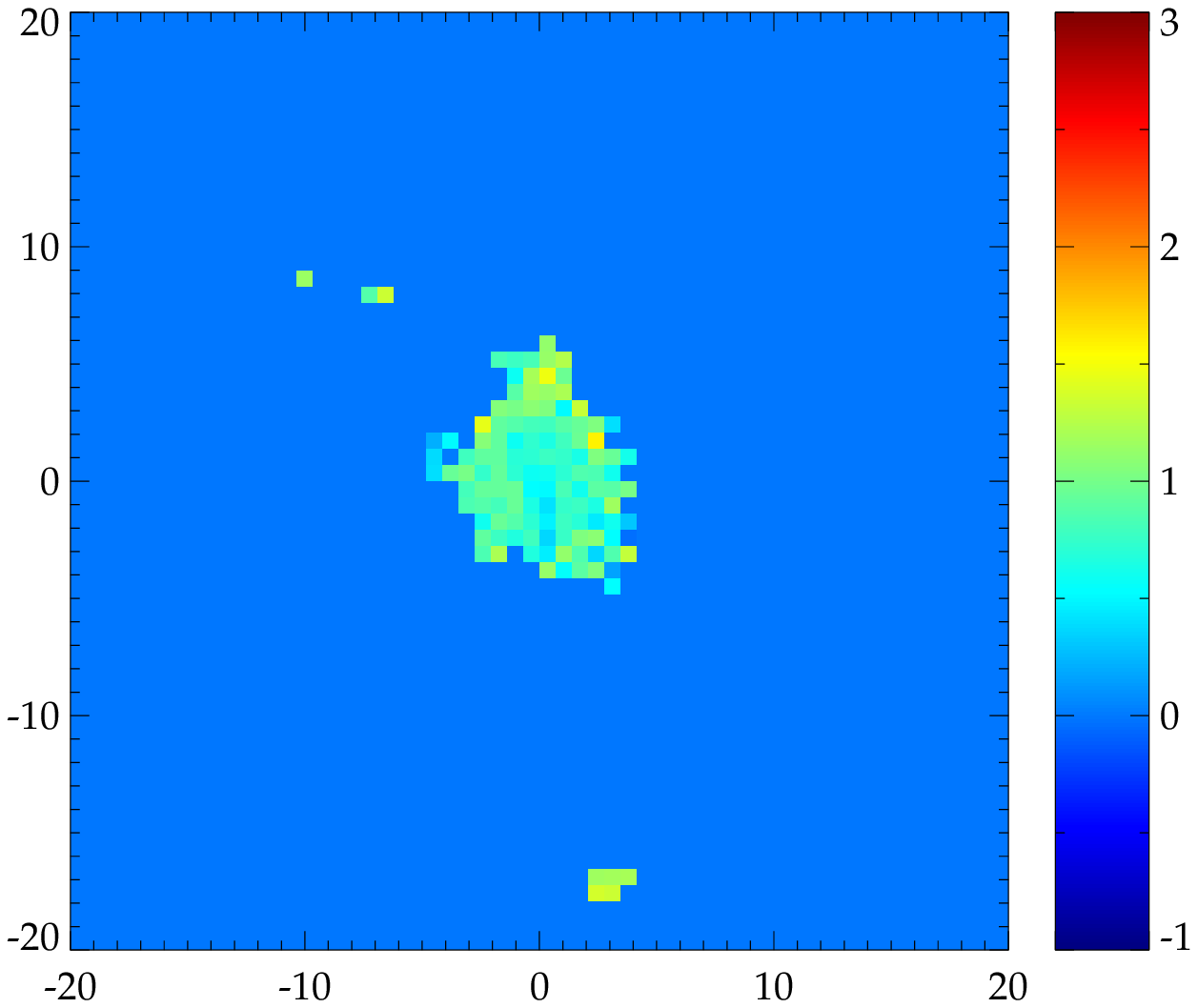}
\includegraphics[height=0.22\textwidth,clip]{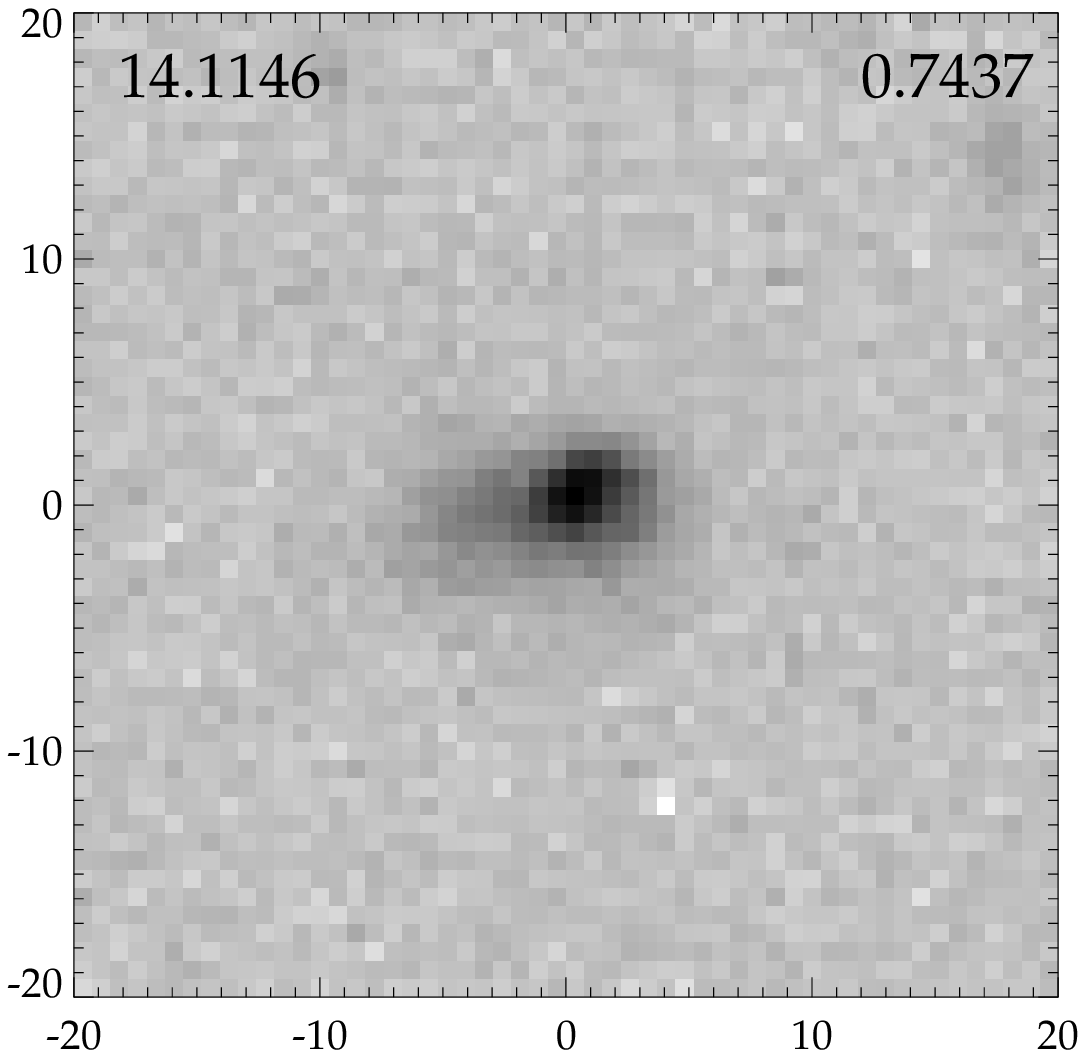} \includegraphics[height=0.22\textwidth,clip]{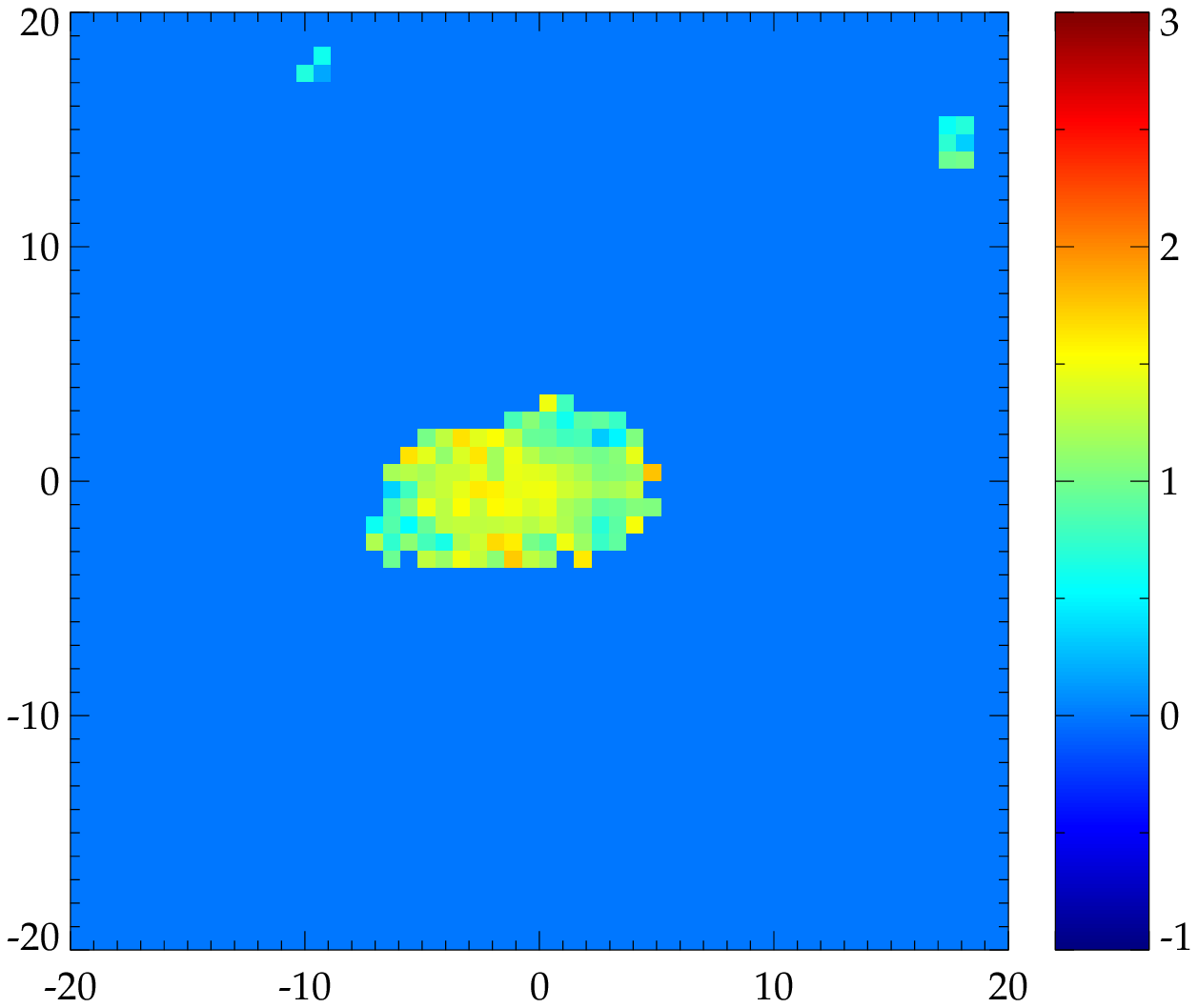}
\includegraphics[height=0.22\textwidth,clip]{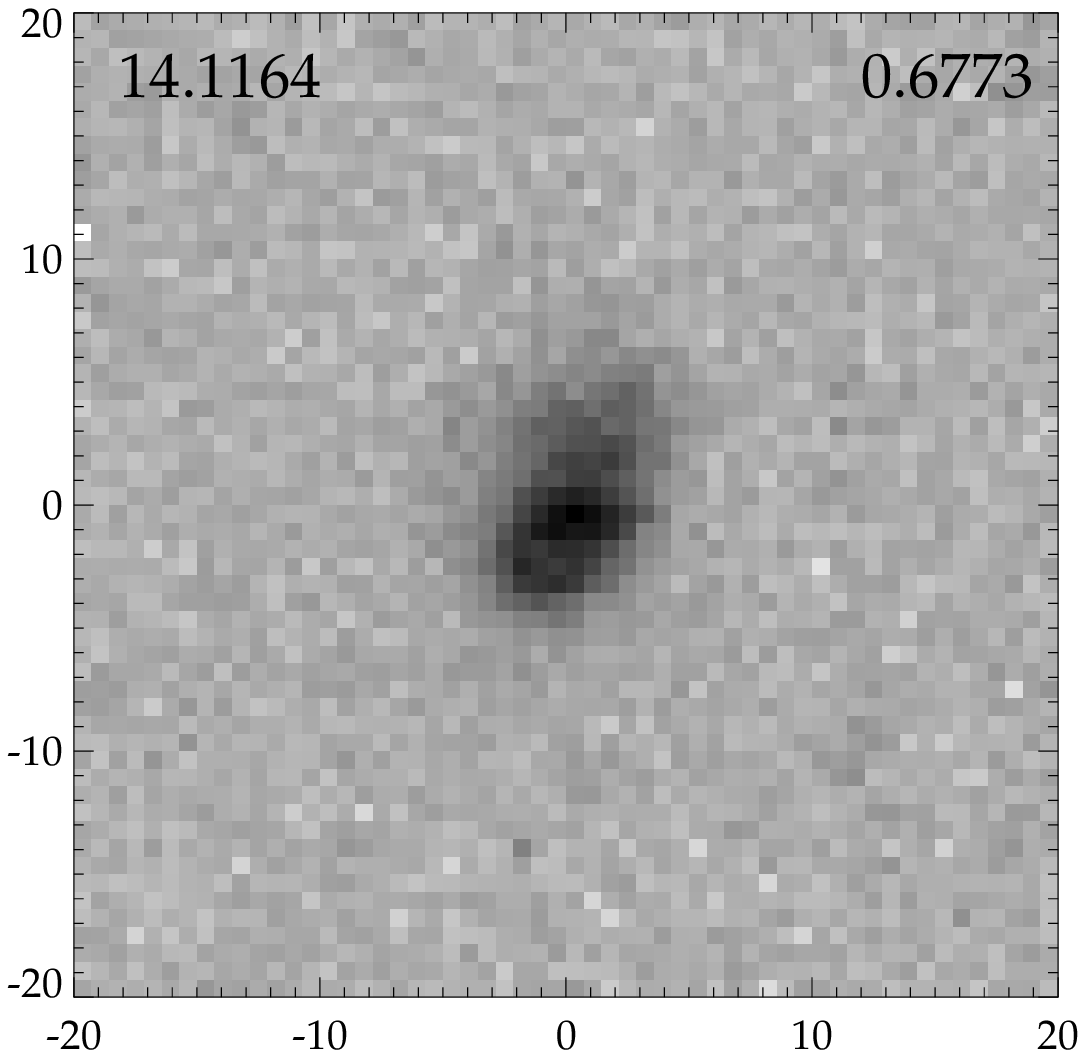} \includegraphics[height=0.22\textwidth,clip]{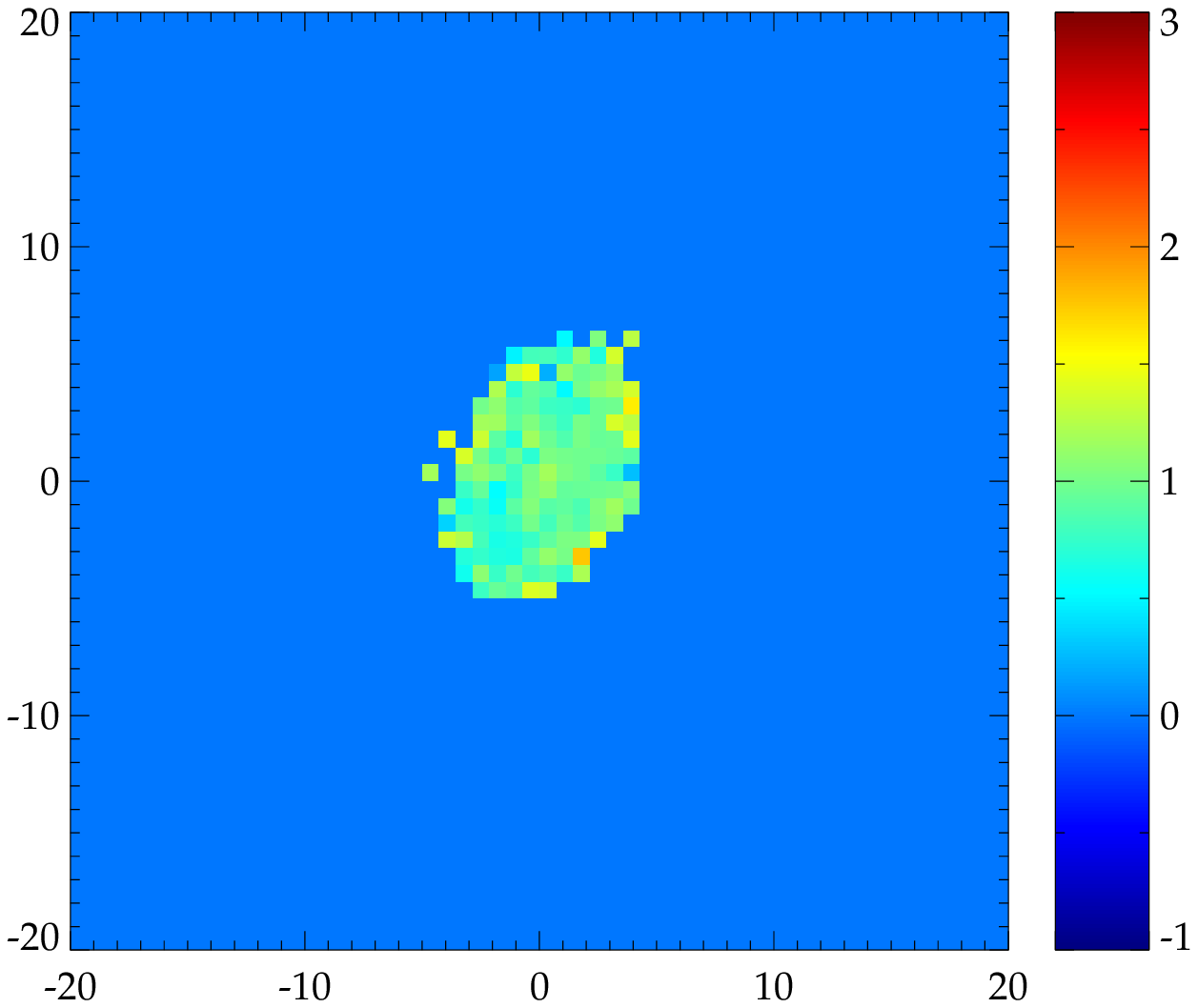}
\includegraphics[height=0.22\textwidth,clip]{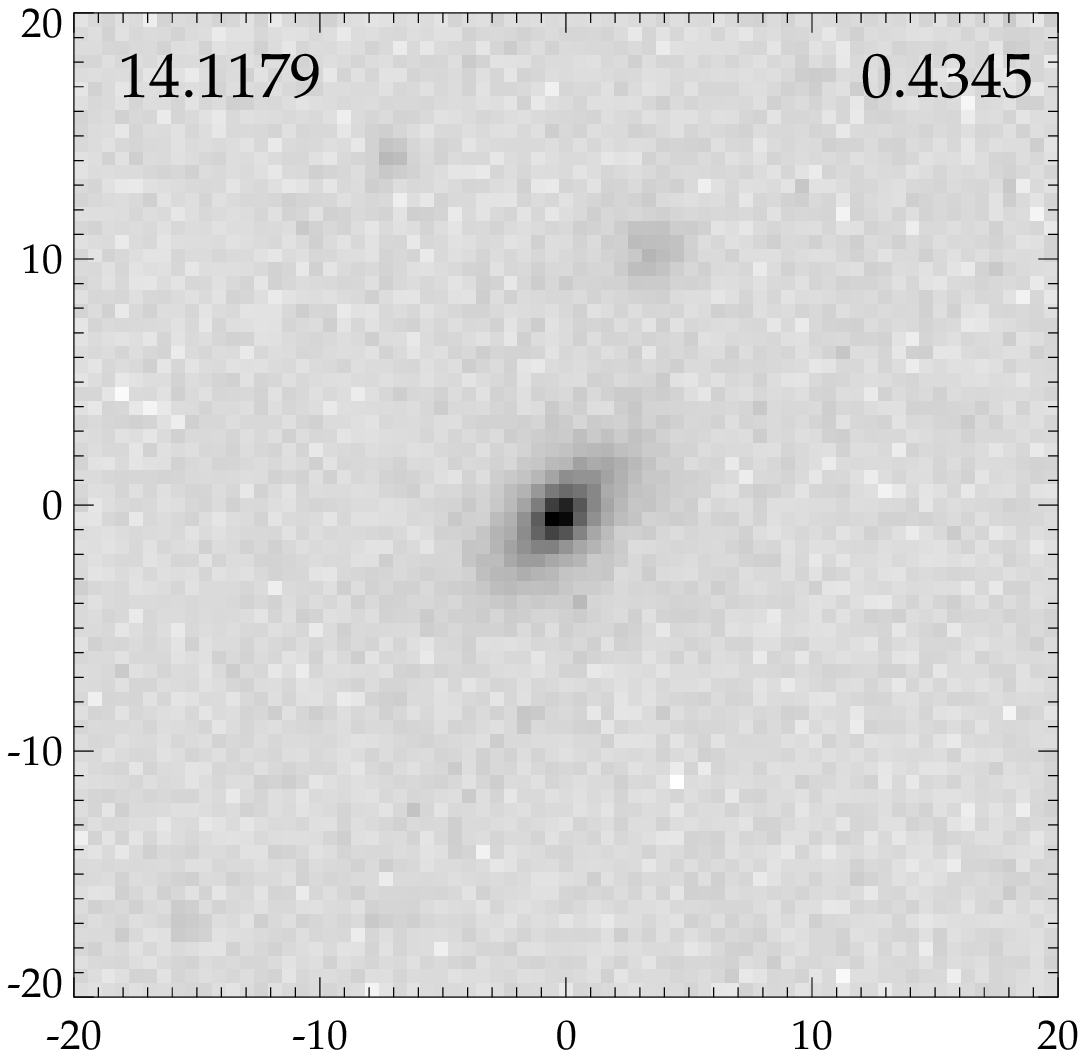} \includegraphics[height=0.22\textwidth,clip]{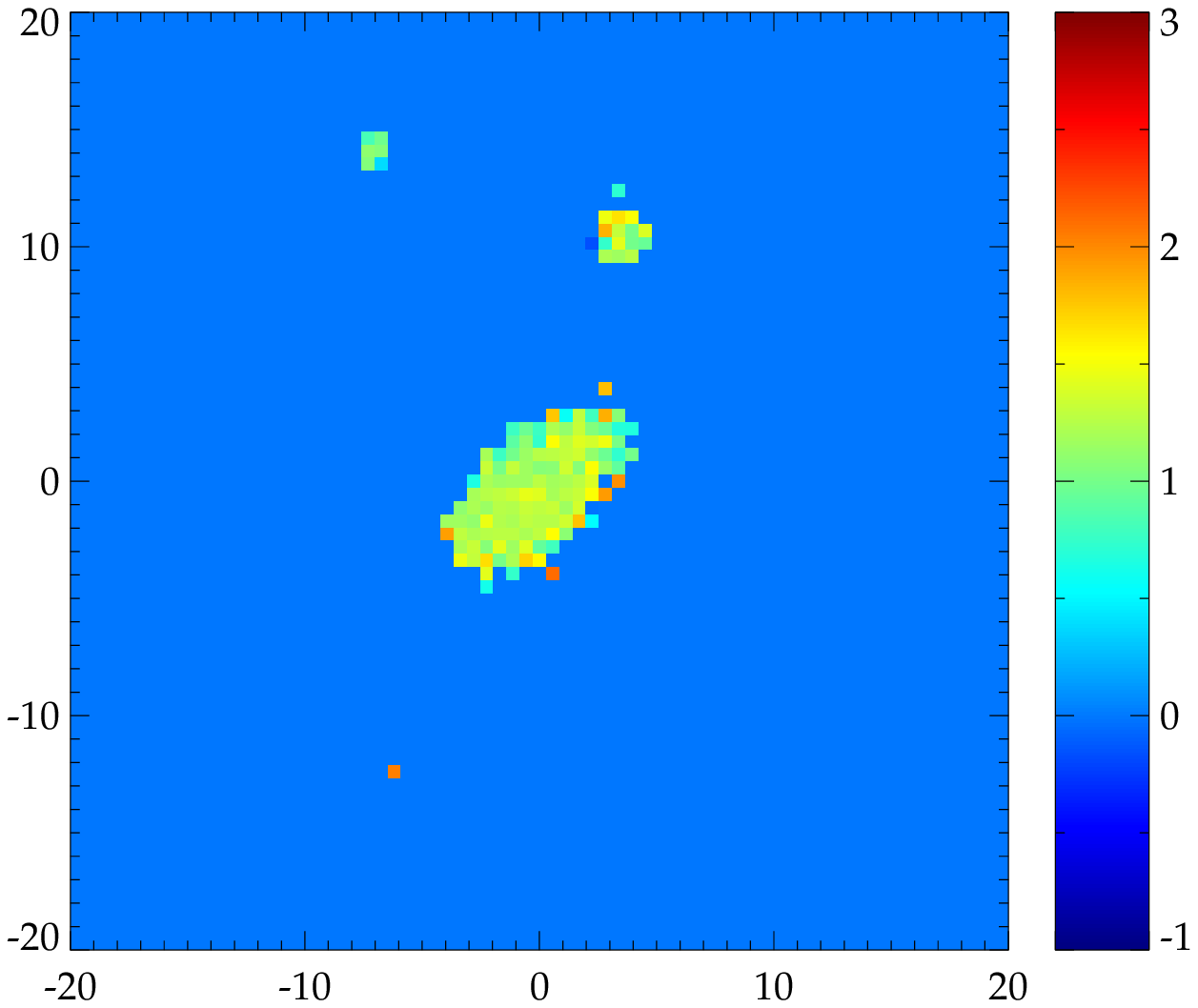}
\includegraphics[height=0.22\textwidth,clip]{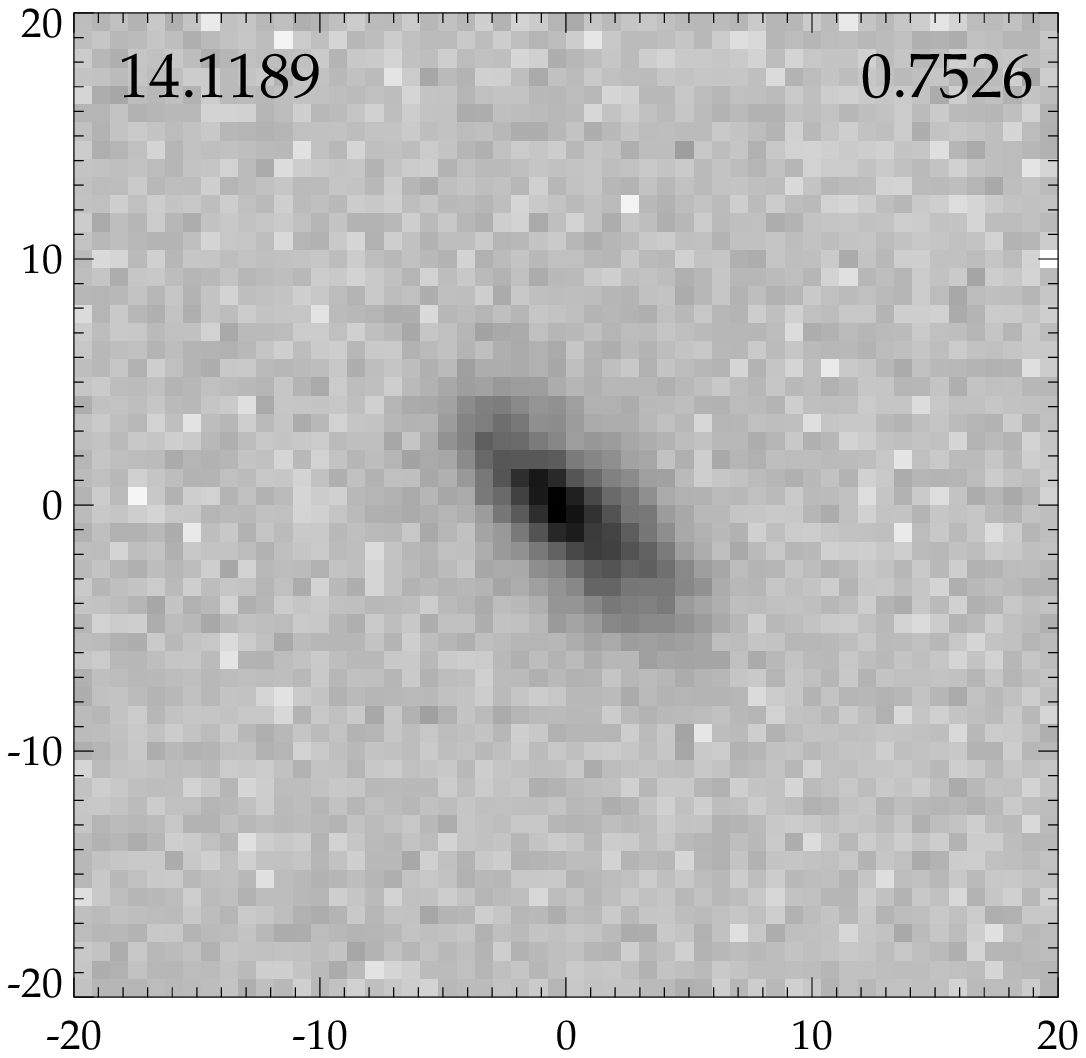} \includegraphics[height=0.22\textwidth,clip]{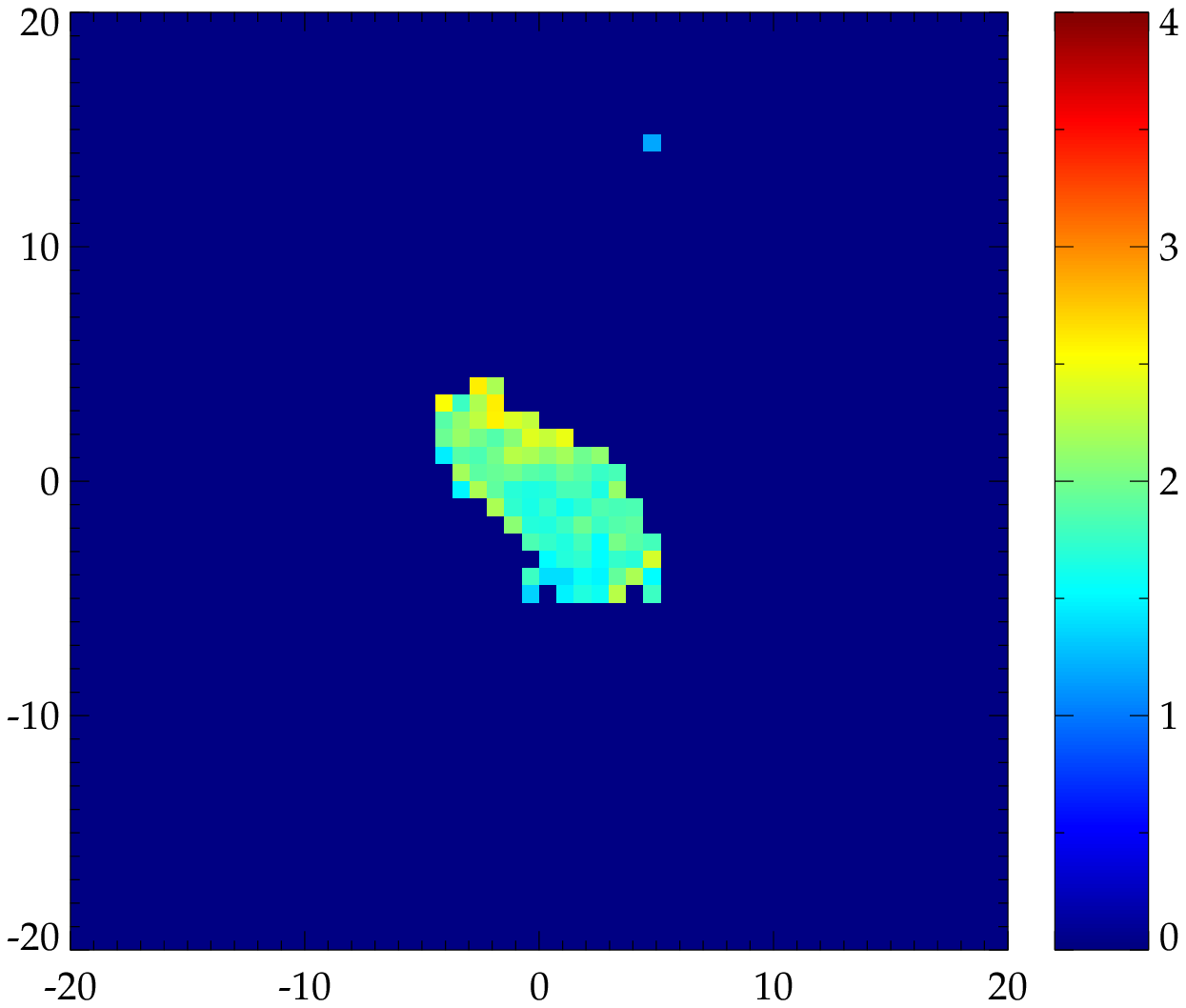}
\includegraphics[height=0.22\textwidth,clip]{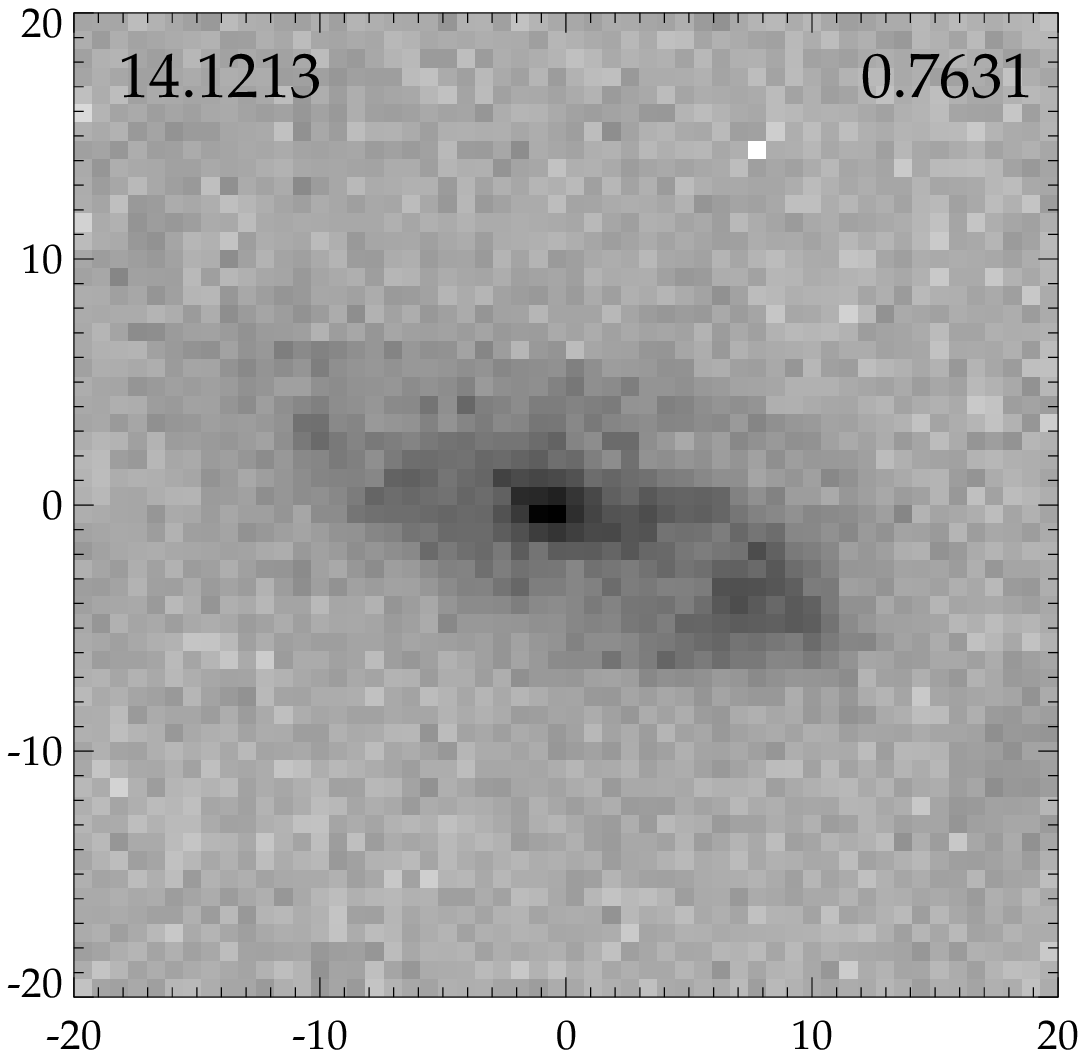} \includegraphics[height=0.22\textwidth,clip]{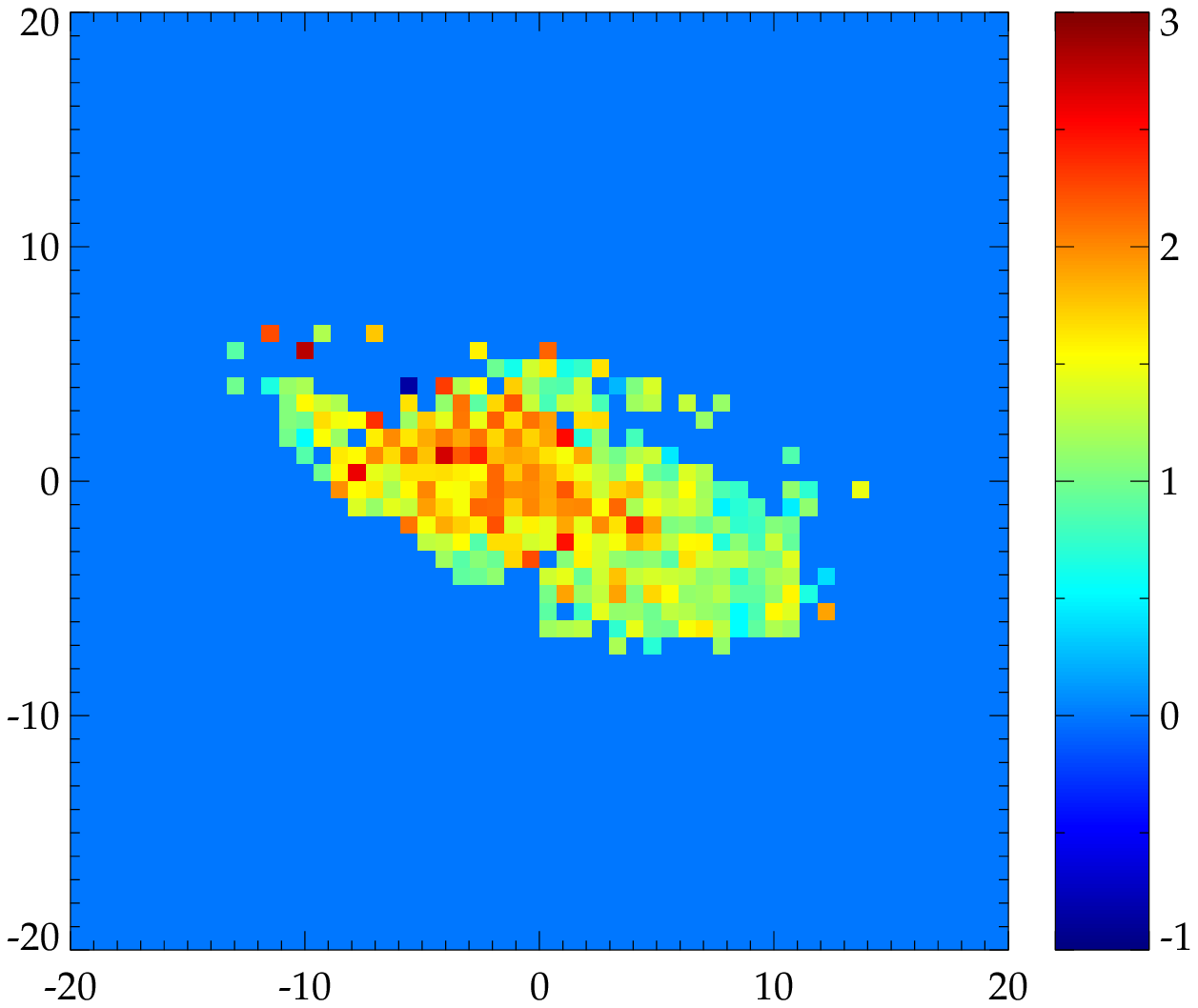}
\includegraphics[height=0.22\textwidth,clip]{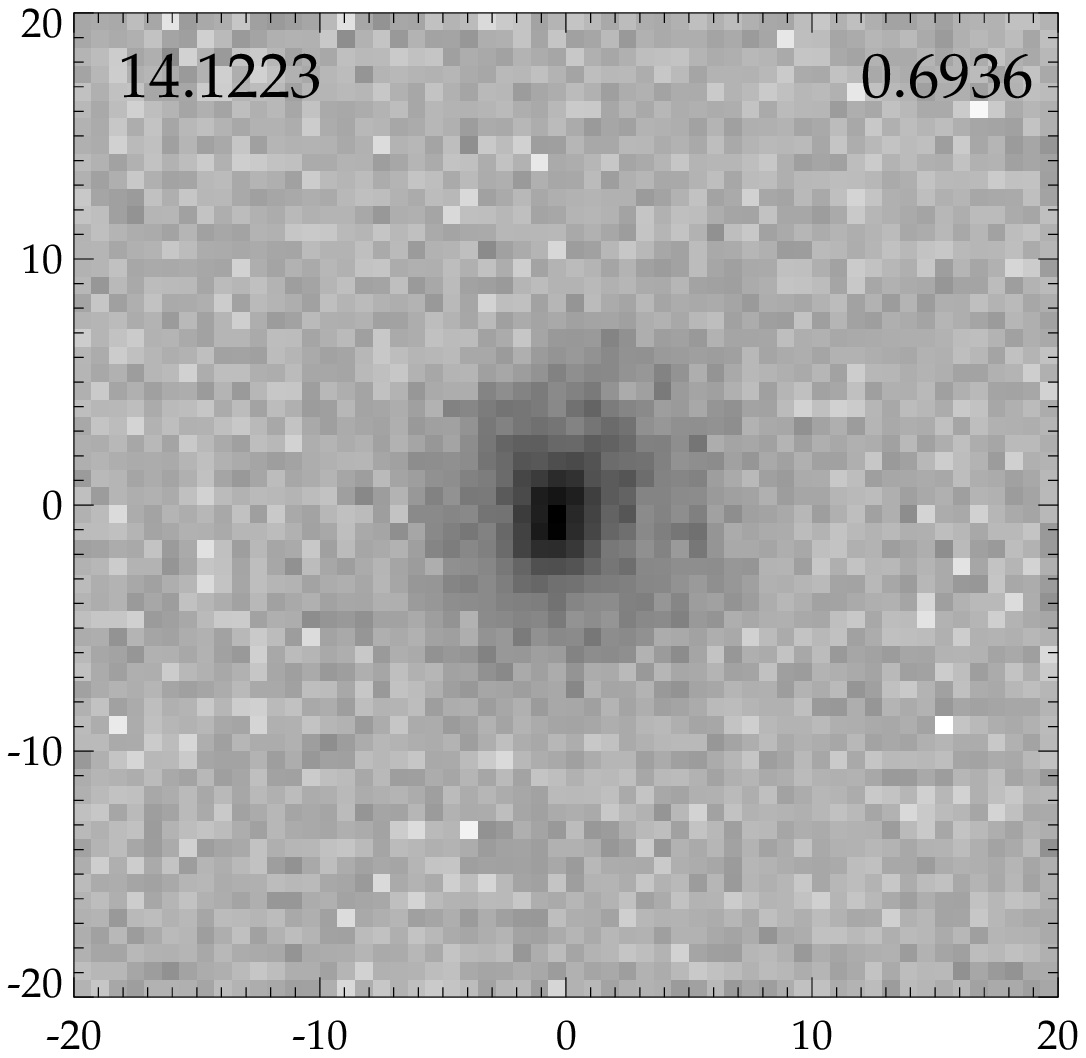} \includegraphics[height=0.22\textwidth,clip]{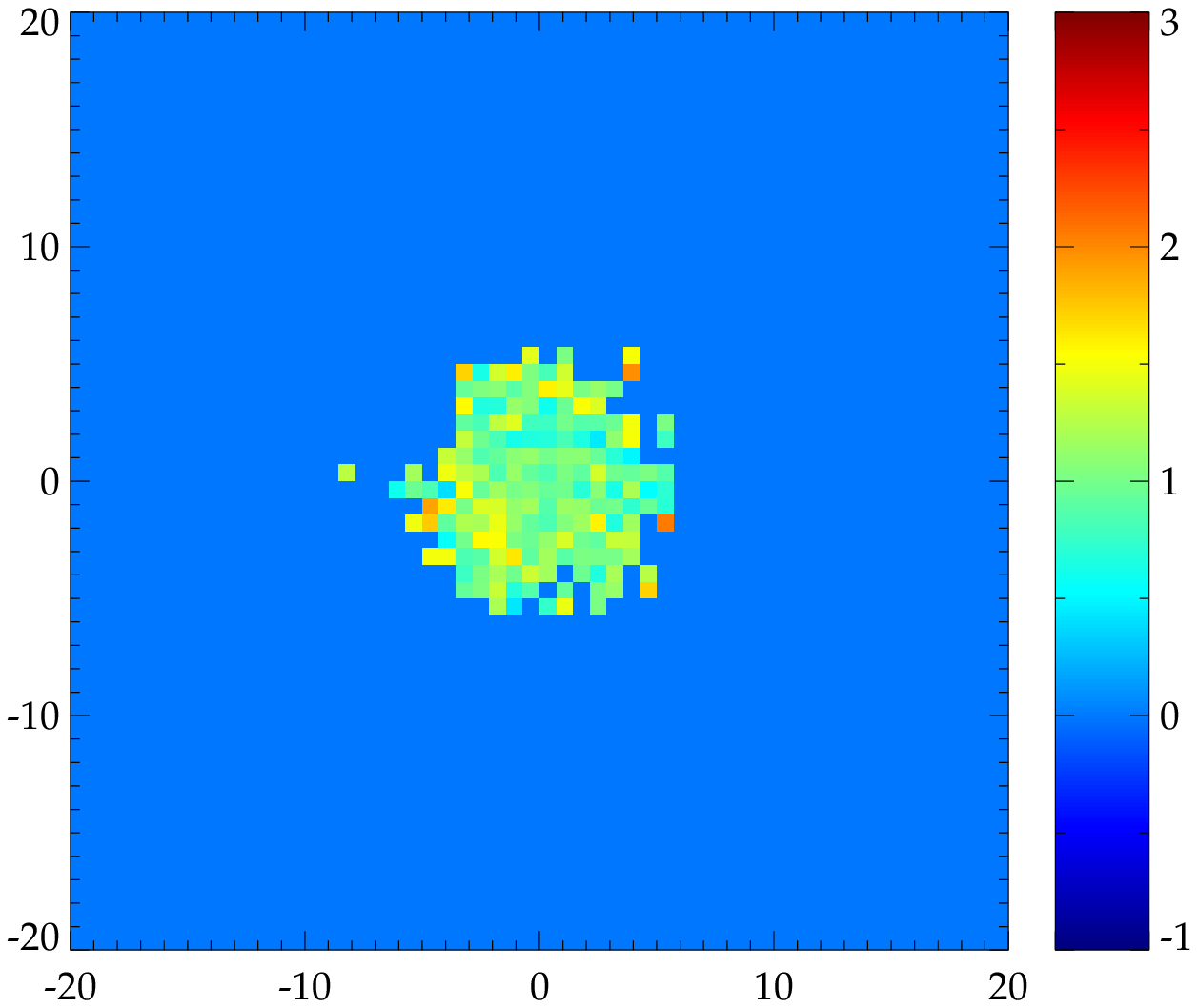}
\includegraphics[height=0.22\textwidth,clip]{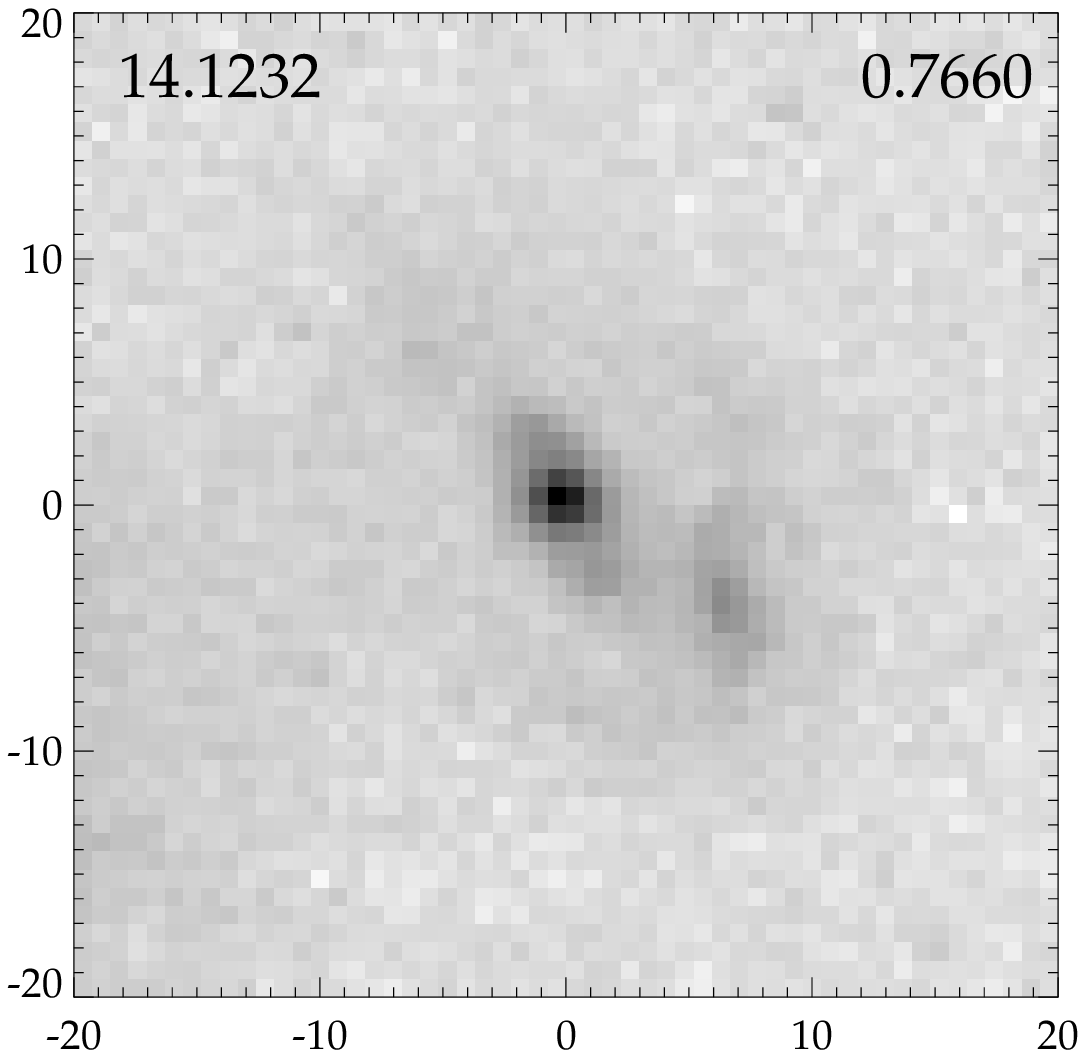} \includegraphics[height=0.22\textwidth,clip]{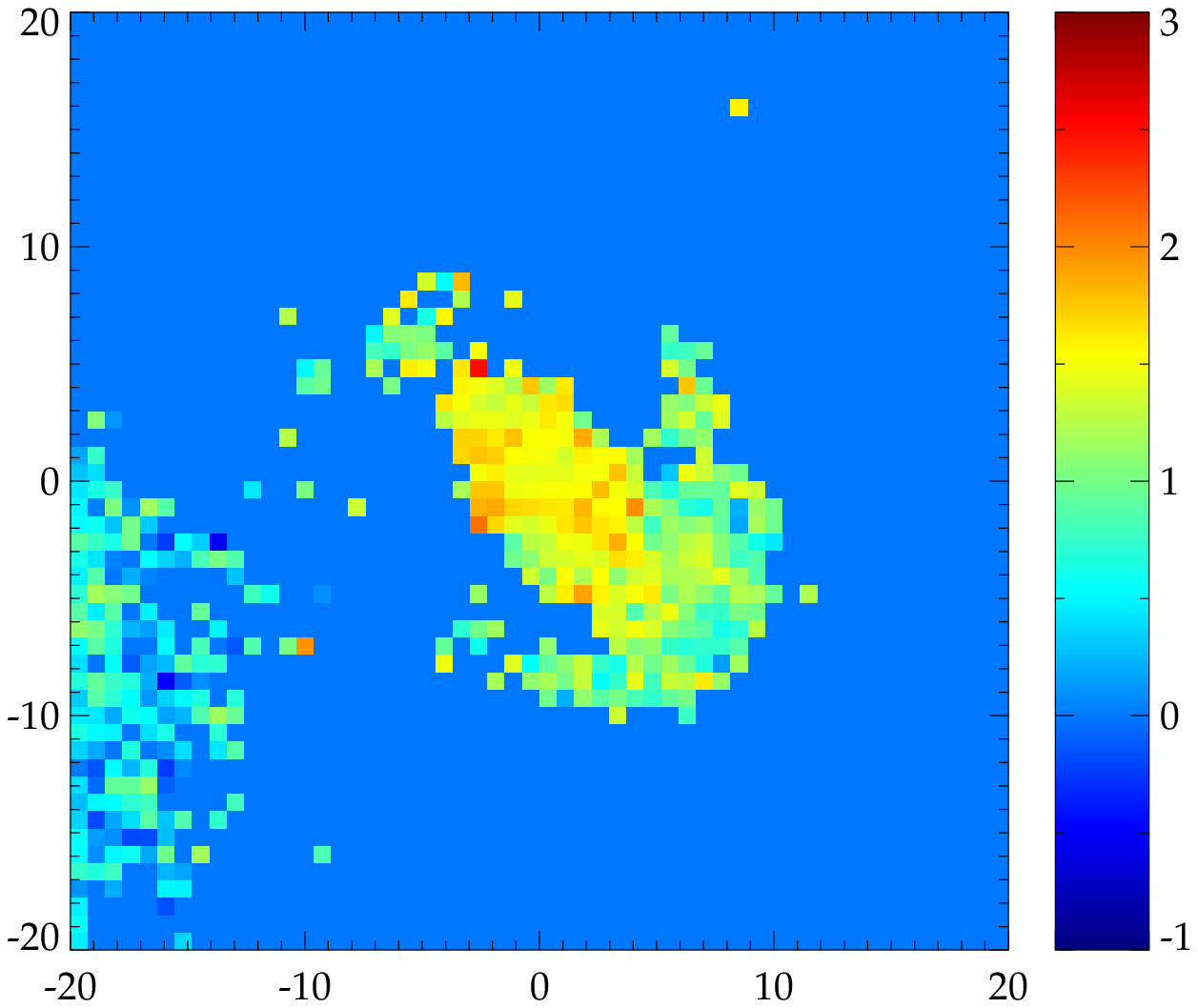}
\includegraphics[height=0.22\textwidth,clip]{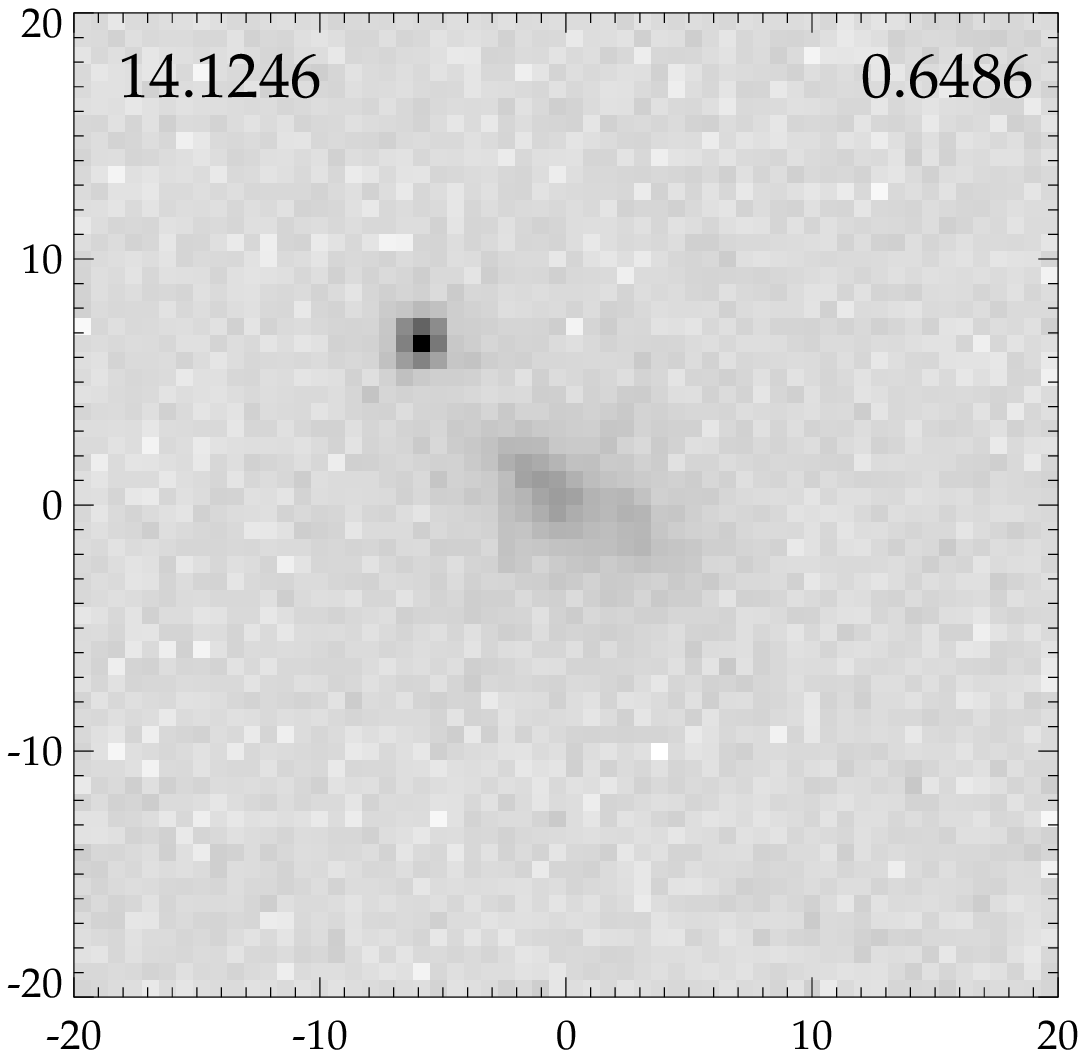} \includegraphics[height=0.22\textwidth,clip]{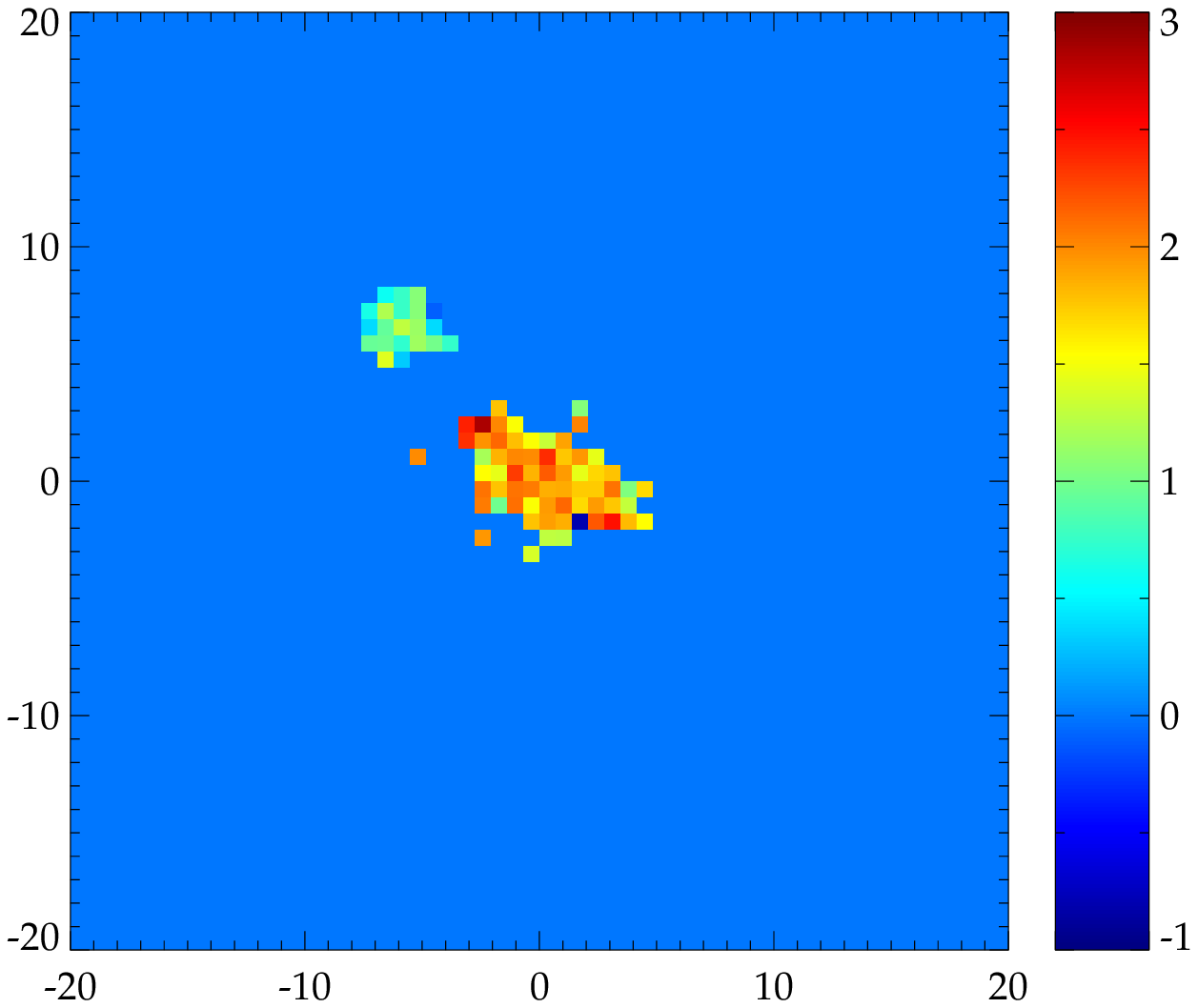}
\includegraphics[height=0.22\textwidth,clip]{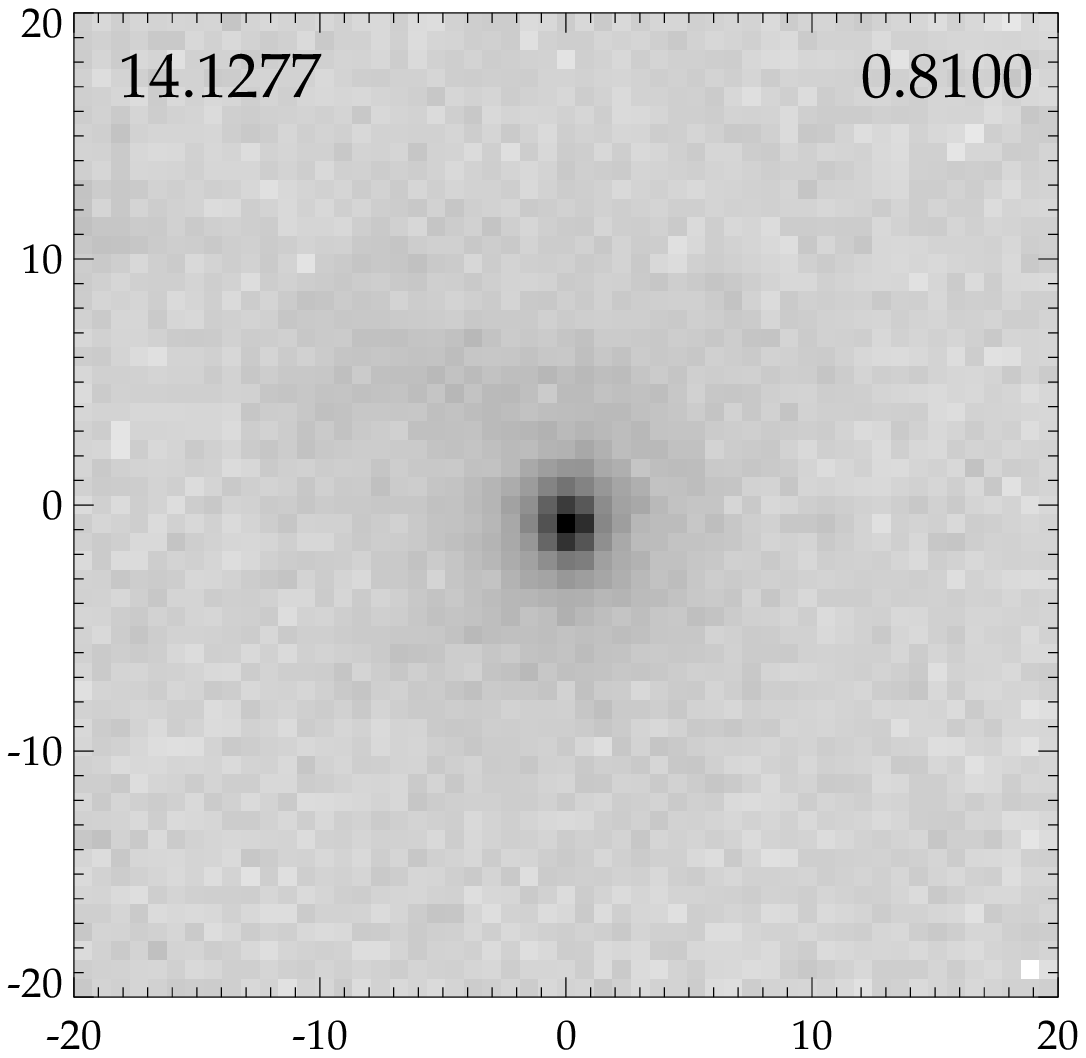} \includegraphics[height=0.22\textwidth,clip]{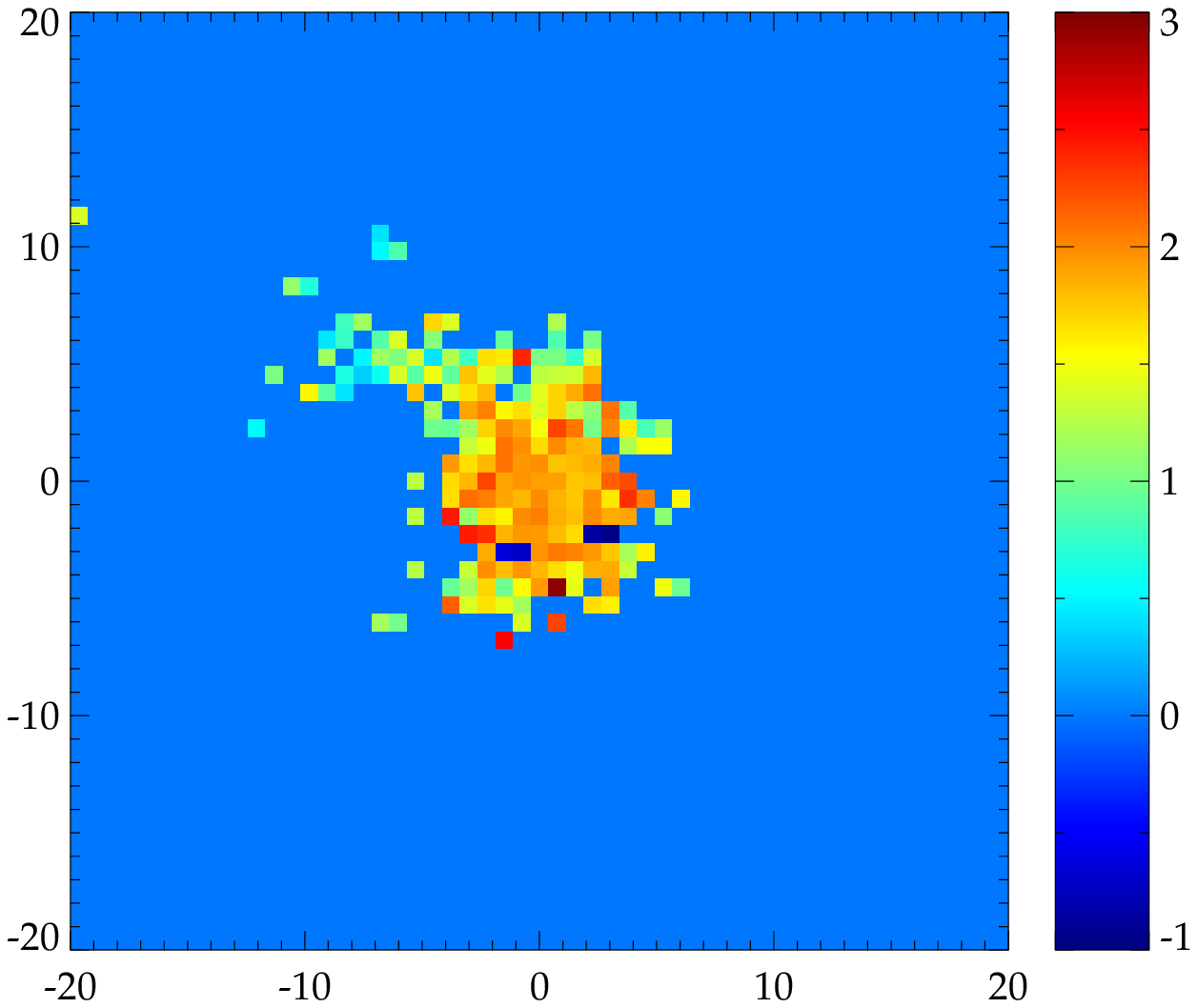}
\includegraphics[height=0.22\textwidth,clip]{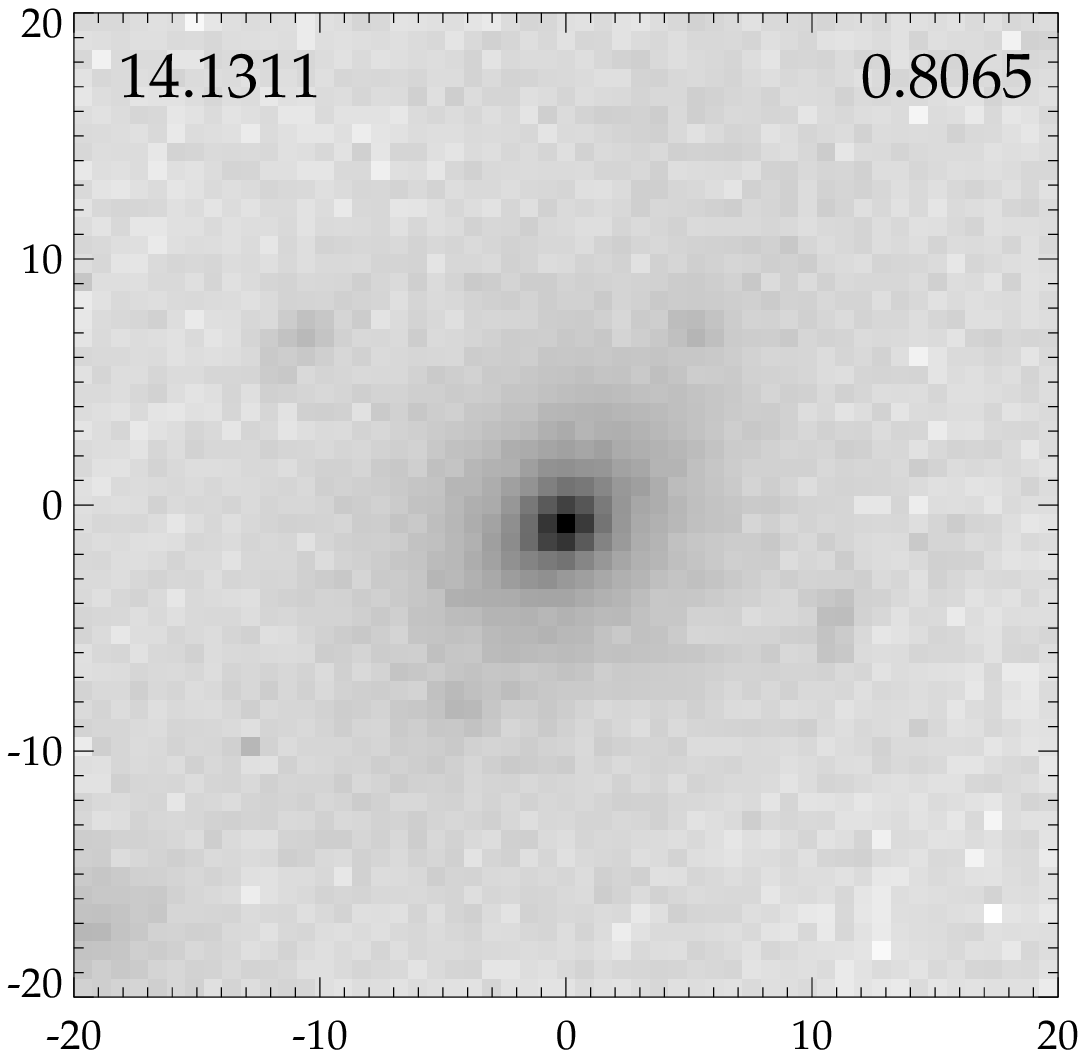} \includegraphics[height=0.22\textwidth,clip]{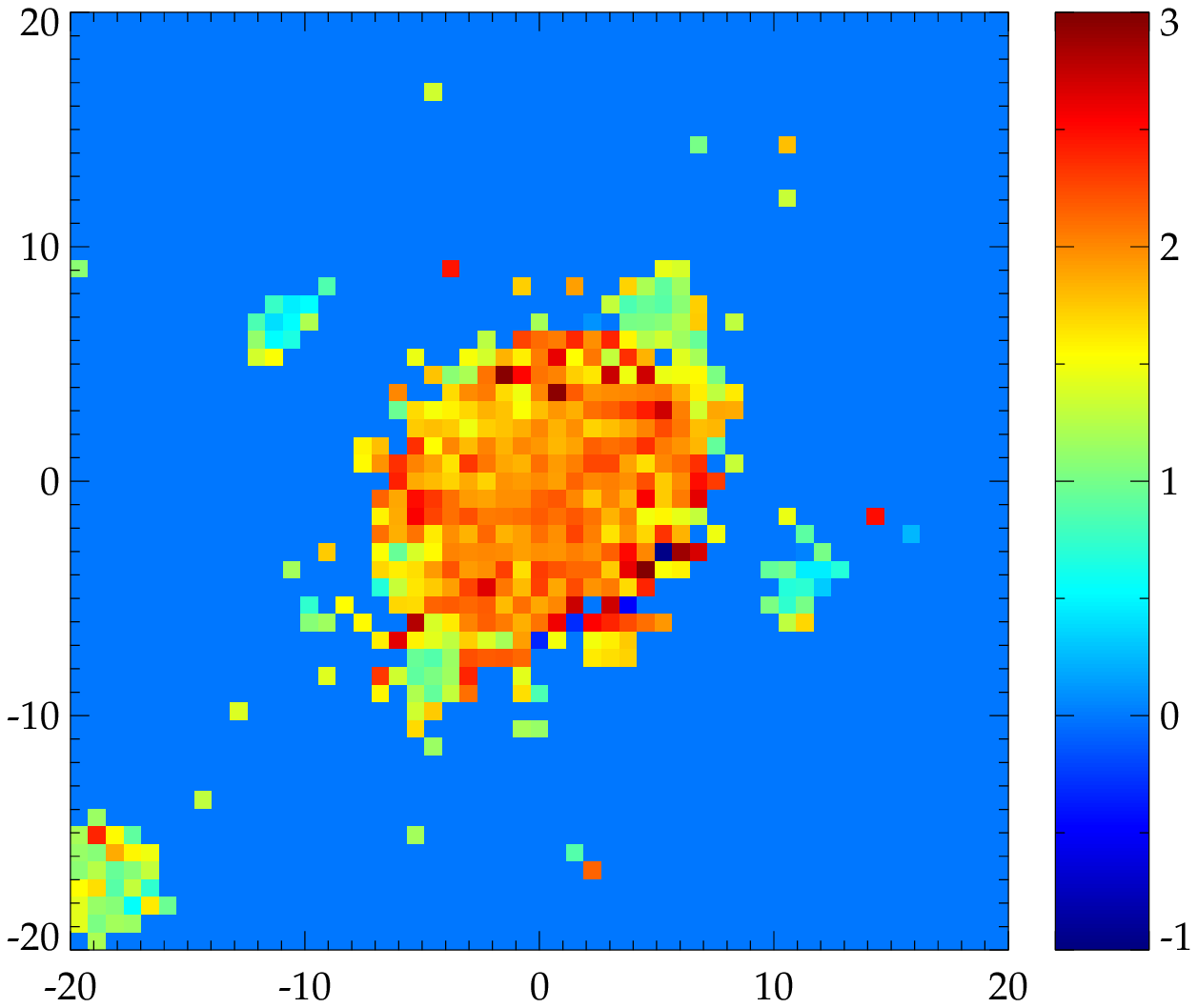}
\caption{Continued.} \end{figure*}

\addtocounter{figure}{-1}
\begin{figure*} \centering

\includegraphics[height=0.22\textwidth,clip]{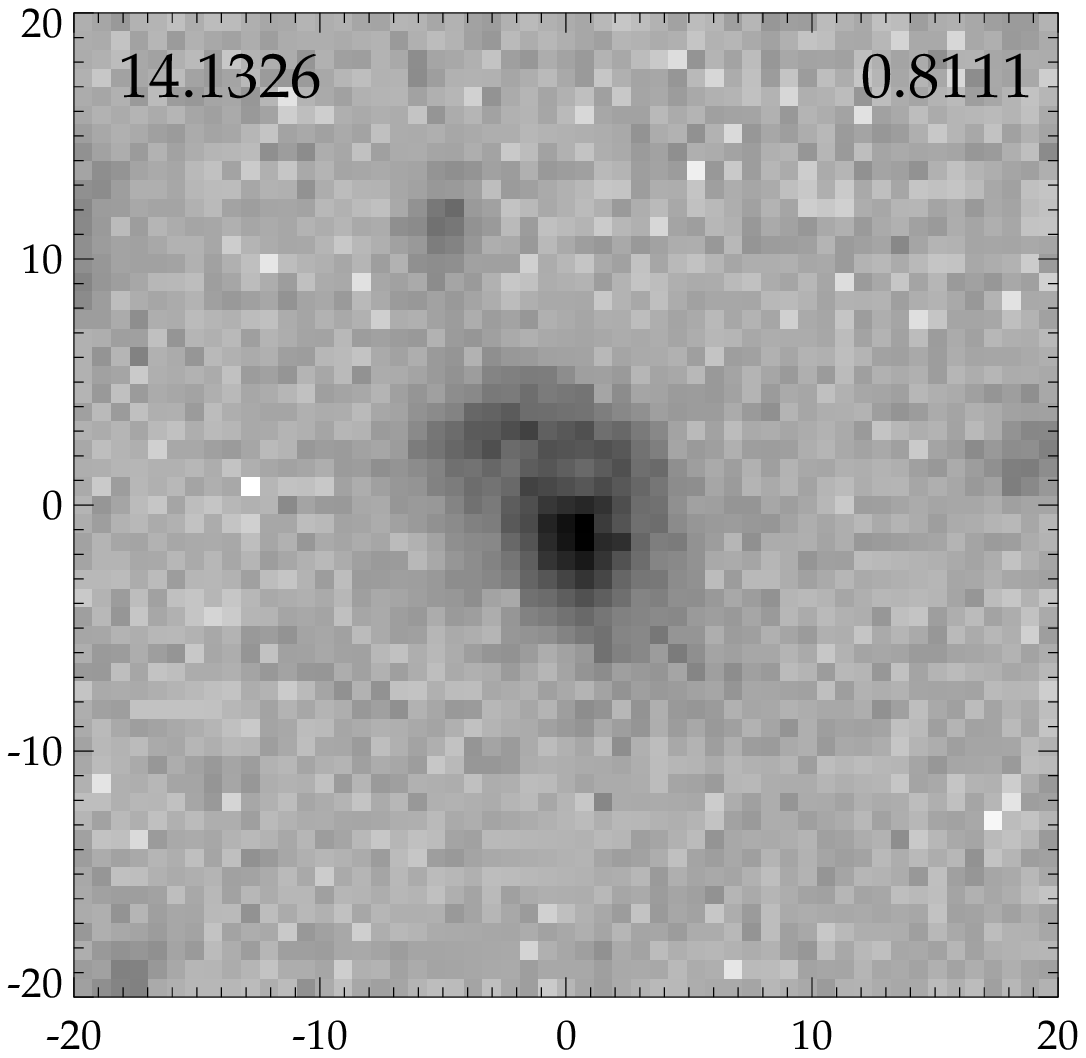} \includegraphics[height=0.22\textwidth,clip]{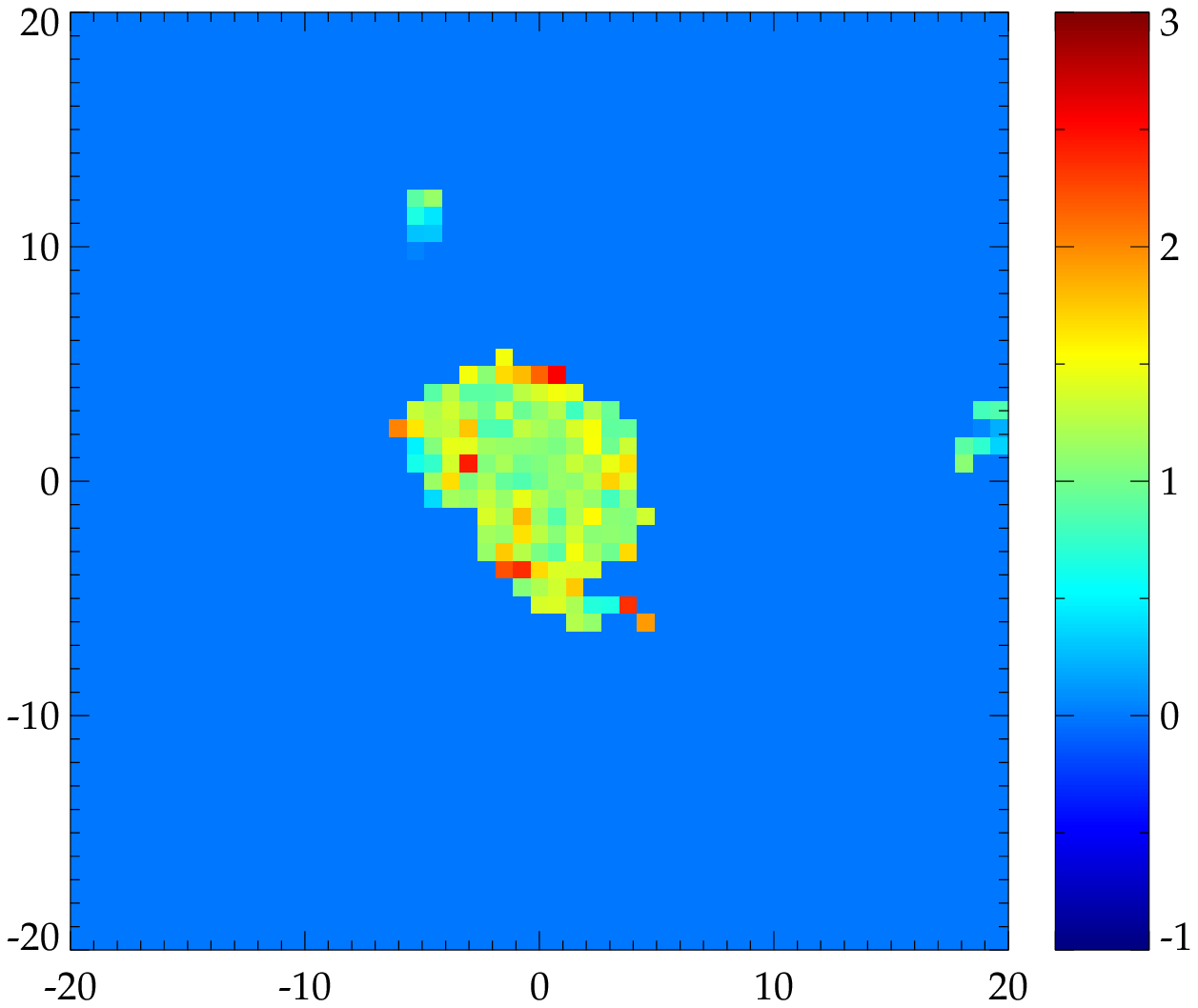}
\includegraphics[height=0.22\textwidth,clip]{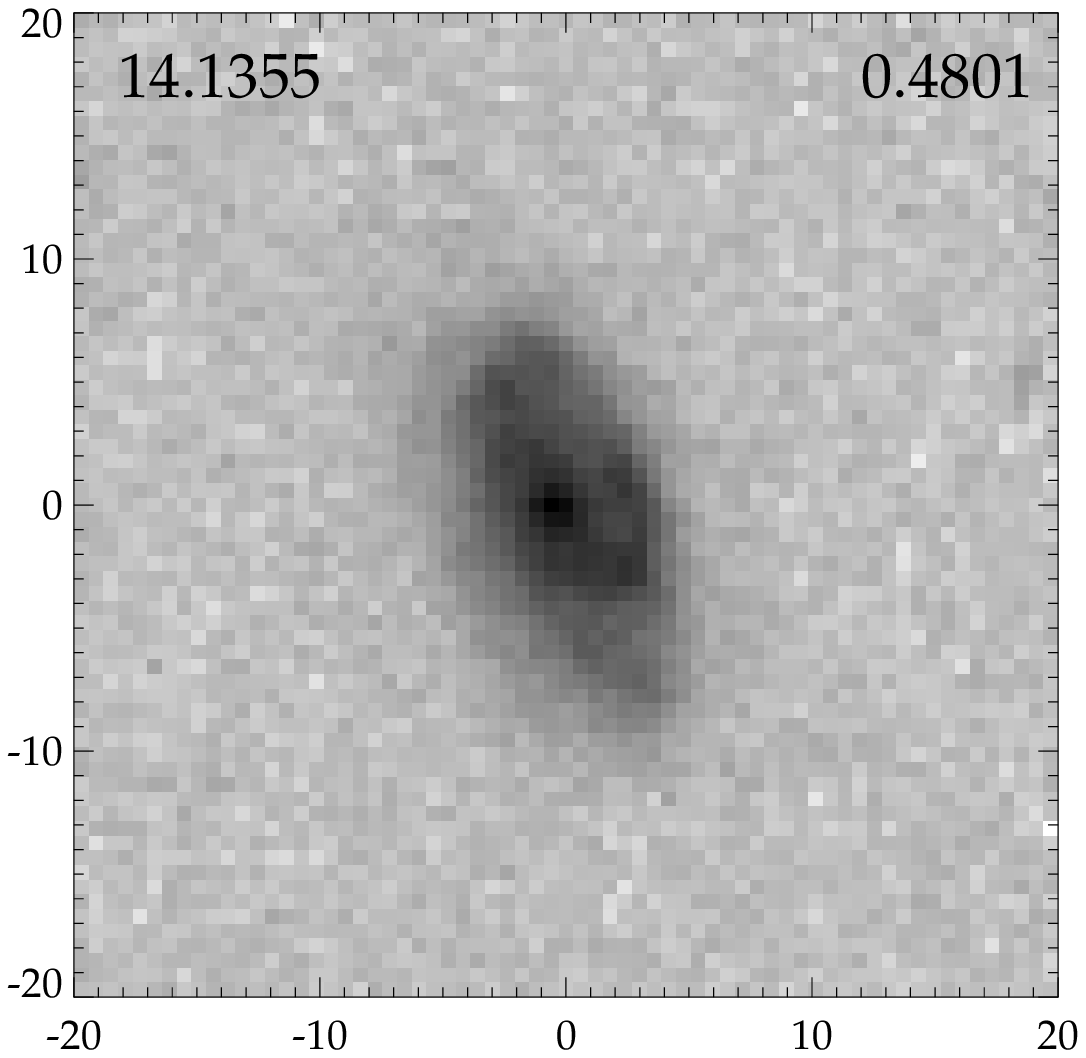} \includegraphics[height=0.22\textwidth,clip]{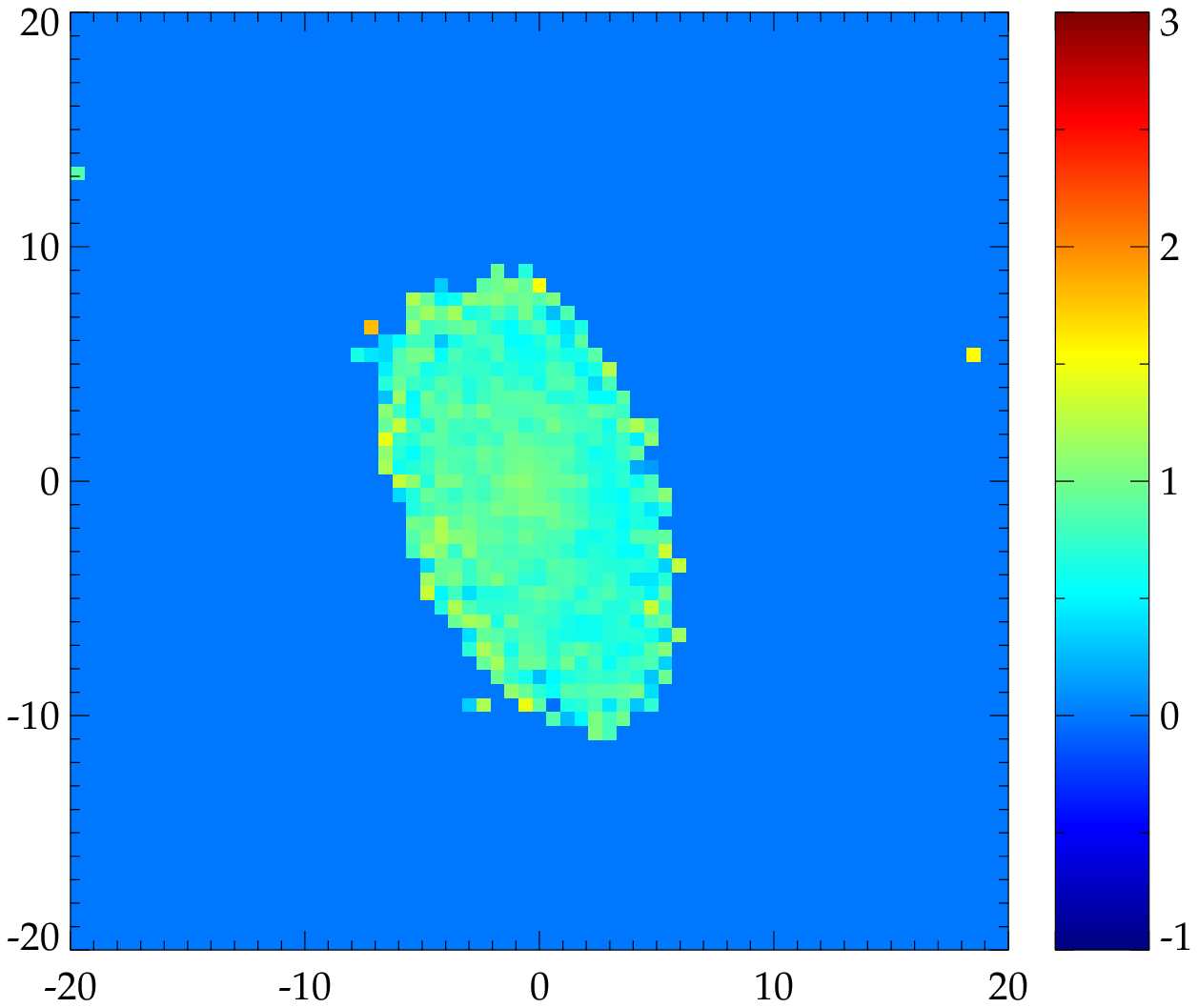}
\includegraphics[height=0.22\textwidth,clip]{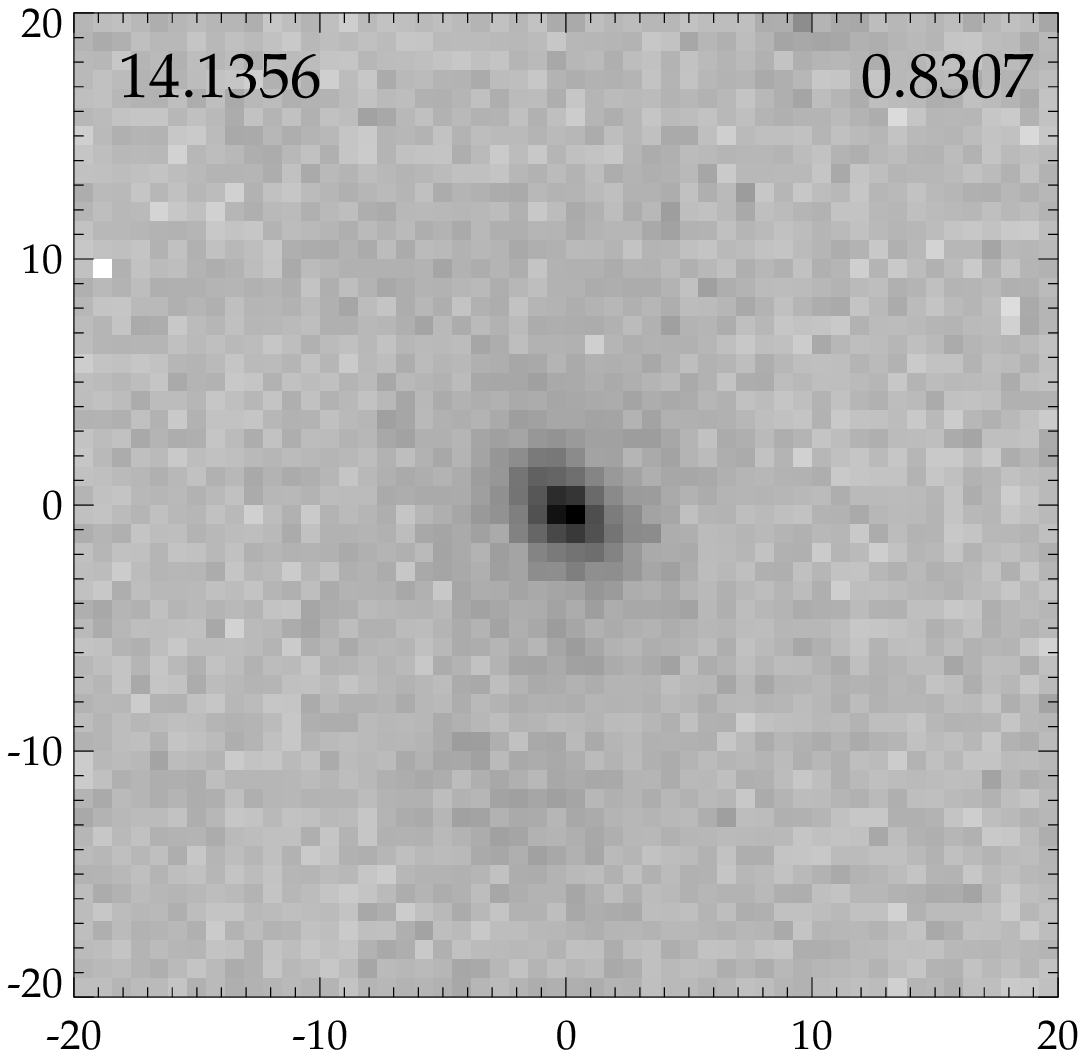} \includegraphics[height=0.22\textwidth,clip]{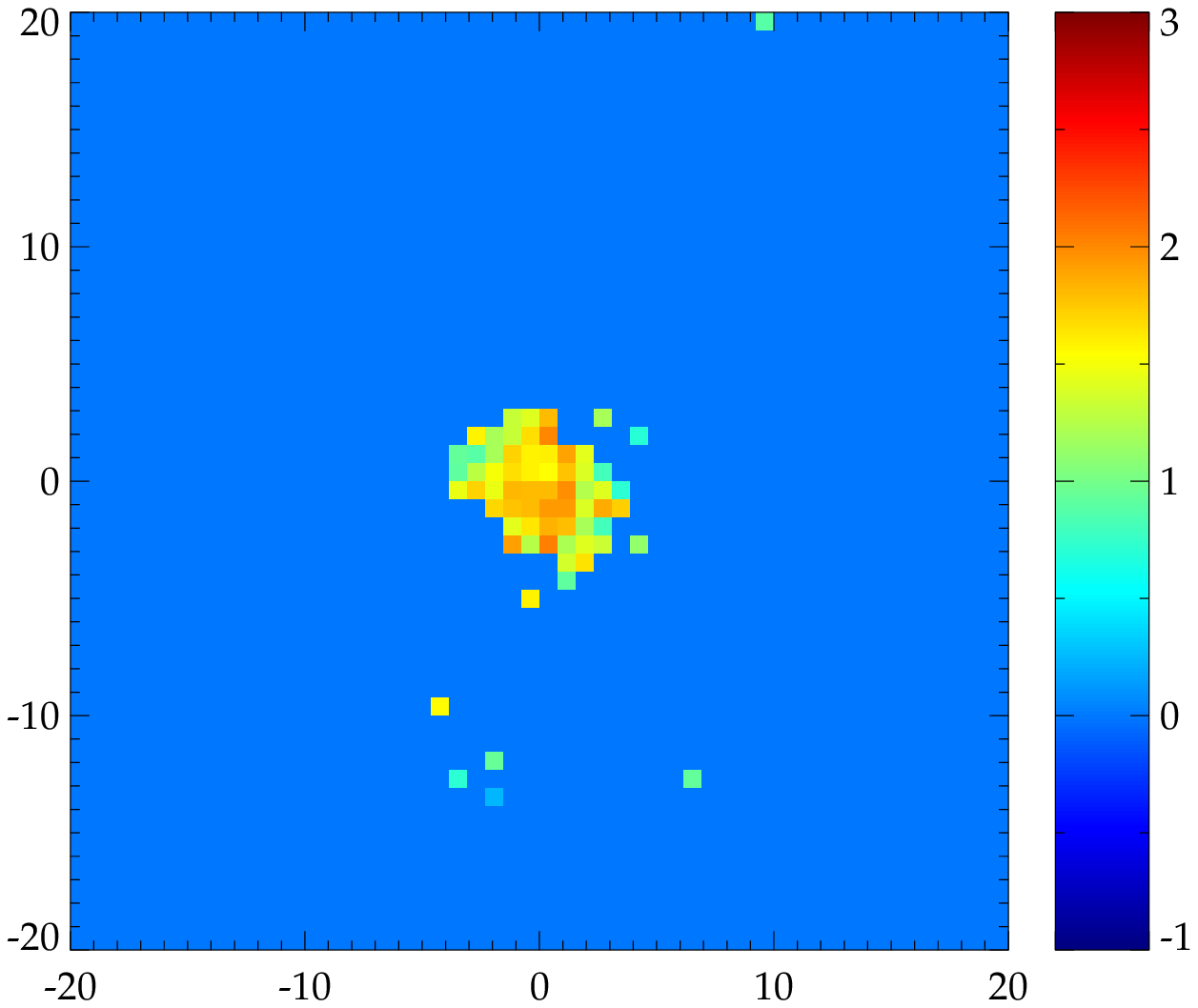}
\includegraphics[height=0.22\textwidth,clip]{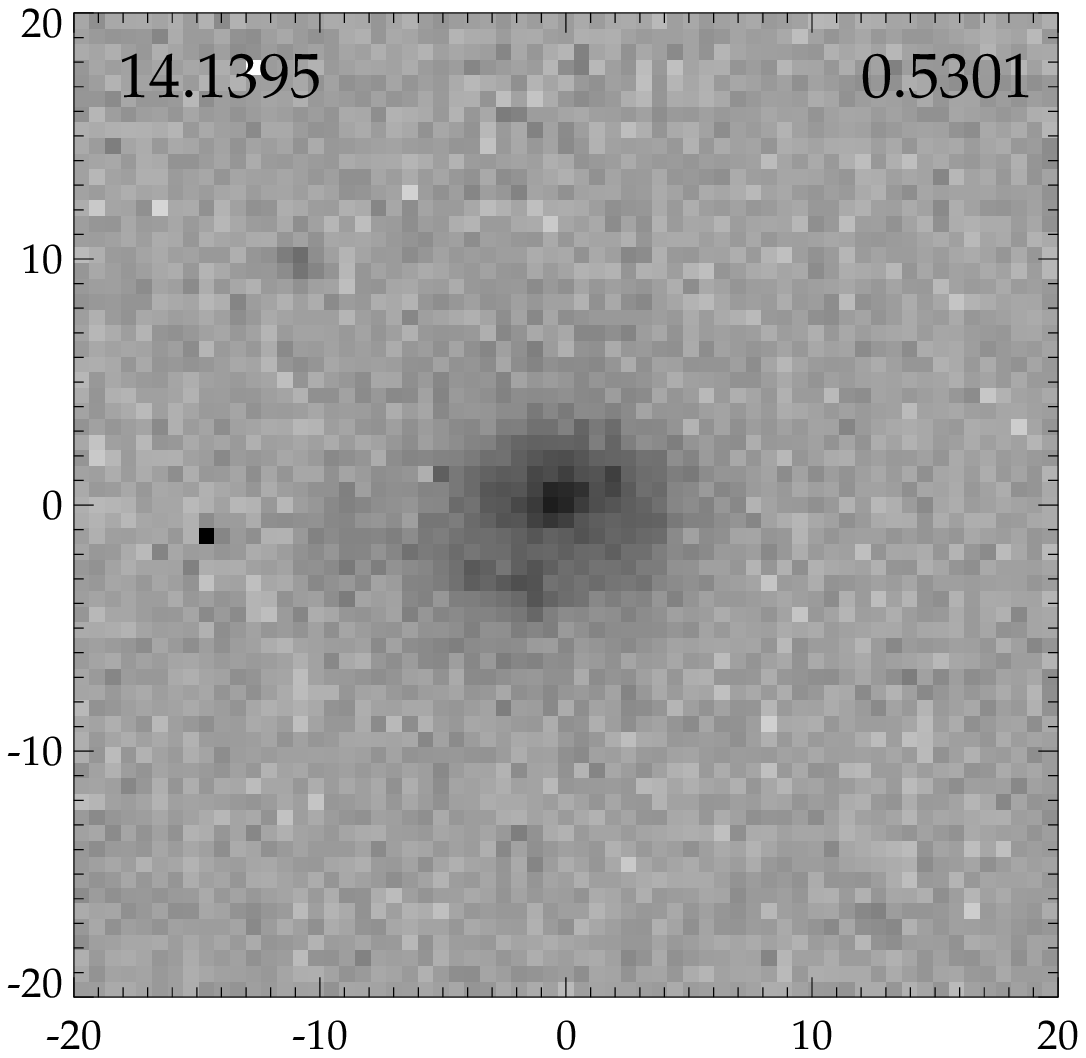} \includegraphics[height=0.22\textwidth,clip]{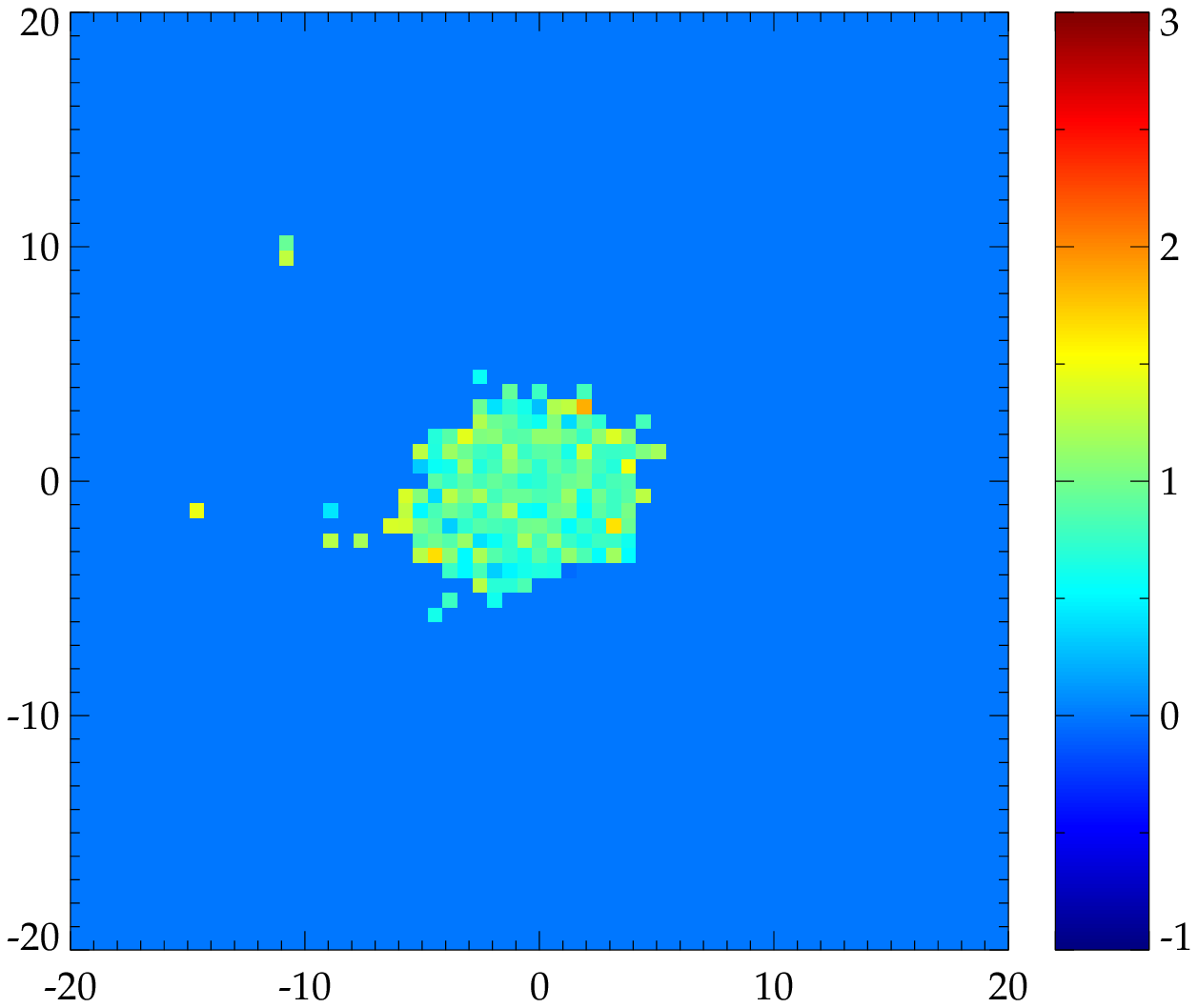}
\includegraphics[height=0.22\textwidth,clip]{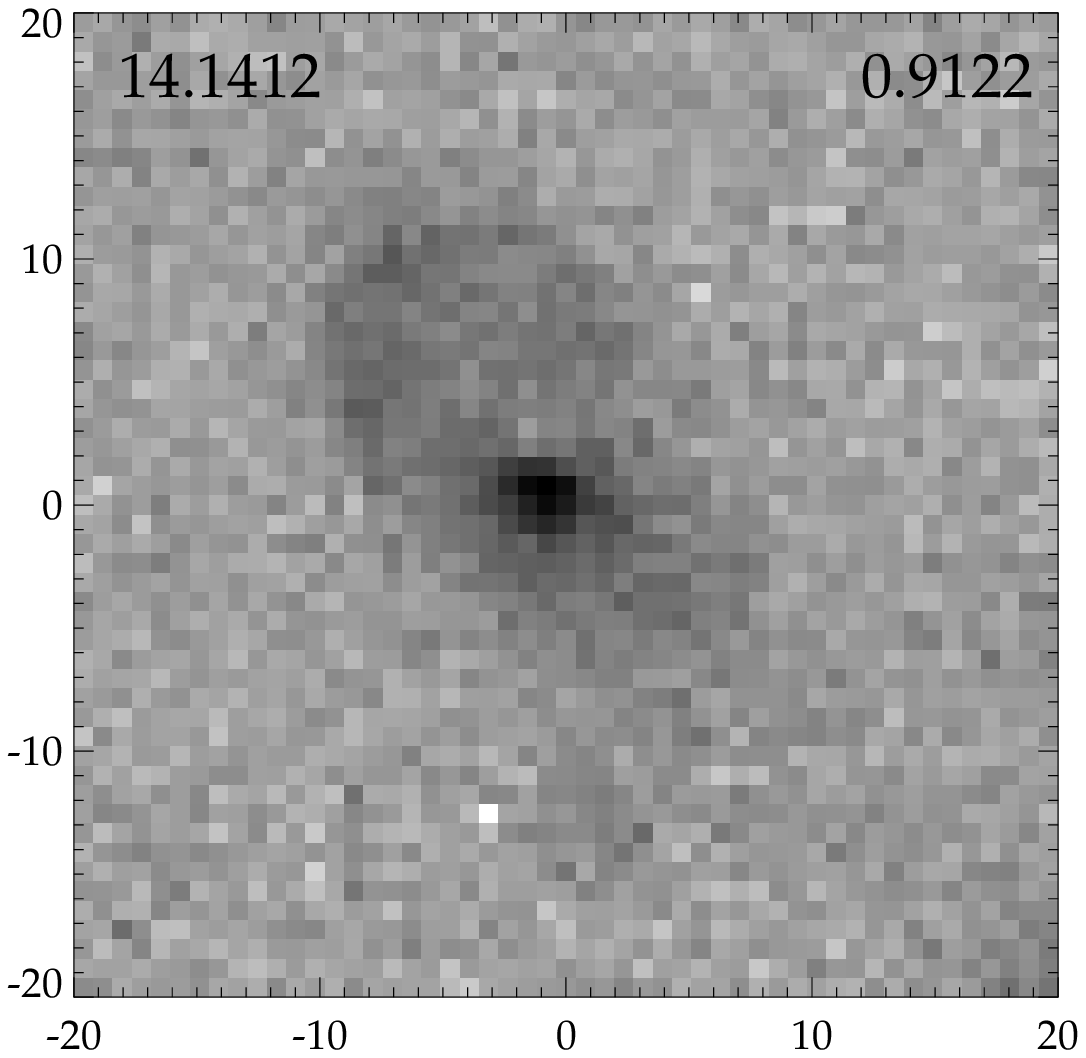} \includegraphics[height=0.22\textwidth,clip]{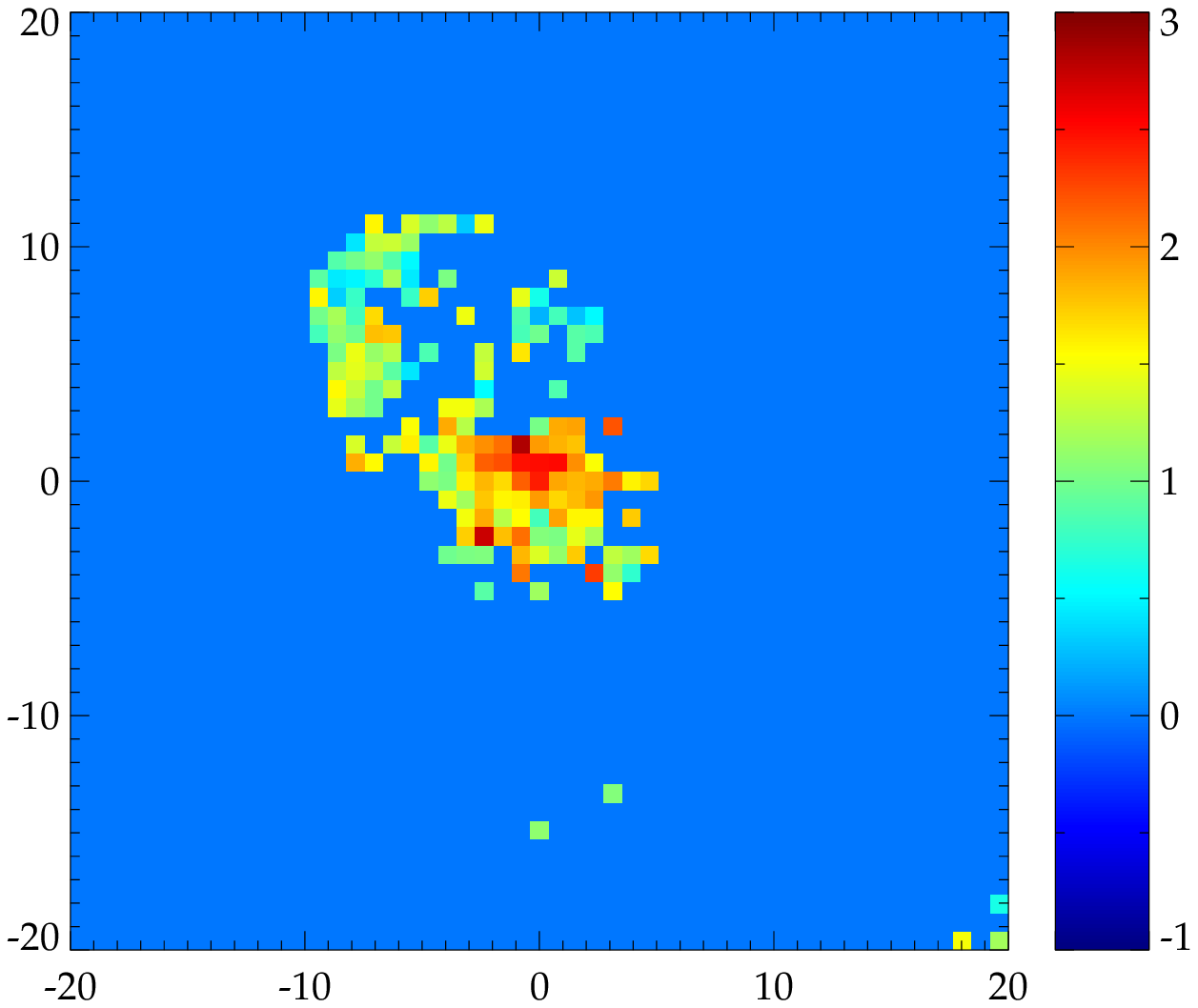}
\includegraphics[height=0.22\textwidth,clip]{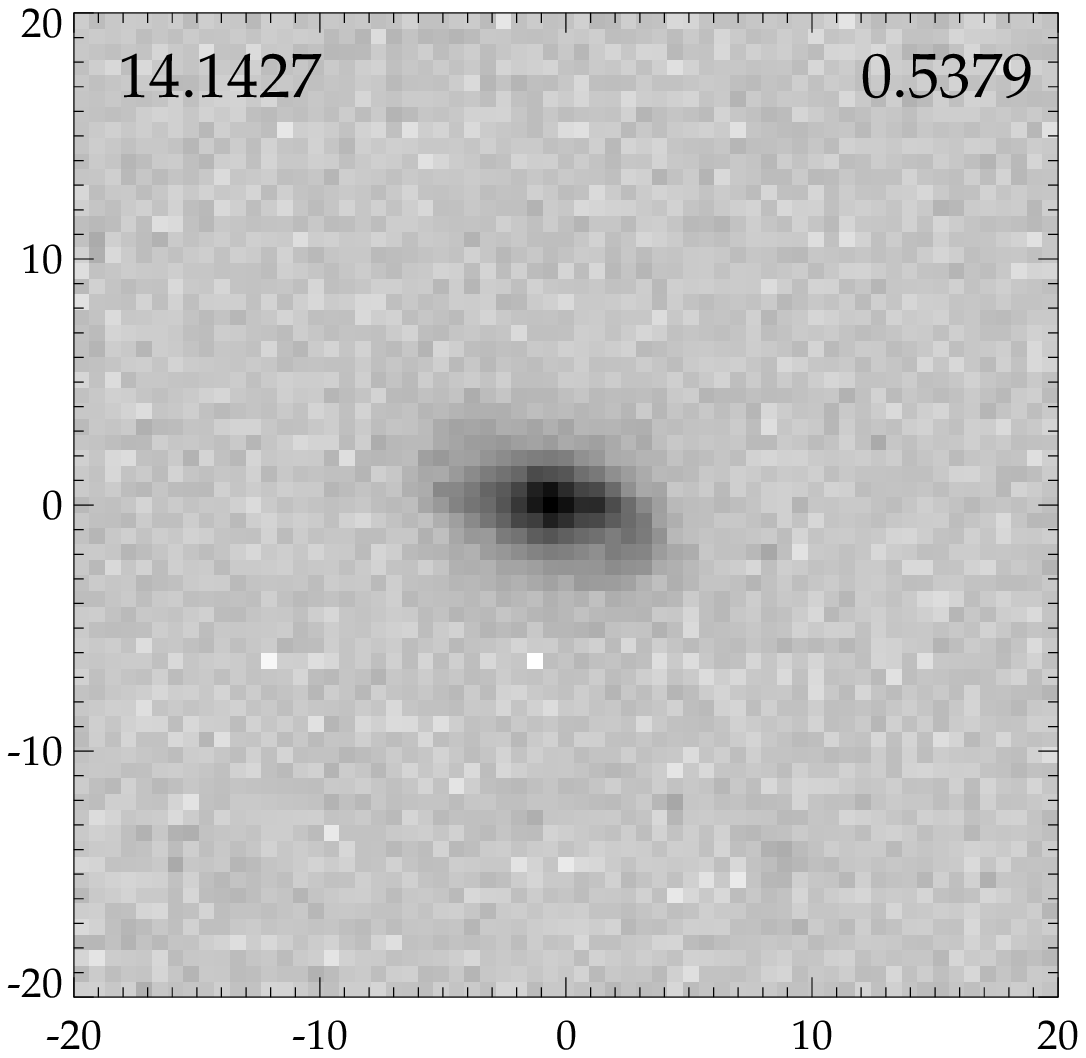} \includegraphics[height=0.22\textwidth,clip]{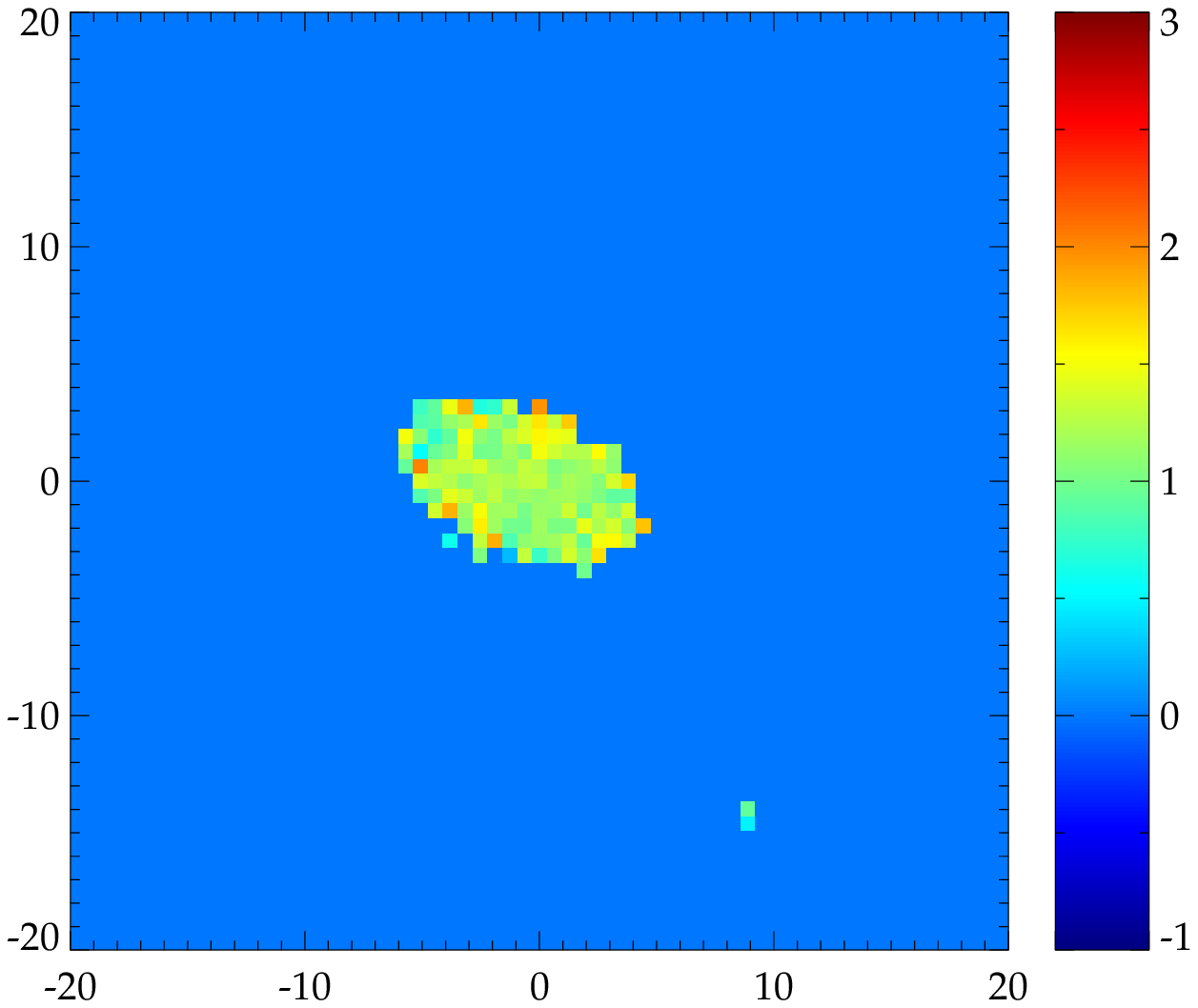}
\includegraphics[height=0.22\textwidth,clip]{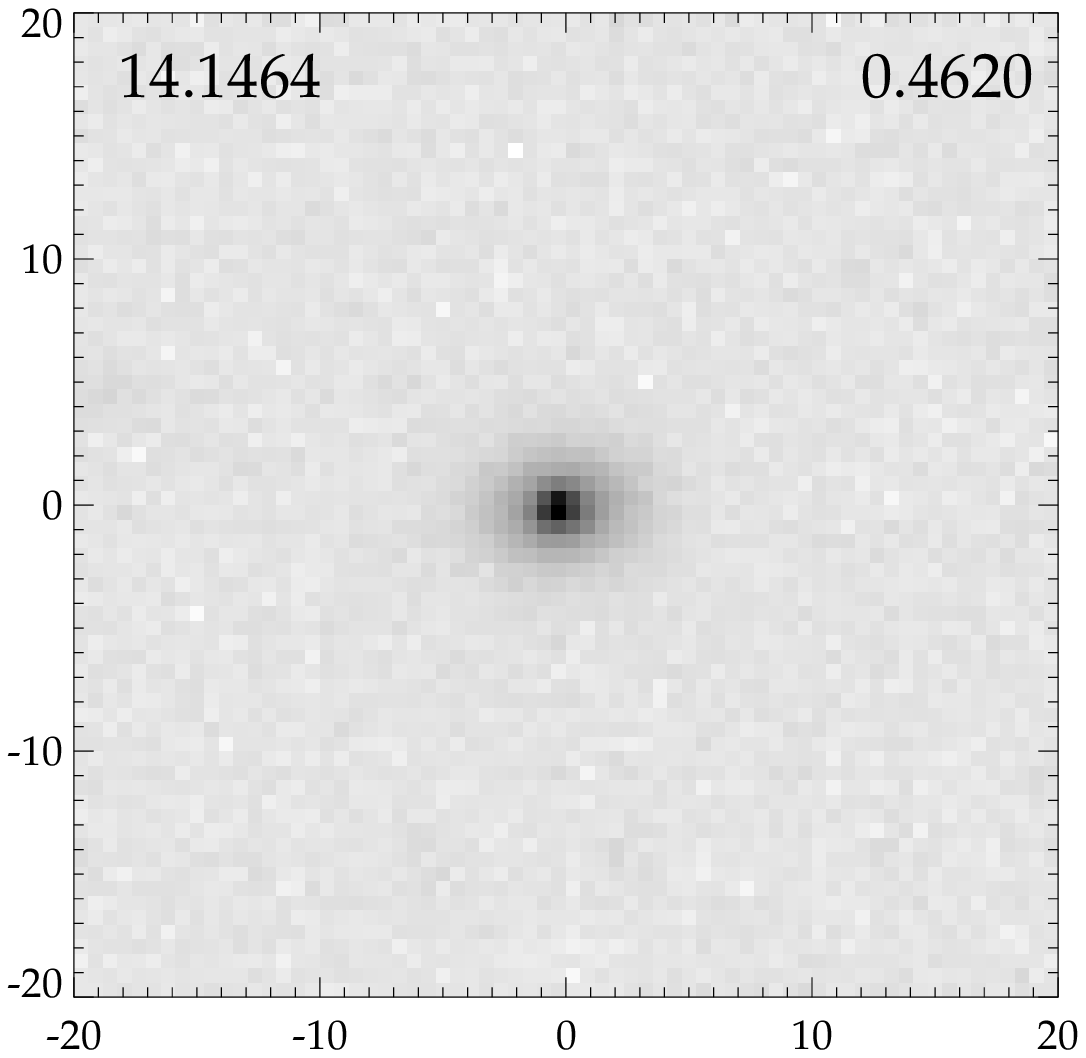} \includegraphics[height=0.22\textwidth,clip]{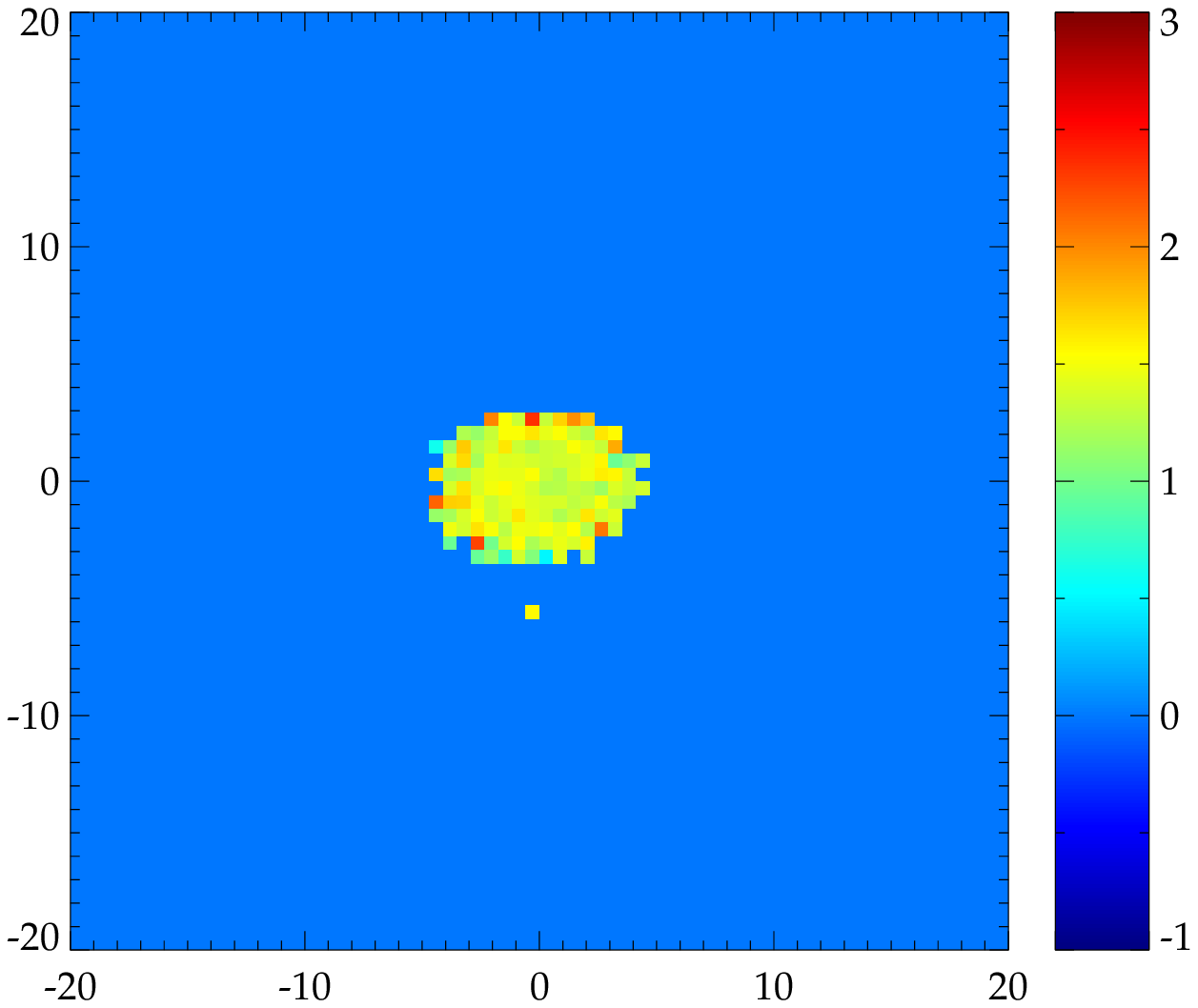}
\includegraphics[height=0.22\textwidth,clip]{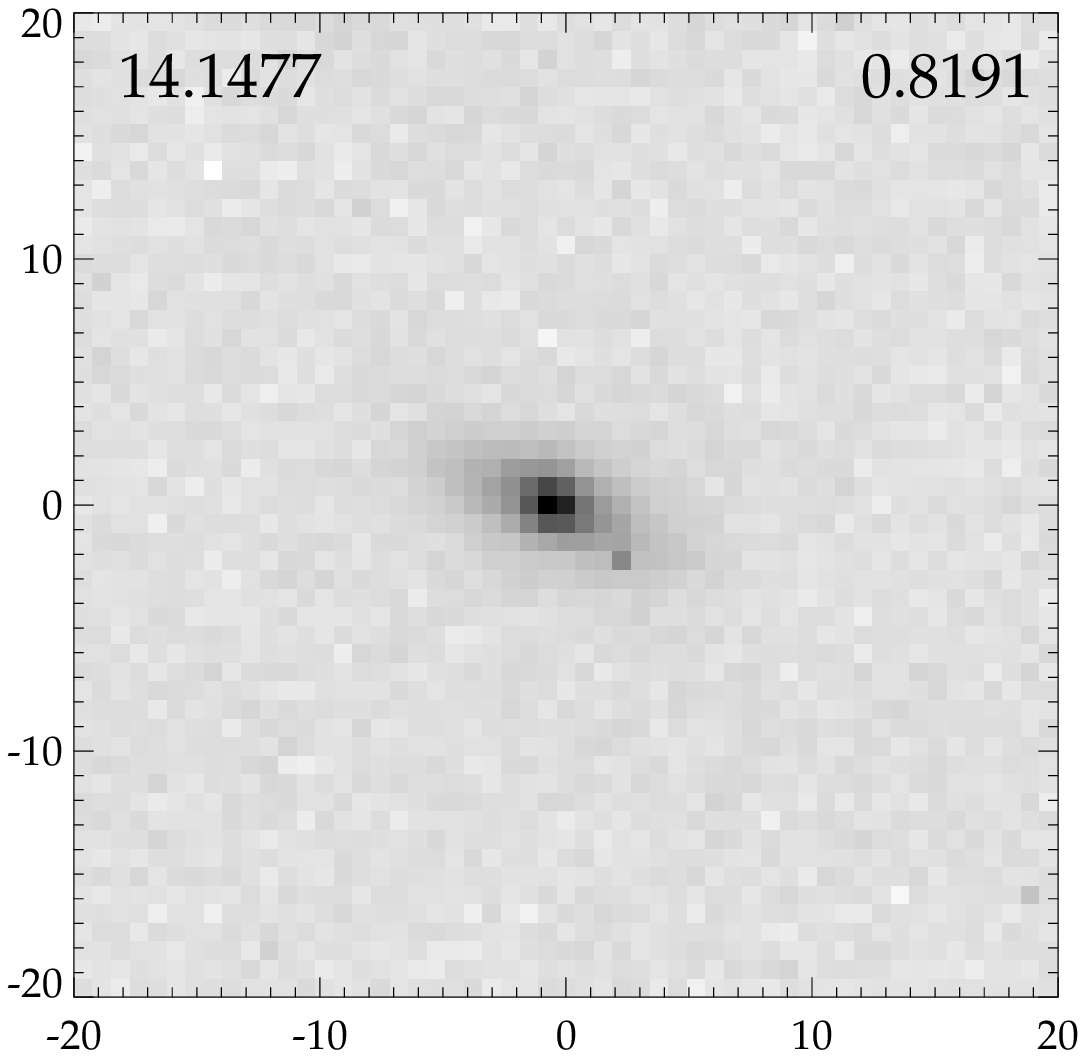} \includegraphics[height=0.22\textwidth,clip]{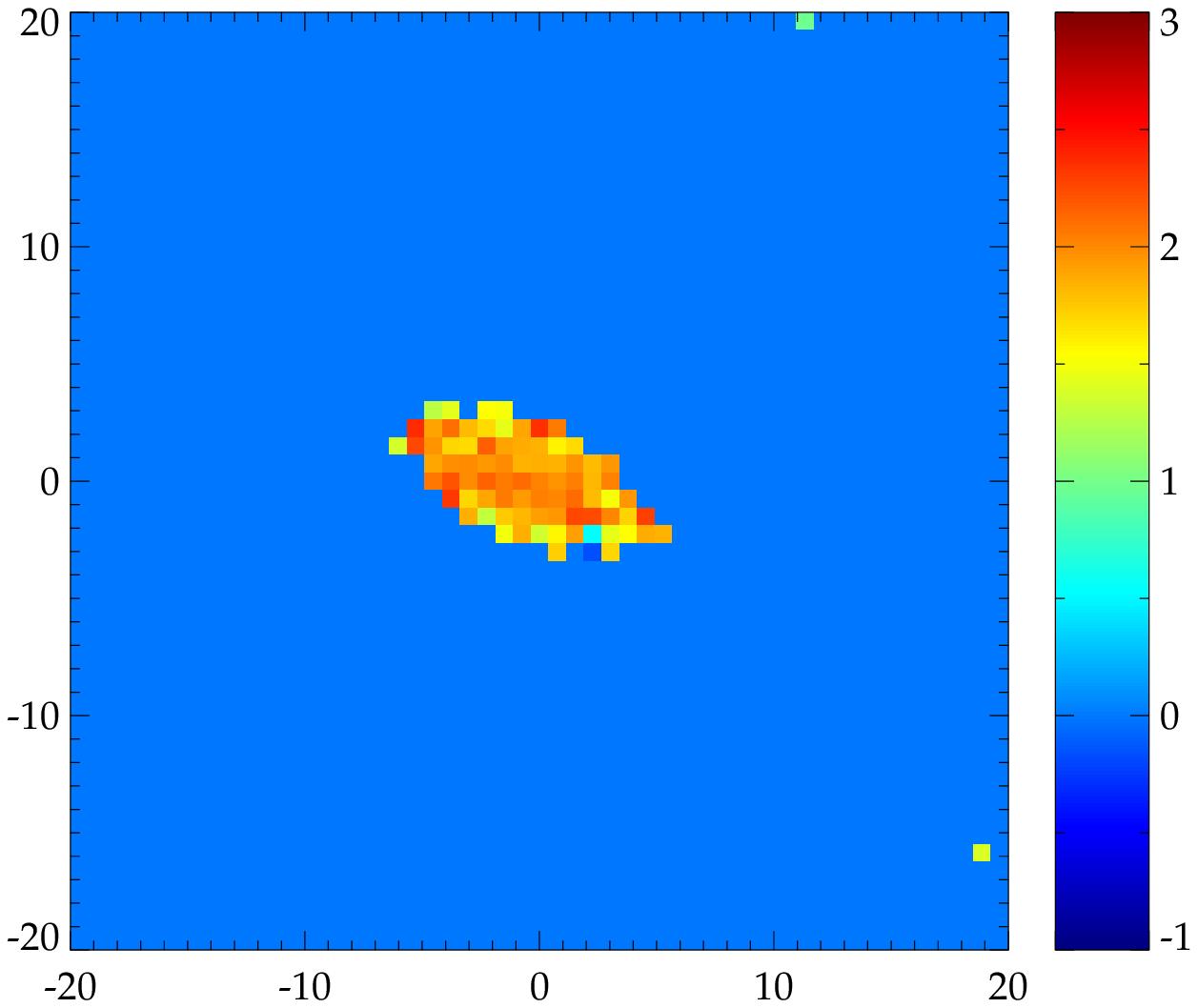}
\includegraphics[height=0.22\textwidth,clip]{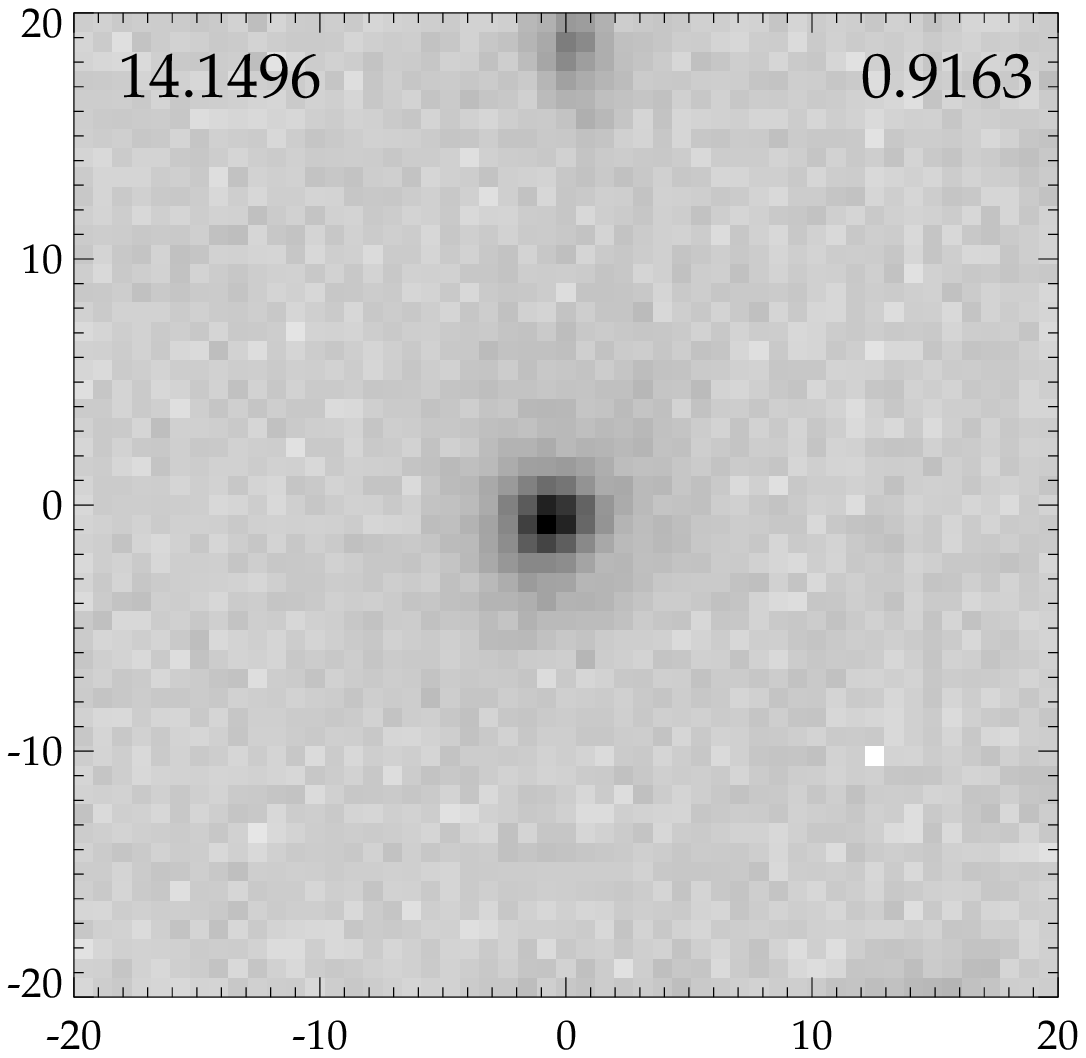} \includegraphics[height=0.22\textwidth,clip]{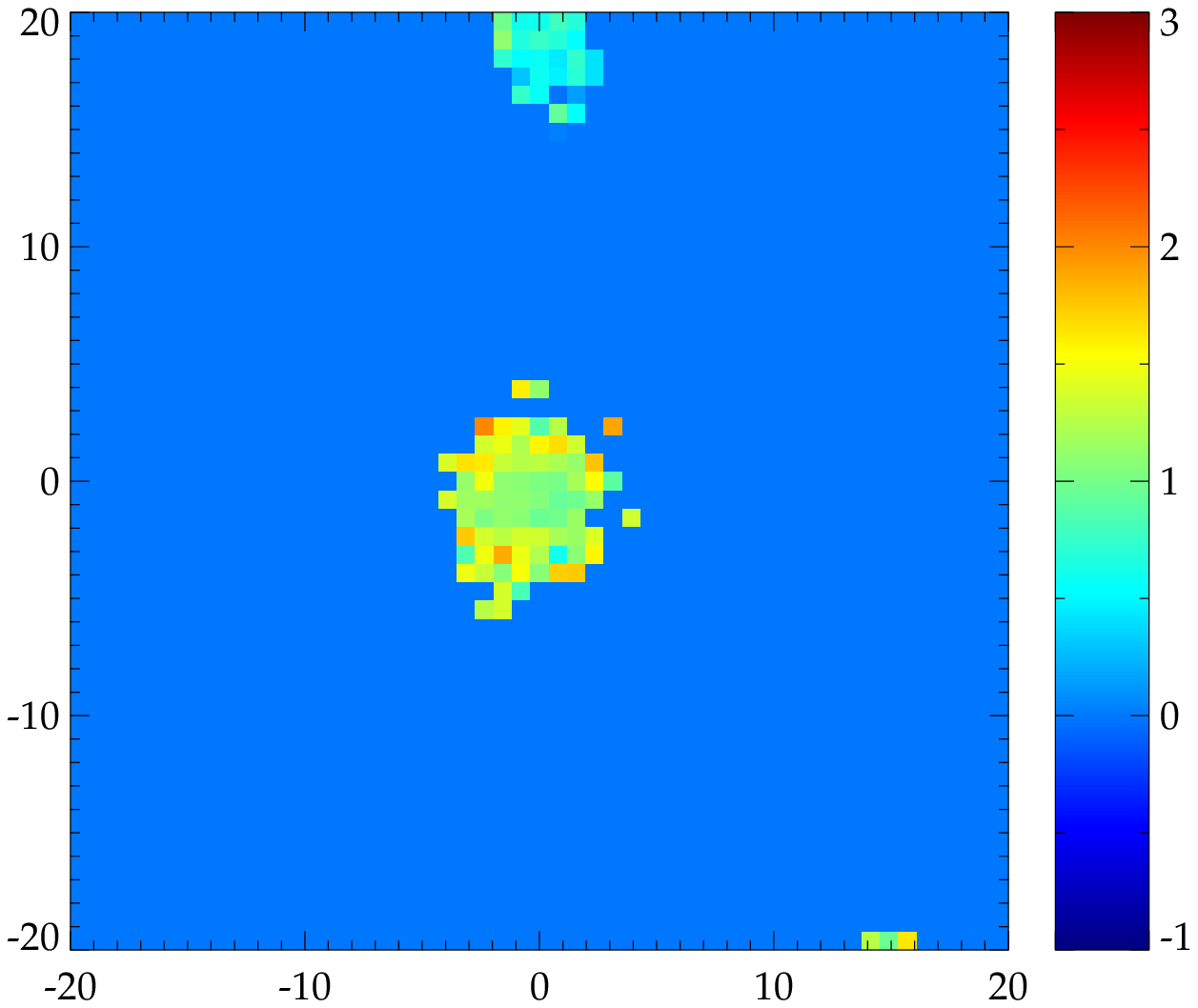}
\includegraphics[height=0.22\textwidth,clip]{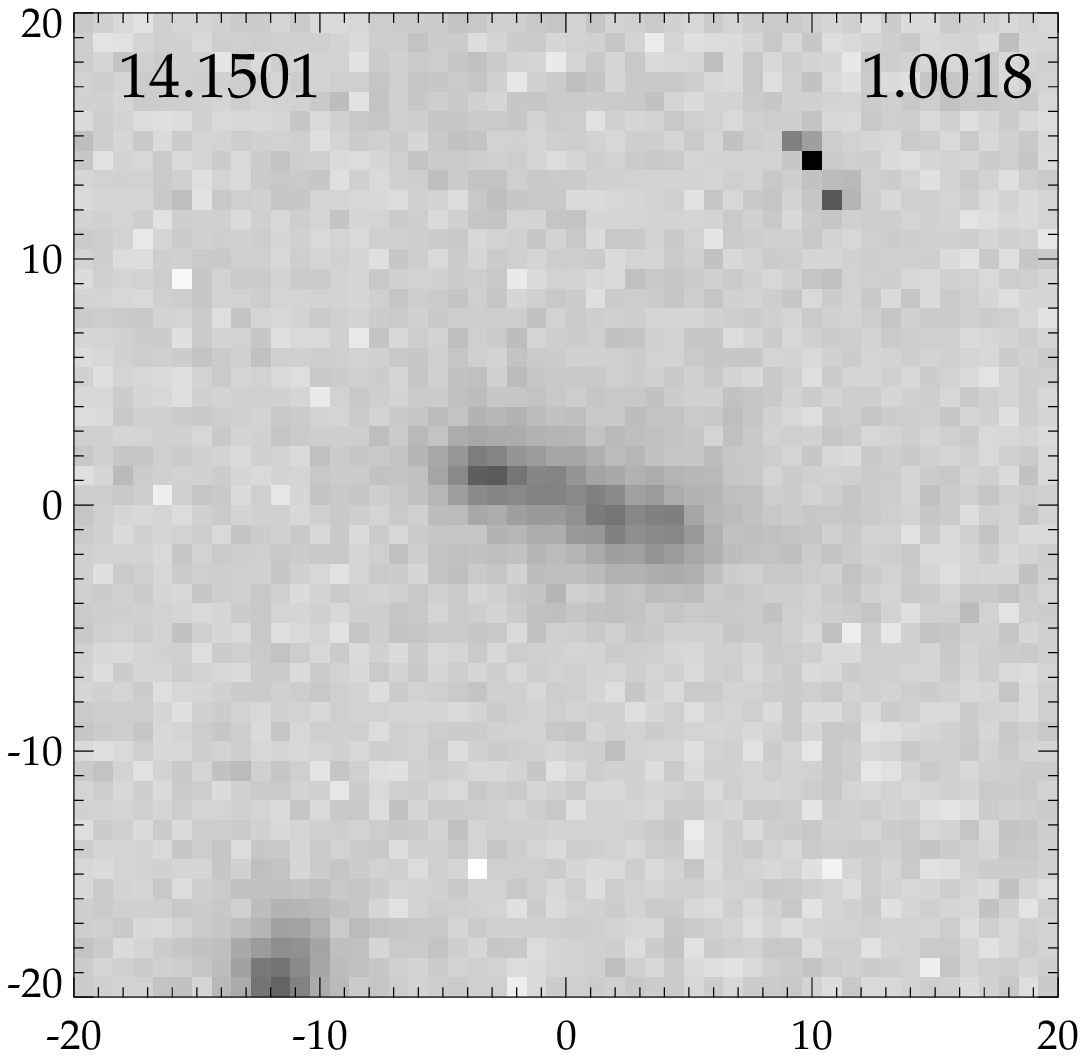} \includegraphics[height=0.22\textwidth,clip]{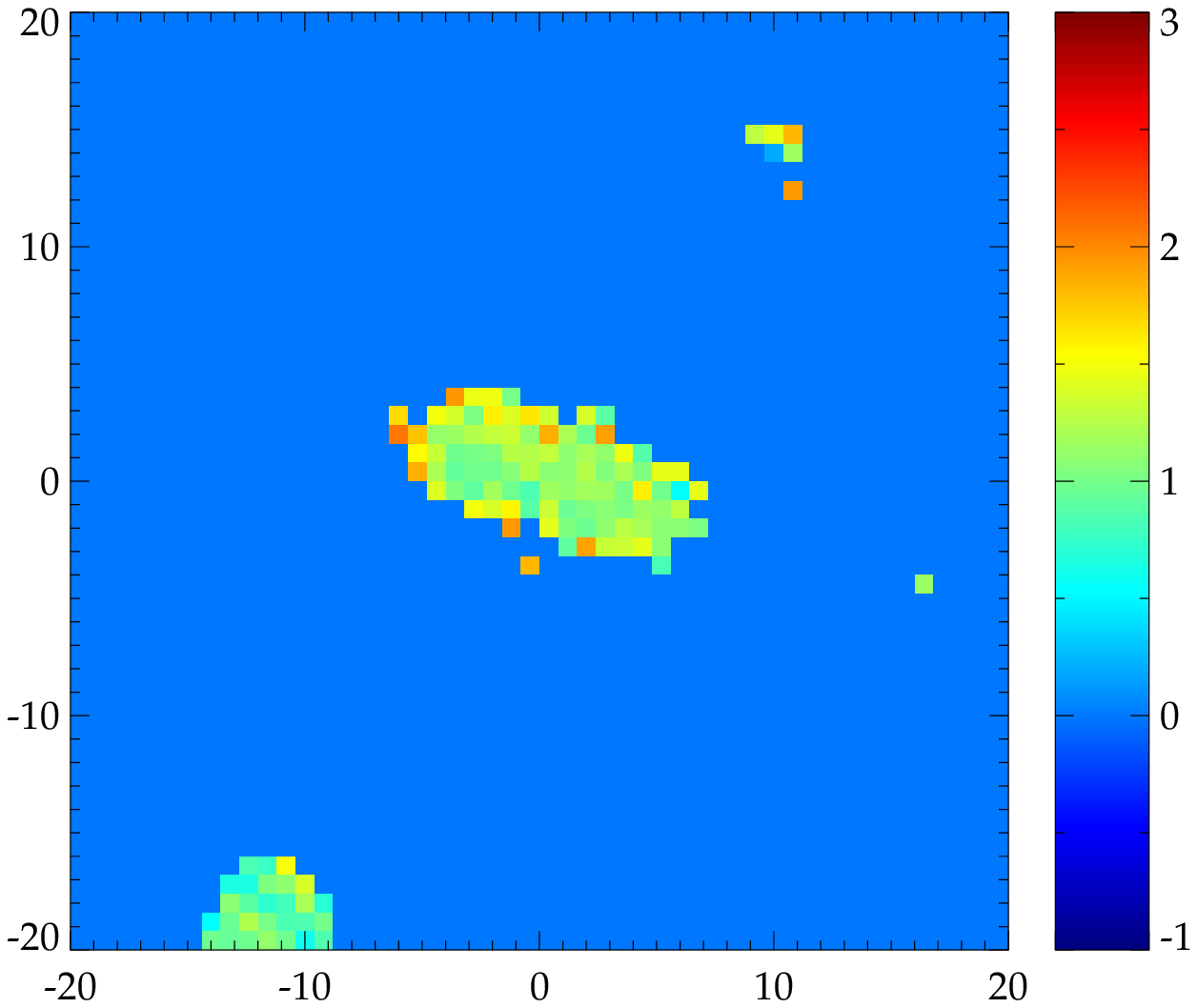}
\includegraphics[height=0.22\textwidth,clip]{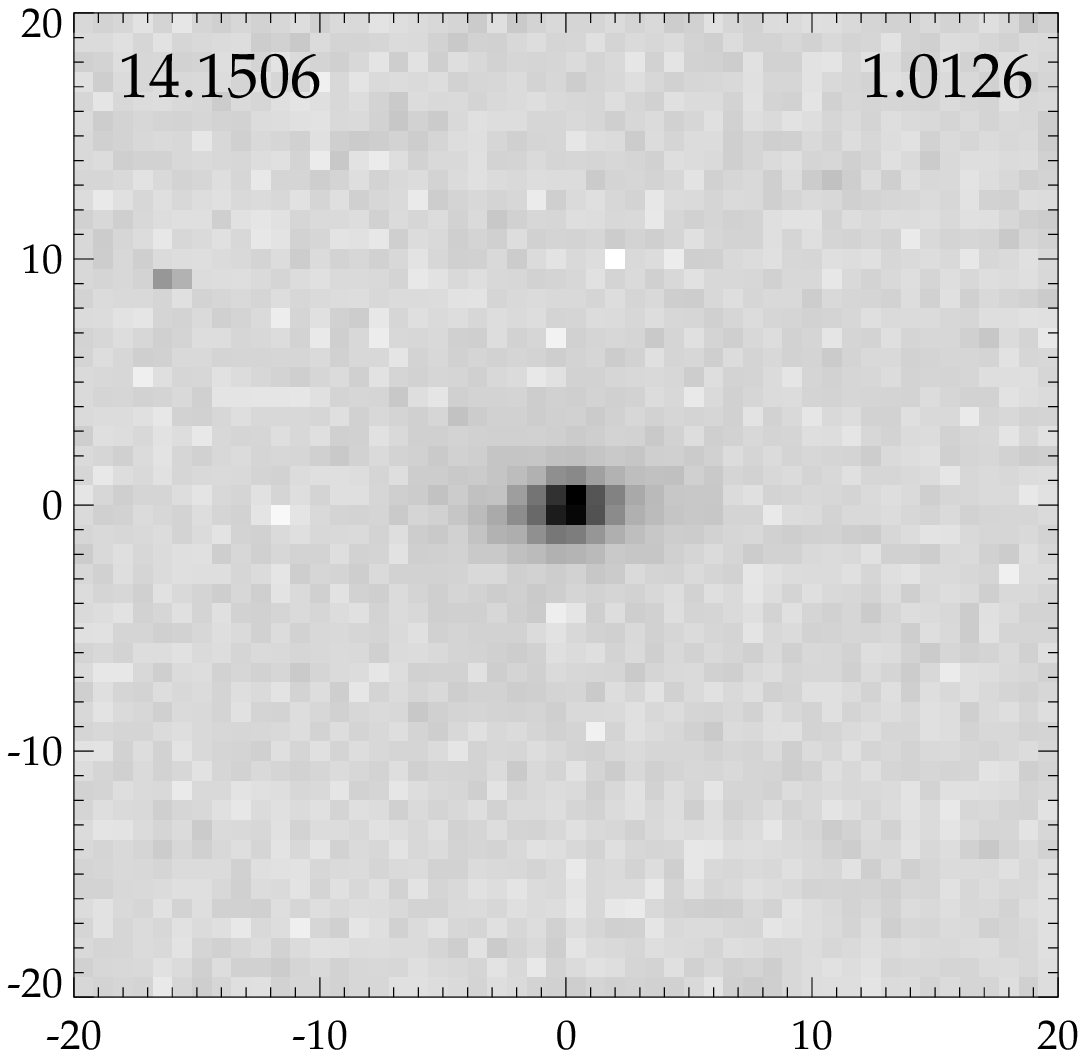} \includegraphics[height=0.22\textwidth,clip]{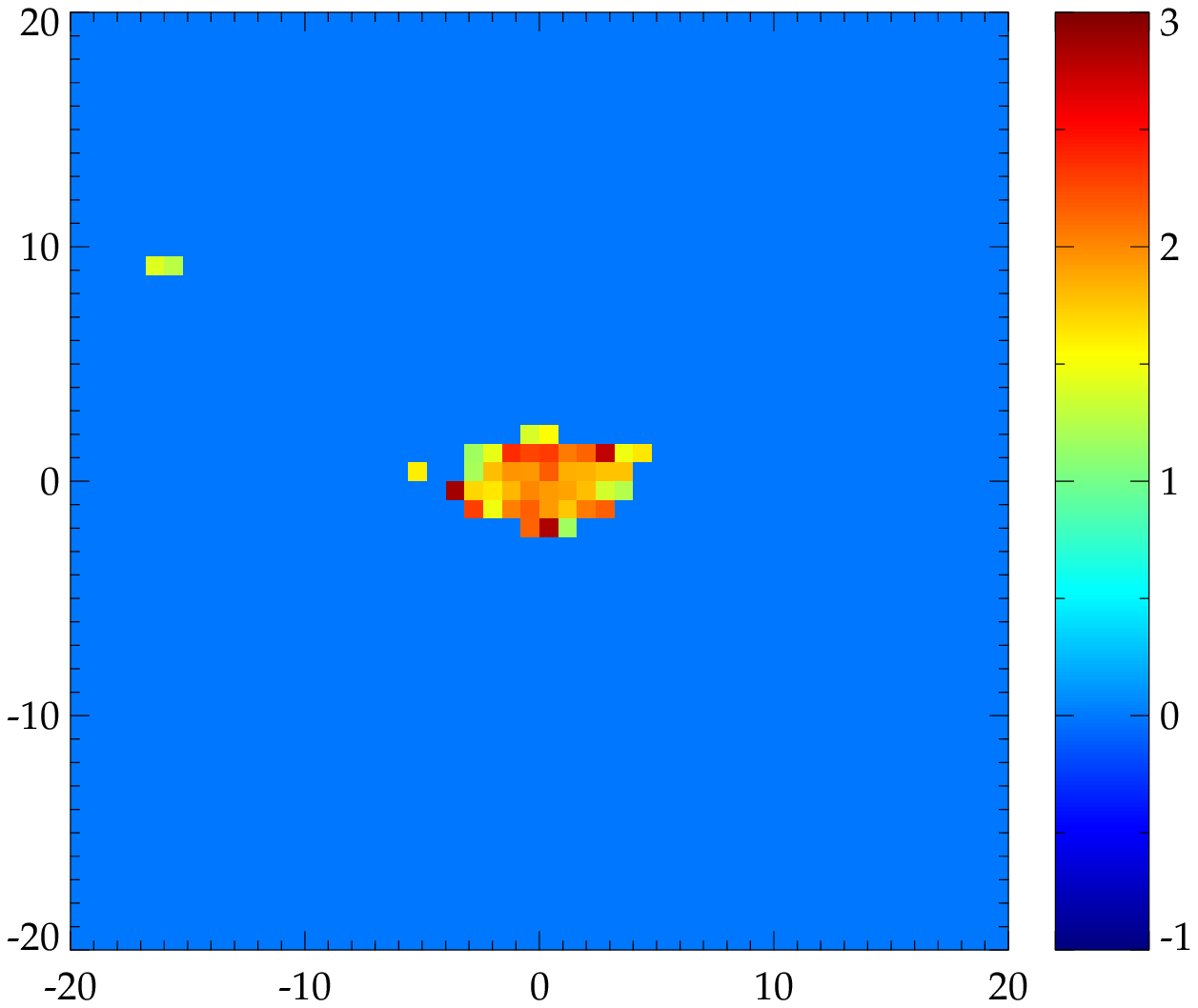}
\includegraphics[height=0.22\textwidth,clip]{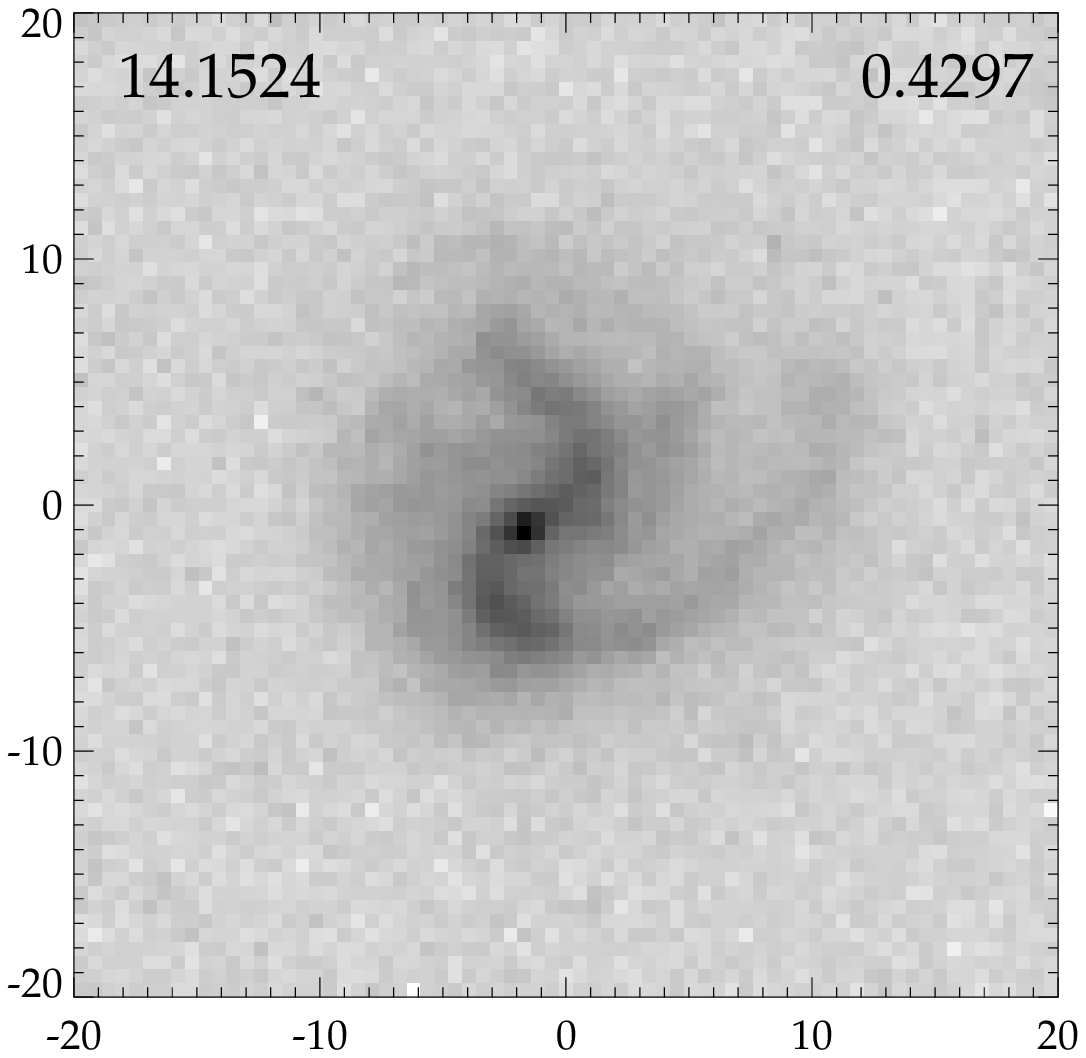} \includegraphics[height=0.22\textwidth,clip]{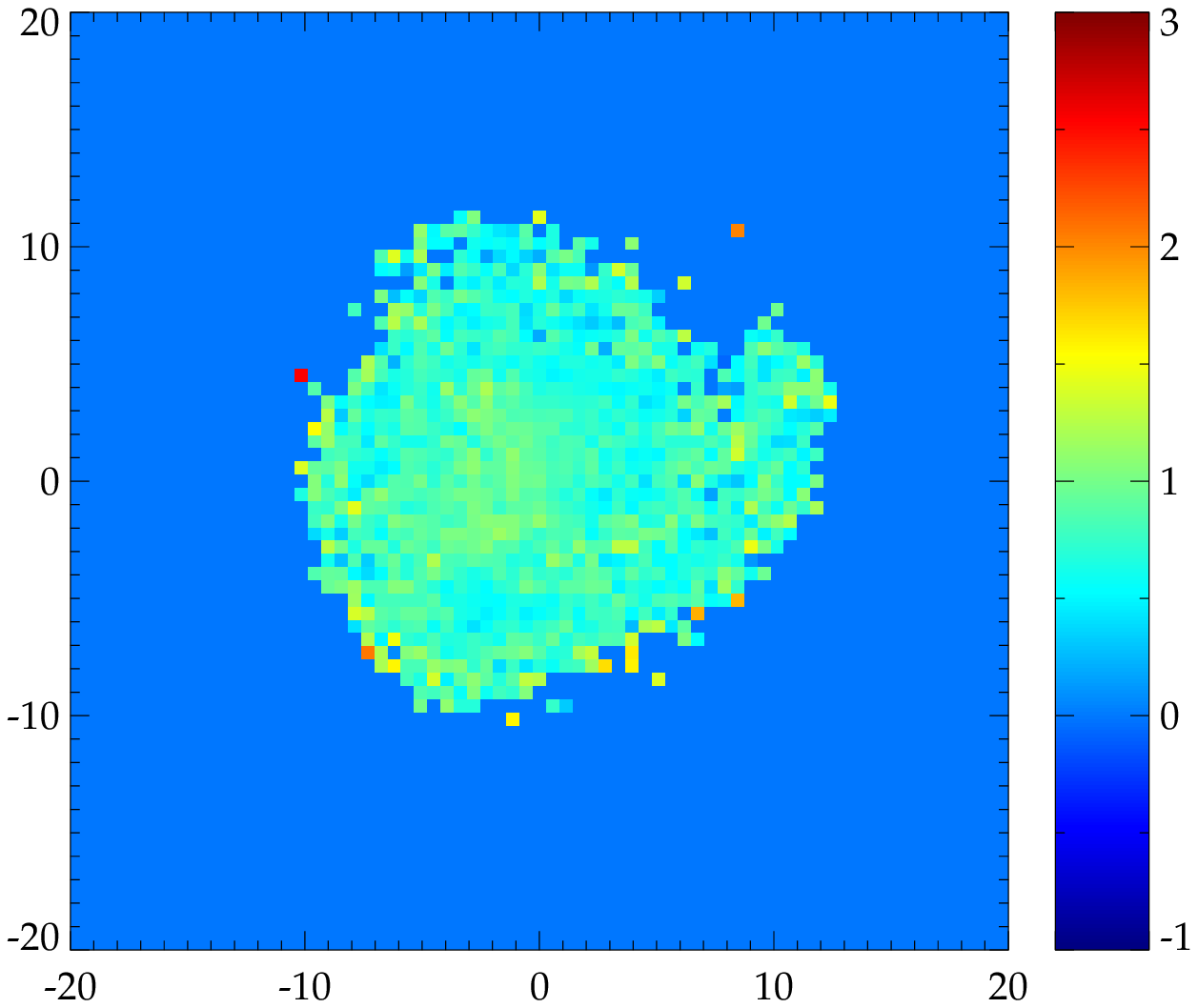}
\caption{Continued.} \end{figure*}

\addtocounter{figure}{-1}
\begin{figure*} \centering
\includegraphics[height=0.22\textwidth,clip]{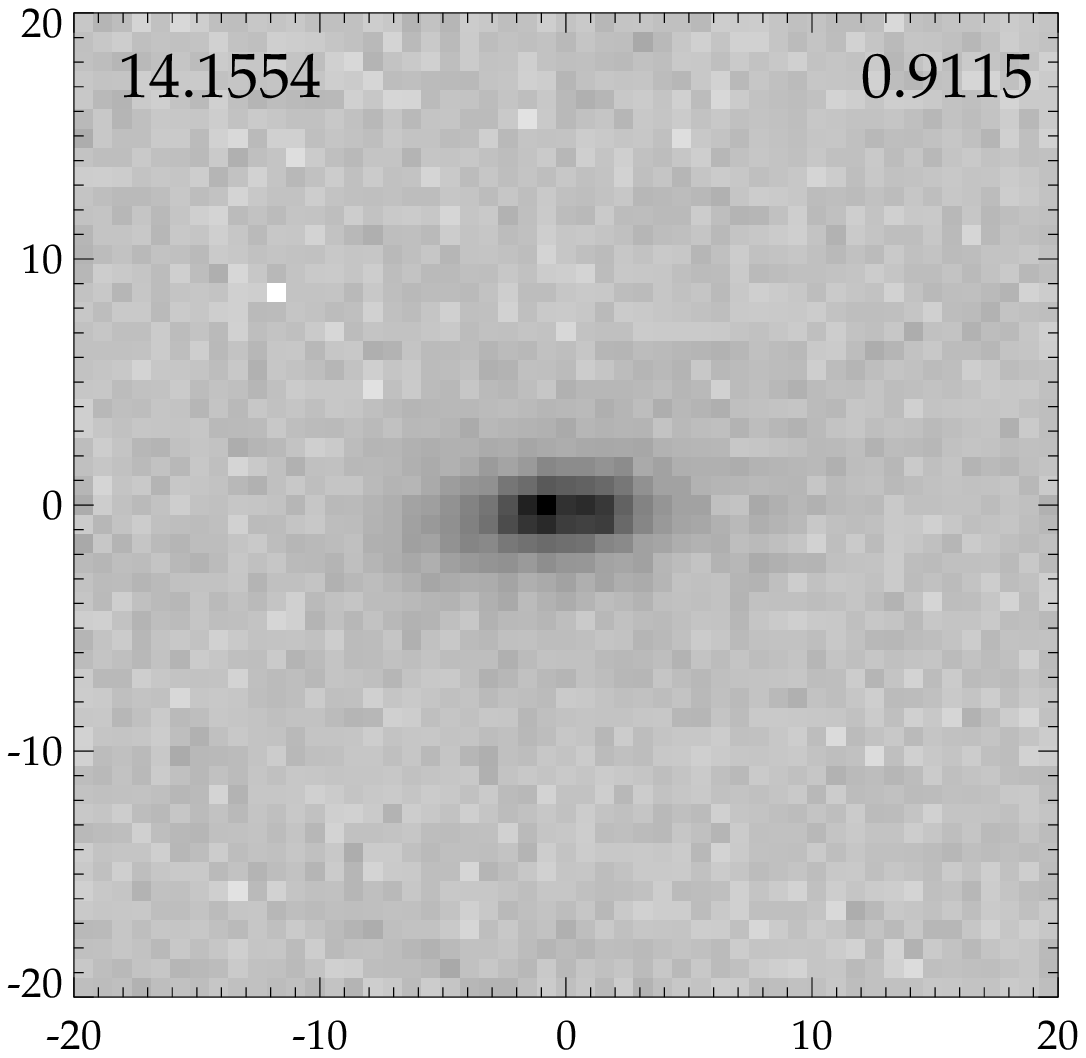} \includegraphics[height=0.22\textwidth,clip]{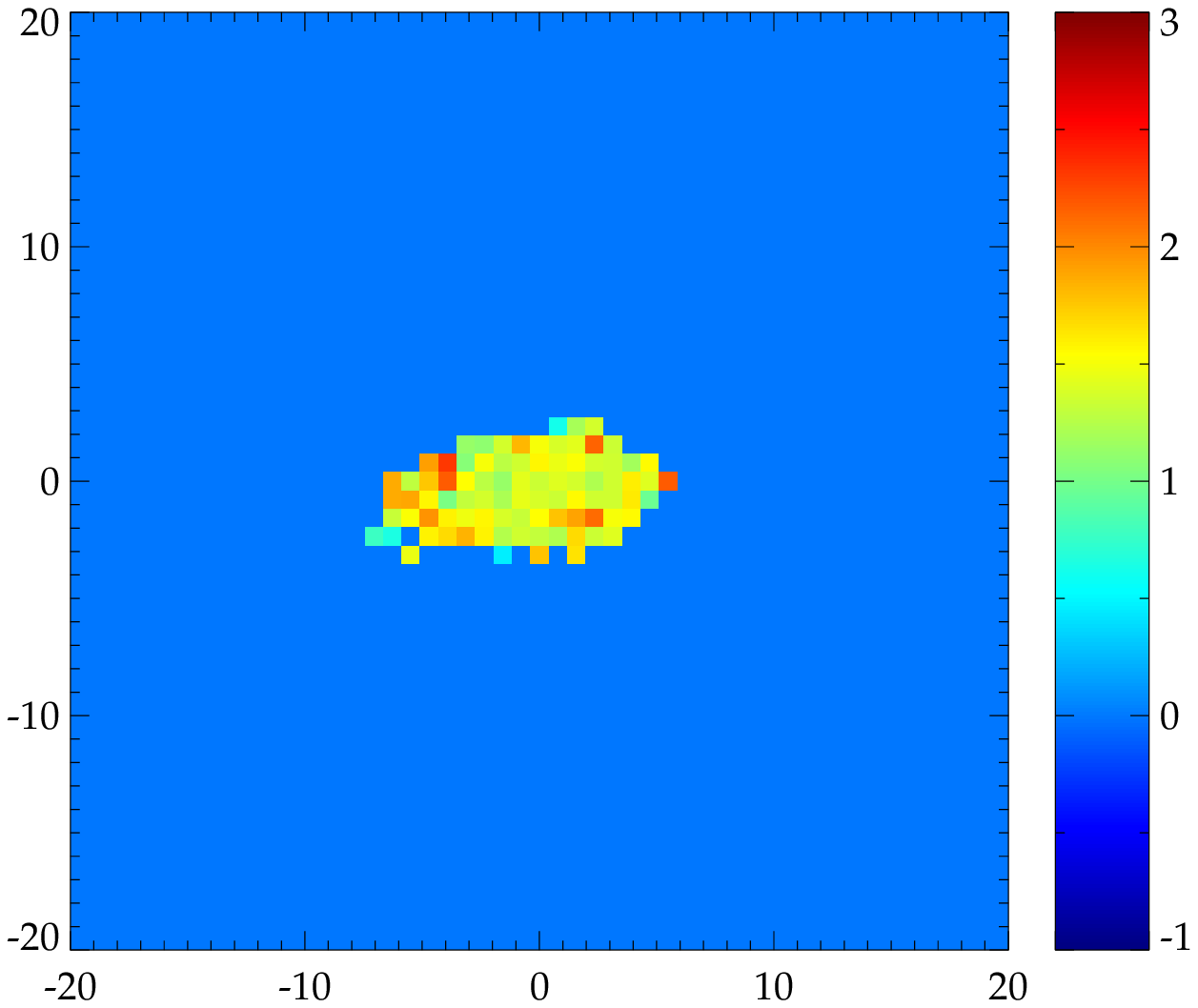}
\includegraphics[height=0.22\textwidth,clip]{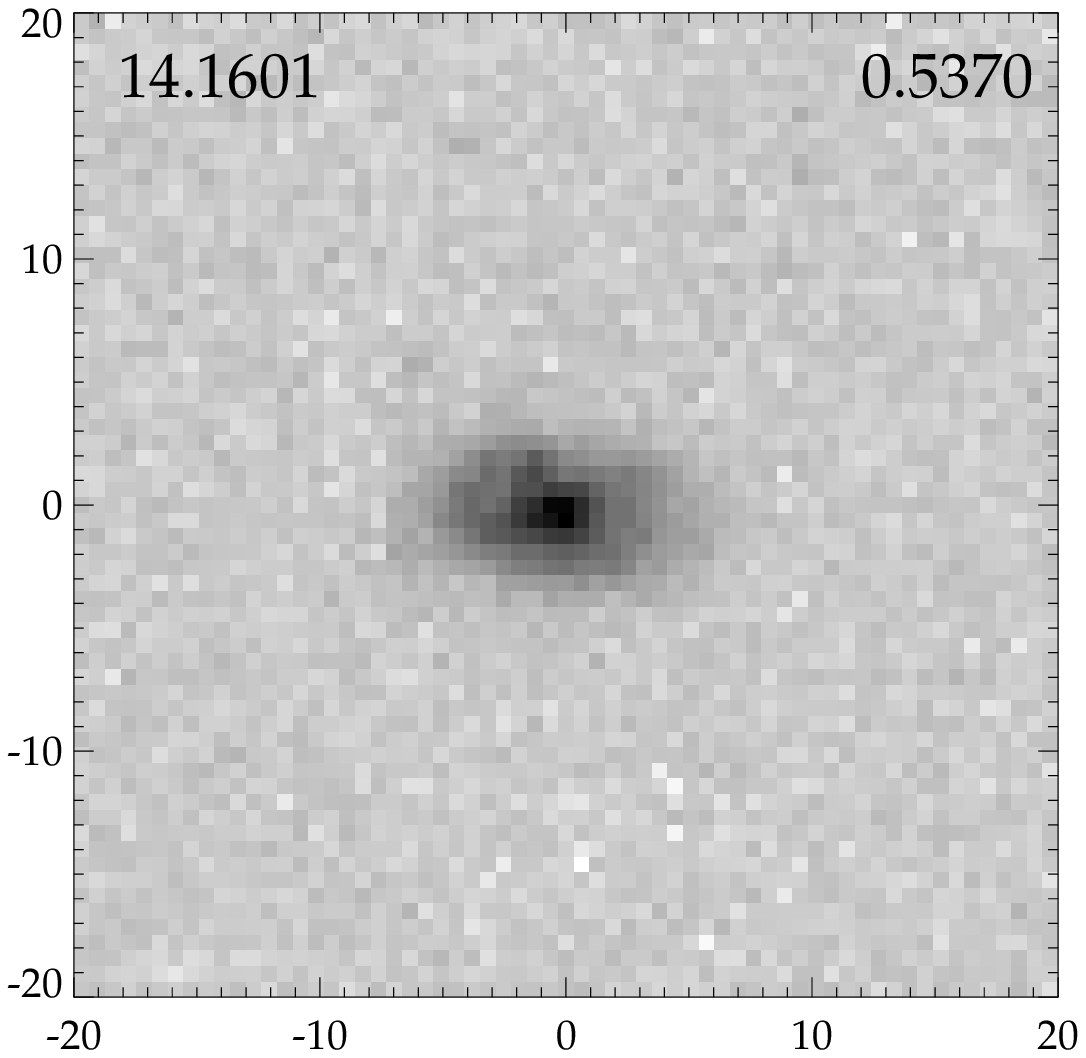} \includegraphics[height=0.22\textwidth,clip]{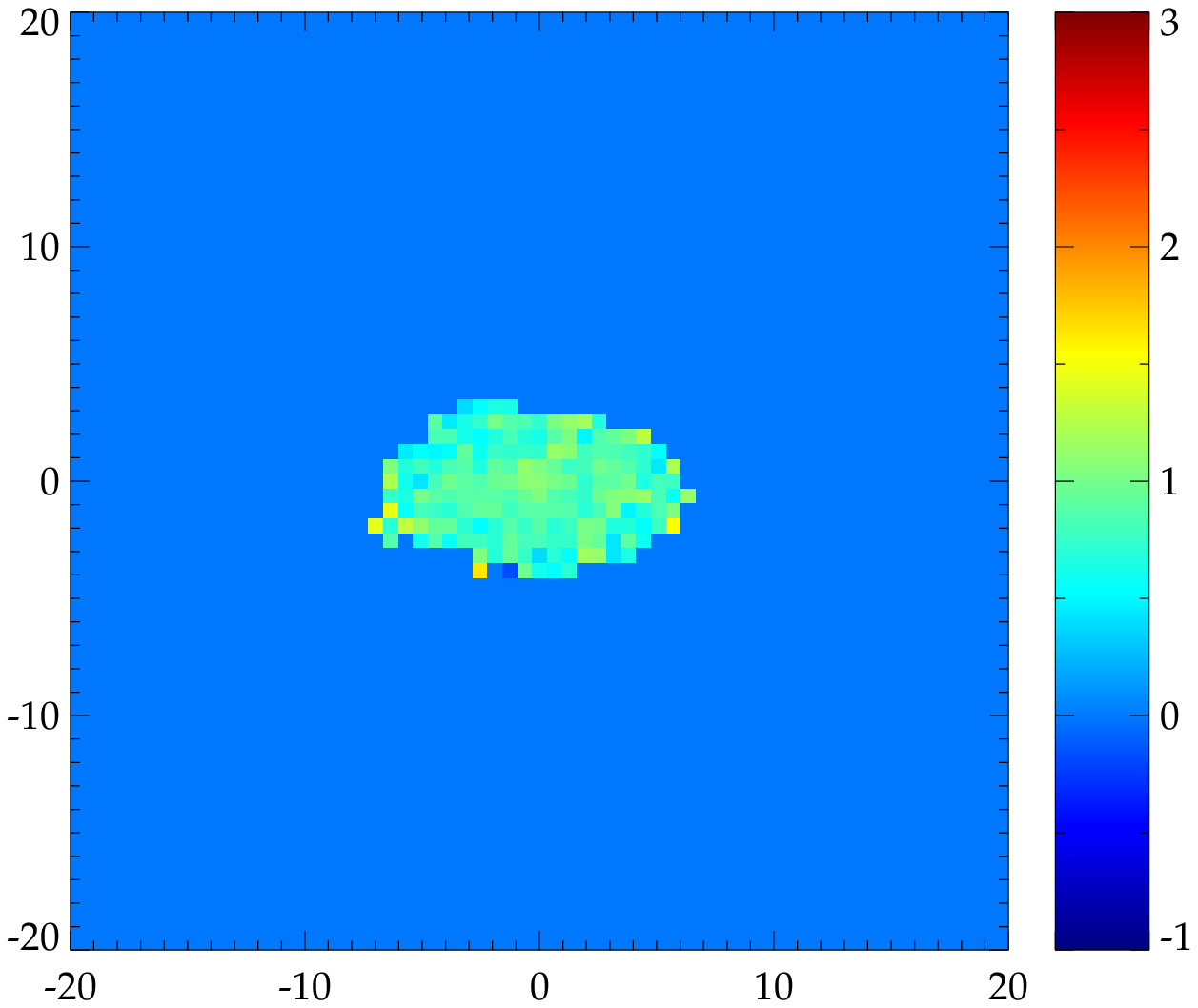}
\includegraphics[height=0.22\textwidth,clip]{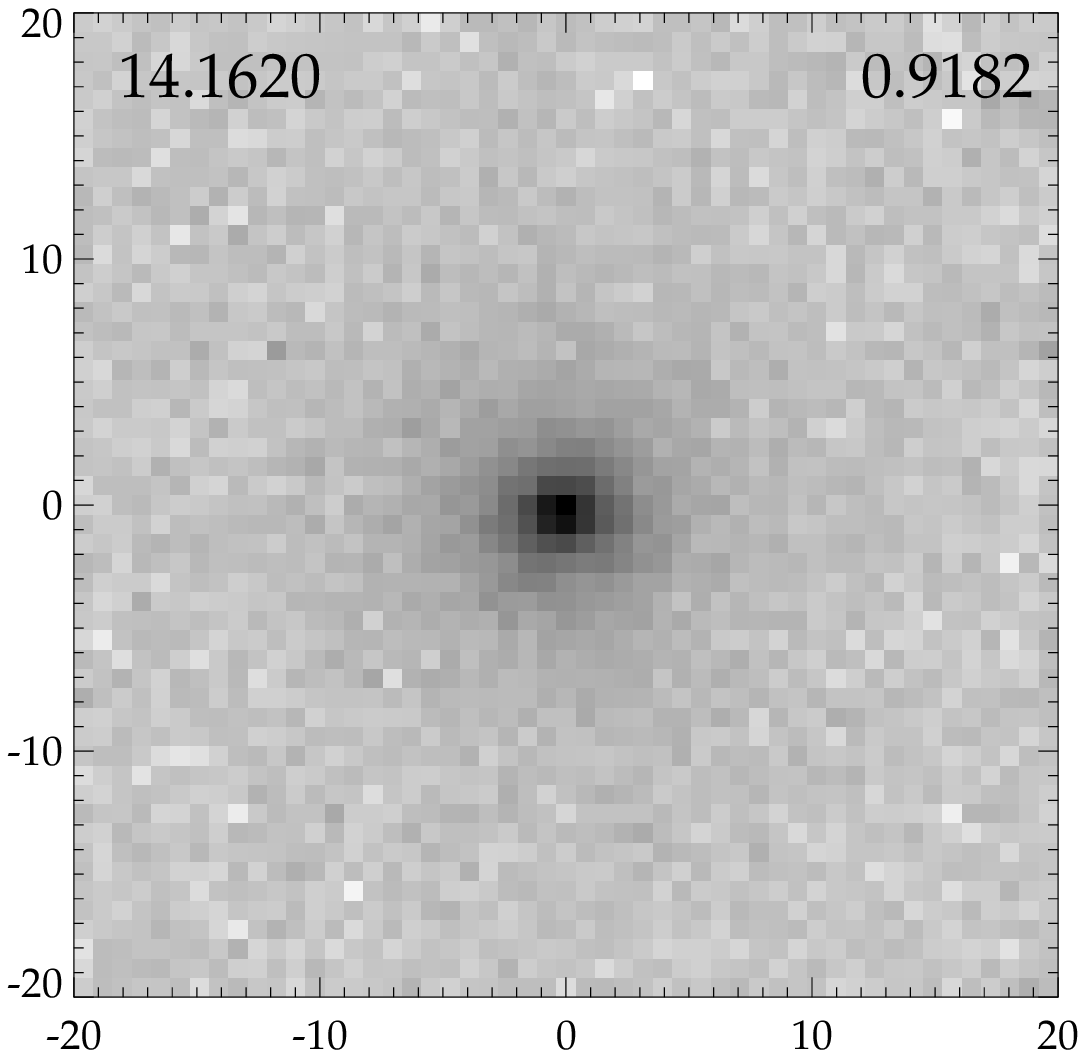} \includegraphics[height=0.22\textwidth,clip]{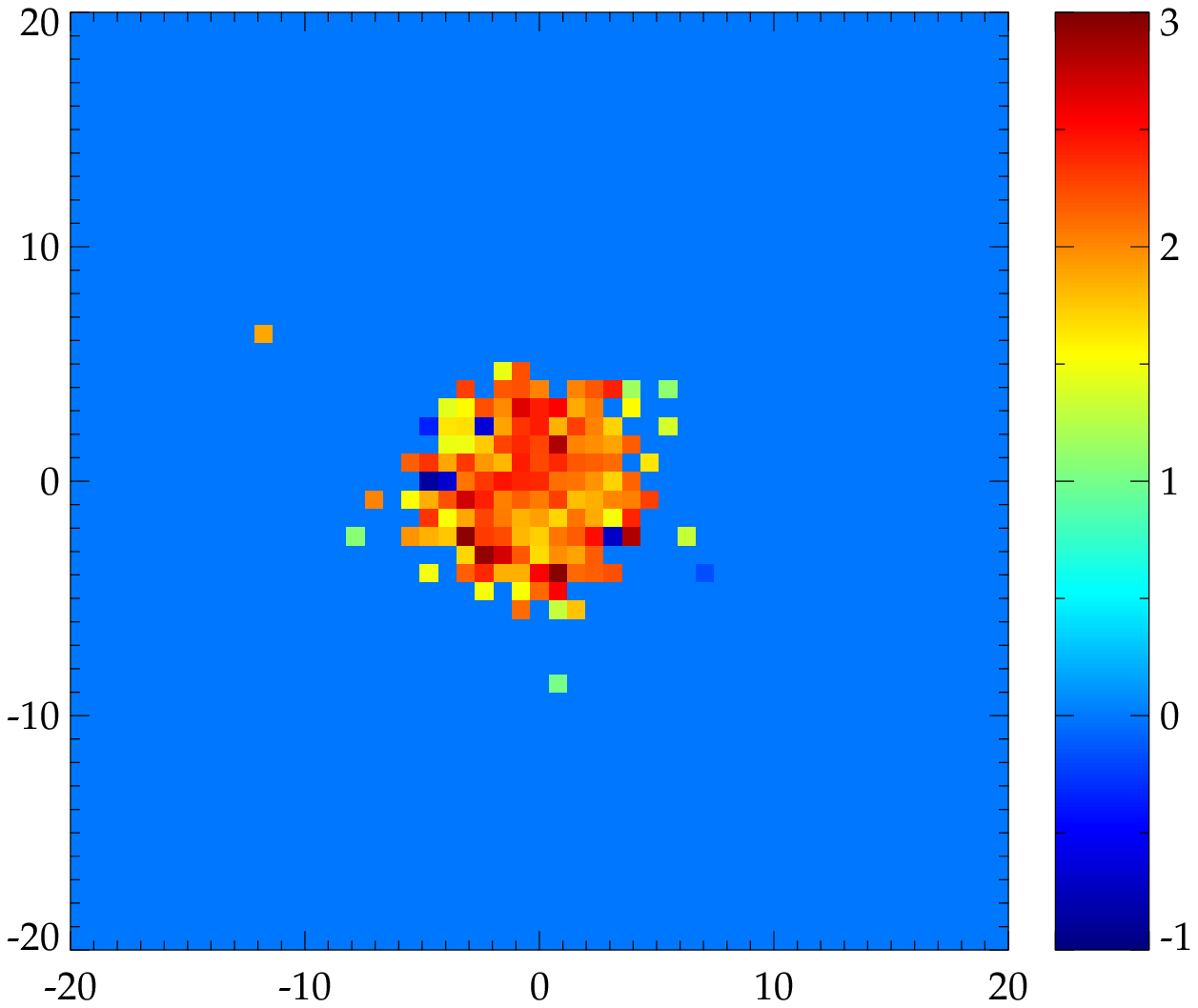} \hspace{0.475\textwidth}
\caption{Continued.
For each target, the name and redshift are labeled in top-left and top-right in the $I_{814}$ image. The color bar shows the color range $-$1 to 3 for the $V_{606}-I_{814}$ color map and 0 to 4 for the $B_{450}-I_{814}$ color map.  The blank in $I_{814}$ imaging was the region beyond the chip border. The size of each image is 40$\times$40\,kpc.[{\it See the electronic edition for a color version of this figure.}]}
\end{figure*}

\begin{figure*}[t] \centering
   \includegraphics[width=0.4\textwidth]{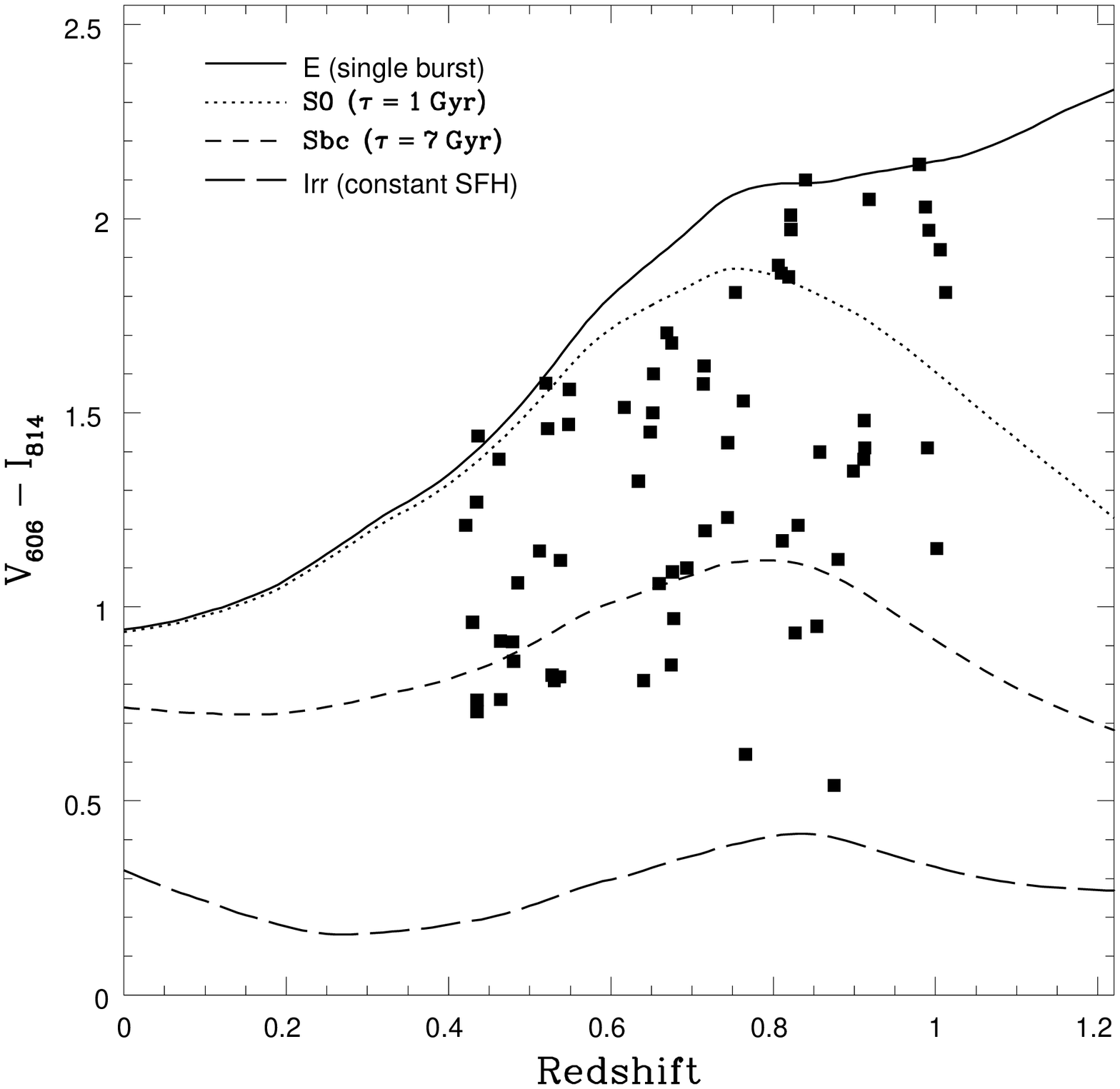}
   \includegraphics[width=0.4\textwidth]{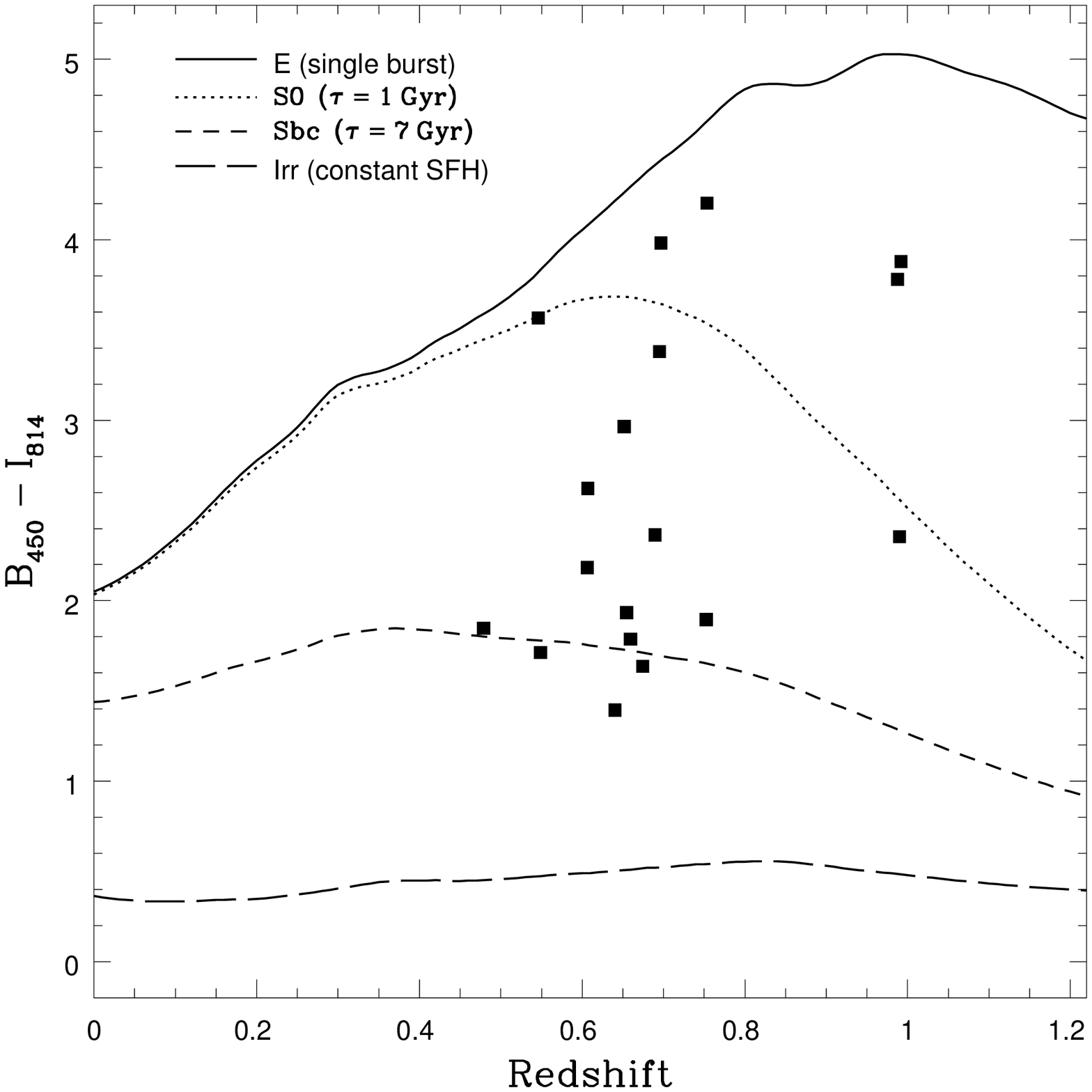}
   \caption{Observed HST colors $V_{606}-I_{814}$ ({\it left panel}) and $B_{450}-I_{814}$ ({\it right panel}) as a function of redshift.}
   \label{modeling}
\end{figure*}

Figure~\ref{hisMB} shows distributions of the absolute $B$ band magnitude of LIRGs and non-LIRGs. Nearly all of our sample galaxies at $z>$\,0.7 are brighter than $M_{\rm AB}(B)$\,=\, $-$20.5 mag and these at $<z\leq$\,0.7 brighter than $M_{\rm AB}(B)$\,=\, $-$19.5 mag. Over the range 0.4\,$<z<$\,1, blue galaxies exhibit a luminosity evolution as significant as $\sim$ 0.8 mag and the luminosity function of red galaxies show little change ($\sim$0.2\,mag, Lilly et al. \cite{Lilly95}; Wolf et al.~\cite{Wolf}). Given the characteristic magnitude at $z\sim$\,0.5 is $M_{\rm AB}(B)$\,$\sim$\,$-$20.5 for both red and blue galaxies (Lilly et al. \cite{Lilly95}), we can see that our sample consist of bright galaxies with $M_{\rm AB}(B)<M_{\rm AB}(B)^{\ast}$+1 with a bias against faint red galaxies at higher redshifts. The effects of sample selection on morphological fractions will be discussed in Section~\ref{dependence}.

\section{Morphological classification}

Marleau \& Simard (\cite{Marleau}) developed a quantitative method to classify galaxy morphologies in the distant universe based on bulge+disk structural parameters. 

With 2-band HST/WFPC2 high resolution observations, we can obtain the spatially resolved color distribution (i.e. color map) for galaxies at $z\sim$1. From a color map of a galaxy, we can see the distribution of stellar populations in the galaxy. Structures such like the red bulge + blue disk clearly seen in nearby galaxies can be recognized. Also, star-forming regions and dusty regions are more noticeable. These physical properties are helpful in recognizing the galaxy. We developed a new reduction pipeline to improve HST/WFPC2 image association and alignment to an accuracy of 0.15 pixel, which allows us to study the properties of the color distribution of galaxies with complex morphologies. We also developed a method to obtain a signal-to-noise ($S/N$) image associated with a color map image. Using the $S/N$ image, a reasonable determination of the target color map area can be done (see Z04 for more details). Figure~\ref{colormap} shows the $I_{814}$ negative greylevel images and the color map images of 75 galaxies in our non-LIRG sample in an order of the CFRS identification (from  left to right, top  to bottom).
With complementary physical properties from the color maps, the morphological classification can be substantially improved. Following Z04, we applied the same classification method to our non-LIRG sample. Here we summarize the morphological classification.

\subsection{Overview of the classification}

We used the software GIM2D (Simard et al. \cite{Simard}) to carry out two-dimensional structural fitting and obtain parameters $B/T$ ratio and $\chi^2$, listed in Table~\ref{table} (Col. 9-14). Visual inspections of the galaxy appearance in HST imaging and color distribution were employed. The observed color was compared to four realistic galaxy models created using GISSEL98 (Bruzual \& Charlot \cite{Bruzual}): Elliptical (single burst), S0 ($\tau\,=\,1$\,Gyr), Sbc ($\tau\,=\,7$\,Gyr) and irregular (constant star  formation rate with  a fixed  age of 0.06\,Gyr). In the models, a formation epoch is adopted at a redshift of $z$\,=\,5. Figure~\ref{modeling} shows modeling HST color $V_{606}-I_{814}$ (left panel) and $B_{450}-I_{814}$ (right panel) as a function of redshift. The integrated color of our sample galaxies is adopted for comparison.

The regular galaxies, i.e. ones well fit by bulge+disk two-dimensional structure, are described by five `Hubble types' mainly based on the $I_{814}$ band $B/T$ ratio: E (0.8\,$<B/T\leq$\,1), S0 (0.5\,$<B/T\leq$\,0.8),  Sab (0.15\,$<B/T\leq$\,0.5), Sbc (0\,$<B/T\leq$\,0.15) and Sd ($B/T$\,=\,0).  An additional two types, compact (C) and irregular (Irr) are included to describe the objects without a clear bulge+disk structure.  Note that the criterion of half-light radius $r_{50}<$\,3.5\,h$^{-1}_{70}$\,kpc is used to identify compacts (Hammer et al. \cite{Hammer01}). A quality  factor is used to give the confidence of our classification. Four labels are taken to classify interacting/merging systems. See the footnotes of Table~\ref{table} for details.

Based on structural parameters and physical properties derived from color maps, morphological classification was applied to our sample by X.Z.Z. and  F.H. independently (same as in Z04). A final version is approved after discussing several inconsistent classifications. Table~\ref{table} lists the results of our morphological classification, including galaxy type (Col. 15), quality factor (Col. 16) and interacting/merging type (Col. 17). 

\subsection{Fractions of different morphological types}

\begin{table*}\centering
  \caption[]{Fractions of different morphological types in different populations. The last column is the upper limit of merger fraction, including objects showing signs of merging/interaction. The fractions of possible merger/interacting galaxies related to each type are given in parentheses.}
  \label{classification}
  \begin{tabular}{ccccccc} 
  \hline
  \noalign{\smallskip}
Galaxy Type & E/S0 & Spiral & Irregular & Compact & Merger & Merger$_{\rm upper}$ \\
     & (\%) &   (\%) &  (\%)     &   (\%)  &  (\%)  &  (\%)  \\
  \noalign{\smallskip}
  \hline
  \noalign{\smallskip}
non-LIRGs     & 27(1)  & 45 (8) & 7 (5)  & 17 (11) & 4  & 29 \\
  \noalign{\smallskip}
  \hline
  \noalign{\smallskip}
non-LIRGs(corrected)     & 32(1)  & 43 (7) & 6 (4)  & 15 (10) & 4  & 26 \\
  \noalign{\smallskip}
  \hline
  \noalign{\smallskip}
LIRGs         &     & 36 (14) & 22 (14) & 25 (14) & 17 & 58 \\
  \noalign{\smallskip}
  \hline
  \noalign{\smallskip}
Field Galaxies& 27  & 42 & 8 & 17 & 6  & 31 \\
  \noalign{\smallskip}
  \hline
  \noalign{\smallskip}
Field Galaxies (w/o Comp.) & 36  & 48 & 10 &  & 6  & 31 \\
  \noalign{\smallskip}
  \hline
  \end{tabular}
\end{table*}

An elliptical galaxy may appear like a compact galaxy. In our classification, however, they are separated due to different surface brightness distributions. While an elliptical galaxy's surface brightness distribution can be well fit by a de Vaucouleurs law (elliptical-type), the distribution of a compact galaxy usually differs from either the de Vaucouleurs law or an exponential function (disk-type). In addition, color maps show that elliptical galaxies usually exhibit a homogeneous red color distribution and compact galaxies show a relatively complex and blue color distribution.

Because of the limited size of our samples, no effort is made to split sample into different redshift bins and address the evolution of the fractions of different morphological types. 
Considering the uncertainty in the $B/T$ ratio determination, we adopted four major morphological types to calculate their frequencies.

 From Table~\ref{table}, we can summarize the frequencies of different types in non-LIRGs. Note that there are 5 insecure classifications, i.e., ones with the quality factor $Q$\,=\,3. Their morphological types are not biased to one or two types and thus the insecure classifications would not considerably affect the fraction distribution. Table~\ref{classification} lists the fraction distributions of the non-LIRGs as well as LIRGs (Z04).

\subsection{Dependence on sample selection}\label{dependence}

In the local universe, morphological fractions are luminosity-dependent in the sense that fainter galaxies include more late-type and irregular galaxies (Marzke et al. \cite{Marzke}; Nakamura et al. \cite{Nakamura}). For bright galaxies, however, their morphology fractions are little dependent on the magnitude threshold used to select them.  Marzke et al. (\cite{Marzke}) measured the dependence of the luminosity function on galaxy morphological type in the $B$ band. They gave the luminosity functions of three galaxy types, i.e., E/S0, Spiral and Irregular/peculiar. The frequency of each type changes only less than one percent when the criterion of bright galaxies is changed from $M_B < M_B^{\ast}+1$ to $M_B<M_B^{\ast}+2$. Irregulars/peculiars start to represent a noticeable fraction at  $M_B>M_B^{\ast}+3$.

Our sample consists of CFRS galaxies with absolute $B$ band magnitude $M_{\rm AB}(B)<M_B^\ast + 1$ at redshifts from 0.4 to 1. The CFRS survey selects galaxies with 17.5\,$\leq I_{\rm AB}\leq$\,22.5. The survey is complete at $M_{\rm AB}(B)$\,$< -$20.4 mag for galaxies up to $z$\,=\,1 and $-$19.6 mag to $z$\,=\,0.4 (Lilly et al. \cite{Lilly95}). Since blue and red galaxies show different amounts of luminosity evolution over the range 0.4\,$<z<$\,1, our sample suffers from an incompleteness of red galaxies at higher redshifts. From the $B$ band luminosity function of red galaxies at $z\sim$0.9 (for Type 1, i.e. E/S0-Sa, in Wolf et al.~\cite{Wolf}), we derived that the red galaxies at $z\sim$\,0.9 are underestimated by $\sim$65\% (alternatively, by a factor of 0.6) due to a higher cut at $M_{\rm AB}(B)<M_B^\ast + 0.2$ than blue galaxies. The red galaxies are mostly related to E/S0 and Sab galaxies. We corrected the underestimate in morphological fractions. Corrected numbers are given in Table~\ref{classification} as well. 

The LIRGs represent 15\% of field galaxies (Hammer et al. \cite{Hammer04a}). Taking LIRGs and non-LIRGs together, the frequencies of different types in the field galaxy population is also listed in Table~\ref{classification}. Because the luminous blue compact galaxies are distinct from regular galaxies and more common at intermediate redshifts (Guzm$\grave{\rm a}$n et al. \cite{Guzman}; Hammer et al. \cite{Hammer01}), we isolate them as a subpopulation. The compact type adopted in this work did not appear in some morphological investigations of distant galaxies in literature. We present corresponding morphological fractions in terms of visual classification for comparison.

Note that we examine the global morphological properties of field galaxies over a redshift range 0.4\,$<z<$\,1. The morphological evolution within this range is smeared. The frequency of irregulars or mergers can alter by a factor of 2 to 3 (e.g. B98; van den Bergh \cite{Bergh01}). This redshift range corresponds to a time span of 3.5\,Gyr. compared to the lookback time of 6.3\,Gyr at $z$\,=\,0.7. Our statistics roughly represent the morphological properties of the field galaxies at $z\sim$\,0.7.

\subsection{Comparison with previous classifications}

Brinchmann et al. (B98) analyzed morphologies of a sample of 341 CFRS and LDSS galaxies with HST imaging data. In the sample, 144 objects satisfy $M_{\rm AB}(B)<-19.5$\,mag and 0.4\,$<z<$\,1 (same selection as our sample). 
In their $A$-$C$ classification, E/S0, spiral and irregular galaxies represent 33.3\%, 46\% and 20\%, respectively. This is in good agreement with our ``visual'' classification. Note that the compact galaxies in their visual classification appear like a star, inconsistent with our criterion. It was pointed out in B98 that 37\% of the $A$-$C$ spirals and 26\% of $A$-$C$ peculiars at $z>$\,0.5 satisfy the criterion for compact galaxies in Guzm$\grave{\rm a}$n et al. (\cite{Guzman}; comparable to our criterion, see Hammer et al. \cite{Hammer01}). This means that in their $A$-$C$ classification, about 22\% objects satisfy the criterion of compacts, which is similar to the fraction we obtained (17\%).

Because of different criteria in identifying compacts, it is not appropriate to compare our fractions of spirals and irregulars with that in B98. For the fraction of E/S0 galaxies, they obtained a value smaller than ours. In addition, van den Bergh (\cite{Bergh01}) performed a visual classification of 229 CFRS galaxies. A constant frequency of about 17\% was suggested for E/S0 galaxies by $z$\,=1.2. We noticed that our fraction of E/S0 will be in good agreement with the fraction of E/S0 and S0/a galaxies in B98. We explain this difference by the limitation of visual classification. Benefiting from a bulge+disk decomposition and color map, our classification is able to better separate E/S0 from early-type spirals.

\section{Detection of bars}

Bar structure can be recognized from characteristic changes of ellipticity and  position angle (P.A.) in isophotes. The existence of a bar will cause an increase of isophote ellipticity and little change of their P.A., and a shape drop of the ellipticity beyond the end of the bar (Regan \& Elmegreen \cite{Regan}; Sheth et al. \cite{Sheth}).

\begin{figure} \centering
   \includegraphics[width=0.4\textwidth]{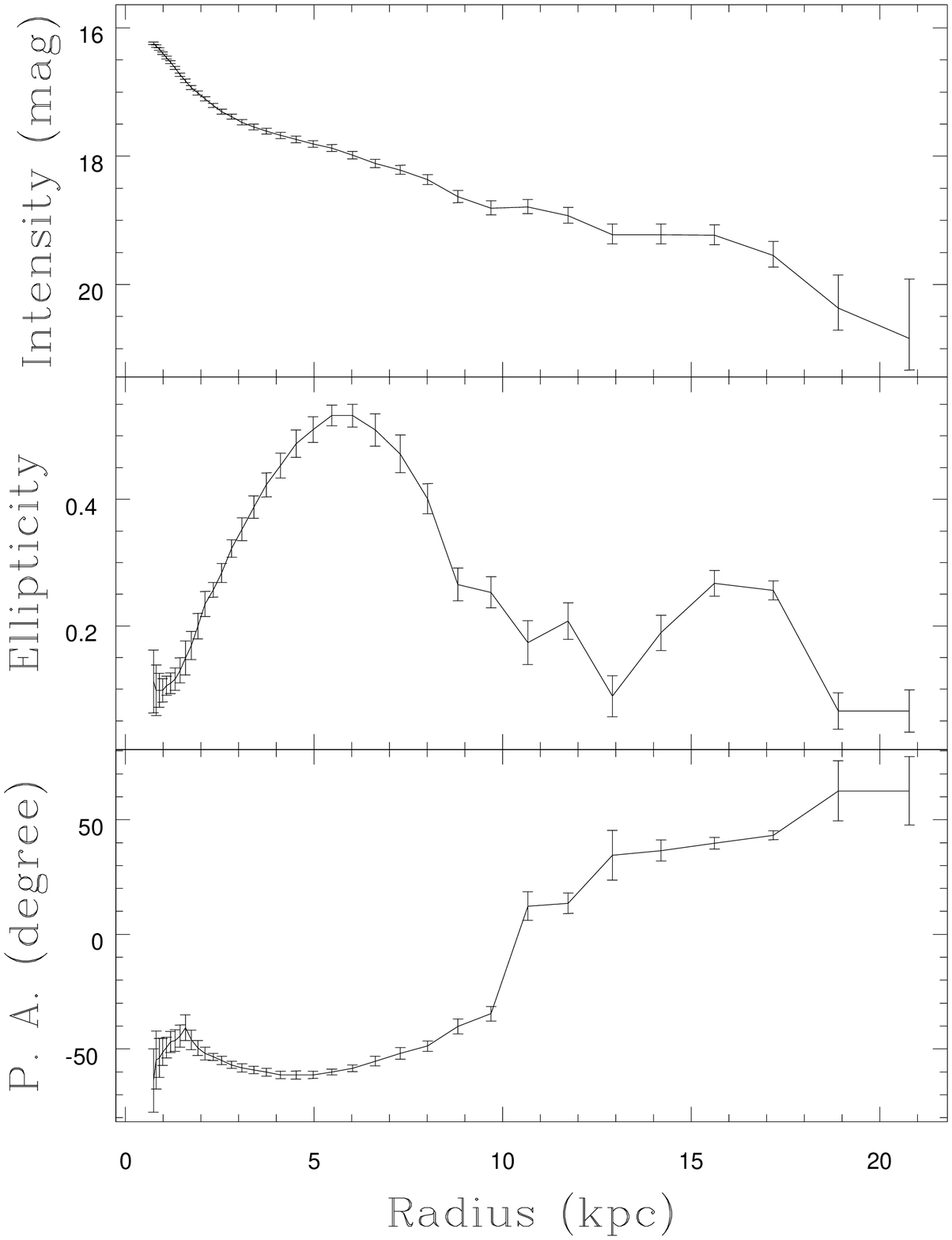}
   \caption{Isophotal analysis of barred spiral galaxy 14.0580. Surface brightness ({\it top panel}, not calibrated), ellipticity ({\it middle panel}) and P.A. ({\it bottom panel}) are shown as a function of semi-major axis length.}
   \label{isophote}
\end{figure}

\begin{figure} \centering
   \includegraphics[height=0.42\textwidth,angle=-90]{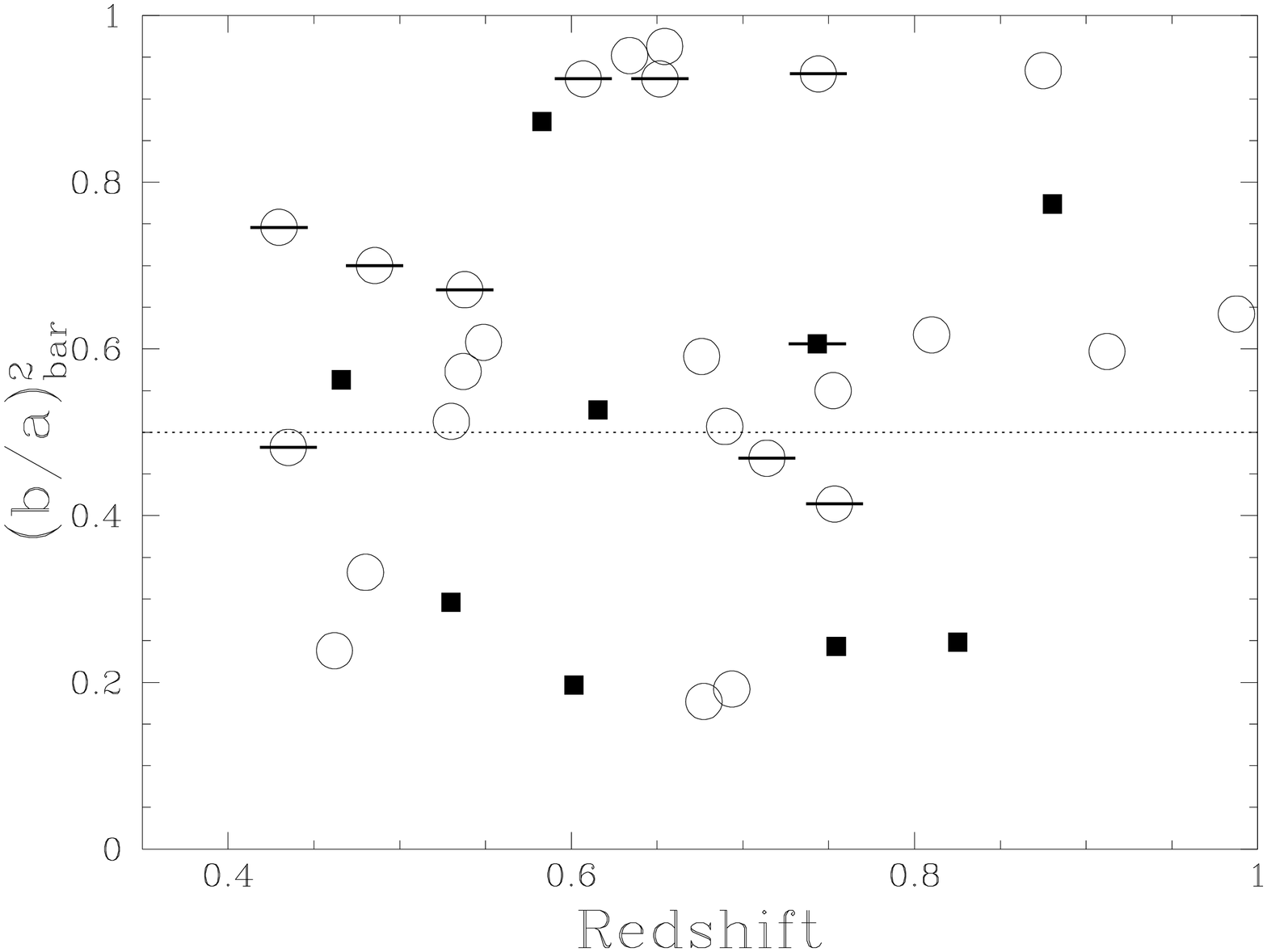}
   \includegraphics[height=0.42\textwidth,angle=-90]{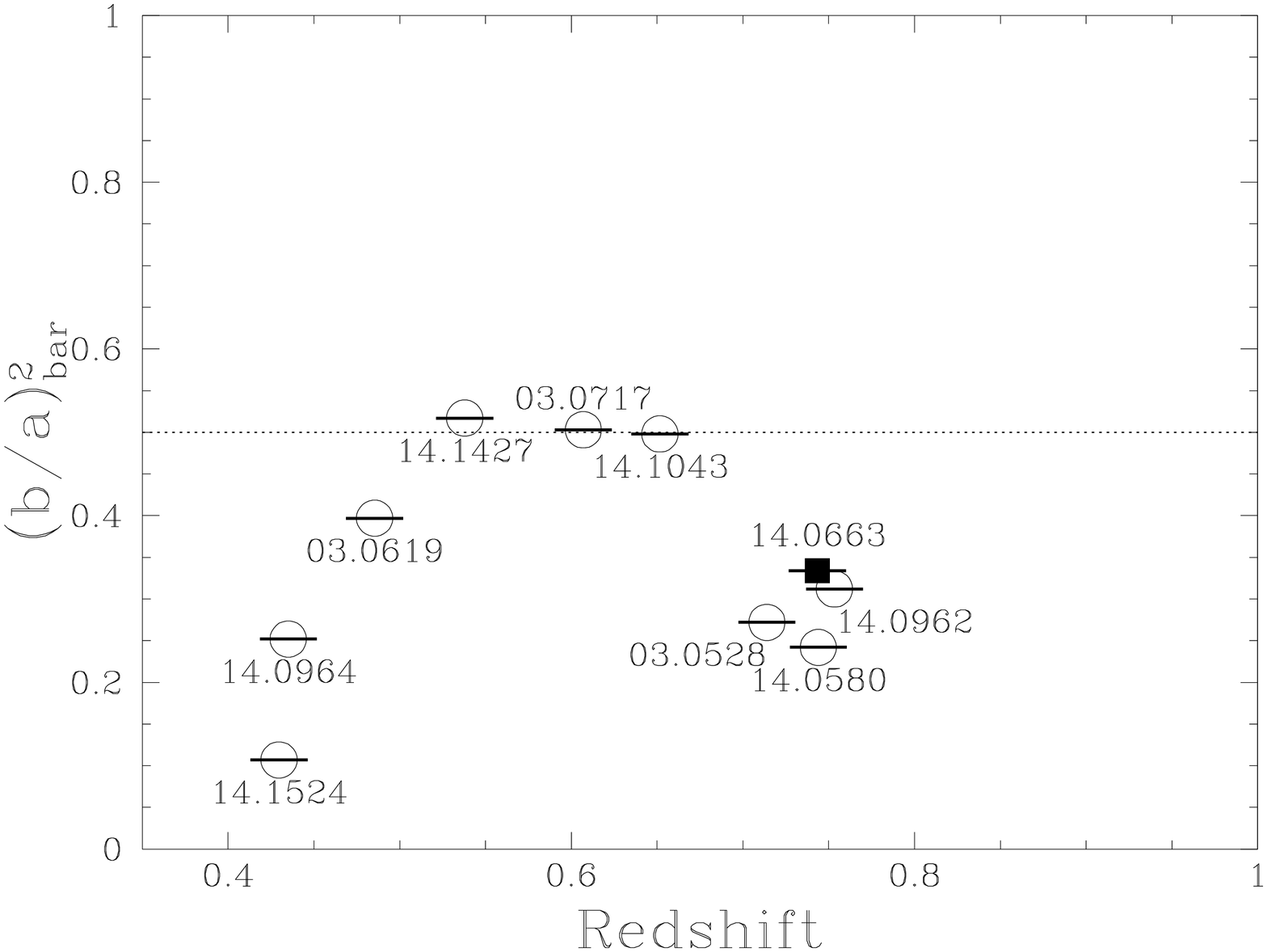}
   \caption{ {\it Top panel:} Bar strength versus redshift. An inner isophote at 85\% and an outer isophote at 1\% of the peak flux are adopted to calculate the bar strength following Abraham et al. (\cite{Abraham99}) for spirals in non-LIRGs (circles) and LIRGs (squares). The {\it bona fide} barred spirals identified in the isophotal analysis are marked with a bar. {\it Bottom panel:} Revised bar strength for the {\it bona fide} barred spirals. See text for more details.}
   \label{bardisk}
\end{figure}

We use the tool IRAF/ELLIPSE to perform the analysis of the surface brightness distribution via fitting the isophotes with ellipses.
An example of the isophotal analysis is shown in Fig.~\ref{isophote}.
This technique is ineffective in detecting bars in highly-inclined spiral galaxies. Ignoring 8 highly-inclined ($i>$60$\degr$) ones (03.0032, 03.0046, 03.1345, 03.1531, 14.0727, 14.0846, 14.1008 and 14.1554), we found that 9 of 26 spiral galaxies have a bar. These barred spirals are 03.0528, 03.0619, 03.0717, 14.0580, 14.0962, 14.0964, 14.1043, 14.1427 and 14.1524.  
Visual examination confirmed the existence of bar structures (see Fig.\ref{colormap}). 
In addition, bar-like structures are also detected in a few objects not classified as spiral galaxies. They are three compacts (03.0327, 03.0508 and 03.0589) and one irregular (03.0645). The bar-like structures in their central regions are elongated light distribution most likely associated with the merging process.

We applied the same isophotal analysis to the companion LIRG sample. Of 36 LIRGs, 13 of them were classified as spirals (see Z04 for details). 4 of the 13 spirals are nearly edge-on. Of 9 low-inclined spiral LIRGs, only one object (14.0663) hosts a bar.  The isophotal analysis revealed bar-like structures in two compact LIRGs 03.0603 and 03.0615.  Both objects show faint diffuse emission surrounding a bright central region. 

To verify our bar detection, we applied another method developed by Abraham et al. (\cite{Abraham99}) to our spiral samples. Abraham et al.'s method is designed to quantitatively measure the bar strength in spiral galaxies. By assuming that the disk of a spiral galaxy is thin and round, and that the bar is characterized by an elliptical light distribution located in the disk plane, the inclination of the galaxy can be corrected to obtain the intrinsic axial ratio of the bar $(b/a)_{bar}$, which is used to measure the strength of the bar (see their paper for a definition). This method has been tested using a local galaxy sample (Frei et al. \cite{Frei}). Strong bar structures are characterized by  $(b/a)^2_{bar}<0.5$ in spiral galaxies with an inclination angle $i < 60\degr$. This method is ineffective in probing the bar structure in the highly-inclined spiral galaxies. 

Following Abraham et al. (\cite{Abraham99}), we took the isophotes at 1\% and 85\% of the peak flux level to measure the entire galaxy and the the inner region, respectively. The parameters of the best-fitting ellipses of the two isophotes are used to calculate the bar strength parameter $(b/a)^2$. Figure~\ref{bardisk} (top panel) shows the distribution of 26 spirals from the non-LIRG sample and 9 spirals from the LIRG sample in the diagram of bar strength versus redshift. The {\it bona fide} barred spirals identified using the technique of isophotal analysis are marked with a bar. We can see from Fig.~\ref{bardisk} that barred spirals are not properly distinguished with a criterion of $(b/a)^2_{bar}<0.5$. Replacing the isophote at 85\% peak flux with the characteristic isophote at the end of each bar, we calculate the bar strength for barred spirals. As Fig.~\ref{bardisk} (bottom panel) shows, the {\it bona fide} barred spirals satisfy the criterion of  $(b/a)^2_{bar}<0.5$. This exercise shows that the bar strength parameter is able to distinguish barred spirals once an inner isophote is properly given. However, the inner isophote at 85\% of peak flux is too arbitrary to reflect the bar structure for our sample of spirals. The brightness of bar characteristic isophotes of the spirals ranges from 34\% to 3\% of their corresponding central surface brightness.
We can conclude that bars in our sample are securely identified using the technique of the isophote analysis. The field galaxies in our sample are bright ($I_{\rm AB}$\,$<$\,22.5 mag) and the characteristic brightness of detected bars varies over a wide range. Note that Abraham et al.'s method is refined in Abraham \& Merrifield (\cite{Abraham00}). An inner isophote is chosen in a series of isophotes so that the bar parameter can be minimized. Our exercise is consistent with this refinement.

On the other hand, observational effects lead to an underestimating of the frequency of bars. Our sample is selected from the CFRS survey with HST 2-band observations. There is no morphological selection effect. Strong bars still can be seen in the $I_{814}$ band when galaxies are redshifted to 0.7, while the weak bar structures become harder to be detected (van den Bergh et al. \cite{Bergh02}). HST $I_{814}$ band imaging used in our examination of bar structures closely corresponds to the rest-frame $B$ band imaging. Figure~\ref{bardisk} shows that the barred spiral galaxies have distributed evenly over 0.4\,$<z<$\,0.8. At redshifts $z>$\,0.8, bandshift effect is no longer negligible as the $I_{814}$ band is bluer than the $B$ band in the rest frame. At the same time, the reduced resolution of WFPC2 imaging makes it difficult to detect a typical 5\,kpc bar (Sheth et al. \cite{Sheth}). The two effects lead to the absence of a bar in spirals at $z>$\,0.8. Our HST imaging data include 7 fields of the Groth Strip Survey (GSS, see Simard et al. \cite{Simard} for more details). The 7 GSS fields are relatively shallow with a detection limit $\sim$24.5\,mag\,arcsec$^{-2}$. In our non-LIRG sample, 40 of 75 objects are from the GSS fields. Considering the cosmological dimming effect, the shallower observations might also contribute to the underestimation of the bar frequency.

In summary, we carefully classified galaxy morphologies in our LIRG and non-LIRG samples and the spiral galaxy subsamples are well characterised. We identified 9 and 1 barred spirals in 26 spiral non-LIRGs and 9 spiral LIRGs, respectively. The probability that LIRGs have significantly fewer bars than non-LIRGs is $\sim$80\% (Student's t Test). This result needs to be confirmed by studying a larger sample.
Taking LIRGs and non-LIRGs together, the bar frequency in the distant spiral galaxies is 31\% if bars are inherently independent of the inclination of their host galaxies on the sky. In the local universe, about 30\% of spiral galaxies host a strong bar seen in optical bands (de Vaucouleurs \cite{Vaucouleurs}) and the bar fraction can increase to 56\% viewed in the near-infrared bands (Eskridge et al. \cite{Eskridge}). Hence, the frequency of bars in spirals at 0.4\,$<\,z\,<$\,1 is similar to or may be higher than that in present-day spirals. This is consistent with the results obtained by Jogee et al. (\cite{Jogee}) based on a much larger sample from the GEMS survey.

\section{Color properties of central regions}

\begin{figure*} \centering
   \includegraphics[width=0.65\textwidth]{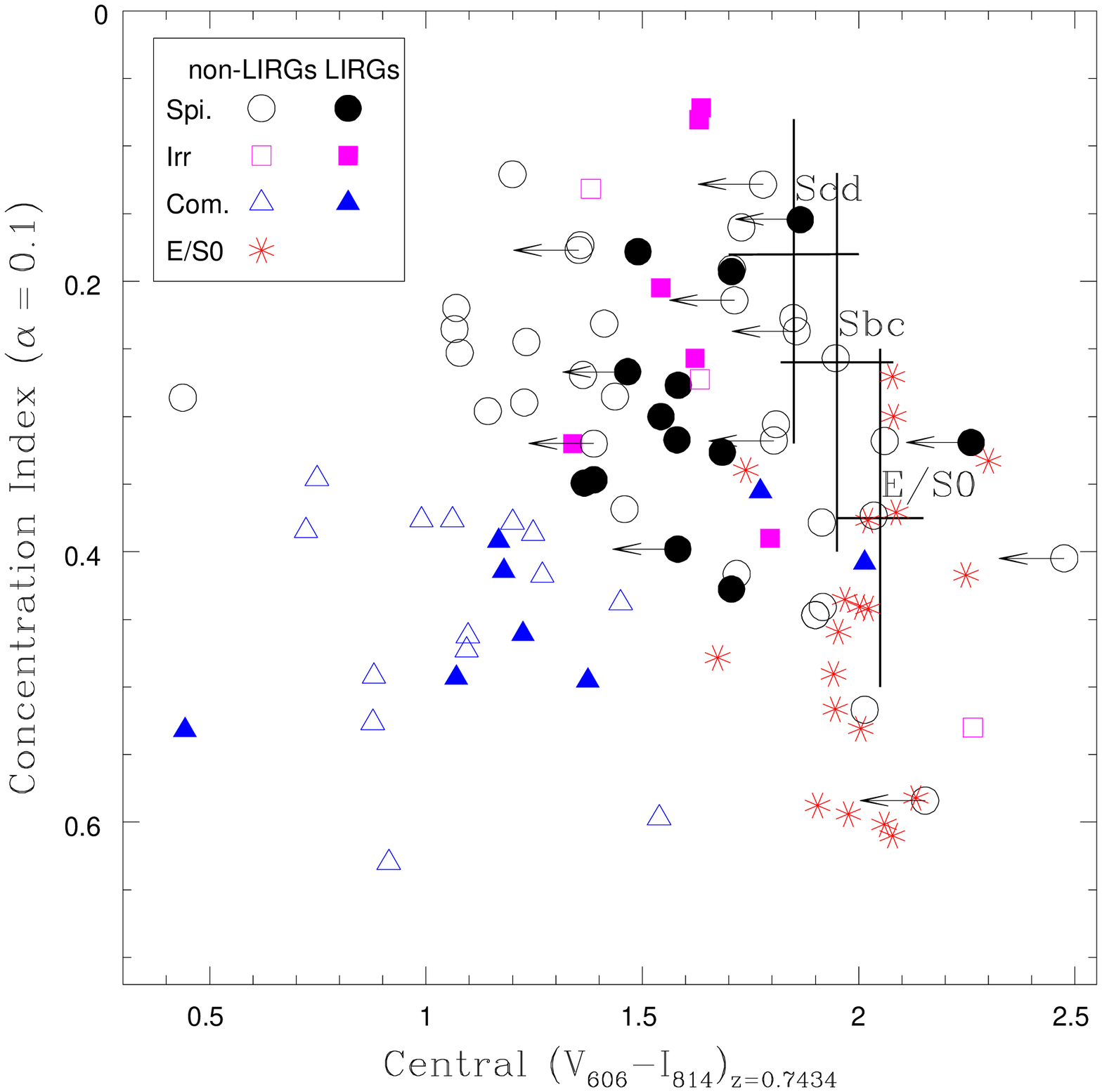}
   \caption{Central color $V_{606}-I_{814}$ versus concentration index. The observed color is corrected to the observed color for objects at redshift 0.7434 (see text for more details). Asterisks are E/S0 galaxies. Squares show irregulars, triangles compacts and circles spirals. Solid symbols represent LIRGs and open symbols non-LIRGs. Compacts show a higher concentration. Note that E/S0 galaxies and some spirals may show a concentrated light distribution (Bershady et al. \cite{Bershady}).}
   \label{colorcompactness}
\end{figure*}

Similar to Z04, we applied an examination of central color versus compactness to our non-LIRG sample.
We adopt an aperture of diameter 2\,kpc to integrate the light centered on the $I_{814}$ brightness peak. We corrected the observed color to a reference frame --- the observed frame for objects at redshift 0.7434. This redshift is the median redshift of our LIRG sample so that we can compare the non-LIRG central color distribution with that of the LIRGs. For the non-LIRGs, the median redshift is 0.6595. At redshift 0.7434, the observed HST color $V_{606}-I_{814}$ corresponds to the color $U$(3440\,\AA)$-B$(4596\,\AA) in the rest frame. Differing from the classical K-correction, our ``relative K-correction'' can remarkably reduce the uncertainties (see Hammer et al. \cite{Hammer01} for more details). In our case, the uncertainty is about $\sim$\,0.1 mag. The concentration index defined in Abraham et al. (\cite{Abraham94}) is used to measure the compactness of a galaxy in the $I_{814}$ band. Here we adopted the ratio of inner radius to outer radius $\alpha$\,=\,0.1 to calculate the concentration index because it is more sensitive than $\alpha$\,=\,0.3 in reflecting the light concentration in the central region. The concentration index is broadly consistent with our criterion to identify compacts. We corrected the effect due to cosmological dimming following B98 (see their Appendix A). A representative sample (Frei et al. \cite{Frei}) is used to derive the compactness distribution of local galaxies. A modeling color of 2.05 is adopted for elliptical galaxies. This color corresponds to the $V_{606}-I_{814}$ color of a single starburst (formed at $z$\,=\,5) at $z$\,=\,0.7434. The bulge color of Sab galaxies is assumed to be bluer by 0.1 mag than that of E/S0 galaxies. Sbc galaxies are bluer than Sab galaxies by 0.1 mag in bulge color. The large crosses show the distribution regions of local galaxies.

Figure~\ref{colorcompactness} shows the compactness as a function of the central color. Non-LIRGs with a visible galactic nucleus (70 of 75) are shown with open symbols and LIRGs with solid ones (see Z04). Different symbols are used to show different morphological types. We point out that the central color should not be directly taken as the bulge color because of the contamination from the disk since several distant spirals show very small bulges or even no bulge. On the other hand, dust extinction may complicate the situation further as a dust lane is often seen in highly-inclined spiral galaxies. We use an arrow to show the high inclination ($i>60\degr$) which often leads to a redder color due to the ``dust-screen''. We can see that three highly-inclined spirals are even redder than an elliptical galaxy in the central color while the others are bluer. 

From Fig.~\ref{colorcompactness}, most distant field galaxies except for E/S0 galaxies are bluer than their local counterparts in the central color. An offset of $\sim$\,0.4 mag exists between the distant and local galaxies. For comparison, the typical color for E, S0, Sbc and Irr galaxies are 2.05, 1.88, 1.10 and 0.38 respectively. Most low-inclined spirals and all compacts host intrinsically blue central regions. The color $U$(3440\,\AA)$-B$(4596\,\AA) is sensitive to young stellar populations. Therefore, the blue central color implies recent star formation relative to the observed epoch in the galactic central region. While spirals distribute close to the local spirals in central color and compactness, compact galaxies are distinctly different from the local galaxies, suggestive of significant evolution in the central color and the compactness.  Nearly all E/S0 galaxies show a central color similar to local counterparts within a considerable scatter. About one quarter of the E/S0 galaxies appear more concentrated than the local E/S0 galaxies. Considering the uncertainties in determining the concentration index, deeper studies are needed to assess this result. If it is confirmed, this would show that the size (mass) growth since $z$\,=\,1 happened to a significant fraction of E/S0 galaxies (see also Ellis et al.~\cite{Ellis}; Drory et al. \cite{Drory}). Most distant E/S0 galaxies have similar compactness to the local ones.

\section{Discussion}

\subsection{Morphological evolution}\label{distribution}

We examined the global morphological properties in the rest-frame $B$ band of field galaxies at 0.4\,$<z<$\,1. Our classification aims to avoid biases from reduced spatial resolution, band-shifting and cosmological dimming effects. Our morphological scheme for regular galaxies is designed to be consistent with the Hubble scheme widely used for local galaxies. We also consider the fact that a substantial fraction of distant galaxies are luminous and compact. However, our analysis is limited by the small size of our sample. Nevertheless, comparison of the morphological frequency distribution of present-day galaxies provides valuable clues to understand galaxy morphological evolution.

Visual examination in previous studies properly classified the morphologies of local galaxies. Nakamura et al. (\cite{Nakamura}) carried out a visual investigation of nearby bright galaxies in the $r^\ast$ band obtained from the SDSS survey. For galaxies brighter than $M_{r^\ast}^\ast$+1, the fractions corresponding to our type are 24\%, 73\% and 3\% for E/S0, Spirals and irregulars respectively. Luminous blue compact galaxies are rare in the present day sample (Garland et al. \cite{Garland}). Note that $M^\ast$ is the same for different types of galaxies in the $B$ band but different in the $r^\ast$ band (Nacamura et al. \cite{Nakamura}). Better determination of the fractions can be derived from a quantitative morphological examination of the SDSS survey.


Compared with the frequency distribution of present-day galaxies, our morphological examination shows that field galaxies at intermediate redshifts include fewer spirals, more compacts, more irregulars and more merging systems. Our results support the conclusion that the irregulars and merging systems have undergone a remarkable evolution since $z=$\,1 (B98; van den Bergh \cite{Bergh01}). Our results show that 27\% of field galaxies are E/S0 galaxies, compared to the present-day fraction of 24\%. We point out that the fraction of E/S0 galaxies could be underestimated in the CFRS survey because of the difficulty of identifing redshifts for elliptical galaxies using low-resolution CFHT spectroscopy (Crampton et al. \cite{Crampton}). Hence we can conclude that the fraction of bright E/S0 galaxies at 0.4\,$<z<$\,1 is comparable to the present day fraction. The identified E/S0 galaxies in our sample show similar colors to their local counterparts (see Fig.~\ref{colorcompactness}), suggesting that they were in place at $z\sim$\,1. This is consistent with previous results indicating that early type galaxies show little/mild evolution since $z=$\,1 (Schade et al. \cite{Schade}; Fontana et al. \cite{Fontana}). A possible evolutionary effect of E/S0 galaxies we found is that $\sim$25\% of the galaxies appear to be more compact than present-day E/S0 galaxies. 

The comparison of the frequency distribution of galaxy types suggests that the morphological evolution since $z=$\,1 most likely concerns spirals.
Hammer et al. (\cite{Hammer04a}) pointed out that most of the field spirals would undergo 4 to 5 violent star formation episodes, i.e., LIRG phases. 
A detailed formation scenario of spiral galaxies is suggested in Hammer et al. (\cite{Hammer04a}) based on a combined study of star formation rate, metal abundance and morphological properties. In the proposed scenario, merger systems, compacts and irregulars are associated with different evolution stages in the assembly of spiral galaxies. Morphological buildup of the spiral galaxies co-occurs with intense star formation related to LIRG phases.
Figure~\ref{colorcompactness} shows that the field galaxies at 0.4\,$<z<$\,1 are distinct from the local ones in compactness and central color.
Z04 reported the distribution of LIRGs along a sequence linking compact objects with blue central color to extended objects with relatively red central color. Roughly speaking, the non-LIRGs except for E/S0 galaxies follow the same sequence with a much larger scatter.  
Z04 discussed the sequence of LIRG distribution in detail. The sequence from compact objects with blue central color to extended objects with red central color is suggested to hint at the formation mechanism for at least part of the present-day spiral galaxies. From the sequence, the disk of a spiral galaxy is built up after the formation of the bulge, accompanied by the morphology evolving from compact into extended as well as the central color evolving from blue to red. We can see from Fig.~\ref{colorcompactness} that most compacts in either non-LIRGs or LIRGs would ultimately evolve into extended spirals, following the sequence of the LIRGs. The compacts are indeed the galaxies forming their bulge. This is also suggested by the spectroscopic properties of the luminous compacts (Hammer et al. \cite{Hammer01}). The morphological diversity of LIRGs implies that massive star formation can be related to different morphologies, supporting the morphological buildup during multiple LIRG episodes.

Moreover, the bulges in spiral galaxies at 0.4\,$<z<$\,1 are expected to be systematically smaller, provided that the present-day massive (early-type) spiral galaxies had mostly assembled since $z$\,=\,1. Ellis et al. (\cite{Ellis}) found that the color of bulges is systematically bluer than that of E/S0 galaxies. This is also suggested by Fig.~\ref{colorcompactness}. It is difficult to derive further constraints to the recent star formation activity superimposed on an old passively-evolving stellar population. A single starburst involving 10\% stellar mass of host galaxies can easily cause a change of 1 mag in the color, intimately depending on the burst's age (see figure 4 in Ellis et al.~\cite{Ellis}). On the other hand, dust extinction may significantly redden the color. The interaction between star formation and dust extinction makes it difficult to quantitatively determine the recent star formation in the central region. Interestingly, the LIRGs appear systematically redder than the non-LIRGs (excluding E/S0 ones) in central color. We know that IR emission is a good indicator of star formation (Kennicutt \cite{Kennicutt}; Flores et al. \cite{Flores04}) and dust plays a key role re-radiating the UV in the infrared band. Dust extinction most likely influences the reddened central color of the LIRG, with respect to that of the non-LIRG. 

Lilly et al. (\cite{Lilly98}) showed that the median type of disk galaxy changes from Sbc at 0.2\,$<z<$\,0.5 to ``Sdm'' at 0.75\,$<z<$\,1, which is confirmed by this study. Due to the uncertainties in determination of the $B/T$ ratio, this result is not so confident. However, we found a link between the $B/T$ determination and the central color. Among 18 spiral non-LIRGs with a central color redder than 1.5, 12 are $B/T\geq$\,0.15. Besides this, among the 15 spiral non-LIRGs with a central color bluer than 1.5, 12 of them have indeed a small or no bulge ($B/T\leq$\,0.03). It is likely that the central color of these spirals without or with only small bulges (partially) suffers from disk contamination. They make up more than one third of the spirals in Fig.~\ref{colorcompactness}. Since nearly all local spirals are Sab and Sbc (Fukugita et al. \cite{Fukugita}), it seems that the $B/T$ ratio of distant spirals is systematically lower than that of the local spirals. Either blue bulges or lower $B/T$ ratios imply a significant evolution of spiral bulges, i.e., the spiral bulges have undergone a significant growth since $z\sim$\,1. In summary our investigation of galaxy central colors suggests a significant evolution of bulge properties, either bluer or smaller than those in present-day spirals. In the following we discuss to which physical processes the bulge evolution is related.

\subsection{Physical processes driving galaxy evolution}

\subsubsection{Implications of the morphological frequency distributions of LIRGs and non-LIRGs}

The morphological classifications for LIRGs and non-LIRGs are based on the same HST imaging observations and same methodology. Hence, direct comparison of the morphological fraction distribution between the LIRGs and non-LIRGs is possible.
LIRGs showing different morphological types hint that the mechanisms igniting the LIRG phase might be diverse. 

From Table~\ref{classification}, we can see that the frequency of ongoing mergers is higher in LIRGs than in non-LIRGs by a factor of 2 to 3. Also LIRGs show more objects with disturbed morphologies and signs of merging/interacting. This suggests the merging process as an important mechanism to trigger violent star formation. We note that our LIRG sample is made up of securely-detected bright sources ($F_{15\mu \rm m}>300\mu$Jy) in deep ISOCAM observations. The LIRG sample is indeed slightly biased to include more ultra-luminous infrared galaxies (ULIRGs, $L_{\rm IR}(8-1000\,\mu$m)\,$\geq$\,$10^{12}\,L_\odot$). In the local universe, the ULIRGs are dominated by major mergers or their remnants (Sanders \& Mirabel \cite{Sanders}). If this is the case in the distant universe, the frequency of mergers in LIRGs might be slightly overestimated.

In addition, other mechanisms seem to substantially contribute to triggering intense star formation as neither companions nor signs of merging/interacting are detected in one third of LIRGs (see Z04). The mechanisms behind the isolated LIRGs (including spirals, irregulars and compacts) are still unclear. Cohen et al. (\cite{Cohen}) indicated that most galaxies in the HDF are in groups. If this is the case for distant field galaxies, the probability of galaxies to pass one another would be higher than that in a random distribution. The interaction caused by nearby galaxies could be one of the major mechanisms to trigger the LIRG phase (Elbaz et al. \cite{Elbaz04}). We investigated the possibility that a LIRG phase could be ignited by internal processes like secular evolution driven by bars.

\subsubsection{Secular evolution versus bulge formation}\label{bars}

Secular evolution driven by bars is proposed to be one of the major mechanisms of bulge formation (Combes \cite{Combes00}). Recent simulations show that round bulges mostly seen in early type spirals cannot be formed via secular evolution while the formation of disk-like bulges in late-type spirals can be satisfactorily explained by this mechanism (Debattista et al. \cite{Debattista}; Shen \& Sellwood \cite{Shen}). It is widely accepted that a bar can transport gas into the central region (e.g. Sakamoto et al. \cite{Sakamoto}) followed by mild circumnuclear starburst (Ho et al. \cite{Ho}). At $z<$\,1, more than {50\% of the star formation density is related to IR radiation (Flores et al. \cite{Flores99}). Our sample galaxies have an intermediate stellar mass and they may double their mass at high star formation rates during multiple LIRG phases (Hammer et al. \cite{Hammer04a}). This strongly hints that the morphological structures of spiral galaxies are mainly assembled in the LIRG phases (see Hammer et al. \cite{Hammer04b} for more discussion). We found that the bar frequency of spiral LIRGs is probably much lower than that of spiral non-LIRGs. This suggests that bar instability is not a favored mechanism to trigger violent star formation. Bar-driven secular evolution is unlikely to be one of the major mechanisms to drive galaxy morphological evolution, especially bulge formation.  The dissolution and weakening of bars may re-distribute stars and contribute to the growth of bulges. However, such bar-driven growth seems negligible (e.g. Shen \& Sellwood \cite{Shen}). 
A similar result is suggested by Kannappan et al. (\cite{Kannappan}), who examined a population of nearby blue-centered spiral galaxies and did not find significant evidence of enhanced central star formation associated with bars. 
In contrast, the significantly higher fraction of merging/interacting systems in LIRGs confirms that the merging process is a major mechanism in transforming distant field galaxies into present-day spirals. Mergers/interactions are very efficient at building up large bulges such as those seen in most present-day spirals (Sa to Sbc, see Hammer et al. \cite{Hammer04a}). Since infrared selection favors violent dynamic events (e.g. merging) and strong bars are often excited by such events, it is surprising that the LIRGs host fewer barred galaxies.

If the bar-driven secular evolution is intimately related to the bulge formation, the bar frequency would be expected as a function of redshift.
Differing from Abraham et al. (\cite{Abraham99}), we did not find an apparent deficiency of barred spiral galaxies at redshifts beyond 0.5. Abraham et al. (\cite{Abraham99}) revealed the deficiency of barred spirals in the HDFs. Since the HDFs cover only 10 square arcminutes, the deficiency can be ascribed to the ``field-to-field'' variance (ven den Bergh et al. \cite{Bergh02}).  Our sample is selected from 87 square arcminute sky areas in two independent fields, and the cosmic variance can be substantially reduced. In addition, the bar detection is related to the sample selection. Compared with the spiral galaxies ($I_{\rm AB}<$\,23.6 mag) in Abraham et al. (\cite{Abraham99}), our sample galaxies are brighter ($I_{\rm AB}<$\,22.5 mag) and bigger (see also Z04). Bars thus are easier to detect (Sheth et al. \cite{Sheth}). The near-infrared view of barred spirals also suggests a similar conclusion (Sheth et al. \cite{Sheth}). It seems unlikely that the bar frequency is luminosity-dependent, i.e., less-luminous spirals show a lower bar frequency than luminous spirals at $z>$\,0.5. Given that bar fraction is not correlated with the redshift, the bar-driven secular process seems not to have a strong impact on bulge formation of most spirals.

About one third (or more) distant spirals are identified as barred ones. This implies that a substantial percentage of spiral galaxies already had a well-built disk to host a bar at intermediate redshifts. If the bar phenomenon is a self-regulated process, their bar occurrence is exclusively related to the spiral galaxy itself (Combes \cite{Combes00}). Simulations suggest that spiral galaxies can form a bar repeatedly, given sufficient gas continuously accreted on the disk (Bournaud \& Combes \cite{Bournaud}). In such a case, the frequency of bars is an indicator of the time scale of a bar, with respect to the lifetime of the disk. The bar frequency of distant spirals then suggests that distant spiral galaxies are comparable to present-day ones. Our current knowledge about the origin and lifetime of bars is insufficient to interpret a constant bar frequency up to $z\sim$\,0.8.

\section{Conclusions}

Distant (0.4\,$<z<\sim$1) field galaxies contain a higher fraction of LIRGs, compared with the local galaxies. ISOCAM deep observations of two CFRS fields 0300+00 and 1415+52 allow us to identify galaxies with luminous infrared emission. From 87 square arcminute sky areas imaged by HST/WFPC2 in $V_{606}$ (or $B_{450}$) and $I_{814}$ filters in the two CFRS fields, we selected a sample of 36 LIRGs at 0.4\,$<z<$\,1.2 and another of 75 non-LIRGs at 0.4\,$<z<$\,1 to study their morphological and color map properties. Zheng et al. (\cite{Zheng}) reported the study of the 36 LIRG sample. In this work, we applied the same methodology to the 75 non-LIRG sample.

We found that the frequency of the E/S0 galaxies at 0.4\,$<z<$\,1 is comparable to the frequency in the local universe. They have a central color and compactness similar to their local counterparts, suggesting that most field E/S0 galaxies were already in place at $z$\,=\,1. The compacts are a distinct sub-population at 0.4\,$<z<$\,1. They show concentrated light distribution and remarkable blue central color. Altogether our morphological examination shows that a significant fraction of the distant field galaxies were undergoing morphological transformation and/or evolution since 0.4\,$<z<$\,1. We suggest that they are associated with the large population of present-day spirals.

The frequency distribution of morphological types is different for LIRGs and non-LIRGs. The LIRGs more often show disturbed morphologies than the non-LIRGs. Thus merging processes still play an important role in triggering star formation. However, one third of the LIRGs are isolated objects without perceptible signs of merging/interacting. This suggests that other physical processes substantially contribute to driving galaxy formation. 

To investigate the roles of the bar-driven secular process in galaxy morphological evolution, we identified the bar structures in spiral galaxies of our samples.Due to bandshifting and reduced resolution effects, strong bars in the rest-frame $B$ band can be detected up to $z$\,=\,0.8 with HST $I_{814}$ band imaging. The detected bars in our sample are equally located at $z\sim$0.5 and $z\sim$0.7. This is inconsistent with the paucity of barred spirals at redshifts beyond 0.5 found in the HDFs by Abraham et al. The difference can be ascribed to cosmic variance. 
We found that spirals selected from LIRGs host much fewer bars than those in non-LIRGs. This suggests that bar instability is not efficient to trigger violent star formation. Star formation since $z$\,=\,1 essentially involves LIRGs. The bar-driven secular evolution is thus excluded from being a major mechanism related to a significant fraction of star formation since $z$\,=\,1. The bar frequency of distant spirals is similar to that of the present-day ones in the rest-frame $B$ band. The absence of bar evolution is strongly suggestive that secular phenomena have a minor impact on the building of present-day spiral morphologies, including their bulges, which is likely related to the strong evolution of merging and LIRG number density, as proposed by Hammer et al. (\cite{Hammer04a}).

\begin{acknowledgements}

X.Z.Z. thanks the Centre National de la Recherche Scientifique (CNRS) and the Minist$\grave{\rm e}$re de l'$\acute{\rm E}$ducation nationale, de la Recherche et de la Technologie (MENRT). We are grateful to the anonymous referee for very helpful comments.

\end{acknowledgements}


\begin{thebibliography}{}
  \bibitem[2000]{Abraham00} Abraham, R. G., \& Merrifield, M. R.
    2000, AJ, 120, 2835
  \bibitem[1999]{Abraham99} Abraham, R. G., Merrifield, M. R., Ellis, R. S., Tanvir, N. R., \& Brinchmann, J.
    1999, MNRAS, 308, 569
  \bibitem[1994]{Abraham94} Abraham, R. G., Valdes, F., Yee, H. K. C., \& van den Bergh, S.
    1994, ApJ, 432, 75
  \bibitem[2004]{Bell} Bell, E. F., Wolf, C., Meisenheimer, K., et al.
    2004, ApJ, 608, 752
  \bibitem[2000]{Bershady} Bershady, M. A., Jangren, A., \& Conselice, C. J.
    2000, AJ, 119, 2645
  \bibitem[2002]{Bournaud} Bournaud, F., \& Combes, F.
    2002, A\&A, 392, 83
  \bibitem[1998]{Brinchmann} Brinchmann, J., Abraham, R., Schade, D., et al.
    1998, ApJ, 499, 112 (B98)
  \bibitem[2000]{BrinchmannEllis} Brinchmann, J., \& Ellis, R. S.
    2000, ApJ, 536, L77
  \bibitem[1993]{Bruzual} Bruzual, A.G., \& Charlot, S.
    1993, ApJ, 405, 538
  \bibitem[2002]{Cimatti} Cimatti, A. et al.
    2002, A\&A, 381, L68
  \bibitem[2000]{Cohen} Cohen, J. G., Hogg, D. W., Blandford, R., et al.
    2000, ApJ, 538, 29
  \bibitem[2000]{Combes00} Combes, F.
    2000, in Building Galaxies: from the Primordial to the Present, ed. F. Hammer et al. (Paris: Editions Fronti$\grave{\rm e}$res)
  \bibitem[1990]{Combes} Combes, F., Debbasch, F., Friedli, D., \&  Pfeffinger. D.
    1990, A\&A, 233, 82
  \bibitem[1995]{Crampton} Crampton, D., Le F$\grave{\rm e}$vre, O., Lilly, S. J., \& Hammer, F.
    1995, ApJ, 455, 96
  \bibitem[2000]{Daddi} Daddi, E., Cimatti, A., \& Renzini, A. 
    2000, A\&A, 362, L45
  \bibitem[2004]{Debattista} Debattista, V. P., Carollo, C. M., Mayer, L., \& Moore, B.
    2004, ApJ, 604, L93
  \bibitem[1963]{Vaucouleurs} de Vaucouleurs, G.
    1963, ApJS, 8, 31
  \bibitem[2004]{Drory} Drory, N., Bender, R., Feulner, G. et al.
    2004, ApJ, 608, 742
  \bibitem[1962]{Eggen} Eggen, O. J., Lynden-Bell, D., \& Sandage, A.
    1962, ApJ, 136, 748
  \bibitem[2002]{Elbaz02} Elbaz, D, Cesarsky, C. J., Chanial, P., et al.
    2002, A\&A, 384, 848
  \bibitem[2003]{Elbaz03} Elbaz, D, \& Cesarsky, C. J.
    2003, Science, 300, 270
  \bibitem[2004]{Elbaz04} Elbaz, D., Marcillac, D., Moy, E.
    2004, astro-ph/0403209
  \bibitem[2001]{Ellis} Ellis, R. S., Abraham, R. G., \& Dickinson, M. 
    2001, ApJ, 551, 111
  \bibitem[2000]{Eskridge} Eskridge, P. B., Frogel, J. A., Pogge, R. W. et al. 
    2000, AJ, 119, 536
  \bibitem[1980]{Fall} Fall, S. M., \& Efstathiou, G.
    1980, MNRAS, 193, 189
  \bibitem[1999]{Flores99} Flores, H., Hammer, F., Thuan, T. X., et al. 
    1999, ApJ, 517, 148
  \bibitem[2004]{Flores04} Flores, H., Hammer, F., Elbaz, D., Cesarsky, C. J., Liang, Y. C., Fadda, D., \& Gruel, N.
    2004, A\&A, 415, 885
  \bibitem[2004]{Fontana} Fontana, A., Pozzetti, L., Donnarumma, I., et al. 
    2004, A\&A, 424, 23
  \bibitem[2003]{Fritze} Fritze-v.Alvensleben, U., \& Gerhard, O. 
    2003, A\&A, 398, 89
  \bibitem[1996]{Frei} Frei, Z., Guhathakurta, P., Gunn, J. E., \& Tyson, J. A.
    1996, AJ, 111, 174
  \bibitem[1998]{Fukugita} Fukugita, M., Hogan, C. J., \& Peebles, P. J. E.
    1998, ApJ, 503, 518
  \bibitem[2003]{Garland} Garland, C. A., Pisano, D. J., Williams, J. P., Guzm$\grave{\rm a}$n, R., \& Castander, F. J.
    2003, AAS, 203, 7002
  \bibitem[2001]{Genzel} Genzel, R., Tacconi, L. J., Rigopoulou, D., Lutz, D., \& Tecza, M.
    2001, ApJ, 563, 527
  \bibitem[1997]{Guzman} Guzm$\grave{\rm a}$n, R., Gallego, J., Koo, D. C., et al.
    1997, ApJ, 489, 559
\bibitem[1997]{Hammer97} Hammer, F., Flores, H., Lilly, S. J., et al.
    1997, ApJ, 481, 49
  \bibitem[2001]{Hammer01} Hammer, F., Gruel, N., Thuan, T. X, Flores, H., \& Infante, L.
    2001, ApJ, 550, 570
  \bibitem[2005]{Hammer04a} Hammer, F., Flores, Elbaz, D., H., Zheng, X. Z., Liang, Y. C., \& Cesarsky, C. 
    2005, A\&A, 430, 115
  \bibitem[2004]{Hammer04b} Hammer, F., Flores, H., Liang, Y. C., Zheng, X. Z., Elbaz, D., \& Cesarsky, C.
    2004, astro-ph/0401246
  \bibitem[2004]{Heavens} Heavens, A., Panter, B., Jimenez, R. \& Duplop, J.
    2004, Nature, 428, 625 
  \bibitem[1997]{Ho} Ho, L. C., Filippenko, A. V., \& Sargent, W. L. W. 
    1997, ApJ, 487, 591
  \bibitem[2001]{Ishida} Ishida, C. M., \& Sanders, D. B.
    2001, BAAS, 198, 3461
  \bibitem[2004]{Jogee} Jogee, S., Barazza, F. D., Rix, H.-W., et al.
    2004, ApJ, 615, L105
  \bibitem[1996]{Kauffmann} Kauffmann, G. 
    1996, MNRAS, 281, 487
  \bibitem[2004]{Kannappan} Kannappan, S. J., Jansen, R. A., \& Barton, E. J.
    2004, AJ, 127, 1371
  \bibitem[1998]{Kennicutt} Kennicutt, R. C. Jr. 
    1998, ARA\&A, 36, 189
  \bibitem[2003]{Kobulnicky} Kobulnicky, H. A., Willmer, C. N. A., Phillips, A. C. et al.
    2003, ApJ, 599, 1006
  \bibitem[1992]{Kormendy} Kormenday, J., \& Sanders, D. B.
    1992, ApJ, 390, L53
  \bibitem[2000]{Fevre} Le F$\grave{\rm e}$vre, O., Abraham, R., Lilly, S. J., et al.
    2000, MNRAS, 311, 565
  \bibitem[2003]{Lilly03} Lilly, S. J., Carollo, C. M., \& Stockton, A. N.
    2003, ApJ, 597, 730
  \bibitem[1998]{Lilly98} Lilly, S. J., Schade, D, Ellis, R., et al.
    1998, ApJ, 500, 75
  \bibitem[1995]{Lilly95} Lilly, S. J., Tresse, L., Hammer, F., Crampton, D.,  \& Le F$\grave{\rm e}$vre, O.
    1995, ApJ, 455, 108
  \bibitem[2004]{Liang} Liang, Y. C., Hammer, F., Flores, H., Elbaz, D., \& Cesarsky, C. J.
    2004, A\&A, 423, 867
  \bibitem[2000]{Madau} Madau, P., \& Pozzetti, L.
    2000, MNRAS, 312, L9
  \bibitem[1998]{Marleau} Marleau, F. R., \& Simard, L.
    1998, ApJ, 507, 585
  \bibitem[1998]{Marzke} Marzke, R. O., da Costa, L. N., Pellegrini, P. S., Willmer, C. N. A., \& Geller, M. J.
    1998, ApJ, 503, 617
  \bibitem[2001]{Menanteau01} Menanteau, F., Abraham, R. G., \& Ellis, R. S.
    2001, MNRAS, 322, 1
  \bibitem[1994]{Mihos} Mihos, J. C., \& Hernquist, L.
    1994, ApJ, 425, L13
  \bibitem[2003]{Nakamura} Nakamura, O., Fukugita, M., Yasuda, N., et al.
    2003, AJ, 125, 1682
  \bibitem[1997]{Regan} Regan, M. W., \& Elmegreen, D. M. 
    1997, AJ, 114, 965
  \bibitem[1999]{Sakamoto} Sakamoto, K., Okumura, S. K., Ishizuki, S., \& Scoville, N. Z.
    1999, ApJ, 525, 691
  \bibitem[1996]{Sanders} Sanders, D. B., \& Mirabel, I. F.
    1996, 34, 749
  \bibitem[1999]{Schade} Schade, D., Lilly, S. J., Crampton, D., et al.
    1999, ApJ, 525, 31
  \bibitem[2004]{Shen} Shen, J., \& Sellwood, J. A.
    2004, ApJ, 604, 614
  \bibitem[2003]{Sheth} Sheth, K., Regan, M. W., Scoville, N. Z., \& Strubbe, L. E.
    2003, ApJ, 592, L13
  \bibitem[2002]{Simard} Simard, L., Willmer, C. N. A., Vogt, N. P., et al.
    2002, ApJS, 142, 1
  \bibitem[2002]{Treu02} Treu, T., Stiavelli, M. Casertano, S., Moller, P., \& Bertin, G.
    2002, ApJ, 564, L13
  \bibitem[2001]{Bergh01} van den Bergh, S.
    2001, AJ, 122, 621
  \bibitem[2000]{Bergh00} van den Bergh, S., Cohen, J. G., Hogg, D. W.,\&  Blandford, R.
    2000, AJ, 120, 2190
  \bibitem[2002]{Bergh02} van den Bergh, S., Abraham, R. G., Whyte, L. F., et al.
    2002, AJ, 123, 2913
  \bibitem[2003]{vanDokkum} van Dokkum, P. G., \& Ellis, R. S.
    2003, ApJ, 592, L53
  \bibitem[2004]{Weiner} Weiner, B. J., Phillips, A. C., Faber, S. M., et al.
    2005, ApJ in press (astro-ph/0411128)
  \bibitem[2004]{Wolf} Walf, C., Meisenheimer, K., Rix, H.-W., Borch, A., Dye, S., \& Kleinheinrich, M.
    2004, A\&A, 401, 73
  \bibitem[2004]{Zheng} Zheng, X. Z., Hammer, F., Flores, H., Ass$\acute{\rm e}$mat, F., \& Pelat, D.
    2004, A\&A, 421, 847 (Z04)

\end{thebibliography}
\end{document}